\documentclass[structabstract]{aa}  
\usepackage{graphicx}
\usepackage{natbib,txfonts}
\usepackage{longtable,lscape}
\usepackage{hyperref}   

\usepackage{amstext} 
\usepackage{subfig}  
\begin{document}
\title{VLBI-selected sample of Compact Symmetric Object candidates
and frequency-dependent position of hotspots}


\author{
   K. V. Sokolovsky\inst{1,2}
   \and
   Y. Y. Kovalev\inst{2,1}
   \and
   A.~B.~Pushkarev\inst{3,4,1}
   \and
   P. Mimica\inst{5}
   \and
   M. Perucho\inst{1,5}
}

\authorrunning{Sokolovsky et al.}
\titlerunning{A new VLBI-selected sample of CSO candidates}

\institute{
   Max--Planck--Institut f\"ur Radioastronomie, Auf dem H\"ugel 69, D-53121 Bonn, Germany
   \and
   Astro Space Center of Lebedev Physical Institute, Profsoyuznaya Str. 84/32, 117997 Moscow, Russia
   \and
   Pulkovo Astronomical Observatory, Pulkovskoe Chaussee 65/1, 196140 St.~Petersburg, Russia
   \and
   Crimean Astrophysical Observatory, 98409 Nauchny, Crimea, Ukraine
   \and 
   Departament d'Astronomia i Astrof\'{\i}sica, Universitat de Val\`encia, C/
   Dr. Moliner 50, 46100 Burjassot (Val\`encia), Spain
}

\offprints{K.~V.~Sokolovsky\\ \email{ksokolov@mpifr-bonn.mpg.de}}

\date{Received July 02, 2011; accepted July 11, 2011}

\abstract
{
The Compact Symmetric Objects (CSOs) are small
($<1$\,kiloparsec) and powerful extragalactic radio sources
showing emission on both sides of an active galactic nucleus
and no signs of strong relativistic beaming. They may be young
radio sources, progenitors of large FR~II radio galaxies.
}
{
We aim to study the statistical properties of CSOs
by constructing and investigating a new large sample of CSO candidates on
the basis of dual-frequency, parsec-scale morphology.
}
{
For the candidate selection we utilized VLBI data for 4170
extragalactic objects obtained simultaneously at 2.3 and
8.6\,GHz ($S$ and $X$ band)
within the VLBA Calibrator Survey 1-6 and the Research and
Development -- VLBA projects. Properties of their broad-band
radio spectra were characterized by using RATAN-600 observations.
Numerical modeling was applied in an attempt to explain the
observed effects.
}
{
A sample of 64 candidate CSOs is identified.
The median two-point $S$--$X$~band spectral index of parsec-scale hotspots
is found to be $-0.52$; with the median brightness temperature
$\sim 10^9$\,K at $X$~band.
Statistical analysis reveals a systematic difference between
positions of brightest CSO components (associated with hotspots) measured 
in the $S$ and $X$~bands.
The distance between these components is found to be on average
$0.32 \pm0.06$\,mas greater at $8.6$\,GHz than at $2.3$\,GHz.
}
{
This difference in distances cannot be explained by different resolutions at
the $S$ and
$X$~bands. It is a manifestation of spectral index gradients across CSO
components, which may potentially provide important physical information about
them. Despite our detailed numerical modeling of a CSO hotspot,
the model was not able to reproduce the magnitude of the observed
positional difference. A more detailed modeling may shed light on the origin
of the effect.
Multifrequency follow-up VLBI observations of the selected sample
are needed to confirm and study newly identified CSOs from the
list presented here.
}

\keywords{
Galaxies: active -- Galaxies: jets -- Radio continuum: galaxies
-- Methods:numerical -- Magnetohydrodynamics (MHD) -- Shock waves
}

\maketitle
%

\section{Introduction}

Compact Symmetric Objects (CSOs) are small ($<1$\,kiloparsec), powerful extragalactic
radio sources that show emission on both sides of an active galactic
nucleus \citep{1994ApJ...432L..87W,1996ApJ...460..612R}. In contrast to the majority of
compact radio sources, relativistic beaming effects are
believed to be small in CSOs owing to their orientation close to the plane of
the sky. The parsec-scale core that marks position of the central engine is often
weak or not detected at all. Kinematic studies of CSOs 
reveal no superluminal motion and suggest source ages 
of a few hundred to a few thousand years
\citep{1998A&A...337...69O,2009AN....330..149P}. CSOs may
be progenitors of large-scale Fanaroff-Riley type II radio galaxies
\citep{1995A&A...302..317F,1996ApJ...460..634R,2002ApJ...568..639P}.
To draw solid conclusions about the properties of CSOs as a class, it is
important to construct a large representative sample of them. 

An interesting side effect of an investigation of a large CSO candidate sample is 
the possibility of finding a supermassive binary black hole pair 
\citep{2004ApJ...602..123M,2006ApJ...646...49R,2009AN....330..206T} that can mimic a CSO.
Such an object would have two compact radio components that share the typical
characteristics of the core: flat spectrum, high brightness temperature
$T_\mathrm{b}$, and complex flux variability pattern.
It has also been suggested that true CSOs are the likely hosts of 
supermassive black hole binaries \citep{2010ApJ...713.1393W}.
Finally, CSOs may be useful as calibrators for continuum observations of other radio
sources thanks to their stable flux density and low fractional
polarization \citep{2003ApJ...597..157T}.

The largest homogeneously selected CSO sample to date is the COINS sample by \cite{2000ApJ...534...90P}.
An initial list of candidates for this sample was selected from
\cite{1988ApJ...328..114P} and Caltech--Jodrell Bank \citep{1994ApJS...95..345T}
VLBI surveys and  from the first VLBA Calibrator Survey \citep{VCS1}. 
Dedicated multifrequency polarimetric VLBA observations of the
candidates have been performed to distinguish between true CSOs and contaminating core-jet type
sources. The sample was further extended to the northern and southern
extremities of the VLBA Calibrator Survey by \cite{2003ApJ...597..157T}.
Another CSO sample based on the VLBA Imaging and Polarization Survey
is being constructed by \cite{2009AN....330..206T}.

In this paper we present a list of CSO candidates selected by using publicly
available\footnote{\url{http://astrogeo.org/vlbi_images/}}
$S$~band (central frequency 2.3\,GHz) and $X$~band (central frequency 8.6\,GHz)
VLBI data collected in the course of the VLBA Calibrator Surveys
(VCS) 1 to 6 by \cite{VCS1}, \cite{VCS2}, \cite{VCS3,VCS4}, \cite{VCS5},
\cite{VCS6}, and the Research
and Development VLBA program 
\citep[RDV, e.g.,][]{1996ApJS..105..299F,1997ApJS..111...95F,2008evn..confE..86P,RDV2009}.
We discuss the basic properties of the compiled sample of CSO candidates
including an unexpected systematic difference in CSO component positions measured at
the lower and higher frequencies and our attempt to reproduce this effect
with detailed numerical modeling.

\section{The VLBI-selected sample of CSO candidates and its basic characteristics}

Among 4170 radio sources observed in the course of VCS and RDV VLBI
experiments, we selected a list of 64 CSO candidates that meet the following
criteria.
\begin{enumerate}
\item S and X band images show two dominating components of comparable
brightness that are presumed to be hotspots on both sides of a two-sided jet.
\item Each component is detected in the $S$ and $X$~bands with signal-to-noise
ratios greater than five.
\item Most of the residual emission, if present, is located between the two brightest
components. This emission may come from the core, mini-lobes, or the jet itself.
\end{enumerate}
The third criterion is meant to distinguish CSO from sources with a bright
core and one-sided jet showing a single jet component comparable in brightness
to the core. 

The list of selected sources is presented in Table~\ref{table:srclist}.
We followed \cite{1999ASPC..180..301F} to estimate errors of the model
component parameters.
The dual-frequency $S$- and $X$-band, naturally weighted CLEAN images of the
sources are shown in Fig.~\ref{fig:images}. 
The lowest contour level was chosen at four times the image rms, 
and plotted contour levels increased by a factors of two.
The beam is shown in the bottom left-hand corner of each image.
Blue and orange spots indicate Gaussian model components 1 and 2 
from Table~\ref{table:srclist}, respectively.
Table~\ref{table:srclist} and Fig.~\ref{fig:images} are 
available in the electronic version of the paper.

The information about broad-band radio spectra of the sources was obtained from
the RATAN-600 multifrequency multi-epoch $1$--$22$\,GHz observations in
1997-2006 \citep{Kovalev_etal99,2002PASA...19...83K} 
and data from the literature collected by means of the CATS database \citep{CATS05}.
Among the selected CSO candidates we identified 30 GHz--Peaked Spectrum (GPS)
sources, 12 steep spectrum sources, and 22 flat spectrum sources. 
For a discussion of GPS sources observed by the RATAN-600, see
\cite{2009AN....330..199S} and \cite{2008evn..confE..96S}.
We consider sources with peaked or steep spectra as the most
promising CSO candidates.

Among 64 selected candidates, 13 are part of the COINS CSO sample
\citep{2000ApJ...534...90P} and three more sources are listed as rejected candidates
for the COINS sample on the basis of their parsec-scale spectra and
polarization properties. Two of them (0357$+$057, 1734$+$063) have flat single-dish
spectra, while the third source (0839$+$187, also known as OJ\,164) shows 
a spectral peak at $\sim1.2$\,GHz. The last source is an important example 
of a kind of GPS quasar with core-jet morphology, which may contaminate CSO
samples, including the one presented in this paper.

Optical identifications and redshifts of 28 CSO candidates are available
from the \cite{2006A&A...455..773V} catalog, while 36 objects have no 
known optical counterpart. Among the identified sources, 19 are quasars,
seven are active galaxies, and two are BL~Lac type objects.
The boundary between quasars and active galaxies (which
include Seyferts and Liners) is somewhat arbitrarily chosen in the \cite{2006A&A...455..773V} 
catalog at an absolute optical magnitude $M_B = -23$. 
Since quasars are intrinsically brighter in the optical band, they are 
more likely to be detected than are radio galaxies. A similar effect for 
GPS sources was reported by \cite{2009AN....330..199S}.
However, we expect that at least half of the CSO
candidates presented in this paper are associated with quasars.
It is remarkable that the compact double radio structure that we
associate with candidate CSOs is found in objects characterized
by a wide range of optical luminosities (quasars, radio galaxies) and
a variety of radio spectrum shapes (steep, GPS, flat).

None of the selected radio sources has a $\gamma$-ray counterpart in the
{\it Fermi} Large Area Telescope first-year catalog (1FGL;
\citealt{2010ApJS..188..405A}), the largest catalog of GeV sources to date.
Since the extragalactic $\gamma$-ray sky is dominated 
by blazars \citep[e.g.,][]{2010ApJ...715..429A,Fermi_VCS,1993ApJ...415L..13T},
this provides additional 
reassurance that the selected sample is not strongly contaminated by 
relativistically beamed objects.

\section{Properties of the dominating parsec-scale components}
\label{s:general_prop}

As described above, we have selected sources that only show two dominant
components both in the $S$ and $X$~bands. We associate these components with hotspots at
opposite ends of a two-sided jet. To quantify the position, flux density, and
size of the hotspots, each source was modeled in the visibility ($uv$)
plane by two circular 
Gaussian components using the {\em Difmap} software \citep{1994BAAS...26..987S}. 
To minimize the effect of the resolution difference between bands, only
$uv$-range covered by both $S$- and $X$-band data was used for the modeling.

\addtocounter{figure}{1} 

\begin{figure}[t]
\centering \includegraphics[width=0.5\textwidth]{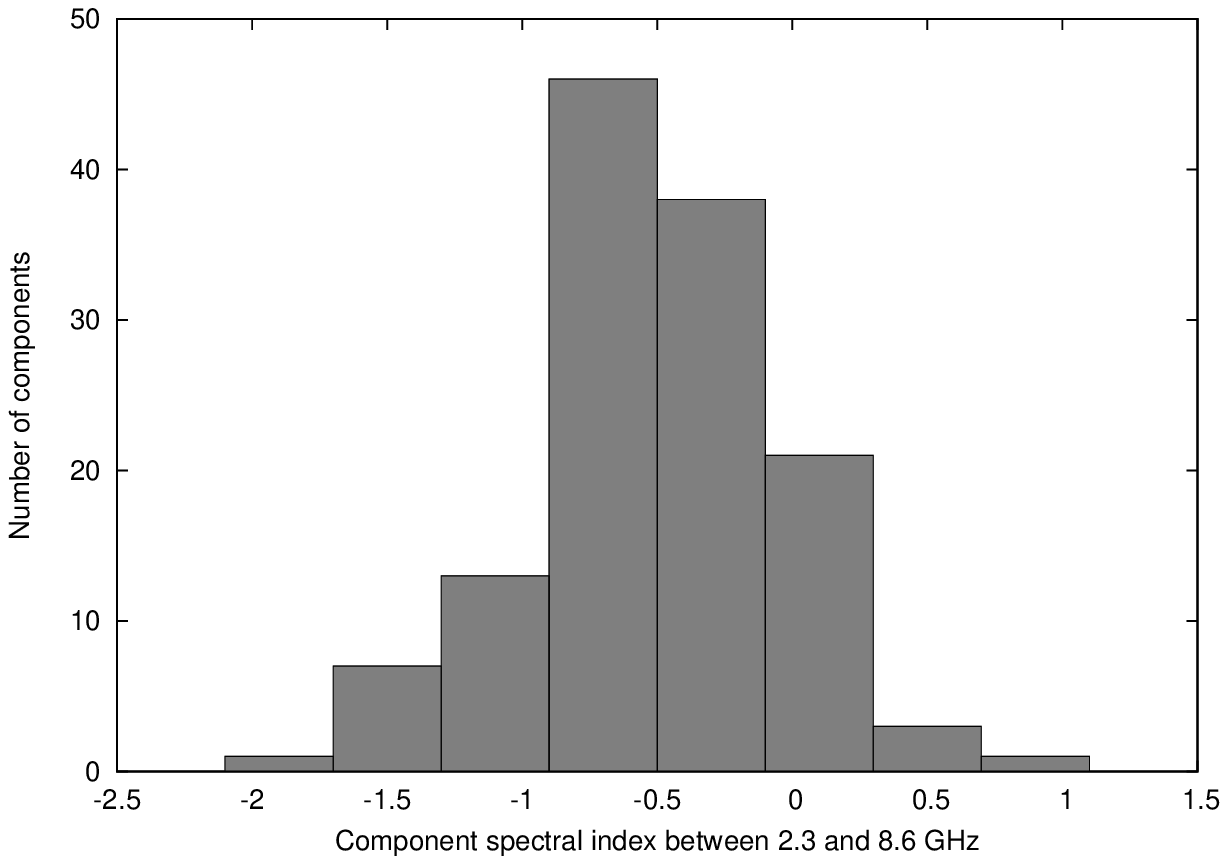}
\caption[]{$S$--$X$~band spectral index, $\alpha$, distribution for bright CSO components
($S\propto\nu^\alpha$).}
\label{fig:v5_a_all}
\end{figure}

The distribution of the two-point, simultaneous $S$--$X$~band spectral index of the components
is presented in Fig.~\ref{fig:v5_a_all}. It shows values in the
range from $-1.83$ to $+0.99$ with the median of $-0.52$. It is evident that 
the majority of the observed components in the selected CSO candidates
have spectral indices that are not typical of flat-spectrum cores.

\begin{figure}[t]
\centering
\includegraphics[width=0.5\textwidth]{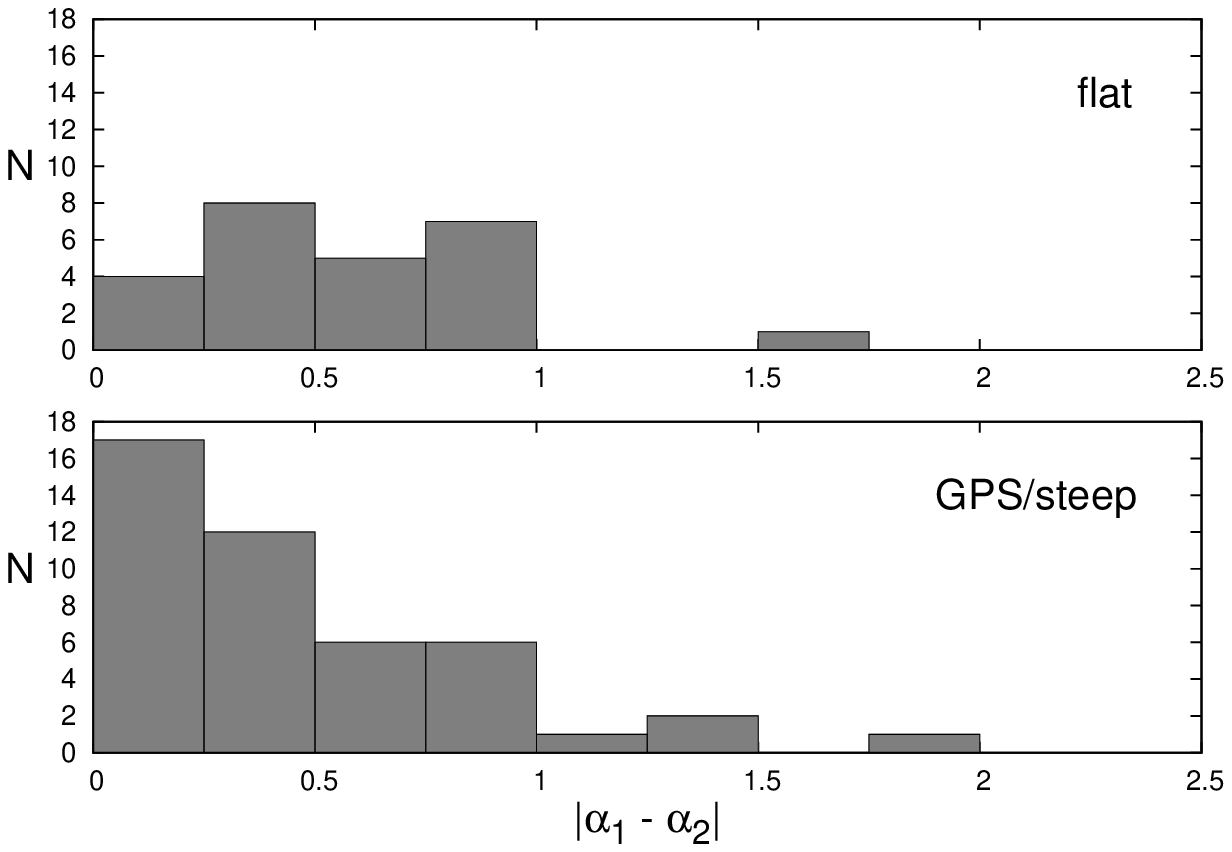}
\caption[]{
Absolute difference between $S$--$X$~band spectral indices of the components~1
($\alpha_1$) and 2~($\alpha_2$)
for objects of various spectral types as presented in Table~\ref{table:srclist}.
The Kolmogorov-Smirnov test indicates at the 94\,\% confidence level
that the distributions are different. }
\label{fig:dA}
\end{figure}

The similarity of broad-band spectra of the two dominant parsec-scale features is expected to 
indicate a true CSO. To establish this similarity, multifrequency VLBI follow-up 
observations of the selected sample of still unconfirmed CSO, similar to the study by
\cite{2000ApJ...534...90P}, are needed. Comparing the two-point simultaneous
$S$--$X$~band spectral 
indices available from our modeling is not sufficient.
However, it may provide a hint that a source 
is probably a true CSO, since a majority of true CSOs (e.g., from the COINS sample)
are found to have similar $S$--$X$~band spectral indices, $\alpha_1$ and $\alpha_2$,
for the two brightest parsec-scale components (Table~\ref{table:srclist}).
On the other hand, a typical core-jet source
is expected to show significantly different spectral indices of
the two dominant components
(absolute difference near to or greater than 0.5) due to the flat radio spectrum
of the synchrotron self-absorbed core and steep-spectrum synchrotron 
radiation of the optically thin jet feature.
Figure~\ref{fig:dA} presents the distribution of the absolute difference
$|\alpha_1 - \alpha_2|$ of the indices for sources in our sample separated
on the basis of their single-dish radio spectrum type --- GPS/steep versus flat.
Figure~\ref{fig:dA} supports the assumption that GPS/steep spectrum
sources tend to have similar spectra of the two brightest parse-scale
components, while flat spectrum sources tend to have dominant parsec-scale
components with different spectra.
The median component spectral index differences are $0.58$ 
for the flat
and $0.34$ for the GPS/steep subsample.
The Kolmogorov-Smirnov test indicates at the 94\,\% confidence level
that the distributions may be different.
Most of the contaminating core-jet sources --- not true CSO --- are expected to be found 
within the flat-spectrum subsample of the CSO candidates list.

\begin{figure}[t]
\centering \includegraphics[width=0.5\textwidth]{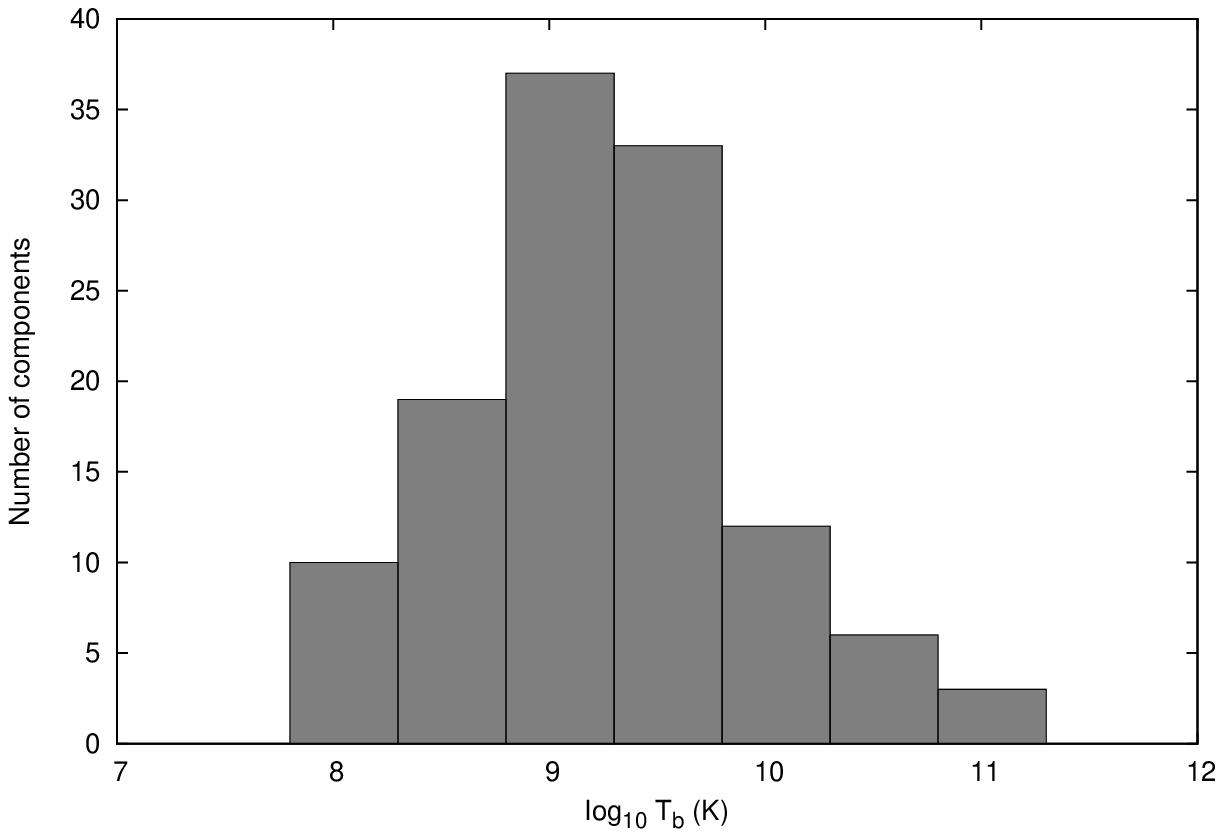}
\caption[]{Brightness temperature distribution (measured at $X$~band) for bright CSO components.}
\label{fig:v5_Tb_X}
\end{figure}

We have also estimated the brightness temperature of the components at
$X$~band, which is found to be
typically $T_\mathrm{b}\sim 10^9$\,K in the observer's frame (see Table~\ref{table:srclist} and
Fig.~\ref{fig:v5_Tb_X}) --- systematically lower than the one measured in
the parsec-scale cores of typical bright extragalactic Doppler-boosted
core-jet sources \citep[e.g.,][]{2005AJ....130.2473K,RDV}.
The values of $T_\mathrm{b}$ presented in Table~\ref{table:srclist}
are computed using component parameters derived from modeling restricted
$uv$-range data for consistency with parameters presented in other
columns of Table~\ref{table:srclist}. Brightness temperatures derived from
the modeling, which takes full $uv$-range available at $X$~band into
account, are
consistent within a factor of two with the values obtained from the
restricted $uv$-range models.

\section{Frequency-dependent component position}
\label{s:pos}

\onlfig{5}{
\begin{figure*}
 \centering
 \subfloat[$2.3$\,GHz]{\label{fig:J1357+4353_S_uv}\includegraphics[width=0.44\textwidth,angle=0,trim=0cm 0cm 0cm 0.6cm,clip]{J1357+4353_S_2002_05_14_yyk_vis_uv.eps}}~~
 \subfloat[$8.6$\,GHz]{\label{fig:J1357+4353_X_uv}\includegraphics[width=0.44\textwidth,angle=0,trim=0cm 0cm 0cm 0.6cm,clip]{J1357+4353_X_2002_05_14_yyk_vis_uv.eps}}
 \caption{$uv$-coverage obtained for the CSO candidate source J1357$+$4353 at $2.3$\,GHz
 (\ref{fig:J1357+4353_S_uv}) and $8.6$\,GHz (\ref{fig:J1357+4353_X_uv})
 observed on 2002 May~14 within the VCS2 \citep{VCS2} VLBA experiment. }
 \label{fig:typicaluv}
\end{figure*}
}

\begin{figure}[t]
\centering \includegraphics[width=0.5\textwidth]{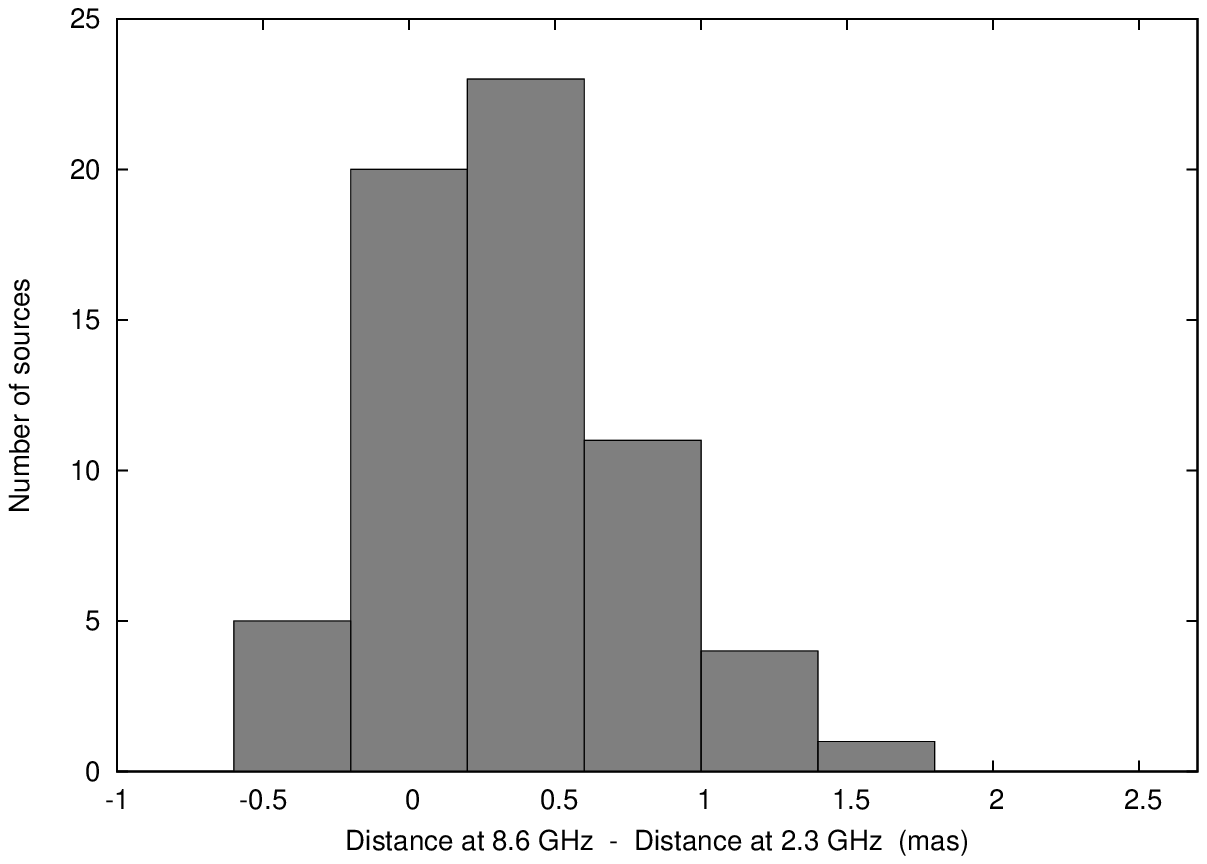}
\caption[]{Difference in distances between two dominant source components
measured in $X$- (8.6\,GHz) and $S$ (2.3\,GHz) bands. 
The median value of the distribution is $0.27$, which is significantly 
($>99.9$\,\% probability according to the sign test) greater than zero.}
\label{fig:dsdx}
\end{figure}

\begin{figure}
\centering \includegraphics[width=0.5\textwidth]{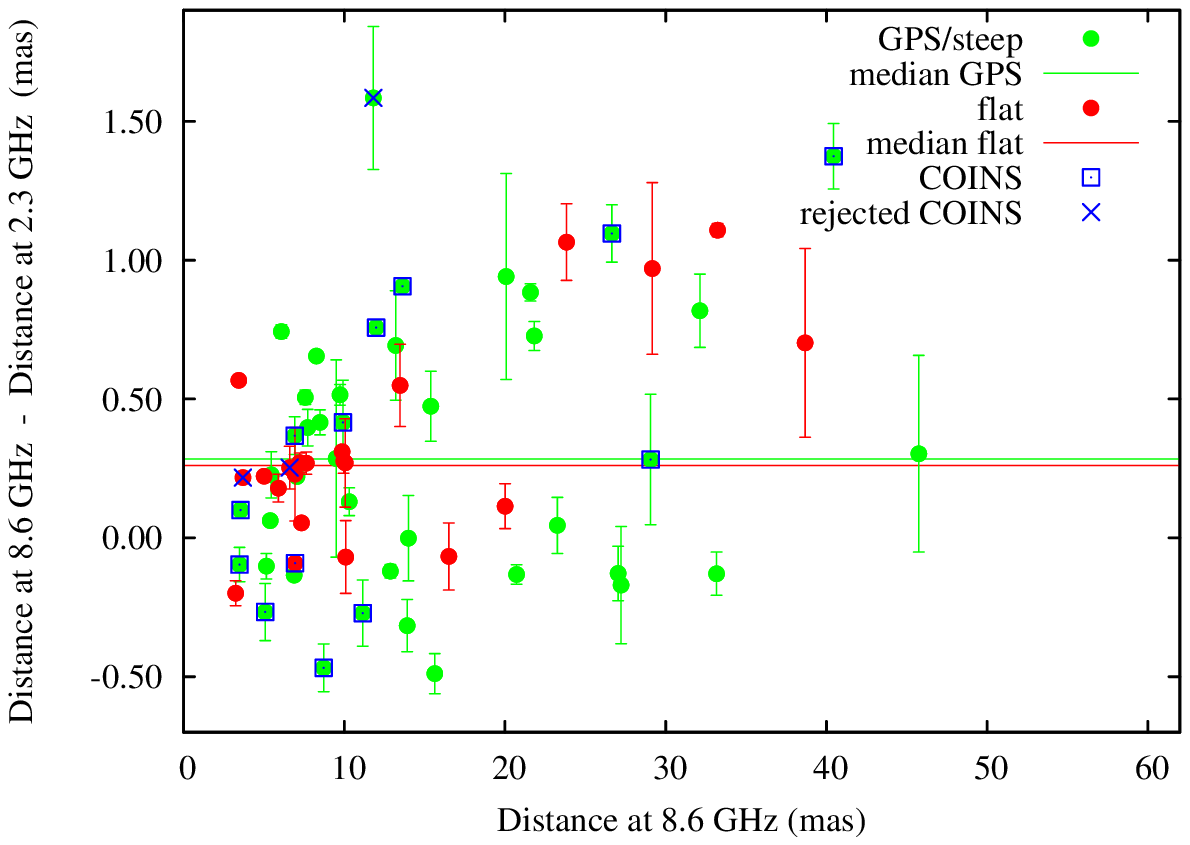}
\caption[]{Difference in distances between two dominant source components measured
in the $S$ and $X$~bands as a function
of distance measured in the $X$~band. Green points represent GPS and steep
spectrum sources, red points represent flat spectrum sources.
Objects included in the COINS sample are
marked with blue squares, while objects rejected from the COINS sample as
probable core-jet sources are marked with blue crosses. The green line
denotes the
median distance difference for GPS/steep spectrum sources, the red line
marks the same for flat spectrum sources.}
\label{fig:distance_difference}
\end{figure}

Interestingly, we have found that distances between the two dominant parsec-scale 
components are systematically greater when measured in the $X$~band
(8.6\,GHz) than those measured in the $S$~band (2.3\,GHz).
An indication was first based on the full $uv$-range
modeling at each frequency. The mean difference in CSO component separation measured 
in the $S$ and $X$~bands was found to be $0.40 \pm 0.07$\,mas (median value
$0.34$\,mas).

The distance between the two outer components of a CSO may appear
different if measured at different frequencies because of {\it (i)}
an observational effect caused by difference in $uv$-coverage, hence the resolution
in $S$ and $X$~bands or {\it (ii)} real spectral index gradients across the
components. The second possibility may potentially provide important
information about physical conditions around the termination shock 
of a CSO jet, particle acceleration efficiency, and cooling rate,
as investigated in Sect.~\ref{sec:discussion}.
To eliminate the former possibility, we redid the model fitting
and component position analysis restricting the used $uv$-range to spatial
frequencies common to $S$- and $X$-band data.
The mean difference was found for the restricted case 
to be $0.32 \pm0.06$\,mas, while
the median value is $0.27$\,mas if all 64 CSO candidates are considered. 
These values are consistent with those obtained using 
the full $uv$-range on the one--$\sigma$ level,
indicating that the difference in $uv$-coverage between
the $S$ and $X$~bands only has a minor effect on the derived component positions.
The magnitude of the difference between the values obtained for the full and
restricted $uv$-range is comparable to the values obtained in our numerical
modeling (Section~\ref{sec:discussion}, Table~\ref{table:1}) where the
effect of resolution difference was simulated in the image plane by
convolving the high-resolution model jet image with Gaussian beams of different
sizes.

In the following discussion and analysis we use the results obtained using the $uv$-range restricted
to the spatial frequencies common to both bands in order to completely eliminate
the systematic effect associated with the difference in
resolution.  A typical $uv$-coverage for the archival $S$- and $X$-band 
VLBI data used to select CSO candidates is presented in Fig.~\ref{fig:typicaluv}
(available in the online version of the paper).
Figure~\ref{fig:dsdx} presents the distribution of the distance differences.
We did not recalculate this value in the source frame because many sources in 
our sample have no redshift information available.
The sign test \citep[e.g.,][]{statTextbook} confirms that the observed median 
difference is indeed greater than zero
with a probability higher than 99.99\,\%. The sign test was
chosen because it is a non-parametric test and makes very few assumptions
about the nature of the distribution under investigation.
Figure~\ref{fig:distance_difference} illustrates the position difference
as a function of component distance measured in the $X$~band for flat- and
GPS/steep-spectrum CSO candidates. The correlation analysis indicates a positive
correlation between 
the $S$--$X$~band distance difference and the distance measured at $X$~band
with the Pearson's correlation coefficient 
$r=0.35$, which corresponds to a 99.4\,\% probability of true correlation at
the given sample size, if all candidates are considered together.

The observed frequency-dependent difference in component
positions could be trivially explained as the well known ``core
shift'' effect
\citep[e.g.,][]{1998A&A...330...79L,2008A&A...483..759K,2011arXiv1103.6032S}
if the presented sample of CSO candidates is heavily
contaminated with core-jet type sources where one of the
two observed bright components is actually a core instead of a
hot spot. 
Spectral index of a component should be flat or inverted
($\alpha \geq 0$) if it is the core, so one may expect a positive
correlation between $\alpha$ and the distance difference.
However, no such correlation is observed. Also, no correlation
was found between $T_\mathrm{b}$ of the first component and the observed distance
difference (Table~\ref{table:srclist}).
Moreover, as shown in \S\ref{s:general_prop}, most of the
dominating parsec-scale features have steep spectral indices
and relatively low brightness temperatures, which is not typical
of Doppler-boosted, opaque parsec-scale cores.

To further test the core shift possibility, we divided the
CSO candidates into two groups based on their single-dish radio
spectra similar to the analysis presented in Fig.~\ref{fig:dA}.
The first group contained flat-spectrum sources that are more
likely to be blazars with core-jet type morphology. The second
group included sources with peaked and steep spectra, which are
more likely to be true CSO. For the flat-spectrum group, the
mean position difference was found to be $0.33\pm0.08$, median
value $0.26$, with $>99.9$\,\% probability that the median
value is greater than zero. For the GPS/steep spectrum group,
the mean difference is $0.31\pm0.08$, the median value $0.28$,
with $>99.9$\,\% probability of it being higher than zero.
A Kolmogorov-Smirnov test can exclude ($p=0.06$) the possibility 
that the frequency-dependent position differences observed in flat-spectrum and
GPS/steep-spectrum sources are drawn from the same parent distribution.

It is widely accepted
\citep[e.g.,][]{1997A&A...325..943S,2010ApJ...715.1071O} that
CSOs are often found among GPS sources identified with galaxies
(as opposed to blazars). In the presented sample of CSO
candidates there are only five GPS sources associated with
galaxies, which is not enough for a statistical analysis.
However, we note that among these five sources (which should be
considered the best CSO candidates), four show a positive
difference between component distances measured at $8.6$ and $2.3$\,GHz.

Overall, it seems highly unlikely that the systematic
difference in component positions at $2.3$ and $8.6$\,GHz in the
selected sources can be attributed to the core shift effect in
core-jet type sources contaminating the CSO candidate sample.
Intrigued by the observed effect, we have turned to detailed
numerical modeling in search of an explanation.

\section{Discussion} 
\label{sec:discussion}

When the jet flow crosses the hotspot, it is deflected and starts to
flow backwards towards the nucleus. This is known as the
backflow. When radio-emitting particles cool in such a backflow, we
expect the higher energy emission (caused by the faster cooling 
high-energy particles) to be concentrated closer to the hotspot, whereas
the lower energy emission (from the slower cooling low energy
particles) should be seen farther away towards the nucleus, while these
particles are being dragged by the backflow. Thus, this should generate a spatial
separation between the peaks of the emission at higher and lower
frequencies such that the high-frequency peak is observed farther away
from the nucleus than the lower frequency one. This is a possible
cause of the observed bias.

We attempted to test this hypothesis by
computing radio emission from simulated relativistic jets. The code
and initial setup is similar to that of \cite{2007MNRAS.382..526P}. 
We performed two simulations, one with a one-sided jet
(M1) and a second one with a two-sided jet (M2), by changing the boundary conditions at the inlet. The radio
emission is computed using the \emph{SPEV} code \citep{2009ApJ...696.1142M}.

The simulated jets are injected with an initial velocity $0.95$~c and
the Mach number $1.66$. The density contrast between the jets and
external medium at the injection point ($10$~pc from the nucleus) is
$10^{-5}$, and their initial radius is $2$~pc. The jet composition is
purely leptonic and the external medium is composed of the neutral
hydrogen. We assume the profile of the external medium in a typical
radio galaxy like 3C~31
\citep{2007MNRAS.382..526P,2002ApJ...581..948H}.

The jets were evolved until they reach the distance of $110$~pc
from the nucleus.  We used these as an input for the \emph{SPEV}
code, and produced synthetic radio images at $2.3$ and $8.6$\,GHz, assuming the
jet angle to the line of sight to be $70^\circ$. Since the simulations
do not include magnetic fields, for the purpose of computing 
the emission, we assumed a randomly oriented magnetic field in a
fraction of equipartition with the thermal fluid, whereby the
ratio of magnetic to internal energy density was fixed to
$\epsilon_B = 10^{-3}$. The relative properties of the emission at
$2.3$ and $8.6$\,GHz are expected to depend only weakly on the value of
$\epsilon_B$ because the radio frequencies are always below the
cooling frequency for the particle distribution. The typical value
of the magnetic field at the injection point is $\approx 20 \mu$G.

\emph{SPEV} code uses Lagrangian particles as markers of the
nonthermal particles responsible for the radio emission (details can
be found in \citealt*{2009ApJ...696.1142M}). In the current work we
assume that these Lagrangian particles are inserted into the jet at
the injection point and their spatial, temporal and spectral evolution
is followed by taking synchrotron and adiabatic losses into account.
We also have the possibility of accelerating additional
particles in strong shock waves (e.g., Mach disk). We denote such
models with a suffix ''s'' (M1s and M2s). For both types of particle
injection (at the jet injection point or in the shocks), we assume
  that the total particle number density is proportional to the
  thermal fluid density and that the total particle energy density is
  proportional to the fluid internal energy density. We fix the
  power-law index of the nonthermal energy distribution at the time
  of injection to $2.25$. These assumptions allow us to determine the
  lower cutoff of the particle injection spectrum (see
  \citealt*{2009ApJ...696.1142M} for more details). To determine the
  upper cutoff we equate the synchrotron cooling time with the
  electron acceleration time (see \citealt*{MGA10} for a more detailed
  discussion of the upper cutoff).

\begin{figure*}[t]
\centering
\includegraphics[scale=0.33,trim=1cm 0cm 0cm 0cm]{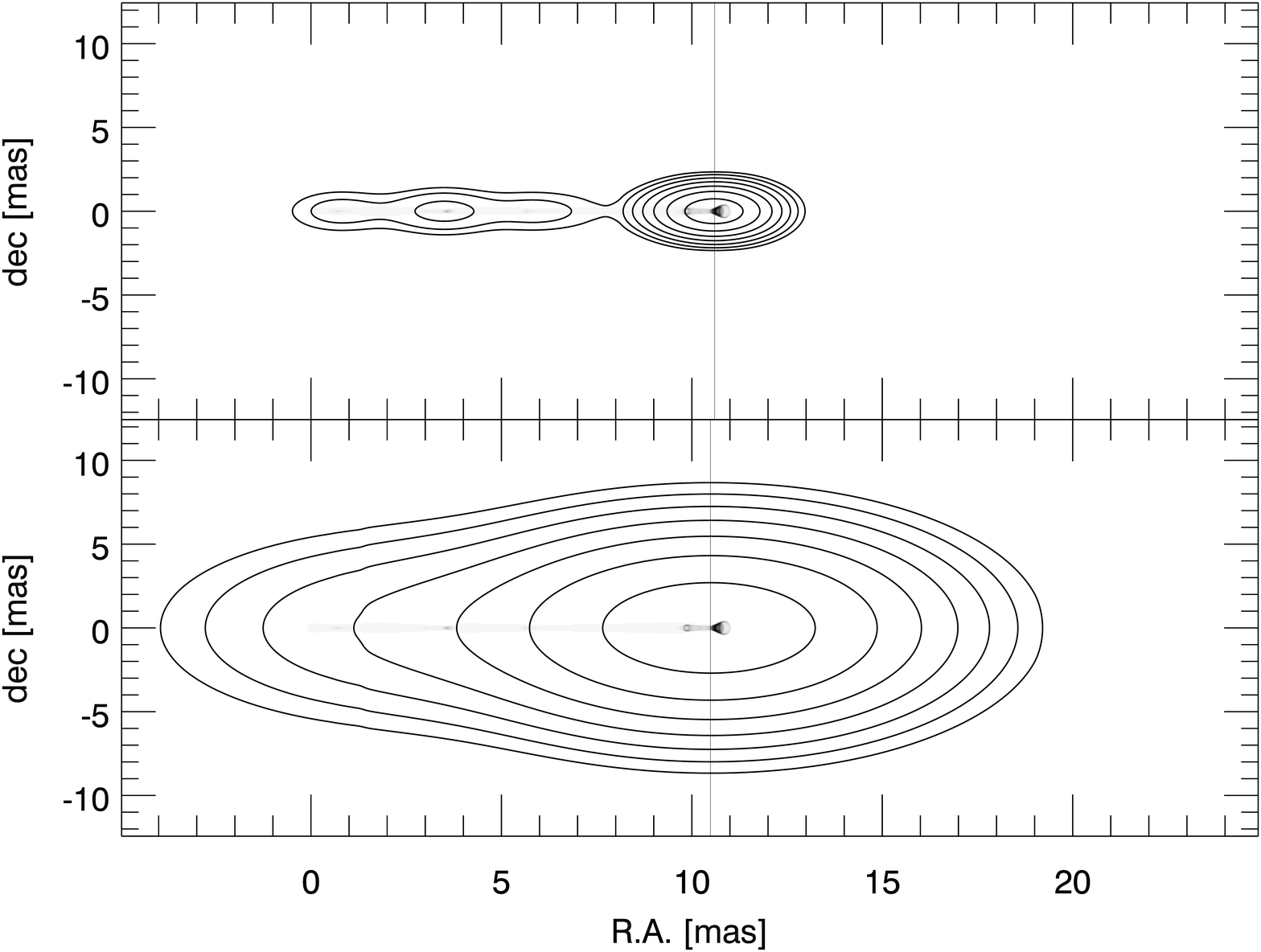}
\caption[]{
Synthetic radio maps at $2.3$\,GHz (lower panel) and $8.6$\,GHz (upper panel) of
the simulated jet model M2s (see Fig.~2 in \citealt*{2007MNRAS.382..526P} for
the typical hydrodynamic structure of such simulated jets). Shades of gray
show the normalized intensity of the unconvolved radio maps. Assuming
that the source is located at $z=1$ contours show the degraded (convolved)
image. Contour levels are 1, 2, 4, 8, 16, 32, and 64\,\% of
the peak intensity. The bright feature in the center of each panel is
the hot spot, and vertical lines denote the longitudinal position of the
peak intensity in the convolved image.
The difference in position of the peak at two frequencies is very small.}
\label{fig:synthetic}
\end{figure*}

The synthetic radio maps obtained from the simulations have much
better resolution than typical VLBI observations. In order to be able
to compare them directly we artificially degrade the quality of
synthetic radio maps to that of the observations. To this end we
compute degraded images assuming the source to be at two redshifts,
$z = 0.04$ and $z = 1$. We note that the angular resolution of the
original synthetic radio maps is $0.14$ and $0.014$ milliarcseconds at
$z=0.04$ and $z=1$, respectively. We convolve the images with a
Gaussian beam whose characteristic half-power size is $6$ and $1.8$
milliarcesonds at $2.3$ and $8.6$\,GHz, respectively. In
Fig.~\ref{fig:synthetic} we show radio maps for model M1.

\addtocounter{table}{1} 
\begin{table}
{
\caption{Difference in positions of the peaks at $8.6$ and $2.3$\,GHz.}
\label{table:1}
\begin{center} 
  \begin{tabular}{lcc}
    \hline\hline
    Model & $\Delta_{0.04}$ & $\Delta_{1}$\\
    index & (mas)           & (mas)       \\
    \hline
    M1s   & $<0.14$ & $0.11$  \\
    M1    & $<0.14$ & $0.11$  \\
    M2s   & $<0.14$ & $0.11$  \\
    M2    & $<0.14$ & $0.11$  \\
    M2cs  & $-0.28$ & $0.055$ \\
    M2c   & $-0.28$ & $0.055$ \\
    \hline
  \end{tabular}
\end{center}
}
\textbf{Note:} $\Delta_{0.04}$ and $\Delta_1$ are the differences for
the source redshift of $0.04$ and $1$, respectively. Positive values
indicate that the peak at the $8.6$\,GHz is farther away from the nucleus
than at $2.3$\,GHz. M1s and M1 stand for the one-sided jet model with and
without shock acceleration, respectively. Similarly, M2s and M2 stand for
the jet of the double-sided jet model, while M2cs and M2c stand for the
counter jet in that same model. When the difference is smaller than the
angular resolution of the unconvolved maps, we give only upper bounds.
\end{table}

In Table~\ref{table:1} we summarize the results of our simulations. We see that the
results do not confirm the hypothesis. Although models M1, M1s, M2, and
M2s, as well as models M2c and M2cs for $z = 1$, show the expected
trend, the differences are too small to be detected, which can also be
seen in Fig.~\ref{fig:synthetic}, where vertical lines denote the position of the
intensity peak. No such trend is observed for models M2c and M2cs for $z = 0.04$.

The population of the nonthermal particles is dominated by the
synchrotron cooling close to the injection site and by the
adiabatic cooling further downstream (see e.g.,
\citealt*{2009ApJ...696.1142M}). In the synchrotron-cooling
region we expect that the higher energy particles
lose their energy faster than the lower energy ones, whereas in
the adiabatic-cooling region they all lose the energy
at the same rate. Nonthermal particles inserted at the jet
injection point reach the adiabatic-cooling region
long before they reach the hotspot. Although there is 
substantial compression of the flow in the shock, if shock
acceleration is not included, this compression alone is not
enough to produce another synchrotron-cooling region.
Therefore, the cooling rate of the high and low energy
particles is the same, and no spatial separation between high
and low frequency emission is expected. If we include shock
acceleration, there is a synchrotron cooling zone
behind the shock, but it is not large enough to reach into the
backflow, because the shock does not accelerate
particles to high enough energy to produce a larger synchrotron
cooling dominated zone. Therefore, the result is qualitatively
the same as when shock acceleration is not taken into account.
This can be seen in the unconvolved radio maps
(Fig.~\ref{fig:synthetic}), where no noticeable difference
between the positions of the peaks can be observed. 

The systematic difference in hotspot positions obtained in our model 
(Table~\ref{table:1}) results fro the fact that the convolution beam at $2.3$\,GHz
is larger than at
$8.6$\,GHz. The former includes more emission from 
jet and backflow regions closer to
the nucleus (blending), and this generates a shift in the
position of the apparent peak towards the nucleus. This results
in the aforementioned apparent position difference in our models, which 
is an effect of model image convolution necessary to make it comparable to
real observations. The same effect should be present in observations of real
sources. However, the predicted magnitude of this
blending effect is still smaller than the observed position difference. 
This is in
accordance with minor difference between the modeling results obtained using
full (different resolution) and restricted (matched resolution) $uv$-range
as discussed in Section~\ref{s:pos}.

In future work we intend to study two possibilities that could
explain the observations. On the one hand, a much more efficient shock
acceleration could produce enough high-energy particles to create a
longer synchrotron cooling zone. On the other hand, the
process of shear layer acceleration (e.g.,
\citealt*{2002ApJ...578..763S,2008ApJ...681...84A})
could produce continuous injection of nonthermal
particles into the backflow. Small velocity gradients present
there would probably result in low acceleration efficiency, lower
emission frequency, and ultimately, a shift in the position of the
peak of emission towards the nucleus for those frequencies.

\section{Summary}

We selected a sample of 64 candidate compact symmetric
objects (CSOs) using simultaneous $S$ (2.3\,GHz) and $X$ (8.6\,GHz) 
band VLBI data for $4170$ compact 
extragalactic radio sources from the VLBA Calibrator Survey 1--6 and
Research and Development VLBA projects. Broad-band radio spectra of the
selected sources were characterized using multifrequency observations with
RATAN--600 and archival data from the literature.
Among the selected CSO candidates, we identified 30 GPS,
12 steep-spectrum, and 22 flat-spectrum
radio sources. The median two-point $S$--$X$~band spectral index for
dominating parsec-scale components was found to be $-0.52$, while
median brightness temperature is on the order of $10^9$\,K.
A multifrequency follow-up VLBI study of the whole selected
sample is needed to confirm true CSO cases.

A systematic difference in distance measured at $2.3$ and $8.6$\,GHz
between the outer CSO components (presumed to be associated
with hotspots) was detected. The $8.6$\,GHz distance was found to be
larger for most sources. 

The analysis excluded the blending effect as the dominant cause for
the observed component position difference.
Even if some CSO candidates in our sample were misclassified and have a core-jet 
parsec-scale morphology, our analysis rules out the apparent 
frequency-dependent shift in the core position (the ``core shift'') as an
explanation for the observed effect. 

Numerical modeling of a CSO hotspot was conducted but failed to
reproduce the magnitude of the observed difference.
The model suggested the observational blending effect as the
dominating mechanism responsible for the position difference, which is
too small to solely account for the observed difference. 
A more detailed modeling that fully accounts for the synchrotron opacity
effects and dedicated simultaneous multifrequency VLBI
observations are needed to determine the exact origin of 
the hotspot position difference effect.
 
\begin{acknowledgements}
The authors are deeply grateful to D.~Kels, who contributed to CSO
candidate selection during his work as a high-school summer
student at the MPIfR in Bonn. We thank A.~P.~Lobanov and T.~Savolainen for
carefully reading the manuscript and providing useful comments. 
We thank the referee for useful comments that helped improve
our analysis and the manuscript. 
KS is supported by the International Max-Planck Research School
(IMPRS) for Astronomy and Astrophysics at the universities of
Bonn and Cologne. YYK was supported in part by the return
fellowship of Alexander von Humboldt foundation and the Russian
Foundation for Basic Research grants 08-02-00545 and 11-02-00368.
PM and MP thankfully acknowledge the computer resources, technical
expertise, and assistance provided by the ``Centre de C\`alcul de la
Universitat de Val\`encia'' through the use of ``Tirant'', the local node of
the Spanish Supercomputation Network.
PM and MP
acknowledge the support from the Spanish Ministry of Education
and Science through grants AYA2007-67626-C03-01 and
CSD2007-00050. MP acknowledges the support from the Spanish
Ministry of Education and Science through grant
AYA2007-67752-C03-02 and a \emph{Juan de la Cierva} fellowship.
The authors made use of the database CATS \citep{CATS05} of the
Special Astrophysical Observatory. This research made use
of the NASA/IPAC Extragalactic Database (NED), which is operated
by the Jet Propulsion Laboratory, California Institute of
Technology, under contract with the National Aeronautics and
Space Administration. This research made use of NASA's Astrophysics
Data System.
\end{acknowledgements}

\onllongtabL{1}{
\small
\begin{landscape}
\renewcommand{\thefootnote}{\Roman{footnote}}
\begin{longtable}{c c rrcrc cc ccl}
\caption{List of CSO candidates selected from VCS and RDV surveys}\label{table:srclist}
\renewcommand{\footnoterule}{}  
 \\
 \hline\hline

Name\footnotemark[1] &
Comp.\footnotemark[2] &
F$_S$\footnotemark[3] (Jy) &
F$_X$\footnotemark[3] (Jy) &
$\alpha$\footnotemark[4] &
R$_X$\footnotemark[5]$^\text{,}$\footnotemark[12] (mas) &
T$_{\mathrm{b}~X}$\footnotemark[6]$^\text{,}$\footnotemark[12] (K) &
D$_X$\footnotemark[7] (mas) &
$D_X-D_S$\footnotemark[8] (mas) & 
opt.\footnotemark[9] &
$z$\footnotemark[10] &
Comments\footnotemark[11]\\

\hline

 J0029+3456  &  1  &  1.168$\pm$0.042  &  0.612$\pm$0.059  &  $-0.50\pm0.08$  &  2.00$\pm$0.18  &     $3\times10^{9}$  &  29.06$\pm$0.23  &  $+0.28\pm0.23$ & a & 0.517 & COINS, GPS $\nu_m$=1.41 \\
             &  2  &  0.835$\pm$0.021  &  0.227$\pm$0.035  &  $-1.00\pm0.12$  &  2.82$\pm$0.42  &     $5\times10^{8}$  &                  &                                  &      &    &          \\
 J0108-0037  &  1  &  0.332$\pm$0.007  &  0.217$\pm$0.004  &  $-0.33\pm0.02$  &  1.13$\pm$0.02  &     $3\times10^{9}$  &   5.15$\pm$0.02  &  $-0.10\pm0.05$ & q & 1.378 & STEEP \\
             &  2  &  0.159$\pm$0.009  &  0.097$\pm$0.003  &  $-0.38\pm0.05$  &  1.24$\pm$0.04  &     $1\times10^{9}$  &                  &                                  &      &    &          \\
 J0111+3906  &  1  &  0.496$\pm$0.005  &  0.540$\pm$0.081  &  $+0.06\pm0.11$  &  1.45$\pm$0.20  &     $4\times10^{9}$  &   5.08$\pm$0.10  &  $-0.27\pm0.10$ & . & ..... & COINS, GPS $\nu_m$=4.64 \\
             &  2  &  0.400$\pm$0.005  &  0.254$\pm$0.009  &  $-0.34\pm0.03$  &  0.24$\pm$0.01  &     $7\times10^{10}$ &                  &                                  &      &    &          \\
 J0127+7323  &  1  &  0.267$\pm$0.006  &  0.096$\pm$0.005  &  $-0.78\pm0.04$  &  1.50$\pm$0.06  &     $7\times10^{8}$  &  13.99$\pm$0.07  &  $-0.00\pm0.15$ & . & ..... & GPS\footnotemark[13]$^,$\footnotemark[14] \\
             &  2  &  0.107$\pm$0.008  &  0.014$\pm$0.002  &  $-1.57\pm0.14$  &  1.11$\pm$0.14  &   $> 2\times10^{8}$  &                  &                                  &      &    &          \\
 J0132+5620  &  1  &  0.458$\pm$0.007  &  0.313$\pm$0.004  &  $-0.29\pm0.01$  &  0.87$\pm$0.01  &     $7\times10^{9}$  &  11.98$\pm$0.01  &  $+0.76\pm0.03$ & . & ..... & COINS, GPS $\nu_m$=3.01 \\
             &  2  &  0.382$\pm$0.011  &  0.042$\pm$0.001  &  $-1.70\pm0.03$  &  1.06$\pm$0.03  &     $7\times10^{8}$  &                  &                                  &      &    &          \\
 J0207+6246  &  1  &  1.085$\pm$0.014  &  0.650$\pm$0.001  &  $-0.39\pm0.01$  &  0.80$\pm$0.00  &     $2\times10^{10}$ &  21.58$\pm$0.00  &  $+0.88\pm0.03$ & . & ..... & GPS $\nu_m$=2.50 \\
             &  2  &  0.309$\pm$0.014  &  0.056$\pm$0.001  &  $-1.30\pm0.03$  &  0.18$\pm$0.00  &   $> 3\times10^{10}$ &                  &                                  &      &    &          \\
 J0209+2932  &  1  &  0.347$\pm$0.008  &  0.133$\pm$0.007  &  $-0.74\pm0.05$  &  2.04$\pm$0.10  &     $6\times10^{8}$  &   7.72$\pm$0.06  &  $+0.40\pm0.07$ & q & 2.195 & STEEP \\
             &  2  &  0.094$\pm$0.005  &  0.074$\pm$0.003  &  $-0.19\pm0.05$  &  1.42$\pm$0.05  &     $6\times10^{8}$  &                  &                                  &      &    &          \\
 J0304+7727  &  1  &  0.320$\pm$0.006  &  0.144$\pm$0.005  &  $-0.61\pm0.03$  &  1.37$\pm$0.04  &     $1\times10^{9}$  &  10.30$\pm$0.05  &  $+0.13\pm0.05$ & . & ..... & GPS\footnotemark[13]$^,$\footnotemark[16] \\
             &  2  &  0.491$\pm$0.007  &  0.119$\pm$0.005  &  $-1.07\pm0.03$  &  2.06$\pm$0.09  &     $5\times10^{8}$  &                  &                                  &      &    &          \\
 J0400+0550  &  1  &  0.338$\pm$0.015  &  0.383$\pm$0.007  &  $+0.10\pm0.04$  &  0.78$\pm$0.01  &     $1\times10^{10}$ &   6.60$\pm$0.01  &  $+0.25\pm0.08$ & q & 0.761 & COINS-REJ, FLAT \\
             &  2  &  0.108$\pm$0.009  &  0.068$\pm$0.002  &  $-0.35\pm0.07$  &  0.60$\pm$0.01  &     $3\times10^{9}$  &                  &                                  &      &    &          \\
 J0424+0204  &  1  &  0.456$\pm$0.019  &  0.518$\pm$0.009  &  $+0.10\pm0.04$  &  0.67$\pm$0.01  &     $2\times10^{10}$ &  15.38$\pm$0.12  &  $+0.47\pm0.13$ & q & 2.056 & STEEP \\
             &  2  &  0.494$\pm$0.025  &  0.044$\pm$0.006  &  $-1.86\pm0.11$  &  1.88$\pm$0.24  &     $2\times10^{8}$  &                  &                                  &      &    &          \\
 J0429+3319  &  1  &  0.450$\pm$0.003  &  0.286$\pm$0.007  &  $-0.34\pm0.02$  &  0.93$\pm$0.02  &     $5\times10^{9}$  &   7.06$\pm$0.01  &  $+0.22\pm0.01$ & . & ..... & STEEP \\
             &  2  &  0.152$\pm$0.003  &  0.082$\pm$0.002  &  $-0.47\pm0.02$  &  0.50$\pm$0.01  &     $5\times10^{9}$  &                  &                                  &      &    &          \\
 J0511+0110  &  1  &  0.132$\pm$0.005  &  0.173$\pm$0.003  &  $+0.21\pm0.03$  &  0.80$\pm$0.01  &     $5\times10^{9}$  &  23.84$\pm$0.13  &  $+1.07\pm0.14$ & . & ..... & FLAT \\
             &  2  &  0.097$\pm$0.005  &  0.040$\pm$0.005  &  $-0.69\pm0.11$  &  2.06$\pm$0.25  &     $2\times10^{8}$  &                  &                                  &      &    &          \\
 J0518+4730  &  1  &  0.561$\pm$0.006  &  0.158$\pm$0.011  &  $-0.98\pm0.06$  &  1.23$\pm$0.08  &     $2\times10^{9}$  &   3.48$\pm$0.06  &  $-0.10\pm0.06$ & . & ..... & COINS, GPS $\nu_m$=2.17 \\
             &  2  &  0.379$\pm$0.005  &  0.146$\pm$0.014  &  $-0.73\pm0.08$  &  1.13$\pm$0.09  &     $2\times10^{9}$  &                  &                                  &      &    &          \\
 J0620+2102  &  1  &  0.417$\pm$0.019  &  0.174$\pm$0.003  &  $-0.67\pm0.04$  &  1.23$\pm$0.02  &     $2\times10^{9}$  &  26.65$\pm$0.09  &  $+1.09\pm0.10$ & . & ..... & COINS, GPS $\nu_m$=1.51 \\
             &  2  &  0.351$\pm$0.016  &  0.087$\pm$0.010  &  $-1.07\pm0.10$  &  1.63$\pm$0.17  &     $6\times10^{8}$  &                  &                                  &      &    &          \\
 J0736+2604  &  1  &  0.107$\pm$0.007  &  0.235$\pm$0.002  &  $+0.60\pm0.05$  &  0.13$\pm$0.00  &   $> 3\times10^{11}$ &   5.89$\pm$0.01  &  $+0.18\pm0.05$ & . & ..... & FLAT \\
             &  2  &  0.193$\pm$0.009  &  0.084$\pm$0.002  &  $-0.64\pm0.04$  &  1.17$\pm$0.03  &     $1\times10^{9}$  &                  &                                  &      &    &          \\
 J0753+4231  &  1  &  0.229$\pm$0.007  &  0.202$\pm$0.004  &  $-0.09\pm0.03$  &  0.54$\pm$0.01  &     $1\times10^{10}$ &   8.72$\pm$0.08  &  $-0.47\pm0.09$ & q & 3.59 & COINS, GPS $\nu_m$=1.01 \\
             &  2  &  0.453$\pm$0.014  &  0.069$\pm$0.007  &  $-1.44\pm0.08$  &  1.75$\pm$0.16  &     $4\times10^{8}$  &                  &                                  &      &    &          \\
 J0814-1806  &  1  &  0.220$\pm$0.004  &  0.112$\pm$0.004  &  $-0.51\pm0.03$  &  1.84$\pm$0.06  &     $5\times10^{8}$  &  23.26$\pm$0.10  &  $+0.04\pm0.10$ & . & ..... & STEEP \\
             &  2  &  0.102$\pm$0.003  &  0.017$\pm$0.003  &  $-1.35\pm0.12$  &  1.61$\pm$0.18  &   $> 1\times10^{8}$  &                  &                                  &      &    &          \\
 J0821-0323  &  1  &  0.336$\pm$0.010  &  0.136$\pm$0.004  &  $-0.69\pm0.03$  &  1.72$\pm$0.05  &     $8\times10^{8}$  &   5.46$\pm$0.06  &  $+0.23\pm0.08$ & q & 2.352 & GPS $\nu_m$=1.46 \\
             &  2  &  0.128$\pm$0.010  &  0.067$\pm$0.006  &  $-0.49\pm0.09$  &  1.29$\pm$0.10  &     $7\times10^{8}$  &                  &                                  &      &    &          \\
 J0842+1835  &  1  &  0.495$\pm$0.021  &  0.429$\pm$0.014  &  $-0.11\pm0.04$  &  0.97$\pm$0.03  &     $7\times10^{9}$  &  11.81$\pm$0.25  &  $+1.58\pm0.26$ & q & 1.272 & COINS-REJ, GPS $\nu_m$=1.17 \\
             &  2  &  0.370$\pm$0.021  &  0.126$\pm$0.026  &  $-0.81\pm0.16$  &  2.46$\pm$0.51  &     $3\times10^{8}$  &                  &                                  &      &    &          \\
 J0935+3633  &  1  &  0.091$\pm$0.002  &  0.121$\pm$0.001  &  $+0.21\pm0.02$  &  0.41$\pm$0.00  &     $1\times10^{10}$ &   5.02$\pm$0.01  &  $+0.22\pm0.01$ & q & 2.835 & FLAT \\
             &  2  &  0.211$\pm$0.003  &  0.079$\pm$0.001  &  $-0.75\pm0.02$  &  0.78$\pm$0.01  &     $2\times10^{9}$  &                  &                                  &      &    &          \\
 J1008-0933  &  1  &  0.504$\pm$0.004  &  0.216$\pm$0.004  &  $-0.65\pm0.02$  &  1.13$\pm$0.02  &     $3\times10^{9}$  &   7.64$\pm$0.01  &  $+0.27\pm0.04$ & . & ..... & FLAT \\
             &  2  &  0.057$\pm$0.004  &  0.068$\pm$0.003  &  $+0.13\pm0.06$  &  0.58$\pm$0.02  &     $4\times10^{9}$  &                  &                                  &      &    &          \\
 J1035+5040  &  1  &  0.234$\pm$0.002  &  0.140$\pm$0.004  &  $-0.39\pm0.02$  &  0.97$\pm$0.02  &     $3\times10^{9}$  &  10.09$\pm$0.13  &  $-0.07\pm0.13$ & . & ..... & FLAT, GPS-REJ \\
             &  2  &  0.099$\pm$0.003  &  0.032$\pm$0.004  &  $-0.87\pm0.11$  &  1.92$\pm$0.25  &     $2\times10^{8}$  &                  &                                  &      &    &          \\
 J1035+5628  &  1  &  0.897$\pm$0.029  &  0.393$\pm$0.023  &  $-0.63\pm0.05$  &  1.54$\pm$0.08  &     $3\times10^{9}$  &  32.13$\pm$0.11  &  $+0.82\pm0.13$ & a & 0.460 & GPS $\nu_m$=1.08 \\
             &  2  &  0.811$\pm$0.048  &  0.269$\pm$0.027  &  $-0.85\pm0.09$  &  2.18$\pm$0.21  &     $1\times10^{9}$  &                  &                                  &      &    &          \\
 J1036-0605  &  1  &  0.146$\pm$0.002  &  0.276$\pm$0.003  &  $+0.49\pm0.02$  &  0.75$\pm$0.01  &     $9\times10^{9}$  &   8.26$\pm$0.01  &  $+0.65\pm0.01$ & . & ..... &  GPS $\nu_m$=4.50\\
             &  2  &  0.306$\pm$0.002  &  0.092$\pm$0.002  &  $-0.92\pm0.01$  &  0.84$\pm$0.01  &     $2\times10^{9}$  &                  &                                  &      &    &          \\
 J1110-1858  &  1  &  0.713$\pm$0.005  &  0.184$\pm$0.003  &  $-1.04\pm0.01$  &  1.53$\pm$0.02  &     $1\times10^{9}$  &  15.63$\pm$0.06  &  $-0.49\pm0.07$ & a & 0.497 & STEEP, GPS-REJ \\
             &  2  &  0.143$\pm$0.005  &  0.065$\pm$0.004  &  $-0.61\pm0.06$  &  2.17$\pm$0.13  &     $2\times10^{8}$  &                  &                                  &      &    &          \\
 J1143+1834  &  1  &  0.189$\pm$0.003  &  0.102$\pm$0.004  &  $-0.47\pm0.04$  &  0.47$\pm$0.01  &     $8\times10^{9}$  &   6.94$\pm$0.03  &  $-0.09\pm0.03$ & . & ..... & COINS, FLAT \\
             &  2  &  0.167$\pm$0.003  &  0.107$\pm$0.005  &  $-0.34\pm0.04$  &  1.44$\pm$0.06  &     $9\times10^{8}$  &                  &                                  &      &    &          \\
 J1207+2754  &  1  &  0.488$\pm$0.018  &  0.340$\pm$0.004  &  $-0.28\pm0.03$  &  0.53$\pm$0.01  &     $2\times10^{10}$ &  10.06$\pm$0.13  &  $+0.27\pm0.16$ & q & 2.177 & FLAT \\
             &  2  &  0.168$\pm$0.015  &  0.044$\pm$0.005  &  $-1.02\pm0.12$  &  2.32$\pm$0.27  &     $1\times10^{8}$  &                  &                                  &      &    &          \\
 J1213-1003  &  1  &  0.115$\pm$0.004  &  0.216$\pm$0.002  &  $+0.49\pm0.03$  &  0.42$\pm$0.00  &     $2\times10^{10}$ &   7.23$\pm$0.02  &  $+0.27\pm0.04$ & b & ..... & FLAT \\
             &  2  &  0.175$\pm$0.004  &  0.075$\pm$0.002  &  $-0.65\pm0.02$  &  2.20$\pm$0.05  &     $3\times10^{8}$  &                  &                                  &      &    &          \\
 J1224+0330  &  1  &  0.262$\pm$0.004  &  0.360$\pm$0.001  &  $+0.24\pm0.01$  &  0.08$\pm$0.00  &   $> 1\times10^{12}$ &   3.43$\pm$0.00  &  $+0.57\pm0.01$ & q & 0.960 & FLAT \\
             &  2  &  1.022$\pm$0.005  &  0.300$\pm$0.002  &  $-0.94\pm0.01$  &  1.10$\pm$0.01  &     $4\times10^{9}$  &                  &                                  &      &    &          \\
 J1247+6723  &  1  &  0.146$\pm$0.003  &  0.091$\pm$0.004  &  $-0.36\pm0.03$  &  0.89$\pm$0.03  &     $2\times10^{9}$  &   7.57$\pm$0.02  &  $+0.51\pm0.03$ & . & ..... & GPS \\
             &  2  &  0.103$\pm$0.003  &  0.054$\pm$0.001  &  $-0.49\pm0.03$  &  1.30$\pm$0.03  &     $5\times10^{8}$  &                  &                                  &      &    &          \\
 J1248-1959  &  1  &  1.245$\pm$0.058  &  0.489$\pm$0.076  &  $-0.72\pm0.12$  &  2.81$\pm$0.43  &     $1\times10^{9}$  &  20.07$\pm$0.32  &  $+0.94\pm0.37$ & q & 1.275 & STEEP \\
             &  2  &  0.825$\pm$0.061  &  0.381$\pm$0.039  &  $-0.59\pm0.10$  &  4.57$\pm$0.46  &     $3\times10^{8}$  &                  &                                  &      &    &          \\
 J1259+5140  &  1  &  0.108$\pm$0.004  &  0.396$\pm$0.001  &  $+0.98\pm0.02$  &  0.06$\pm$0.00  &   $> 2\times10^{12}$ &  20.00$\pm$0.06  &  $+0.11\pm0.08$ & . & ..... & FLAT \\
             &  2  &  0.060$\pm$0.004  &  0.027$\pm$0.002  &  $-0.60\pm0.07$  &  2.09$\pm$0.12  &     $1\times10^{8}$  &                  &                                  &      &    &          \\
 J1302+6902  &  1  &  0.127$\pm$0.005  &  0.160$\pm$0.003  &  $+0.17\pm0.04$  &  0.83$\pm$0.01  &     $4\times10^{9}$  &  16.51$\pm$0.07  &  $-0.07\pm0.12$ & a & 0.57 & FLAT \\
             &  2  &  0.051$\pm$0.005  &  0.023$\pm$0.003  &  $-0.59\pm0.11$  &  1.39$\pm$0.13  &     $2\times10^{8}$  &                  &                                  &      &    &          \\
 J1311+1417  &  1  &  0.454$\pm$0.003  &  0.136$\pm$0.003  &  $-0.92\pm0.02$  &  1.02$\pm$0.02  &     $2\times10^{9}$  &   3.54$\pm$0.02  &  $+0.10\pm0.03$ & q & 1.952 & COINS, GPS $\nu_m$=1.17 \\
             &  2  &  0.383$\pm$0.003  &  0.115$\pm$0.004  &  $-0.92\pm0.03$  &  1.31$\pm$0.04  &     $1\times10^{9}$  &                  &                                  &      &    &          \\
 J1319-0049  &  1  &  0.245$\pm$0.005  &  0.215$\pm$0.006  &  $-0.10\pm0.02$  &  0.75$\pm$0.01  &     $6\times10^{9}$  &  45.75$\pm$0.35  &  $+0.30\pm0.35$ & q & 0.892 & STEEP \\
             &  2  &  0.151$\pm$0.005  &  0.046$\pm$0.010  &  $-0.90\pm0.17$  &  3.21$\pm$0.70  &     $7\times10^{7}$  &                  &                                  &      &    &          \\
 J1320+0140  &  1  &  0.259$\pm$0.011  &  0.346$\pm$0.005  &  $+0.22\pm0.03$  &  0.64$\pm$0.01  &     $1\times10^{10}$ &  29.17$\pm$0.28  &  $+0.97\pm0.31$ & q & 1.232 & FLAT \\
             &  2  &  0.189$\pm$0.011  &  0.096$\pm$0.015  &  $-0.52\pm0.13$  &  3.59$\pm$0.57  &     $1\times10^{8}$  &                  &                                  &      &    &          \\
 J1324+4048  &  1  &  0.369$\pm$0.004  &  0.131$\pm$0.002  &  $-0.80\pm0.01$  &  0.81$\pm$0.01  &     $4\times10^{9}$  &   5.39$\pm$0.01  &  $+0.06\pm0.01$ & q & 0.495 & GPS $\nu_m$=2.79 \\
             &  2  &  0.284$\pm$0.003  &  0.124$\pm$0.001  &  $-0.64\pm0.01$  &  0.79$\pm$0.01  &     $4\times10^{9}$  &                  &                                  &      &    &          \\
 J1335+5844  &  1  &  0.197$\pm$0.005  &  0.453$\pm$0.004  &  $+0.64\pm0.02$  &  0.72$\pm$0.01  &     $2\times10^{10}$ &  12.87$\pm$0.01  &  $-0.12\pm0.02$ & . & ..... & GPS $\nu_m$=6.45 \\
             &  2  &  0.493$\pm$0.005  &  0.174$\pm$0.004  &  $-0.80\pm0.02$  &  1.26$\pm$0.03  &     $2\times10^{9}$  &                  &                                  &      &    &          \\
 J1350-2204  &  1  &  0.526$\pm$0.009  &  0.209$\pm$0.005  &  $-0.71\pm0.02$  &  1.57$\pm$0.04  &     $1\times10^{9}$  &  27.23$\pm$0.21  &  $-0.17\pm0.21$ & . & ..... & GPS $\nu_m$=0.65 \\
             &  2  &  0.402$\pm$0.007  &  0.062$\pm$0.008  &  $-1.44\pm0.10$  &  3.27$\pm$0.41  &     $1\times10^{8}$  &                  &                                  &      &    &          \\
 J1357+4353  &  1  &  0.425$\pm$0.015  &  0.223$\pm$0.013  &  $-0.49\pm0.05$  &  1.83$\pm$0.10  &     $1\times10^{9}$  &  11.15$\pm$0.08  &  $-0.27\pm0.12$ & . & ..... & COINS, GPS $\nu_m$=1.75 \\
             &  2  &  0.281$\pm$0.018  &  0.101$\pm$0.009  &  $-0.77\pm0.08$  &  1.55$\pm$0.12  &     $7\times10^{8}$  &                  &                                  &      &    &          \\
 J1358+4737  &  1  &  0.454$\pm$0.003  &  0.174$\pm$0.004  &  $-0.73\pm0.02$  &  1.19$\pm$0.02  &     $2\times10^{9}$  &   6.08$\pm$0.02  &  $+0.74\pm0.02$ & a & 0.230 & GPS $\nu_m$=2.41 \\
             &  2  &  0.159$\pm$0.003  &  0.059$\pm$0.002  &  $-0.75\pm0.03$  &  0.83$\pm$0.02  &     $1\times10^{9}$  &                  &                                  &      &    &          \\
 J1443+6332  &  1  &  0.191$\pm$0.007  &  0.322$\pm$0.002  &  $+0.40\pm0.03$  &  0.26$\pm$0.00  &     $8\times10^{10}$ &   8.49$\pm$0.04  &  $+0.42\pm0.04$ & q & 1.380 & GPS $\nu_m$=1.53 \\
             &  2  &  0.535$\pm$0.009  &  0.067$\pm$0.004  &  $-1.59\pm0.05$  &  1.46$\pm$0.08  &     $6\times10^{8}$  &                  &                                  &      &    &          \\
 J1451+1343  &  1  &  0.173$\pm$0.007  &  0.086$\pm$0.002  &  $-0.53\pm0.03$  &  1.07$\pm$0.02  &     $1\times10^{9}$  &  27.04$\pm$0.02  &  $-0.13\pm0.10$ & . & ..... & STEEP \\
             &  2  &  0.192$\pm$0.017  &  0.082$\pm$0.002  &  $-0.64\pm0.07$  &  1.45$\pm$0.03  &     $6\times10^{8}$  &                  &                                  &      &    &          \\
 J1503+0917  &  1  &  0.336$\pm$0.010  &  0.189$\pm$0.006  &  $-0.44\pm0.03$  &  0.98$\pm$0.03  &     $3\times10^{9}$  &  13.20$\pm$0.13  &  $+0.69\pm0.20$ & . & ..... & STEEP \\
             &  2  &  0.064$\pm$0.006  &  0.011$\pm$0.003  &  $-1.32\pm0.21$  &  1.16$\pm$0.25  &     $1\times10^{8}$  &                  &                                  &      &    &          \\
 J1558-1409  &  1  &  0.309$\pm$0.005  &  0.160$\pm$0.003  &  $-0.51\pm0.02$  &  0.94$\pm$0.02  &     $3\times10^{9}$  &   6.89$\pm$0.01  &  $-0.13\pm0.02$ & a & 0.097 & GPS $\nu_m$=2.39 \\
             &  2  &  0.228$\pm$0.005  &  0.140$\pm$0.003  &  $-0.37\pm0.02$  &  1.02$\pm$0.02  &     $2\times10^{9}$  &                  &                                  &      &    &          \\
 J1602+2418  &  1  &  0.068$\pm$0.002  &  0.094$\pm$0.001  &  $+0.25\pm0.02$  &  0.66$\pm$0.01  &     $4\times10^{9}$  &   7.34$\pm$0.01  &  $+0.05\pm0.02$ & . & ..... & FLAT \\
             &  2  &  0.174$\pm$0.002  &  0.070$\pm$0.002  &  $-0.69\pm0.03$  &  1.11$\pm$0.03  &     $9\times10^{8}$  &                  &                                  &      &    &          \\
 J1734+0926  &  1  &  0.904$\pm$0.009  &  0.319$\pm$0.005  &  $-0.80\pm0.01$  &  1.15$\pm$0.01  &     $4\times10^{9}$  &  13.62$\pm$0.02  &  $+0.91\pm0.02$ & . & ..... & COINS, GPS $\nu_m$=2.18 \\
             &  2  &  0.636$\pm$0.005  &  0.215$\pm$0.006  &  $-0.83\pm0.02$  &  1.41$\pm$0.03  &     $2\times10^{9}$  &                  &                                  &      &    &          \\
 J1737+0621  &  1  &  0.821$\pm$0.011  &  0.724$\pm$0.003  &  $-0.10\pm0.01$  &  0.51$\pm$0.00  &     $5\times10^{10}$ &   3.69$\pm$0.01  &  $+0.22\pm0.02$ & q & 1.207 & COINS-REJ, FLAT \\
             &  2  &  0.519$\pm$0.006  &  0.183$\pm$0.004  &  $-0.80\pm0.02$  &  1.52$\pm$0.03  &     $1\times10^{9}$  &                  &                                  &      &    &          \\
 J1742-1517  &  1  &  0.056$\pm$0.013  &  0.109$\pm$0.001  &  $+0.51\pm0.17$  &  0.20$\pm$0.00  &   $> 5\times10^{10}$ &  38.68$\pm$0.06  &  $+0.70\pm0.34$ & . & ..... & FLAT\footnotemark[13]$^,$\footnotemark[16] \\
             &  2  &  0.144$\pm$0.018  &  0.059$\pm$0.003  &  $-0.68\pm0.10$  &  2.07$\pm$0.11  &     $2\times10^{8}$  &                  &                                  &      &    &          \\
 J1921+4333  &  1  &  0.112$\pm$0.002  &  0.114$\pm$0.005  &  $+0.01\pm0.04$  &  1.18$\pm$0.05  &     $1\times10^{9}$  &   3.25$\pm$0.04  &  $-0.20\pm0.04$ & . & ..... & FLAT \\
             &  2  &  0.150$\pm$0.003  &  0.080$\pm$0.005  &  $-0.47\pm0.05$  &  1.23$\pm$0.07  &     $9\times10^{8}$  &                  &                                  &      &    &          \\
 J1935-1602  &  1  &  0.061$\pm$0.003  &  0.140$\pm$0.001  &  $+0.63\pm0.04$  &  0.35$\pm$0.00  &     $2\times10^{10}$ &   9.96$\pm$0.01  &  $+0.28\pm0.04$ & q & 1.460 & FLAT \\
             &  2  &  0.175$\pm$0.003  &  0.138$\pm$0.002  &  $-0.18\pm0.02$  &  1.84$\pm$0.03  &     $7\times10^{8}$  &                  &                                  &      &    &          \\
 J1935+8130  &  1  &  0.355$\pm$0.007  &  0.159$\pm$0.010  &  $-0.62\pm0.05$  &  1.21$\pm$0.07  &     $2\times10^{9}$  &   9.74$\pm$0.03  &  $+0.51\pm0.04$ & . & ..... & GPS $\nu_m$=3.41 \\
             &  2  &  0.180$\pm$0.005  &  0.088$\pm$0.003  &  $-0.55\pm0.03$  &  0.69$\pm$0.02  &     $3\times10^{9}$  &                  &                                  &      &    &          \\
 J1944+5448  &  1  &  0.932$\pm$0.059  &  0.338$\pm$0.016  &  $-0.78\pm0.06$  &  1.39$\pm$0.06  &     $3\times10^{9}$  &  40.45$\pm$0.04  &  $+1.37\pm0.12$ & . & ..... & COINS, GPS $\nu_m$=1.06 \\
             &  2  &  0.456$\pm$0.027  &  0.099$\pm$0.007  &  $-1.17\pm0.07$  &  1.02$\pm$0.06  &     $2\times10^{9}$  &                  &                                  &      &    &          \\
 J1950+0807  &  1  &  0.888$\pm$0.014  &  0.485$\pm$0.005  &  $-0.46\pm0.01$  &  0.64$\pm$0.01  &     $2\times10^{10}$ &  21.83$\pm$0.04  &  $+0.73\pm0.05$ & . & ..... & GPS $\nu_m$=2.43 \\
             &  2  &  0.640$\pm$0.028  &  0.244$\pm$0.016  &  $-0.74\pm0.06$  &  1.25$\pm$0.07  &     $3\times10^{9}$  &                  &                                  &      &    &          \\
 J2022+6136  &  1  &  1.238$\pm$0.051  &  2.208$\pm$0.034  &  $+0.44\pm0.03$  &  0.76$\pm$0.01  &     $7\times10^{10}$ &   6.91$\pm$0.01  &  $+0.37\pm0.07$ & a & 0.227 & COINS, GPS $\nu_m$=5.91 \\
             &  2  &  1.480$\pm$0.064  &  0.879$\pm$0.013  &  $-0.40\pm0.04$  &  0.78$\pm$0.01  &     $3\times10^{10}$ &                  &                                  &      &    &          \\
 J2120+6642  &  1  &  0.164$\pm$0.004  &  0.100$\pm$0.001  &  $-0.38\pm0.02$  &  1.03$\pm$0.01  &     $2\times10^{9}$  &   9.87$\pm$0.01  &  $+0.31\pm0.02$ & . & ..... & FLAT \\
             &  2  &  0.117$\pm$0.004  &  0.054$\pm$0.001  &  $-0.58\pm0.03$  &  0.92$\pm$0.02  &     $1\times10^{9}$  &                  &                                  &      &    &          \\
 J2123+1007  &  1  &  0.076$\pm$0.008  &  0.117$\pm$0.003  &  $+0.33\pm0.08$  &  0.49$\pm$0.01  &     $8\times10^{9}$  &   9.49$\pm$0.34  &  $+0.29\pm0.35$ & q & 0.932 & STEEP \\
             &  2  &  0.393$\pm$0.014  &  0.118$\pm$0.026  &  $-0.91\pm0.17$  &  3.17$\pm$0.69  &     $2\times10^{8}$  &                  &                                  &      &    &          \\
 J2131+8430  &  1  &  0.333$\pm$0.007  &  0.137$\pm$0.004  &  $-0.67\pm0.03$  &  1.02$\pm$0.03  &     $2\times10^{9}$  &  13.92$\pm$0.09  &  $-0.32\pm0.09$ & . & ..... & STEEP\footnotemark[13]$^,$\footnotemark[16] \\
             &  2  &  0.223$\pm$0.004  &  0.031$\pm$0.004  &  $-1.48\pm0.11$  &  1.46$\pm$0.18  &     $2\times10^{8}$  &                  &                                  &      &    &          \\
 J2137+3455  &  1  &  0.210$\pm$0.006  &  0.089$\pm$0.005  &  $-0.65\pm0.05$  &  1.05$\pm$0.05  &     $1\times10^{9}$  &   6.94$\pm$0.17  &  $+0.23\pm0.17$ & . & ..... & FLAT \\
             &  2  &  0.127$\pm$0.005  &  0.030$\pm$0.005  &  $-1.10\pm0.13$  &  2.11$\pm$0.33  &     $1\times10^{8}$  &                  &                                  &      &    &          \\
 J2203+1007  &  1  &  0.150$\pm$0.009  &  0.157$\pm$0.006  &  $+0.03\pm0.06$  &  1.36$\pm$0.05  &     $1\times10^{9}$  &   9.92$\pm$0.04  &  $+0.42\pm0.15$ & . & ..... & COINS, GPS $\nu_m$=4.58 \\
             &  2  &  0.100$\pm$0.011  &  0.057$\pm$0.004  &  $-0.43\pm0.09$  &  1.32$\pm$0.07  &     $6\times10^{8}$  &                  &                                  &      &    &          \\
 J2253+0236  &  1  &  0.193$\pm$0.002  &  0.094$\pm$0.002  &  $-0.55\pm0.02$  &  1.02$\pm$0.02  &     $1\times10^{9}$  &  33.24$\pm$0.01  &  $+1.11\pm0.02$ & . & ..... & FLAT \\
             &  2  &  0.047$\pm$0.002  &  0.061$\pm$0.003  &  $+0.19\pm0.04$  &  0.52$\pm$0.01  &   $> 4\times10^{9}$  &                  &                                  &      &    &          \\
 J2254+0054  &  1  &  0.205$\pm$0.009  &  0.221$\pm$0.005  &  $+0.06\pm0.04$  &  1.03$\pm$0.02  &     $4\times10^{9}$  &  13.47$\pm$0.13  &  $+0.55\pm0.15$ & b & ..... & FLAT \\
             &  2  &  0.180$\pm$0.009  &  0.082$\pm$0.009  &  $-0.60\pm0.09$  &  2.51$\pm$0.26  &     $2\times10^{8}$  &                  &                                  &      &    &          \\
 J2347-1856  &  1  &  0.497$\pm$0.010  &  0.195$\pm$0.008  &  $-0.72\pm0.04$  &  1.90$\pm$0.07  &     $1\times10^{9}$  &  33.17$\pm$0.08  &  $-0.13\pm0.08$ & . & ..... & GPS\footnotemark[17] $\nu_m$=2.07 \\
             &  2  &  0.328$\pm$0.006  &  0.140$\pm$0.011  &  $-0.66\pm0.06$  &  1.84$\pm$0.13  &     $7\times10^{8}$  &                  &                                  &      &    &          \\
 J2355-2125  &  1  &  0.303$\pm$0.004  &  0.231$\pm$0.002  &  $-0.21\pm0.01$  &  0.84$\pm$0.01  &     $5\times10^{9}$  &  20.72$\pm$0.03  &  $-0.13\pm0.04$ & . & ..... & GPS\footnotemark[13]$^,$\footnotemark[15] \\
             &  2  &  0.216$\pm$0.005  &  0.071$\pm$0.003  &  $-0.84\pm0.04$  &  1.58$\pm$0.06  &     $5\times10^{8}$  &                  &                                  &      &    &          \\

\hline
\end{longtable}
\footnotetext[1]{Source J2000 epoch name. Precise VLBI positions of these sources may be obtained from
\url{http://astrogeo.org/rfc/} }
\footnotetext[2]{Component number. Each source was modeled with two circular
 Gaussian components. The first component is the one that is brighter at
 8\,GHz.}
\footnotetext[3]{F$_X$ (F$_S$) is the component flux in Jansky measured at 8\,GHz (2\,GHz).}
\footnotetext[4]{$\alpha$ is the component spectral index between 2 and
 8\,GHz.}
\footnotetext[5]{R$_X$ is the component size (FWHM of the Gaussian model) in
 milliarcseconds measured at 8\,GHz.}
\footnotetext[6]{T$_{\mathrm{b}~X}$ is the component brightness temperature in Kelvin
 measured at 8\,GHz.}
\footnotetext[7]{$D_X$ is the distance in milliarcseconds between CSO
 components measured at 8\,GHz.}
\footnotetext[8]{$D_X-D_S$ is the distance difference (milliarcseconds)
 between CSO components measured at 8 and 2\,GHz.}
\footnotetext[9]{opt. --- optical classification according to
 \cite{2006A&A...455..773V}: 'q' stands for quasar, 'a' is an active
 galaxy, 'b' is a BL~Lacertae type object.}
\footnotetext[10]{$z$ --- redshift from \cite{2006A&A...455..773V}.}
\footnotetext[11]{Comments: COINS --- the source is part of the COINS 
\citep{2000ApJ...534...90P};
COINS-REJ --- the source was considered as a candidate for the COINS sample
but was rejected \citep{2000ApJ...534...90P};
GPS --- GPS source, part of the RATAN-600 GPS sample
\cite{2009AN....330..199S} unless indicated otherwise, for these sources
approximate spectral peak
frequency (in GHz) is indicated; GPS-REJ the source was reported in the
literature as
a GPS candidate, but was not confirmed by RATAN-600 observations; STEEP ---
steep spectrum source, FLAT --- flat spectrum source.}
\footnotetext[12]{To distinguish between resolved and unresolved components we use the
criterium proposed by \citealt*{2005AJ....130.2473K}:
$\psi > \text{HPBW} \sqrt{\frac{4
\ln{2}}{\pi}\ln{\frac{\text{SNR}}{\text{SNR}-1}}}$
where $\psi$ is the component best-fit angular size, $\text{HPBW}$ is the
Half
Power Beam Width, $\text{SNR}$ is the signal to noise ratio of the
component.}
\footnotetext[13]{No RATAN-600 observations available for this source.}
\footnotetext[14]{Reported as GPS by \cite{1999A&AS..135..273M}.}
\footnotetext[15]{Reported as GPS by \cite{Volmer}.}
\footnotetext[16]{Spectral classification based on nonsimultaneous literature data
collected by the CATS database \citep{CATS05}.}
\footnotetext[16]{Listed as the confirmed CSO by \cite{2003ApJ...597..157T}.}
\renewcommand{\thefootnote}{\arabic{footnote}}
\end{landscape}
}


\Online

\clearpage
\setcounter{figure}{0} 
\begin{figure*}
 \centering

 \includegraphics[width=0.44\textwidth,angle=0,trim=0 3cm 1cm 5cm,clip]{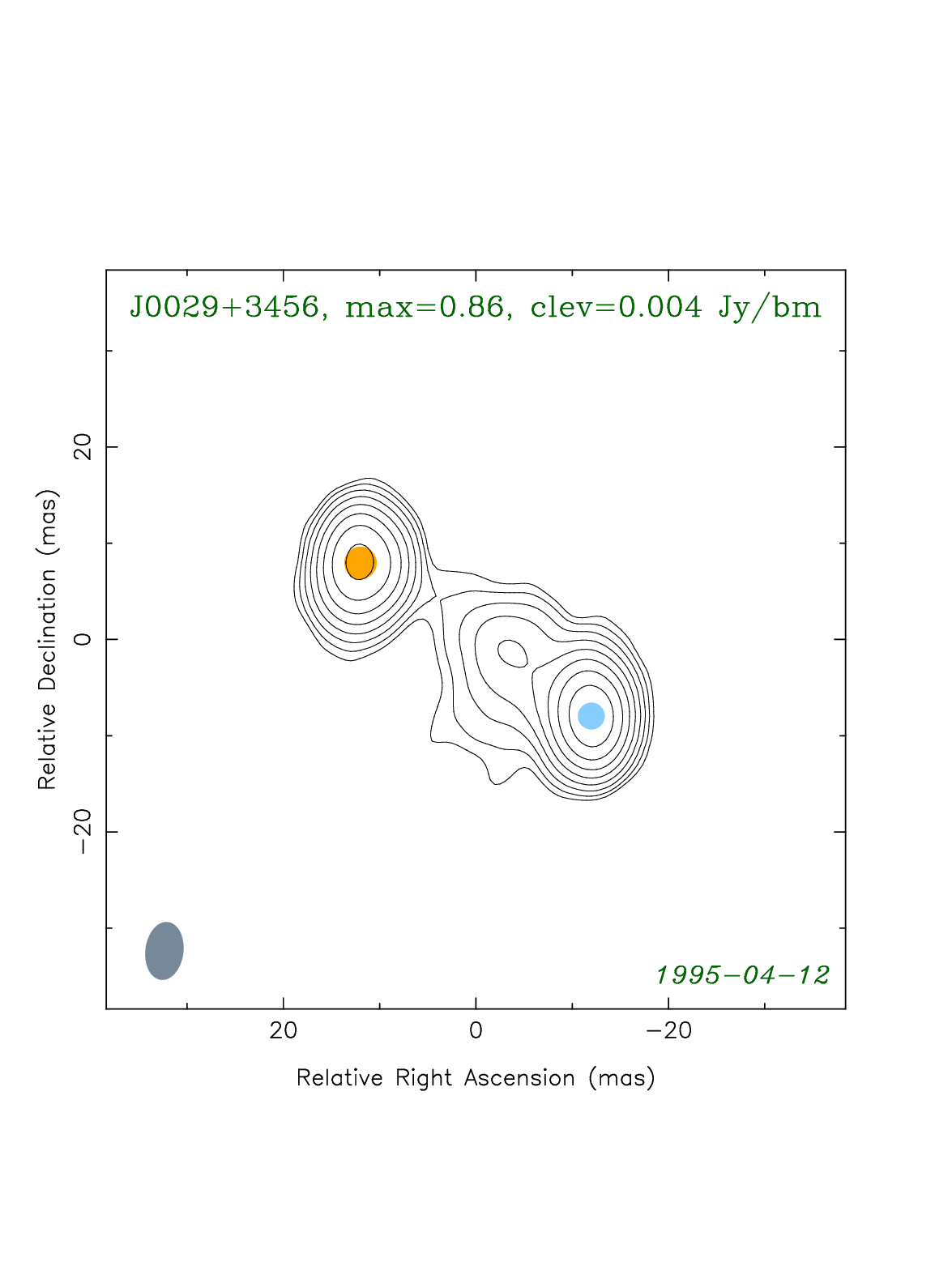}
 \includegraphics[width=0.44\textwidth,angle=0,trim=0 3cm 1cm 5cm,clip]{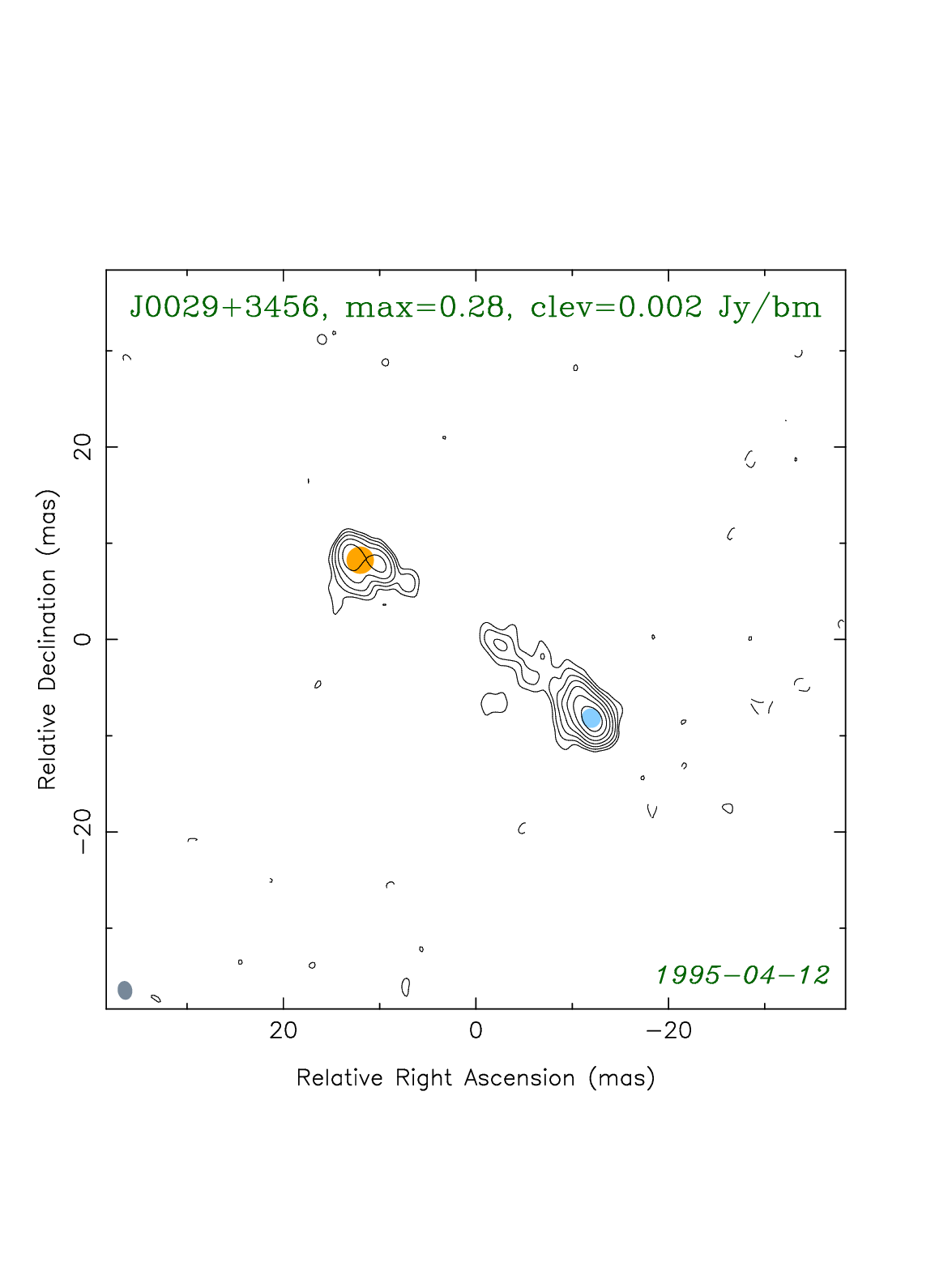}
\\
 \includegraphics[width=0.44\textwidth,angle=0,trim=0 3cm 1cm 5cm,clip]{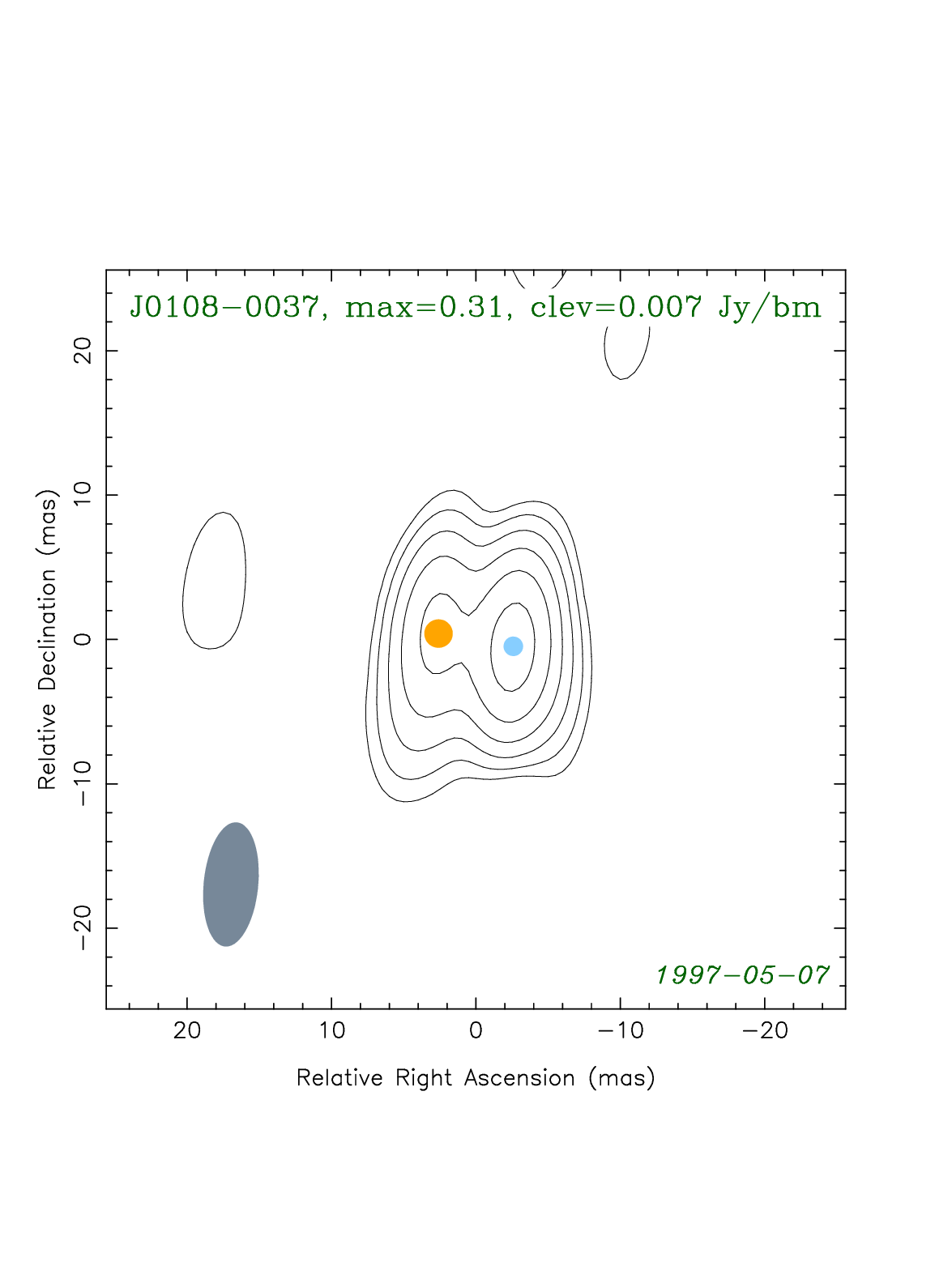}
 \includegraphics[width=0.44\textwidth,angle=0,trim=0 3cm 1cm 5cm,clip]{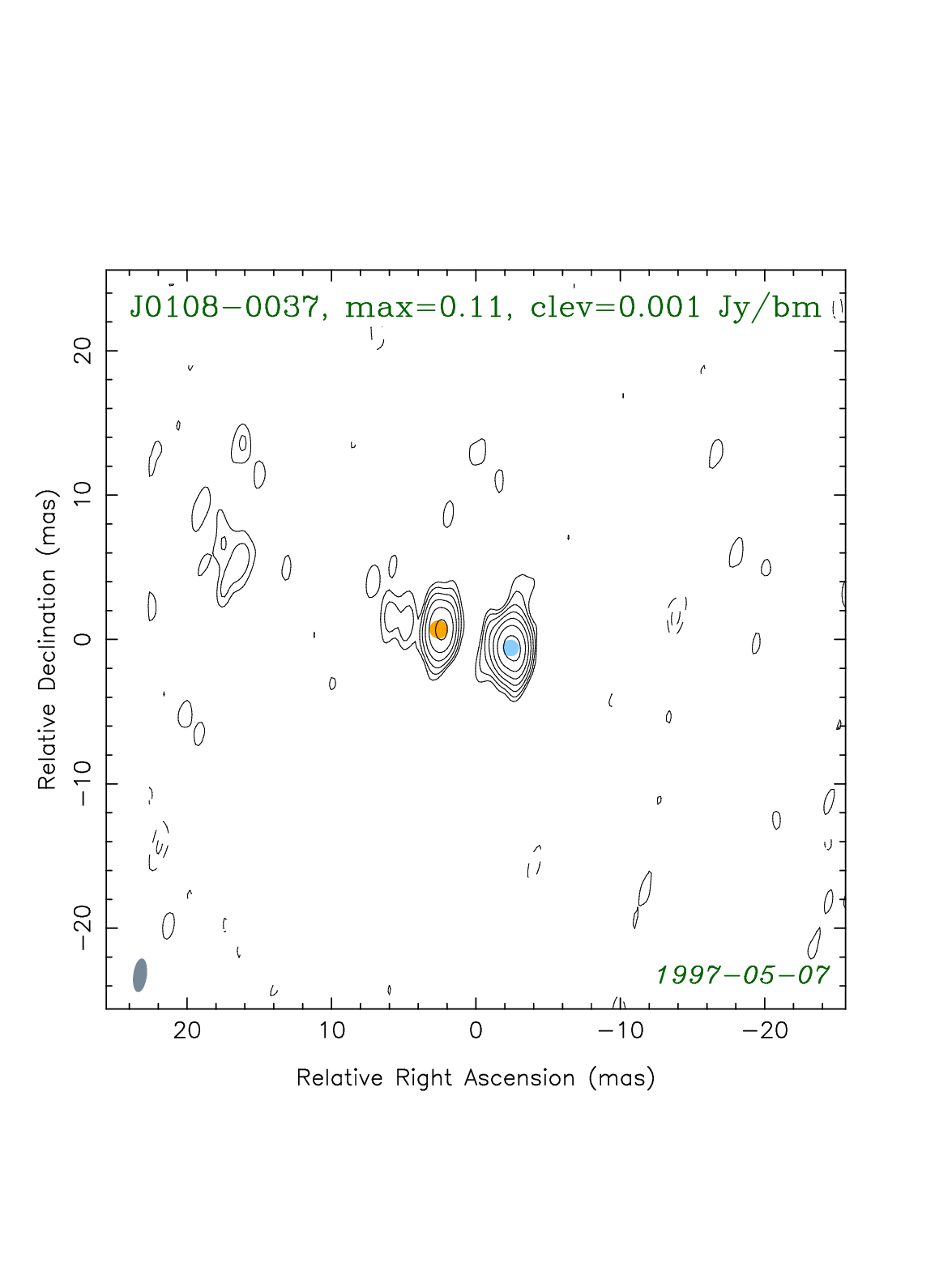}
\\
 \includegraphics[width=0.44\textwidth,angle=0,trim=0 3cm 1cm 5cm,clip]{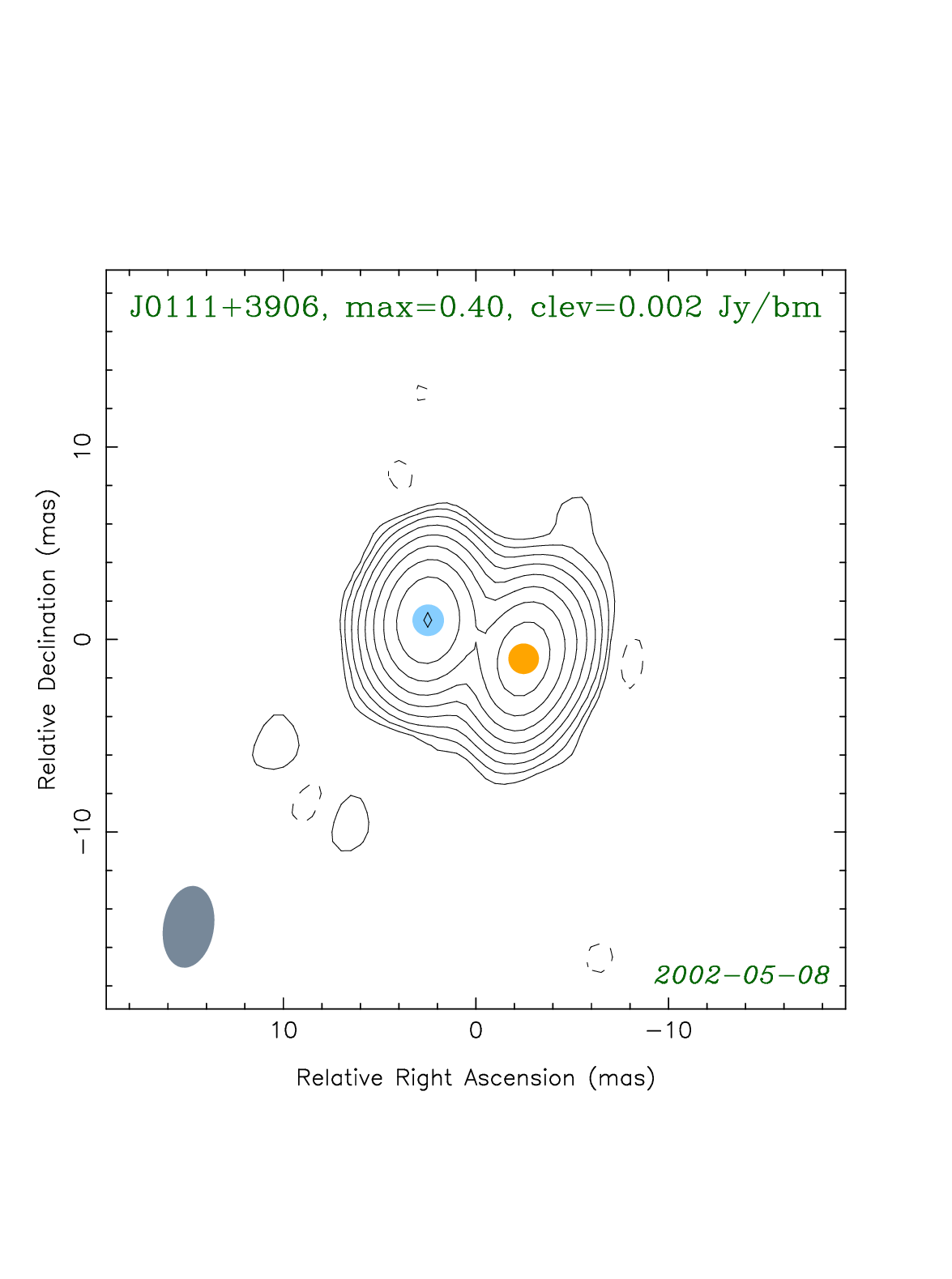}
 \includegraphics[width=0.44\textwidth,angle=0,trim=0 3cm 1cm 5cm,clip]{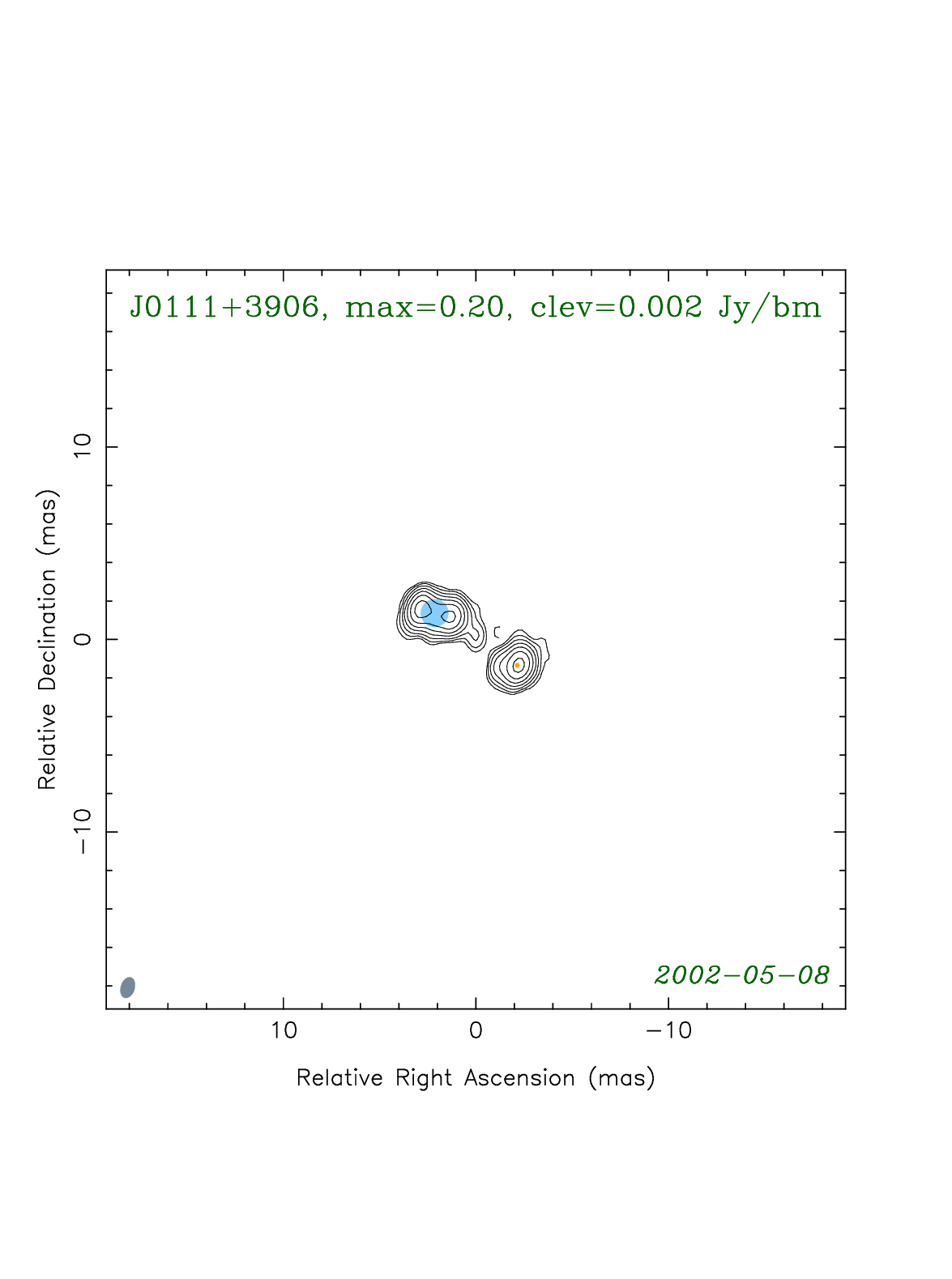}

\caption{
Dual-frequency simultaneous
2 and 8 GHz CLEAN images. The lowest contour
value `clev' is chosen at four times the rms level, 
and the peak brightness is given by `max' (Jy/beam). The contour levels
increase by factors of two. The dashed contours indicate negative flux.
The beam is shown in the bottom left corner of the images at the half-power level.
An epoch of observation is shown in the bottom right corner. 
Blue and orange spots indicate Gaussian model components 1 and 2, respectively.
The brightness distribution is shown at the full resolution of naturally weighted
data, while the model components are fitted in a restricted $uv$-range that is common
to both bands, see \S\,\ref{s:pos} for details.
}
\label{fig:images}
\end{figure*}

\clearpage
\addtocounter{figure}{-1}
\begin{figure*}
 \centering

 \includegraphics[width=0.44\textwidth,angle=0,trim=0 3cm 1cm 5cm,clip]{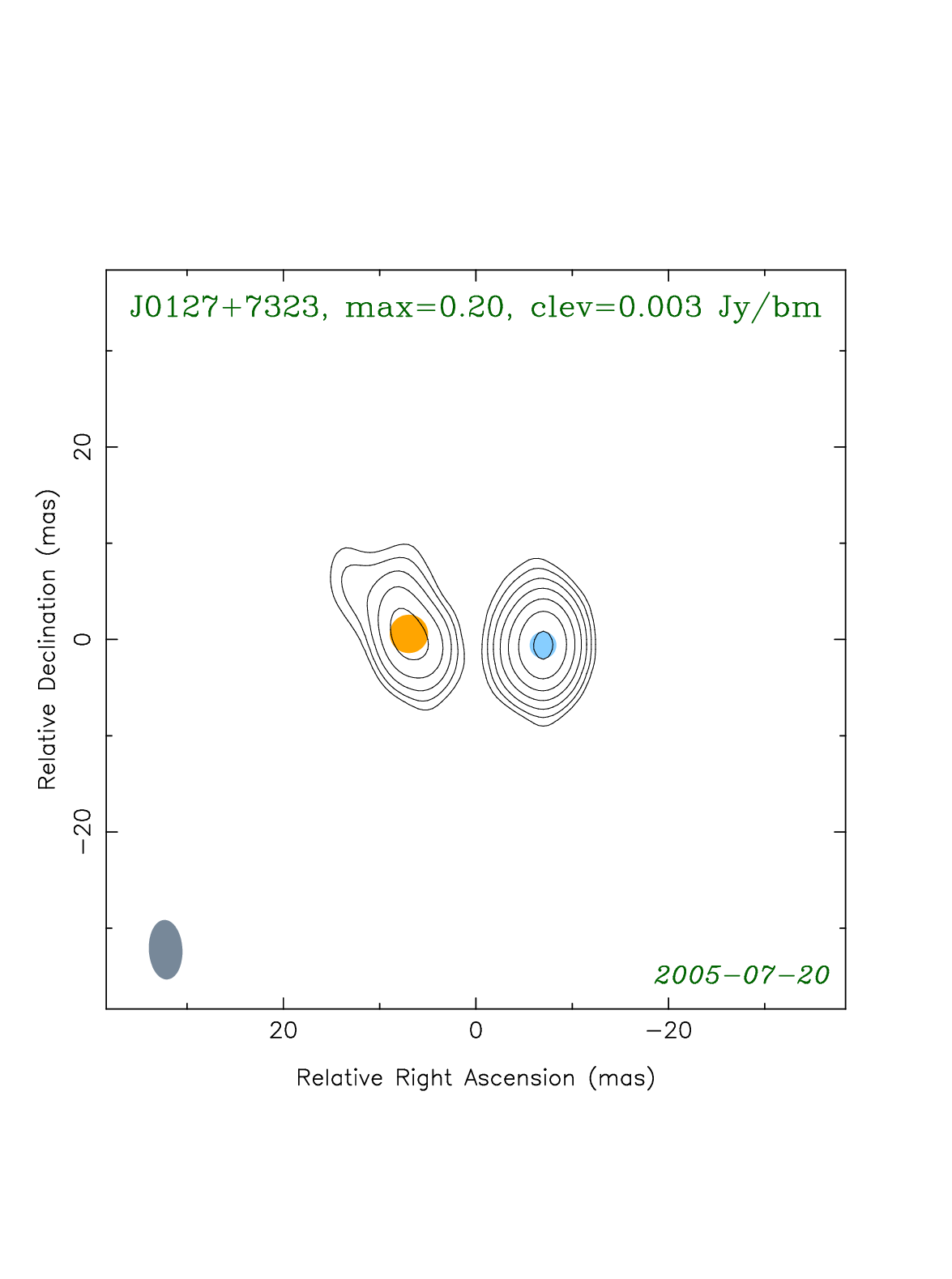}
 \includegraphics[width=0.44\textwidth,angle=0,trim=0 3cm 1cm 5cm,clip]{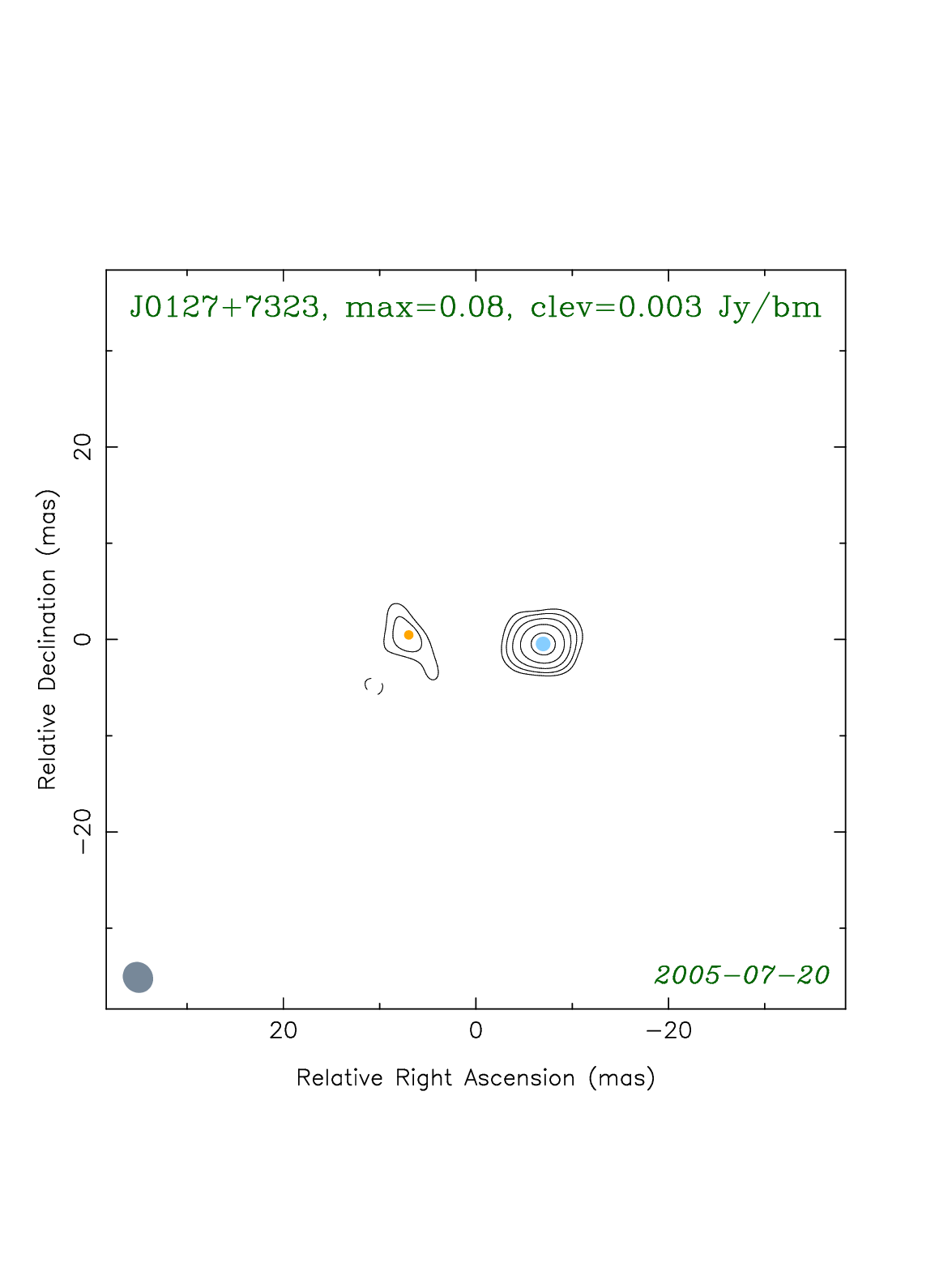}

 \includegraphics[width=0.44\textwidth,angle=0,trim=0 3cm 1cm 5cm,clip]{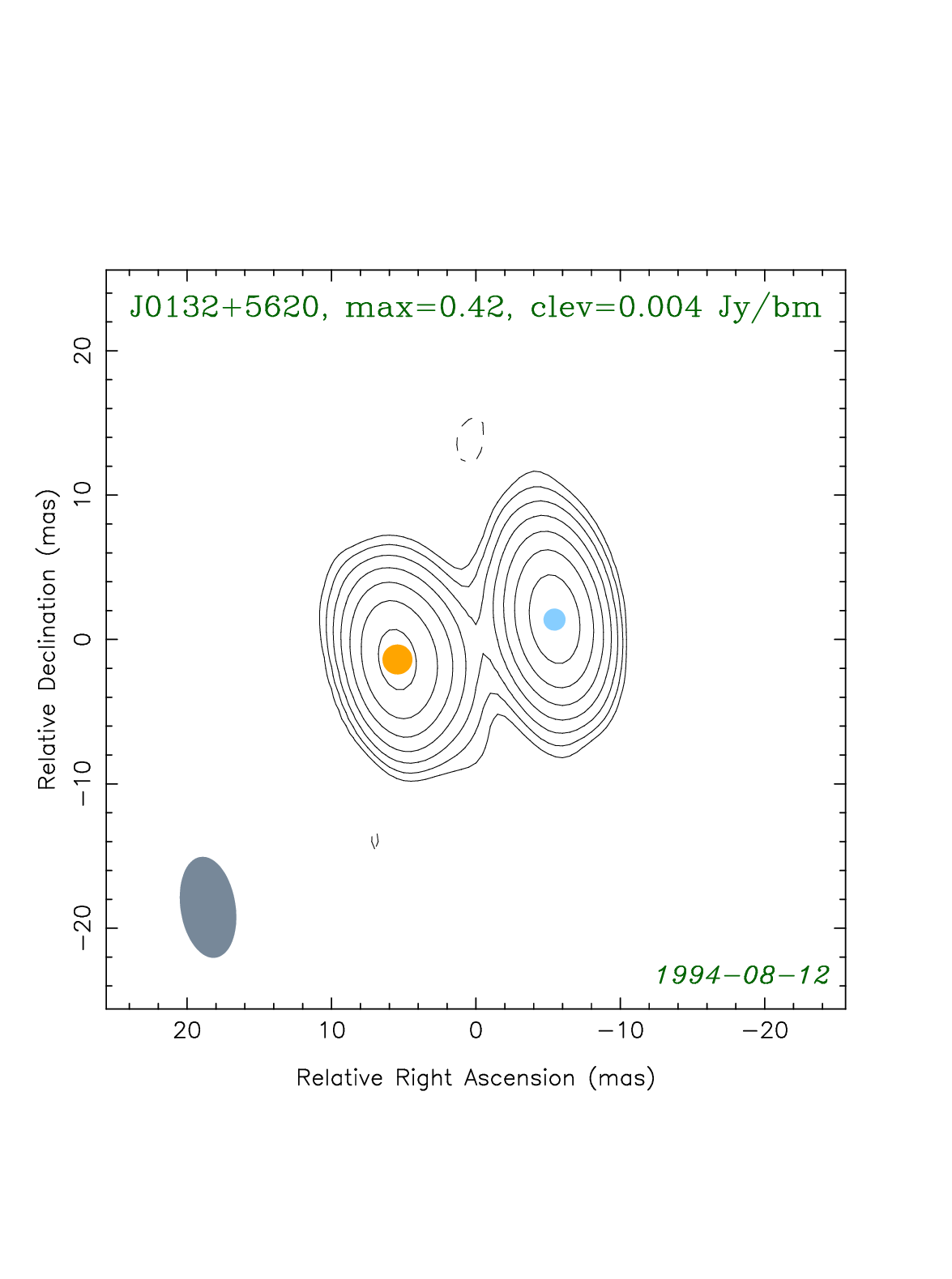}
 \includegraphics[width=0.44\textwidth,angle=0,trim=0 3cm 1cm 5cm,clip]{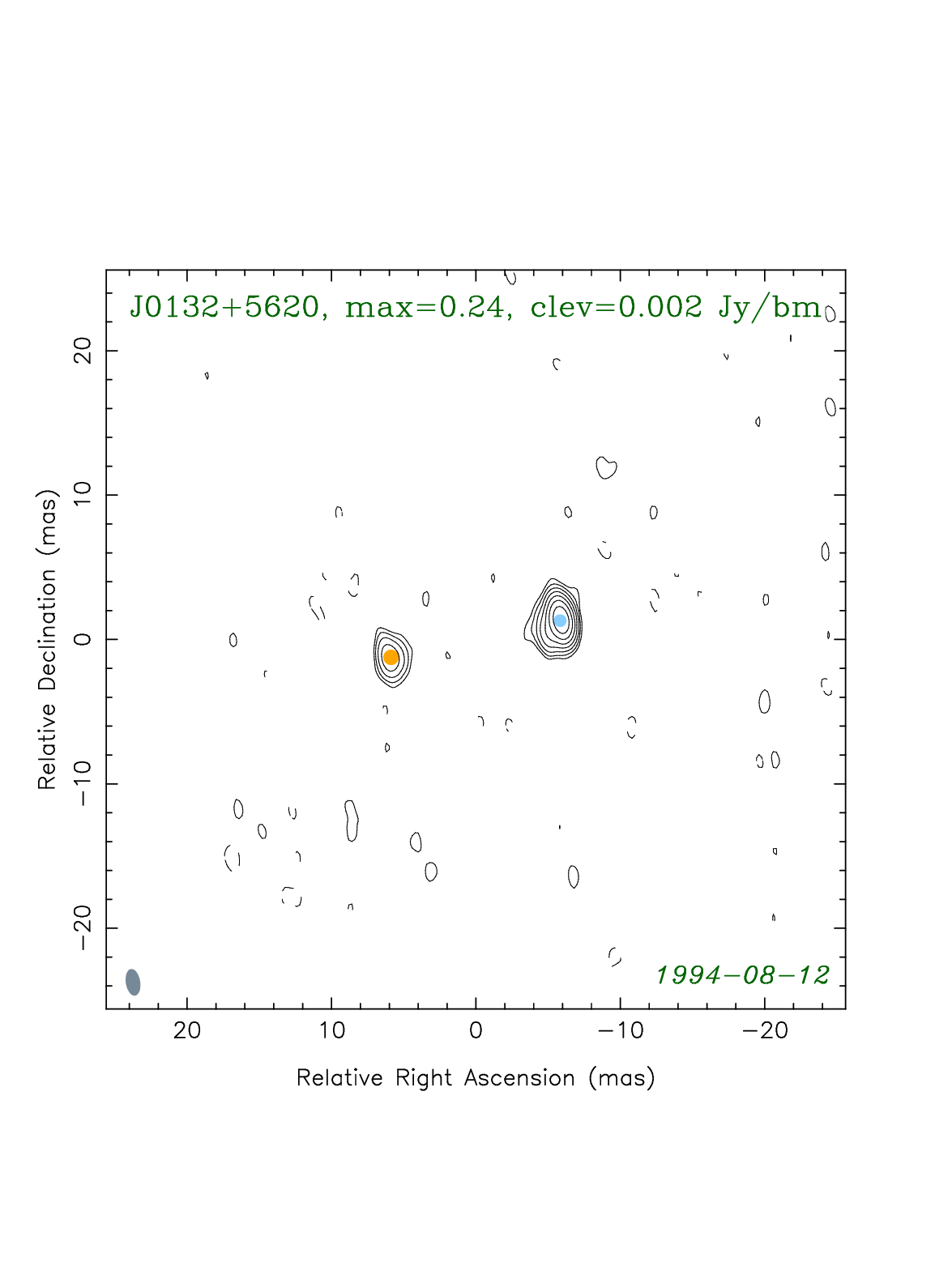}

 \includegraphics[width=0.44\textwidth,angle=0,trim=0 3cm 1cm 5cm,clip]{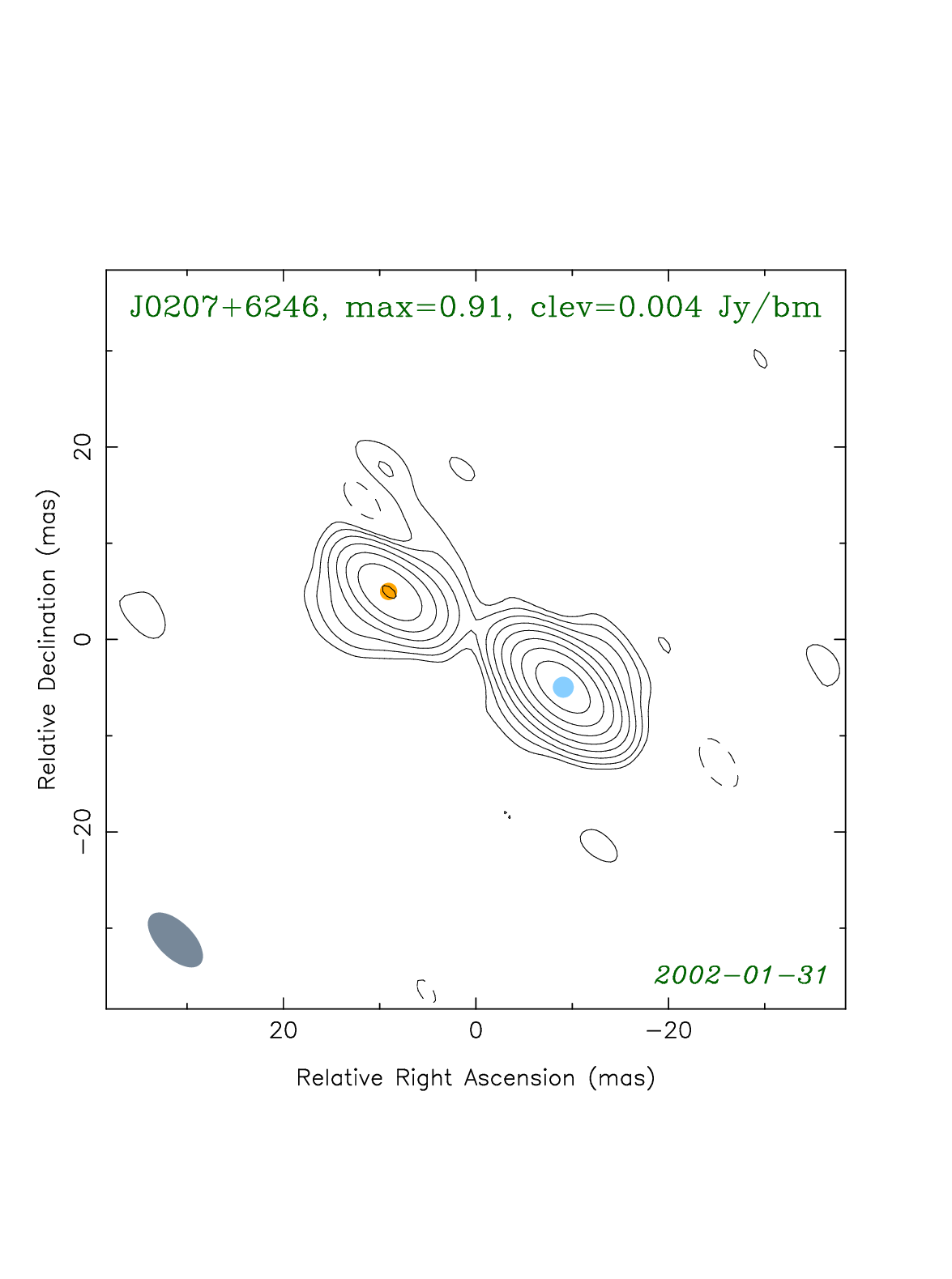}
 \includegraphics[width=0.44\textwidth,angle=0,trim=0 3cm 1cm 5cm,clip]{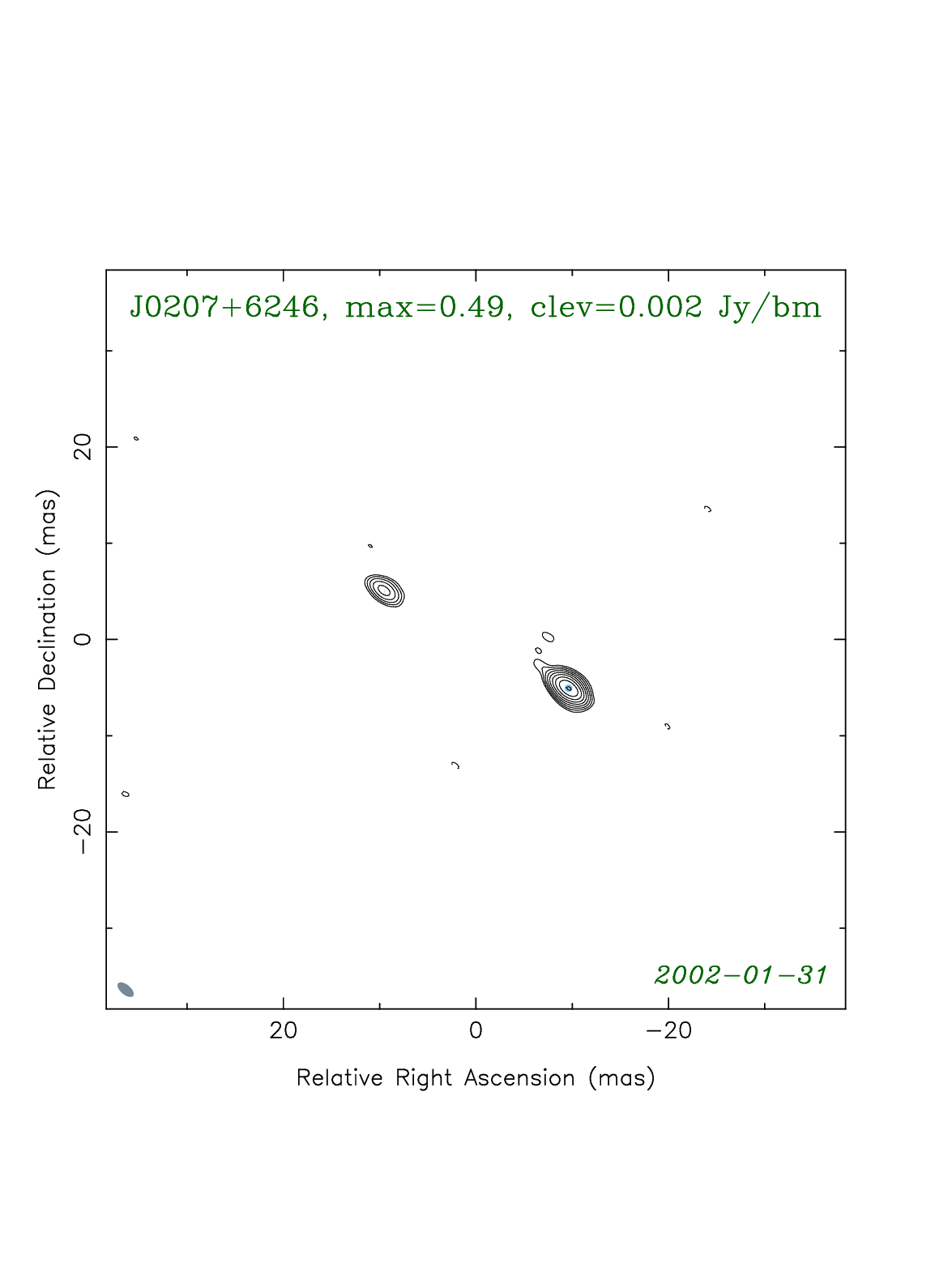}

 \caption{continued}
\end{figure*}

\clearpage
\addtocounter{figure}{-1}
\begin{figure*}
 \centering

 \includegraphics[width=0.44\textwidth,angle=0,trim=0 3cm 1cm 5cm,clip]{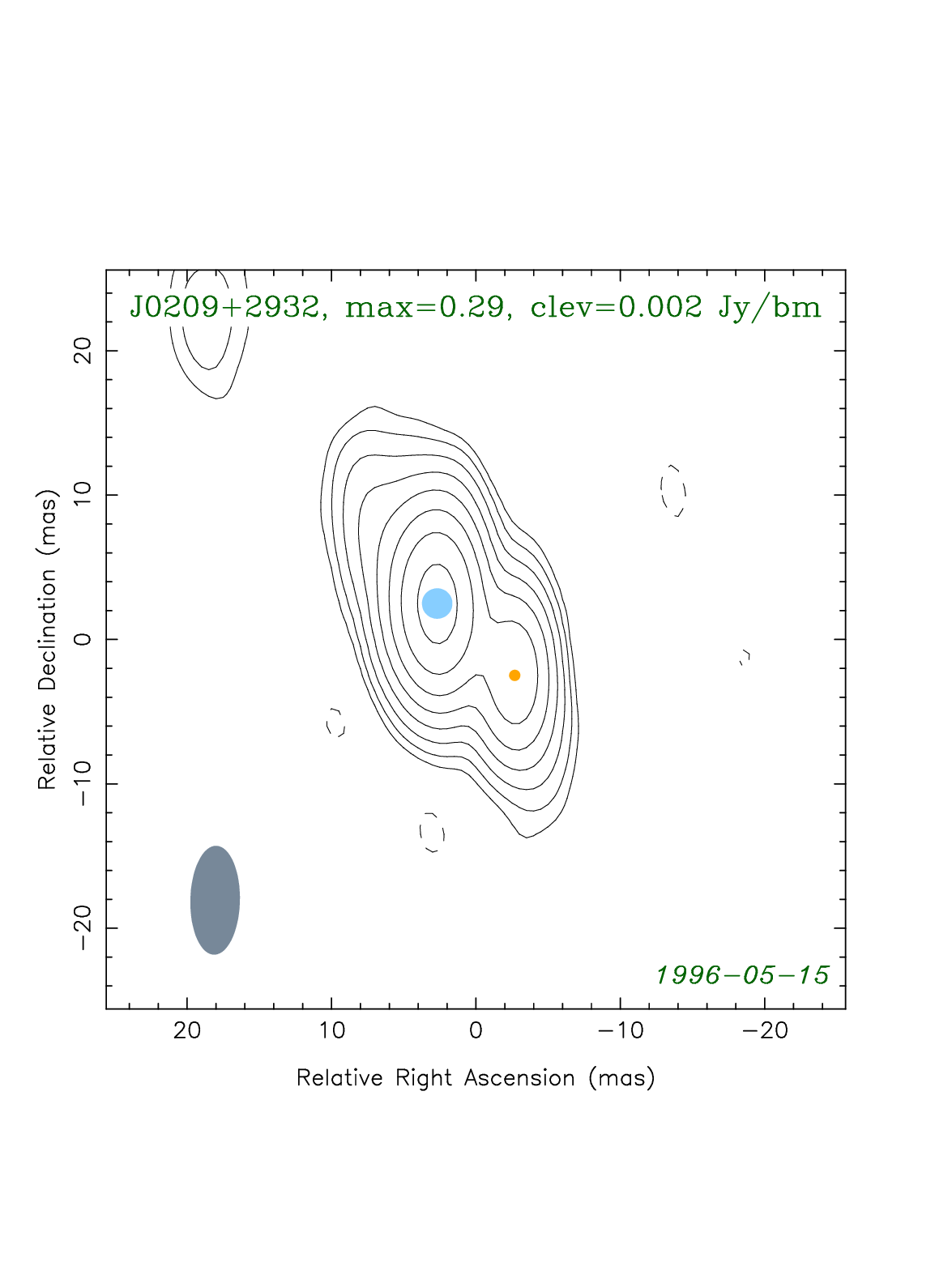}
 \includegraphics[width=0.44\textwidth,angle=0,trim=0 3cm 1cm 5cm,clip]{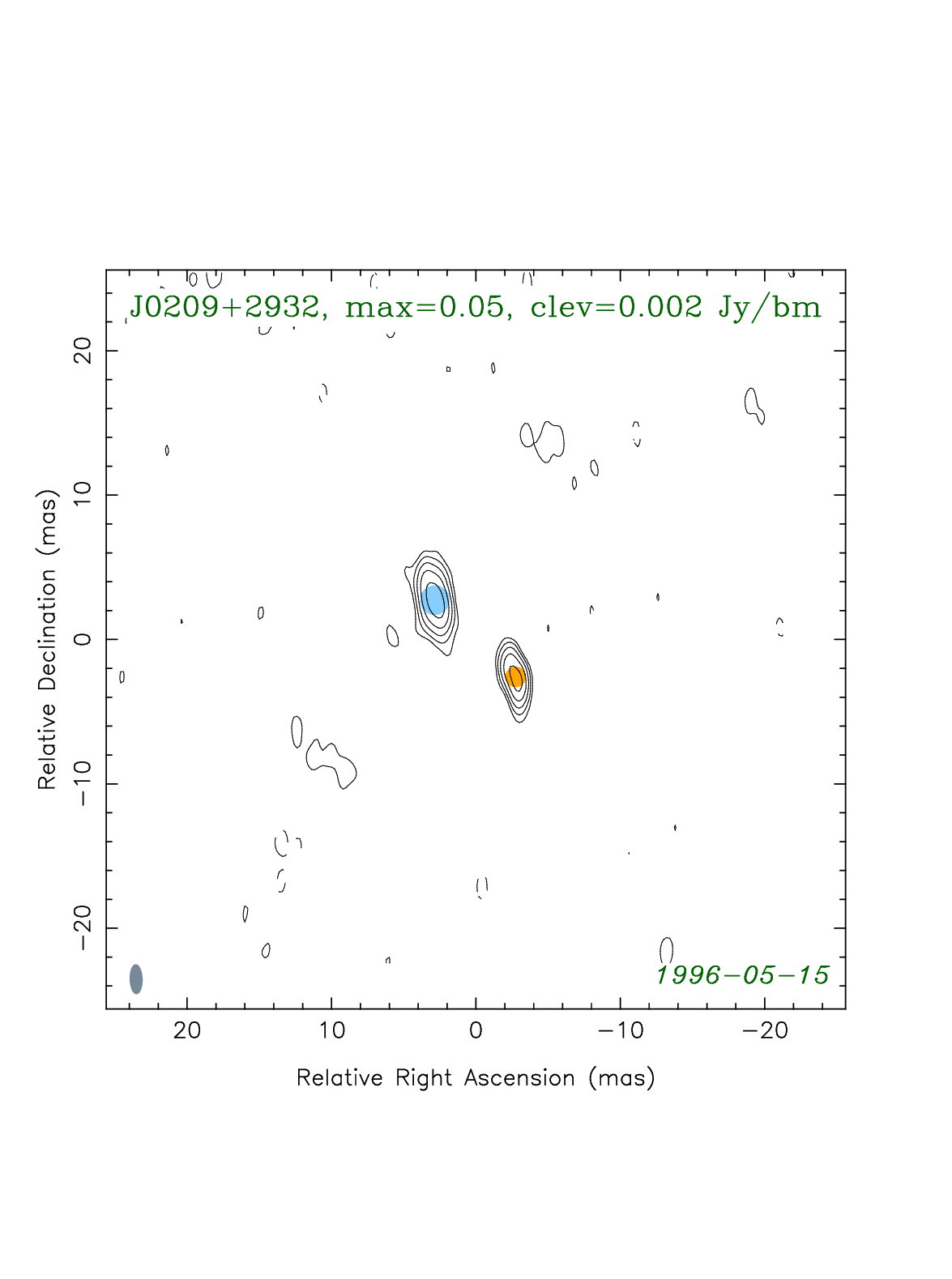}

 \includegraphics[width=0.44\textwidth,angle=0,trim=0 3cm 1cm 5cm,clip]{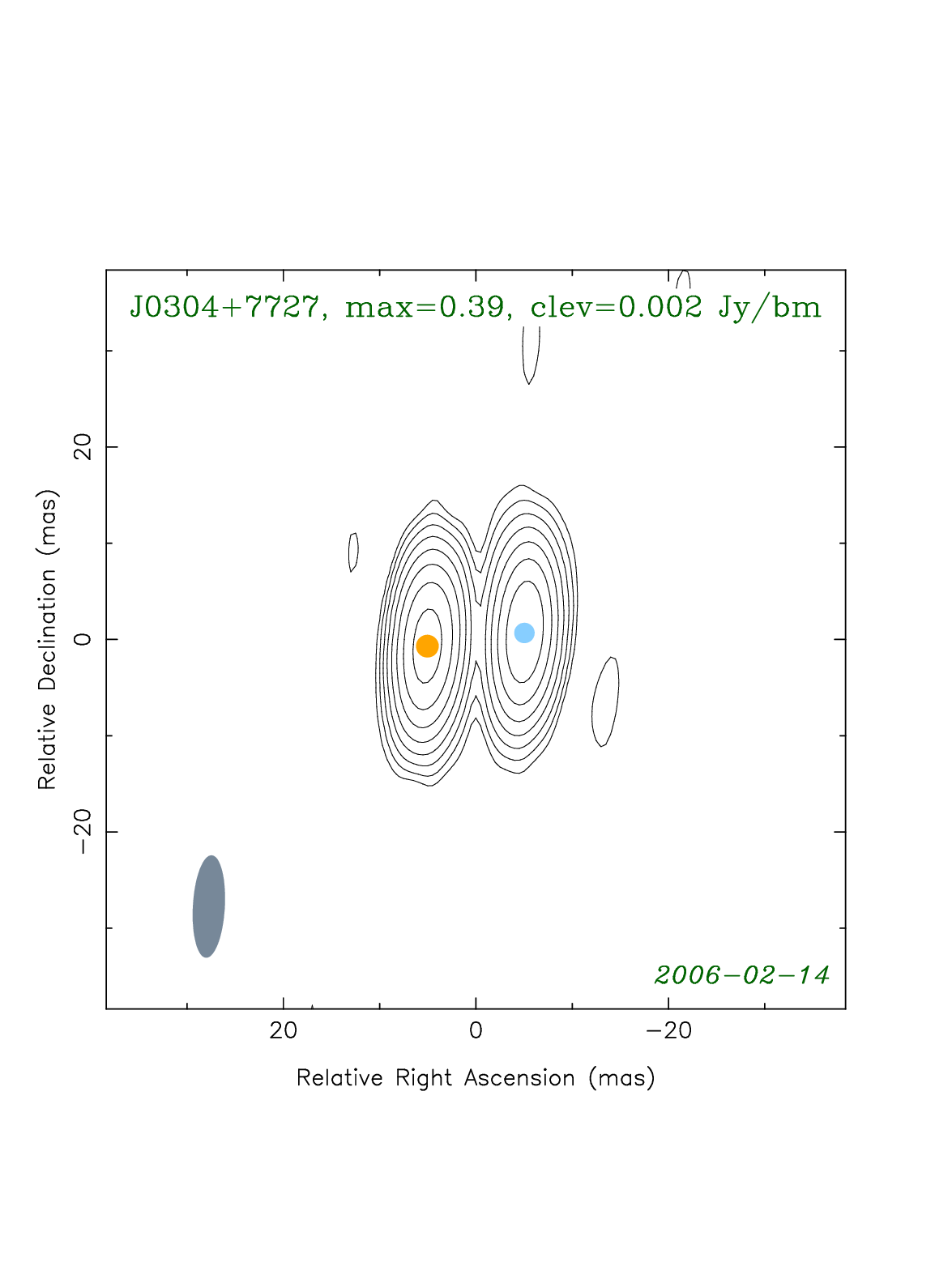}
 \includegraphics[width=0.44\textwidth,angle=0,trim=0 3cm 1cm 5cm,clip]{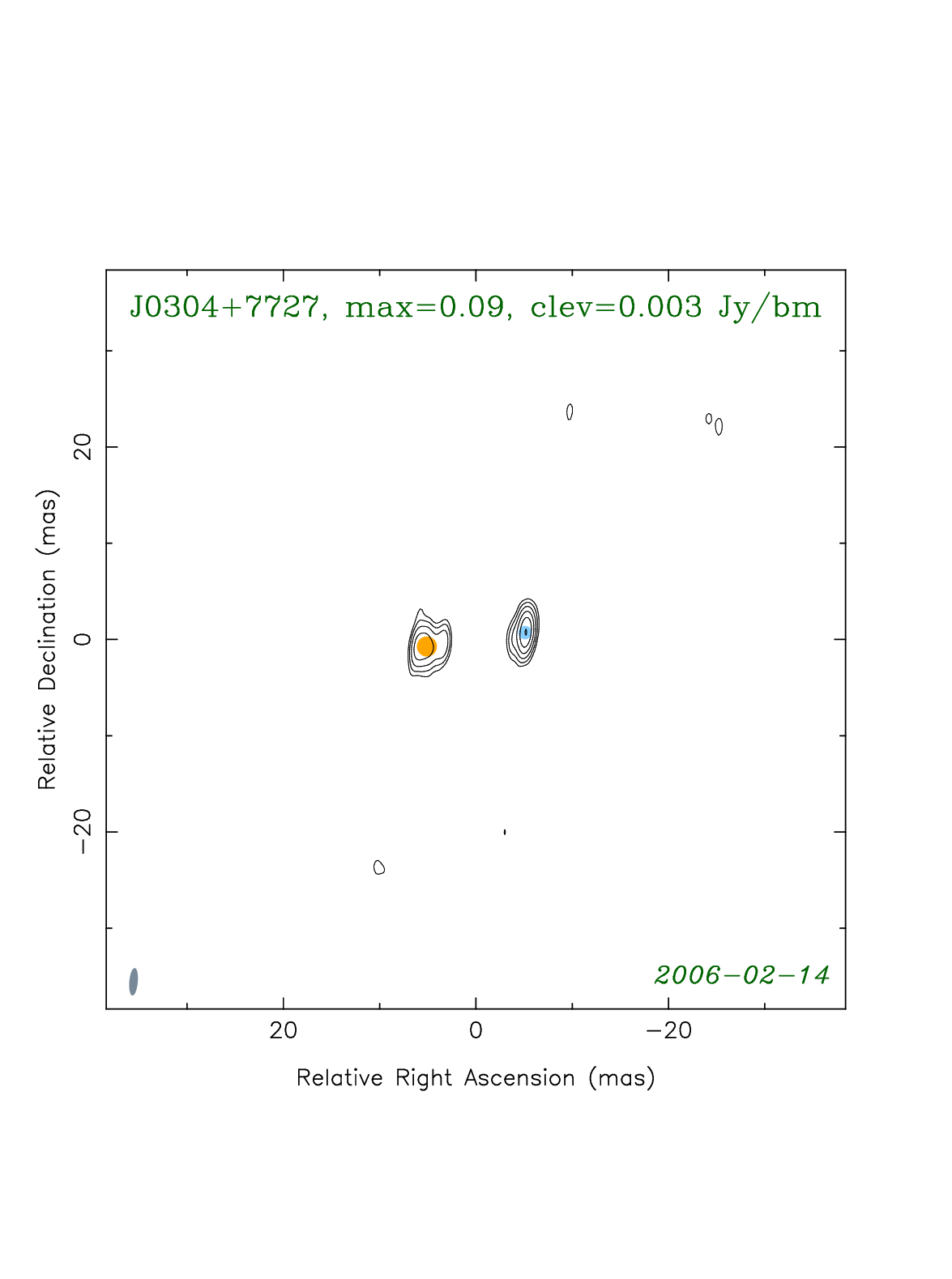}

 \includegraphics[width=0.44\textwidth,angle=0,trim=0 3cm 1cm 5cm,clip]{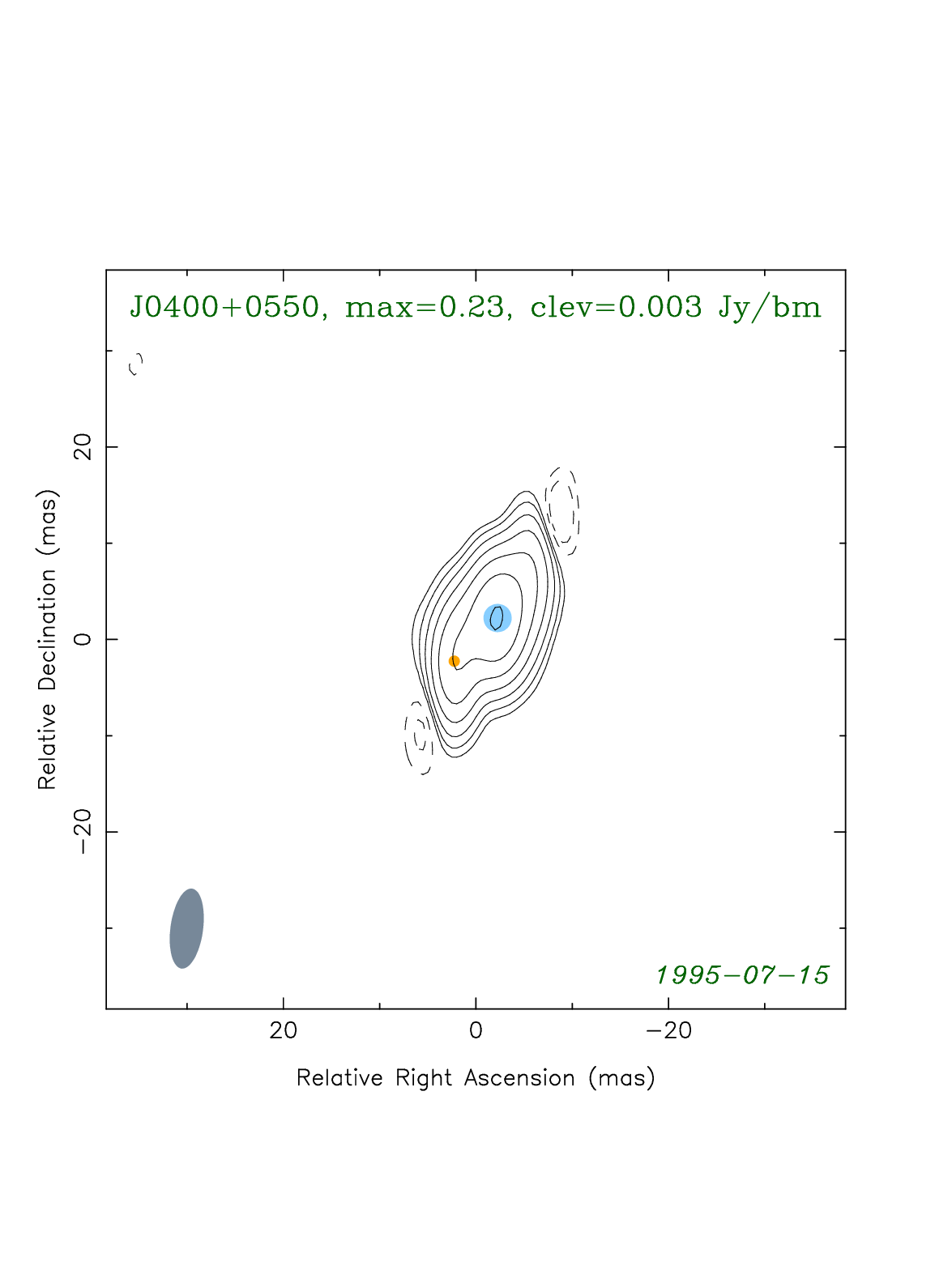}
 \includegraphics[width=0.44\textwidth,angle=0,trim=0 3cm 1cm 5cm,clip]{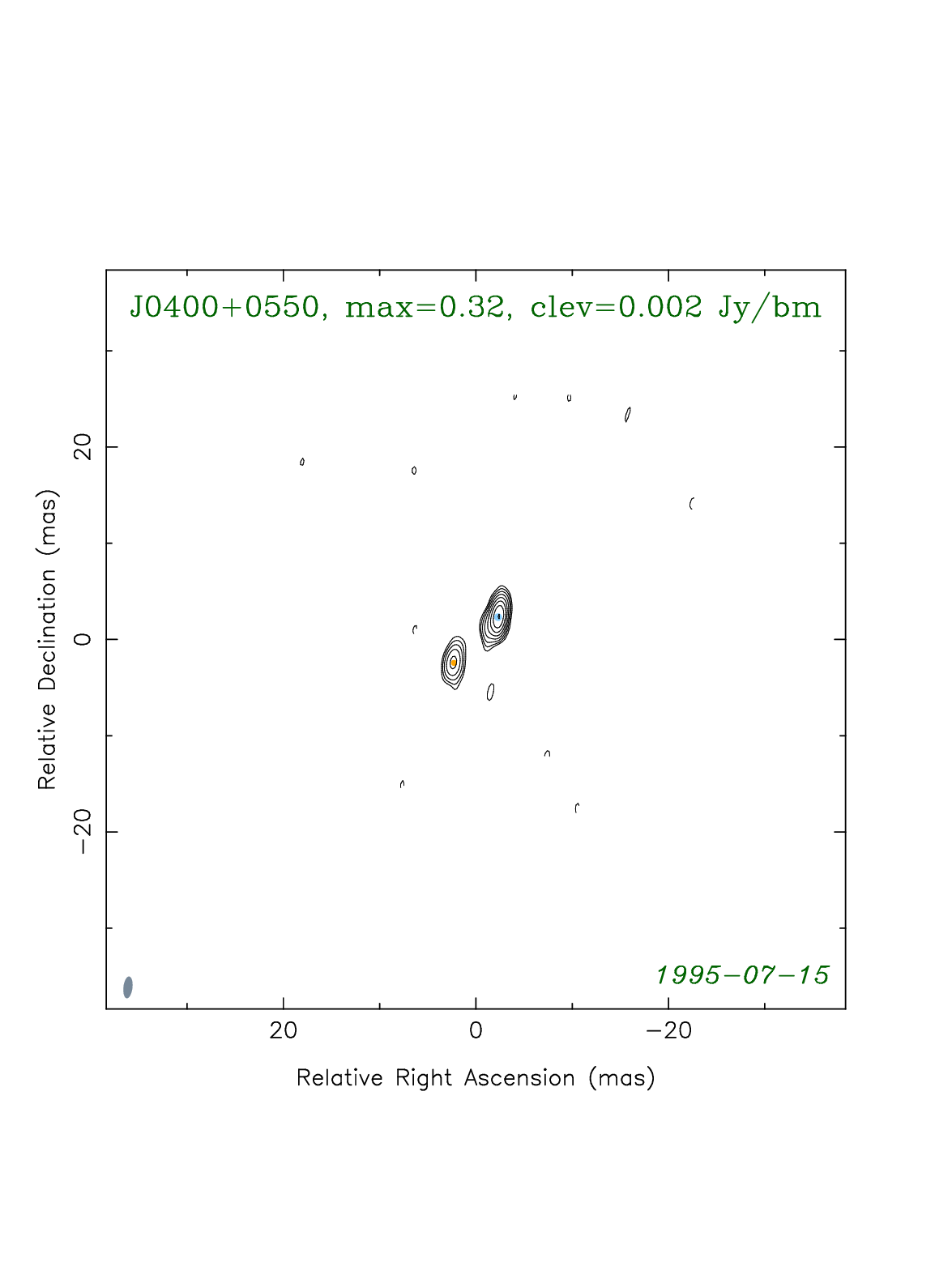}

 \caption{continued}
\end{figure*}

\clearpage
\addtocounter{figure}{-1}
\begin{figure*}
 \centering

 \includegraphics[width=0.44\textwidth,angle=0,trim=0 3cm 1cm 5cm,clip]{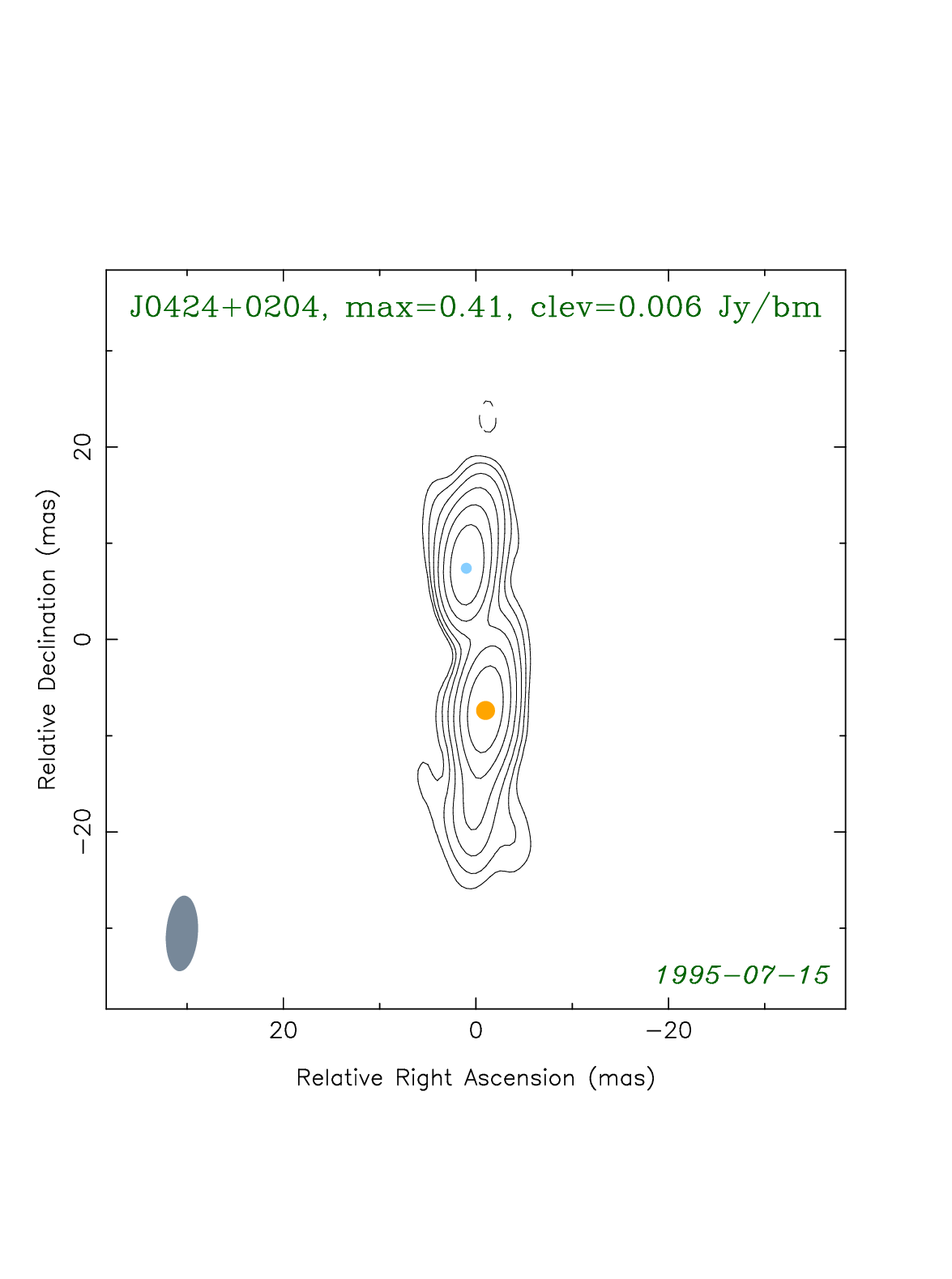}
 \includegraphics[width=0.44\textwidth,angle=0,trim=0 3cm 1cm 5cm,clip]{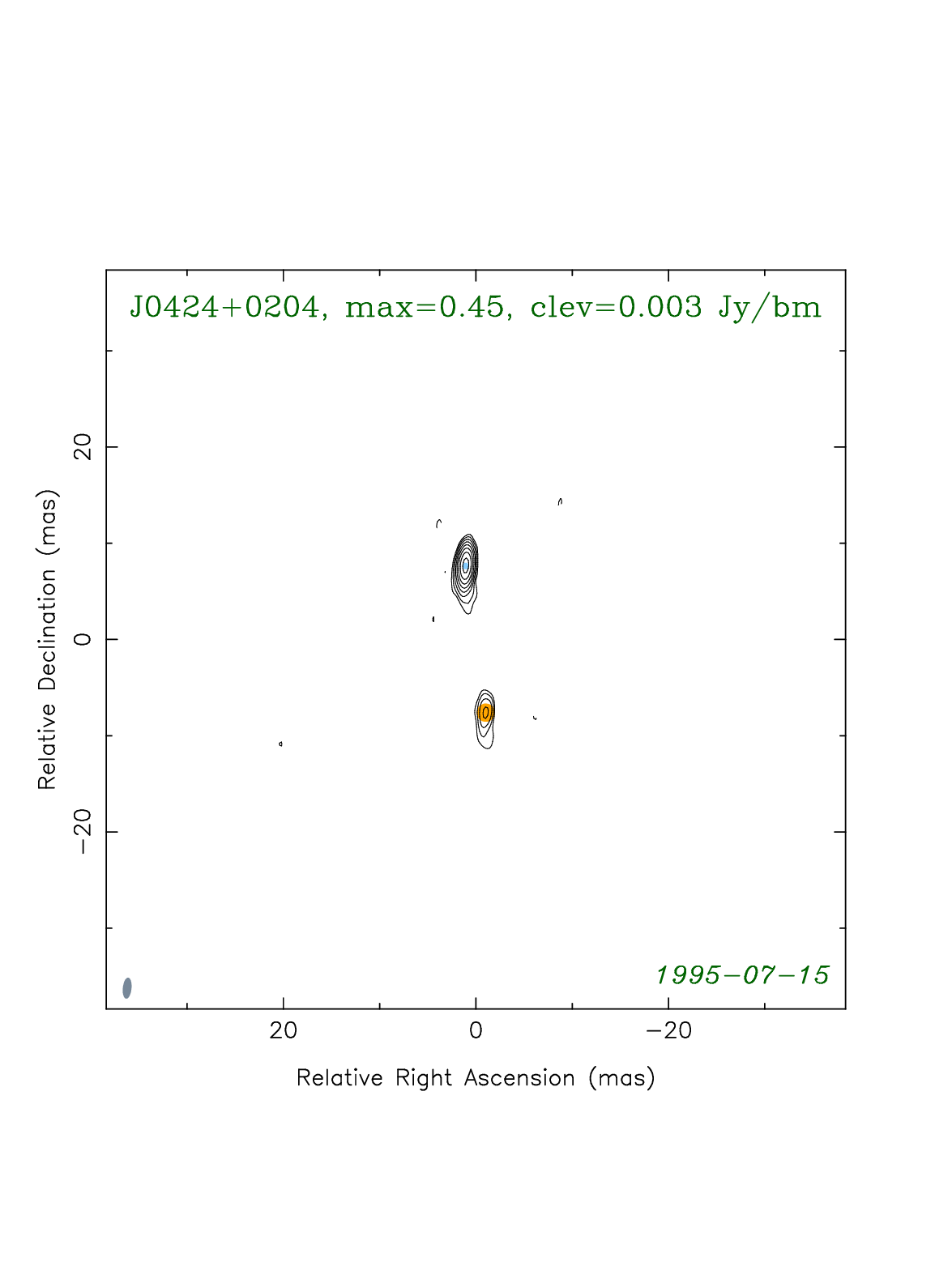}

 \includegraphics[width=0.44\textwidth,angle=0,trim=0 3cm 1cm 5cm,clip]{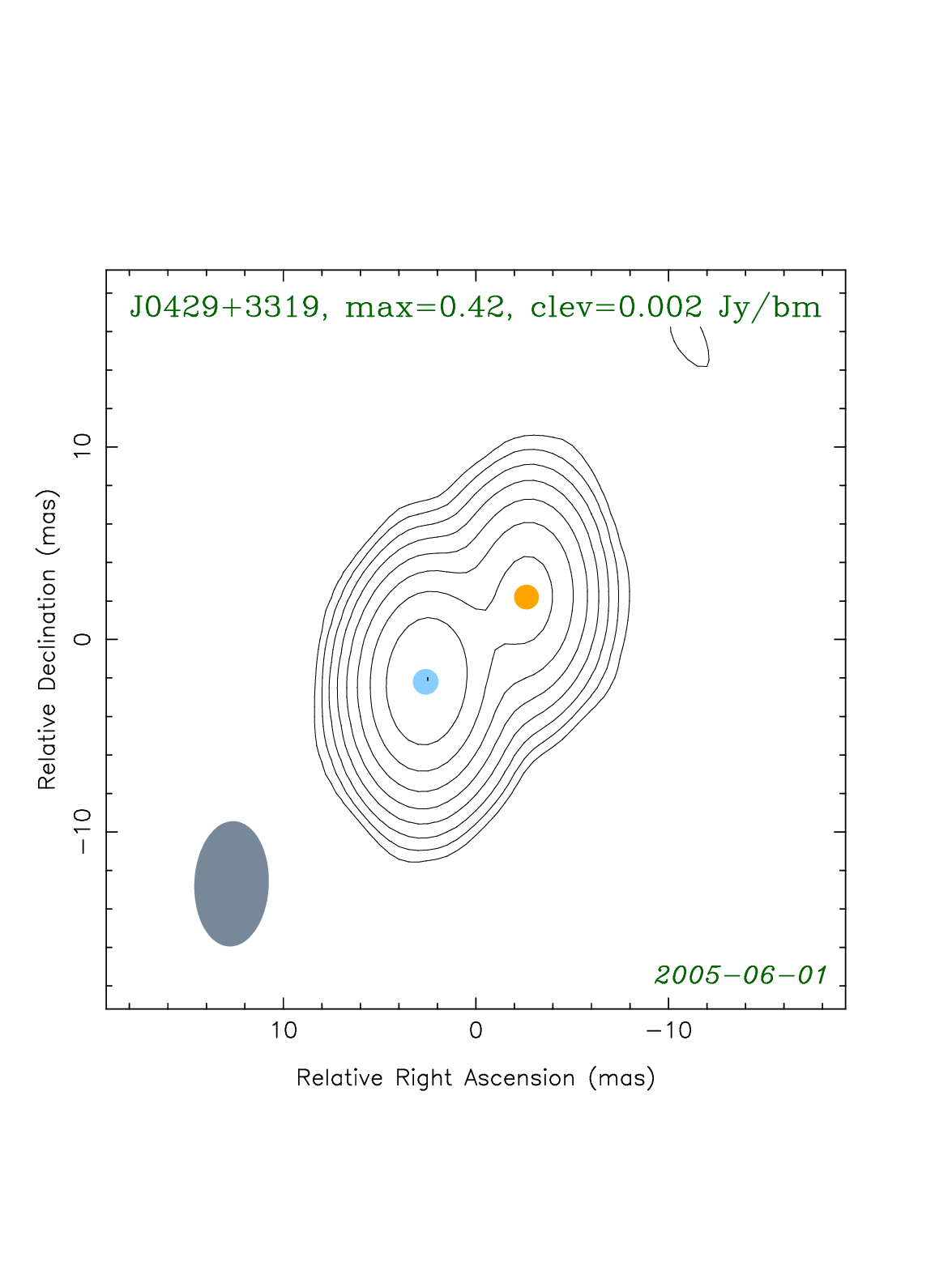}
 \includegraphics[width=0.44\textwidth,angle=0,trim=0 3cm 1cm 5cm,clip]{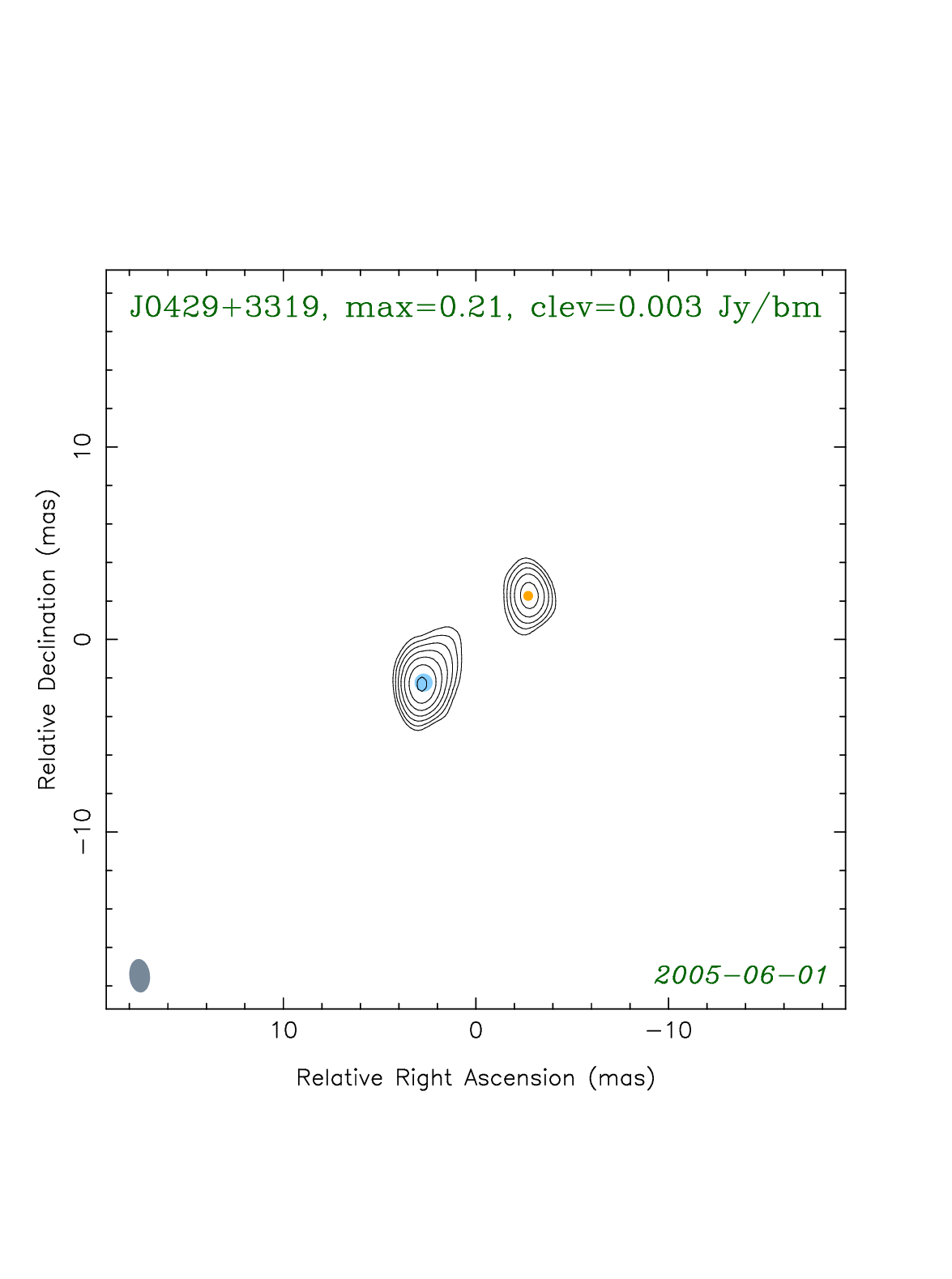}

 \includegraphics[width=0.44\textwidth,angle=0,trim=0 3cm 1cm 5cm,clip]{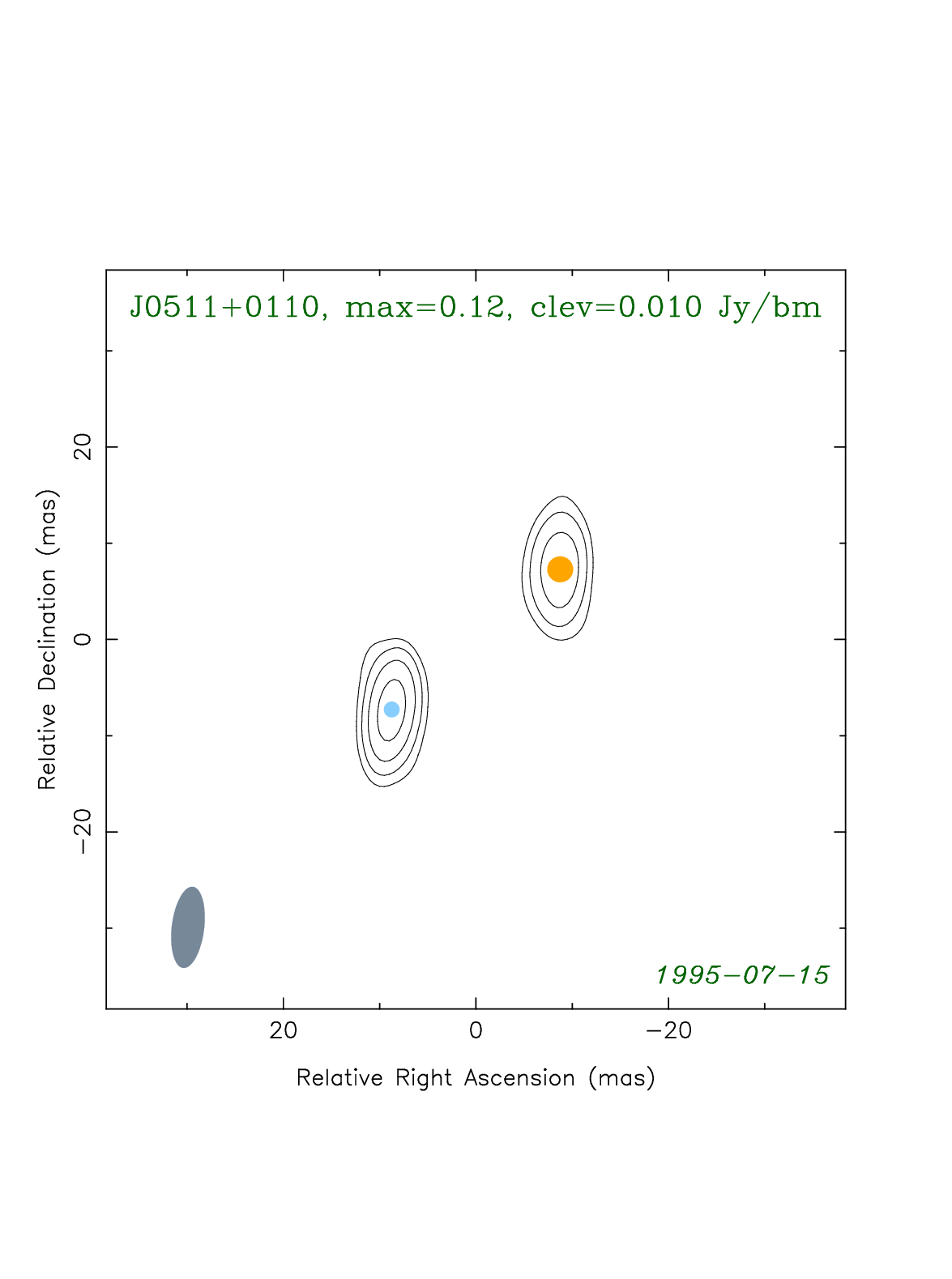}
 \includegraphics[width=0.44\textwidth,angle=0,trim=0 3cm 1cm 5cm,clip]{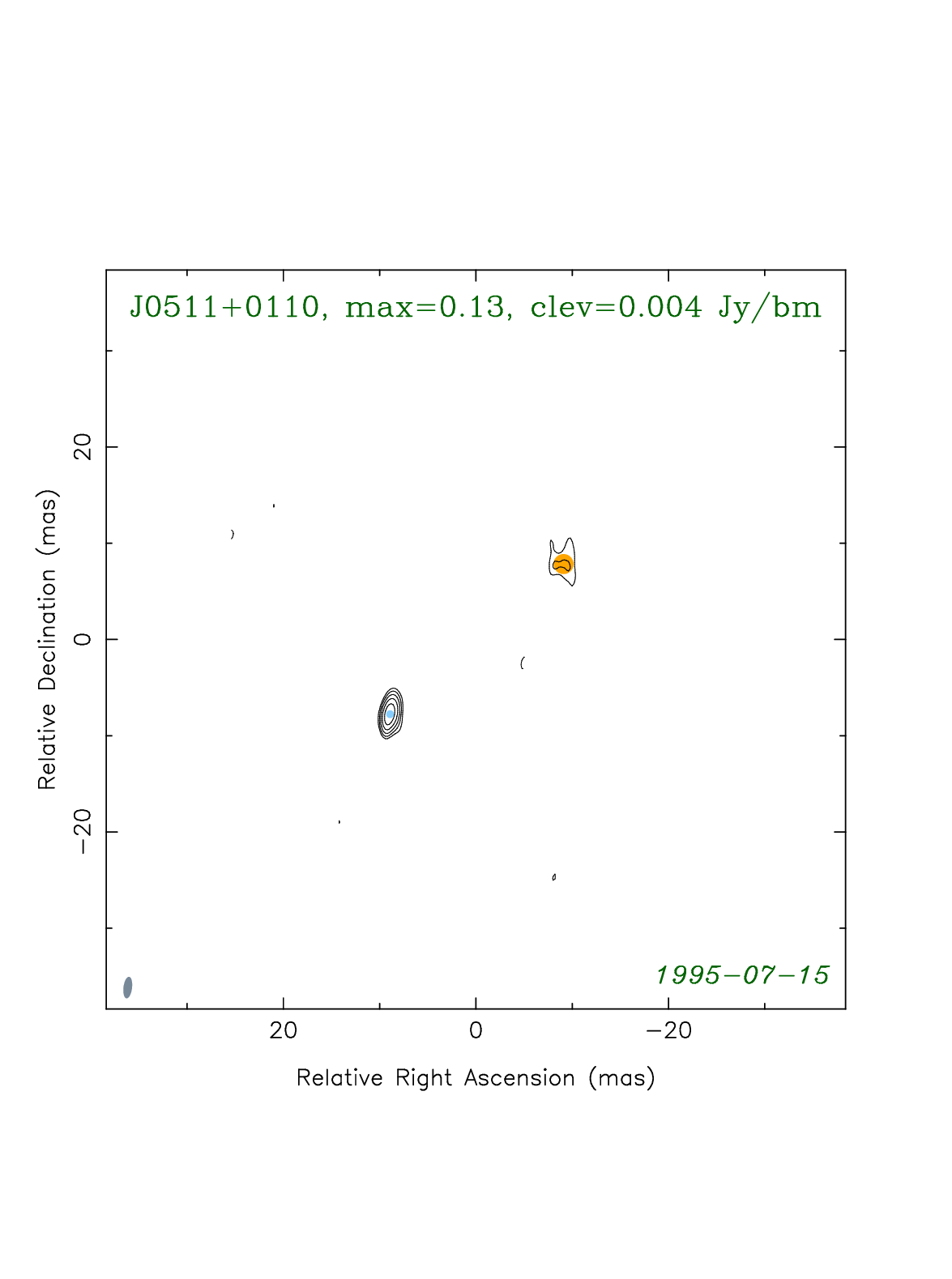}

 \caption{continued}
\end{figure*}

\clearpage
\addtocounter{figure}{-1}
\begin{figure*}
 \centering

 \includegraphics[width=0.44\textwidth,angle=0,trim=0 3cm 1cm 5cm,clip]{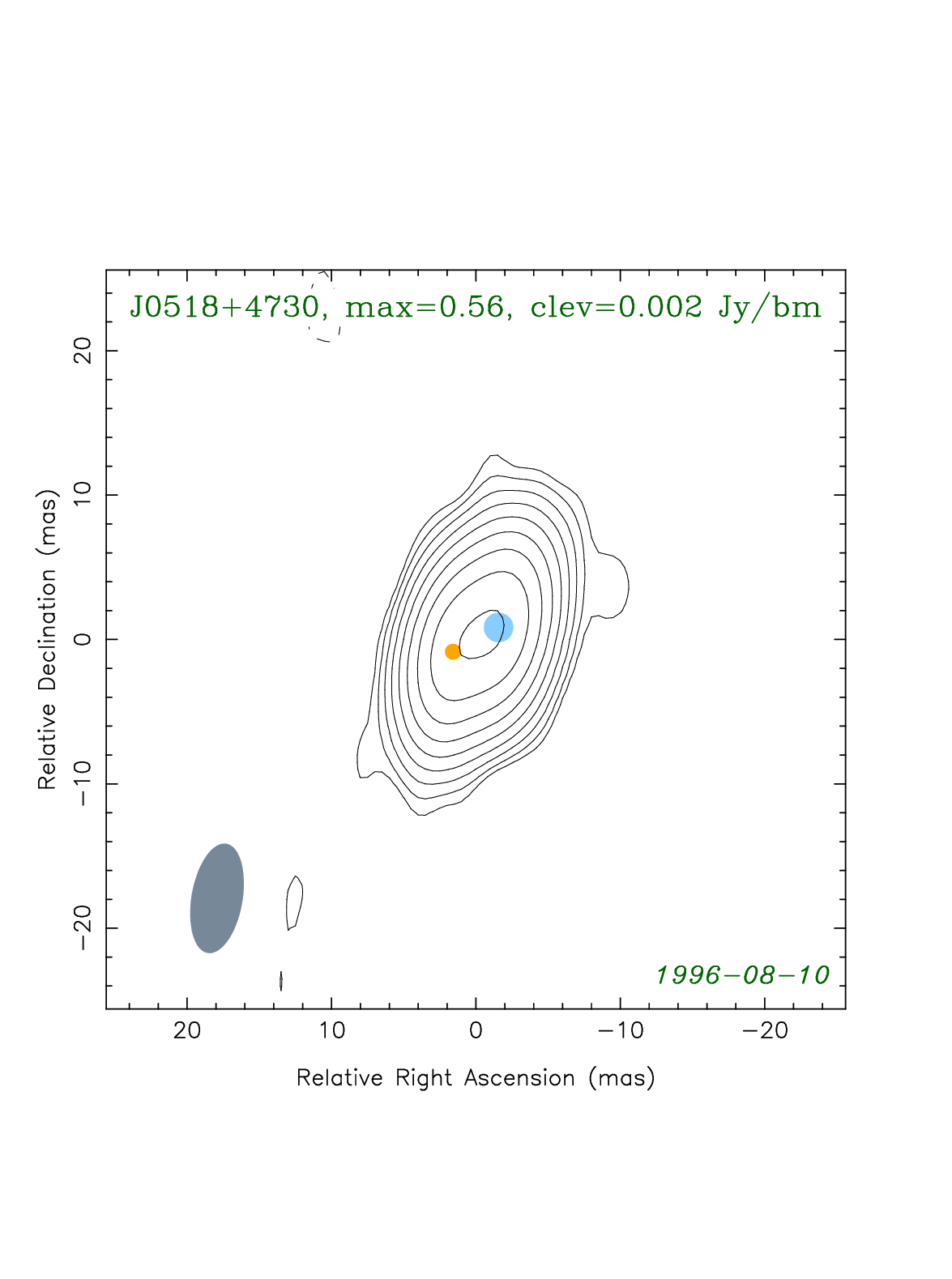}
 \includegraphics[width=0.44\textwidth,angle=0,trim=0 3cm 1cm 5cm,clip]{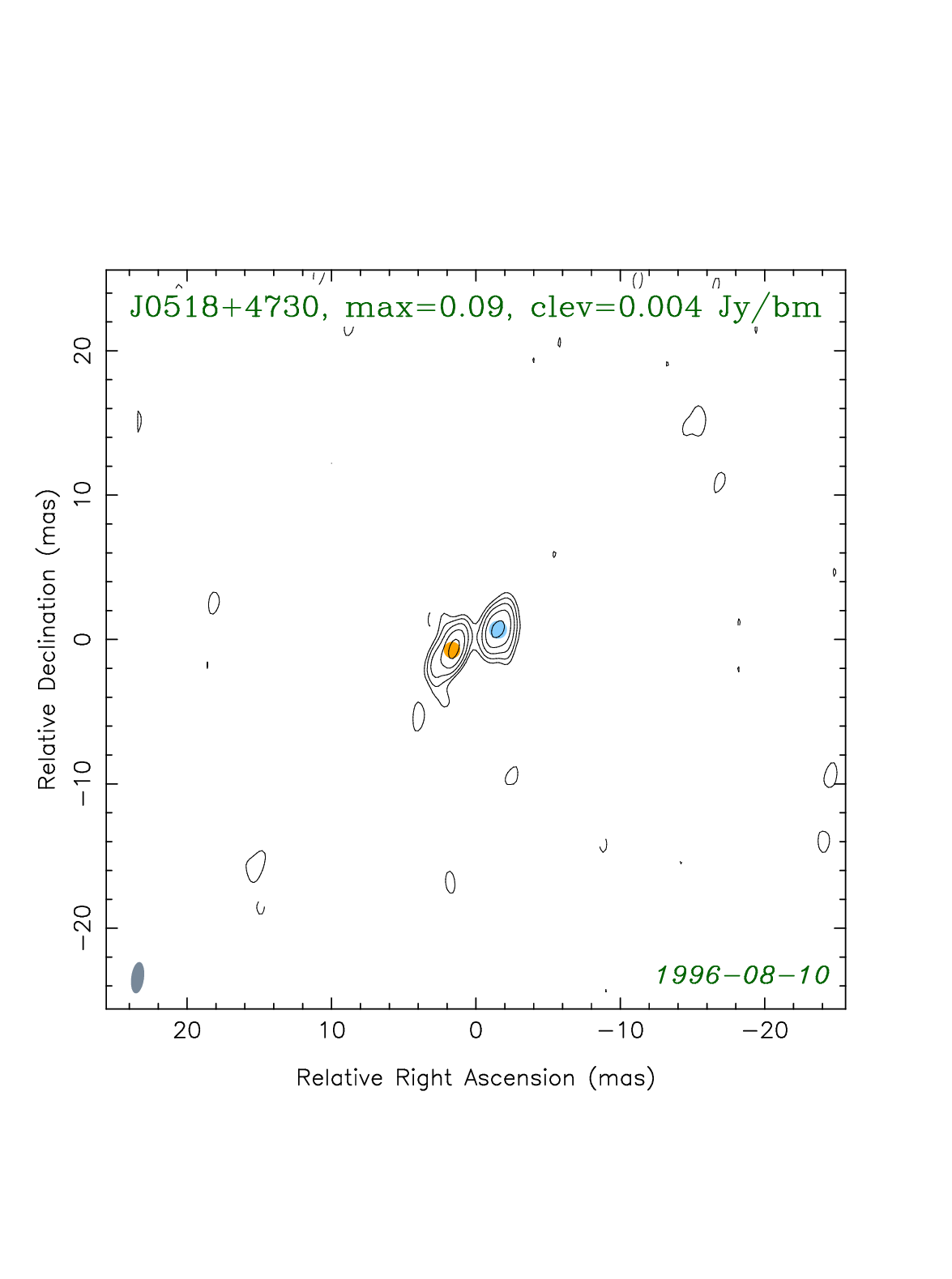}

 \includegraphics[width=0.44\textwidth,angle=0,trim=0 3cm 1cm 5cm,clip]{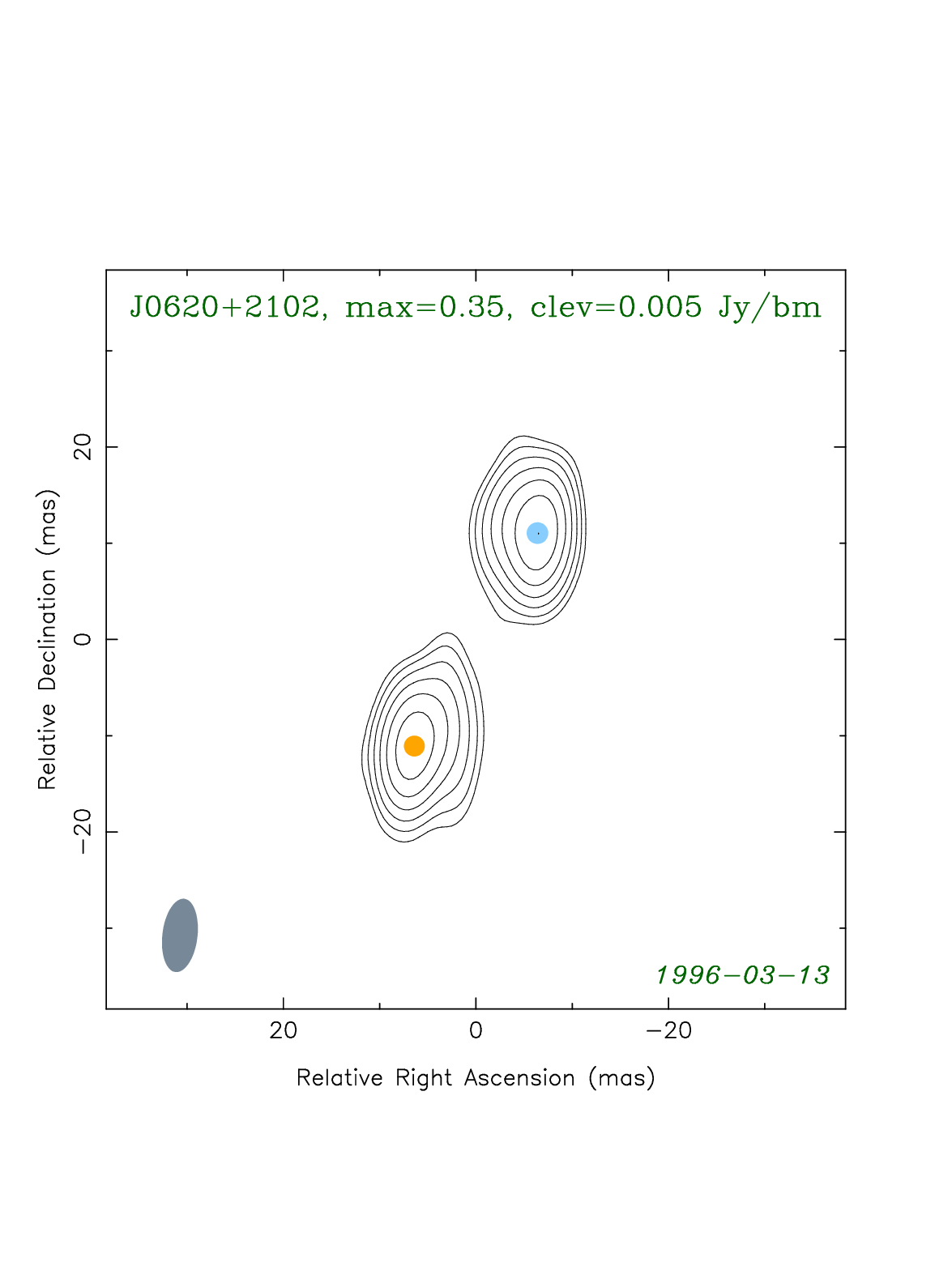}
 \includegraphics[width=0.44\textwidth,angle=0,trim=0 3cm 1cm 5cm,clip]{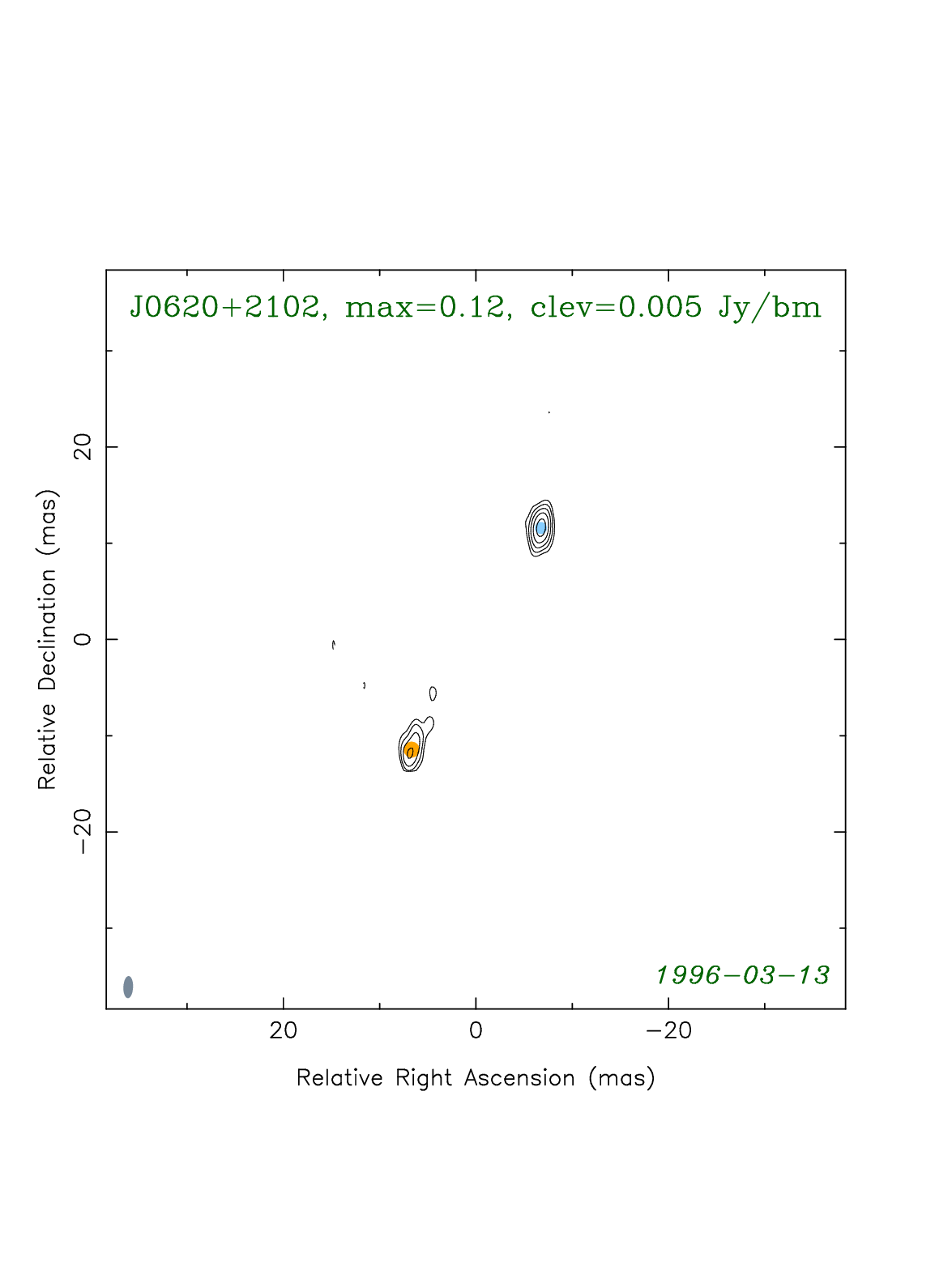}

 \includegraphics[width=0.44\textwidth,angle=0,trim=0 3cm 1cm 5cm,clip]{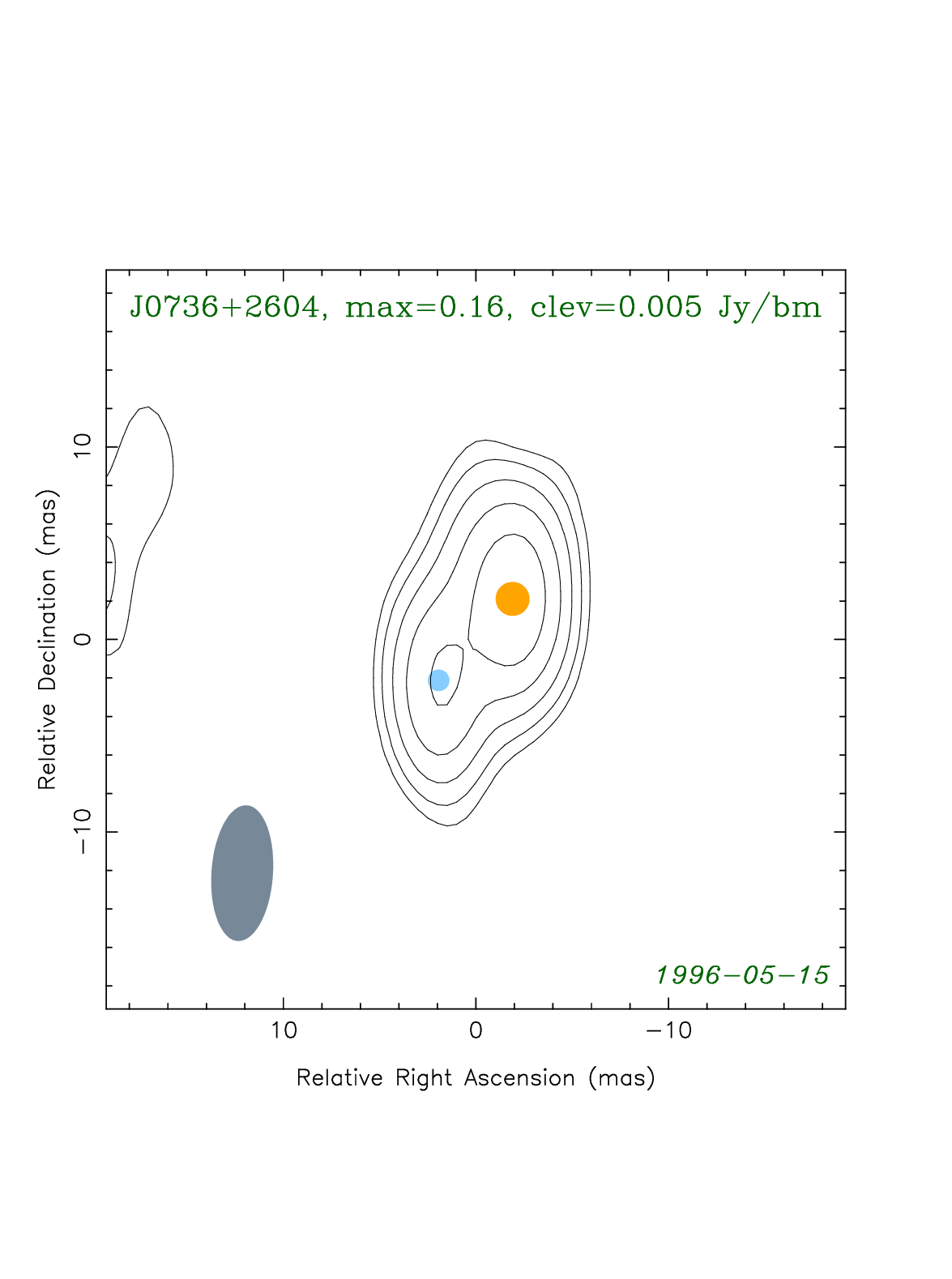}
 \includegraphics[width=0.44\textwidth,angle=0,trim=0 3cm 1cm 5cm,clip]{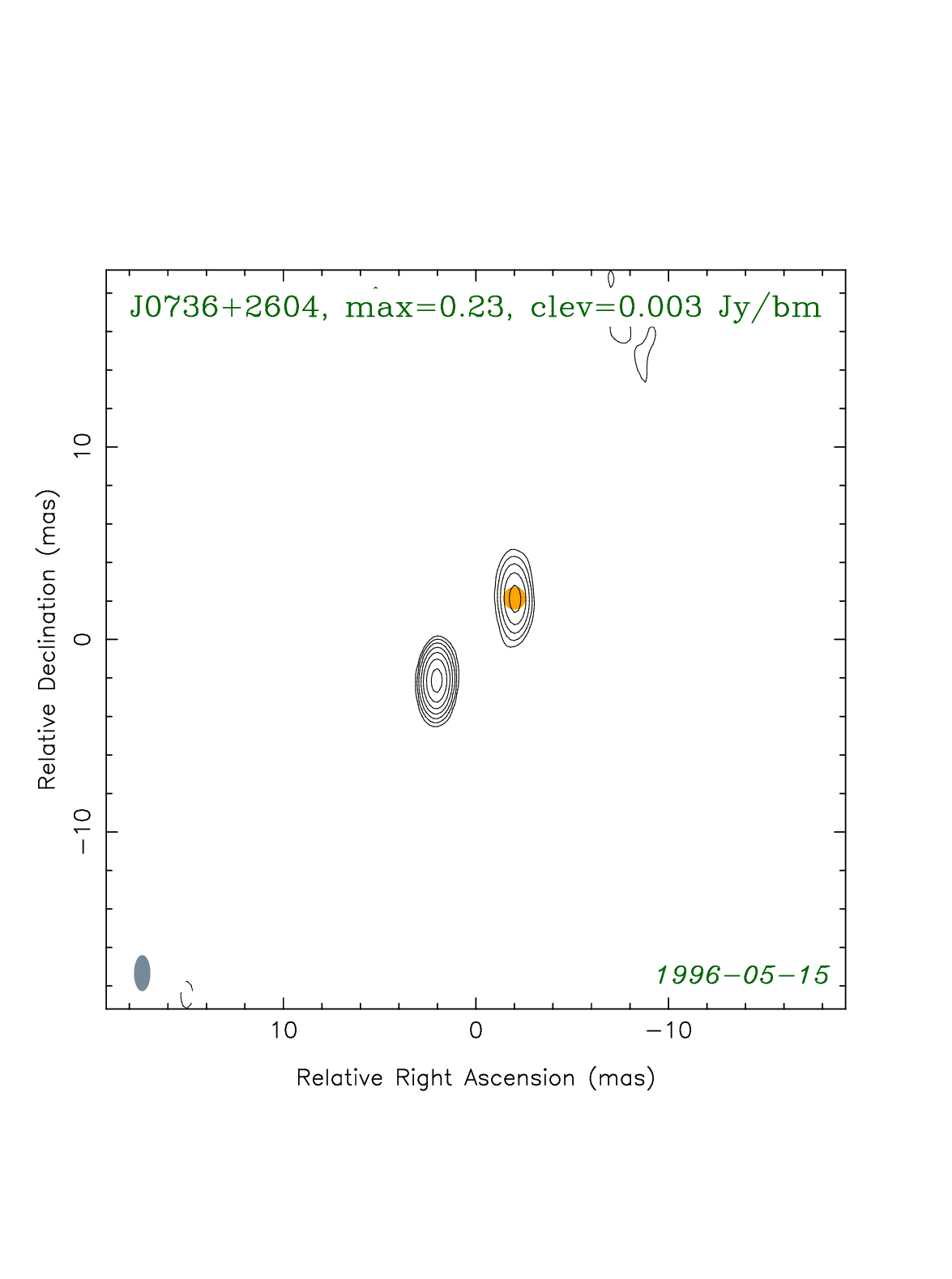}

 \caption{continued}
\end{figure*}

\clearpage
\addtocounter{figure}{-1}
\begin{figure*}
 \centering

 \includegraphics[width=0.44\textwidth,angle=0,trim=0 3cm 1cm 5cm,clip]{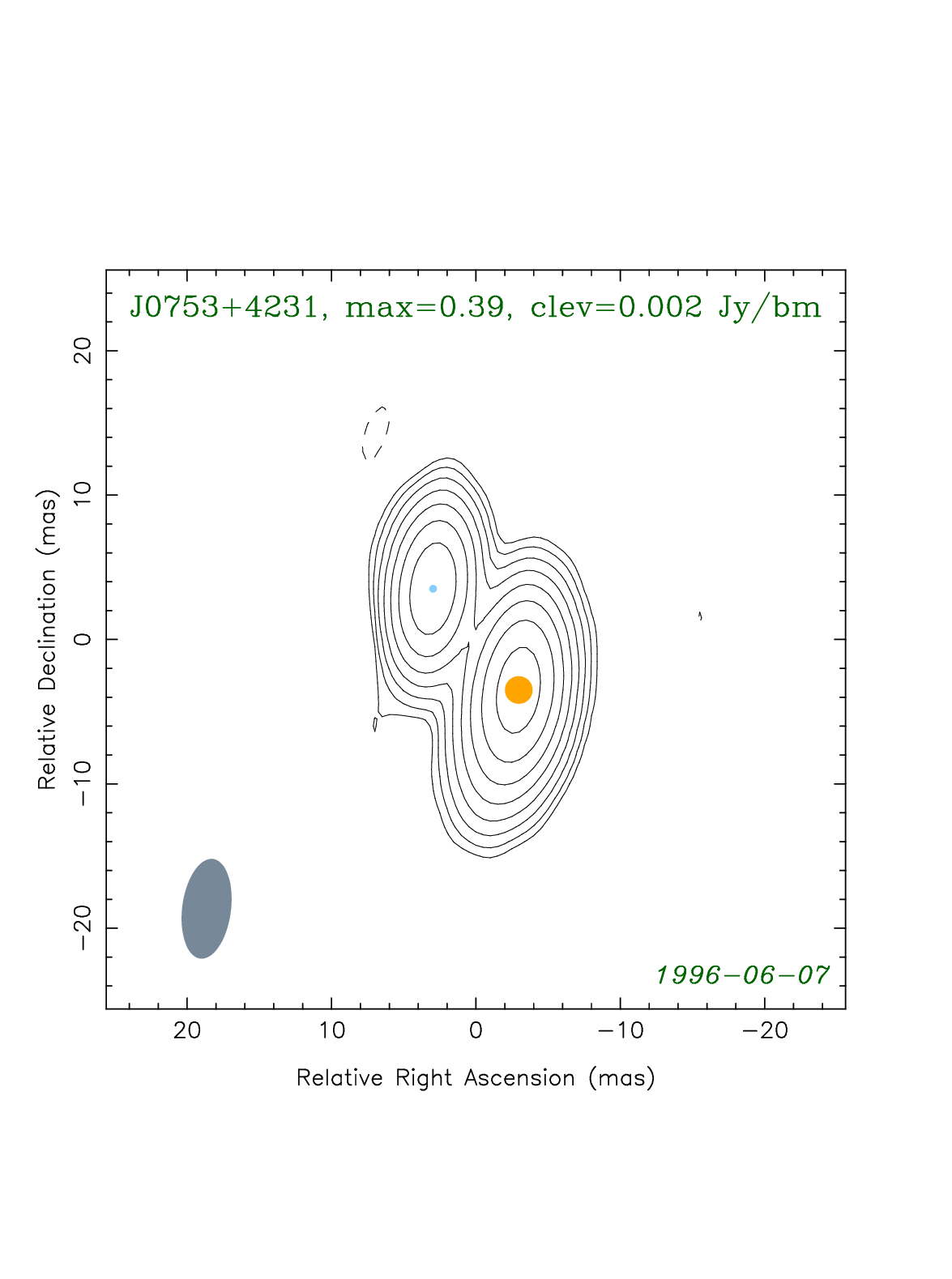}
 \includegraphics[width=0.44\textwidth,angle=0,trim=0 3cm 1cm 5cm,clip]{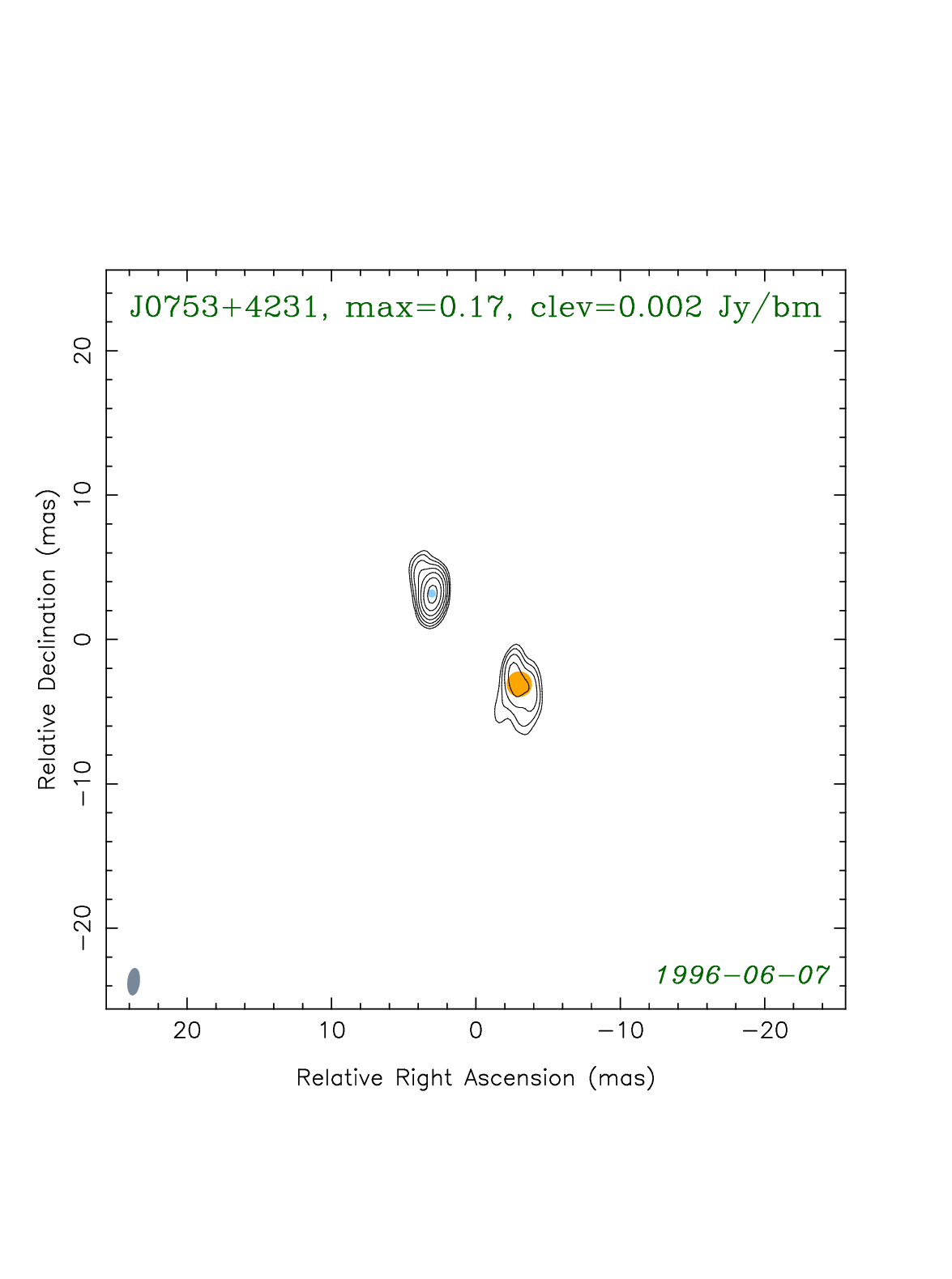}

 \includegraphics[width=0.44\textwidth,angle=0,trim=0 3cm 1cm 5cm,clip]{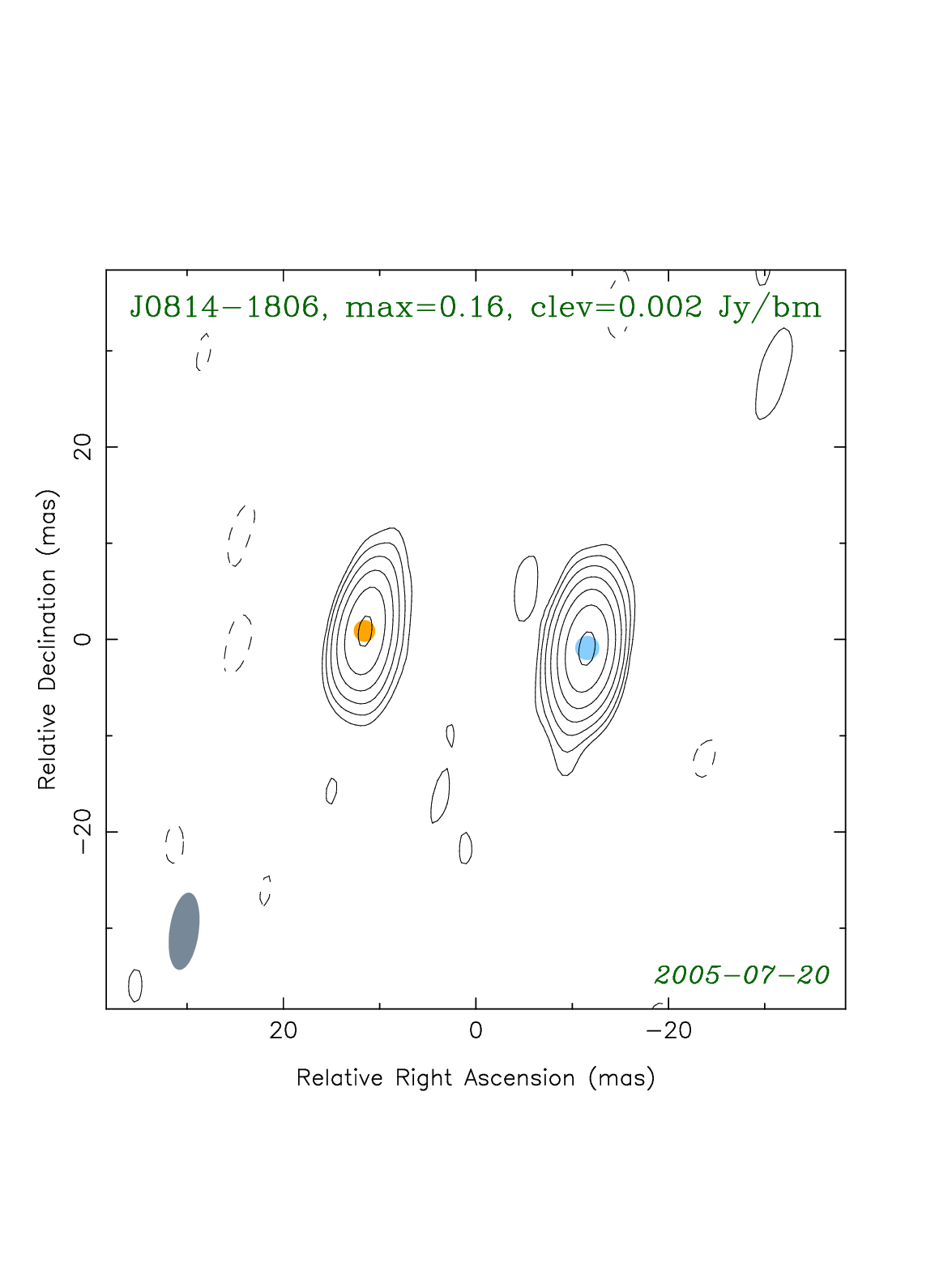}
 \includegraphics[width=0.44\textwidth,angle=0,trim=0 3cm 1cm 5cm,clip]{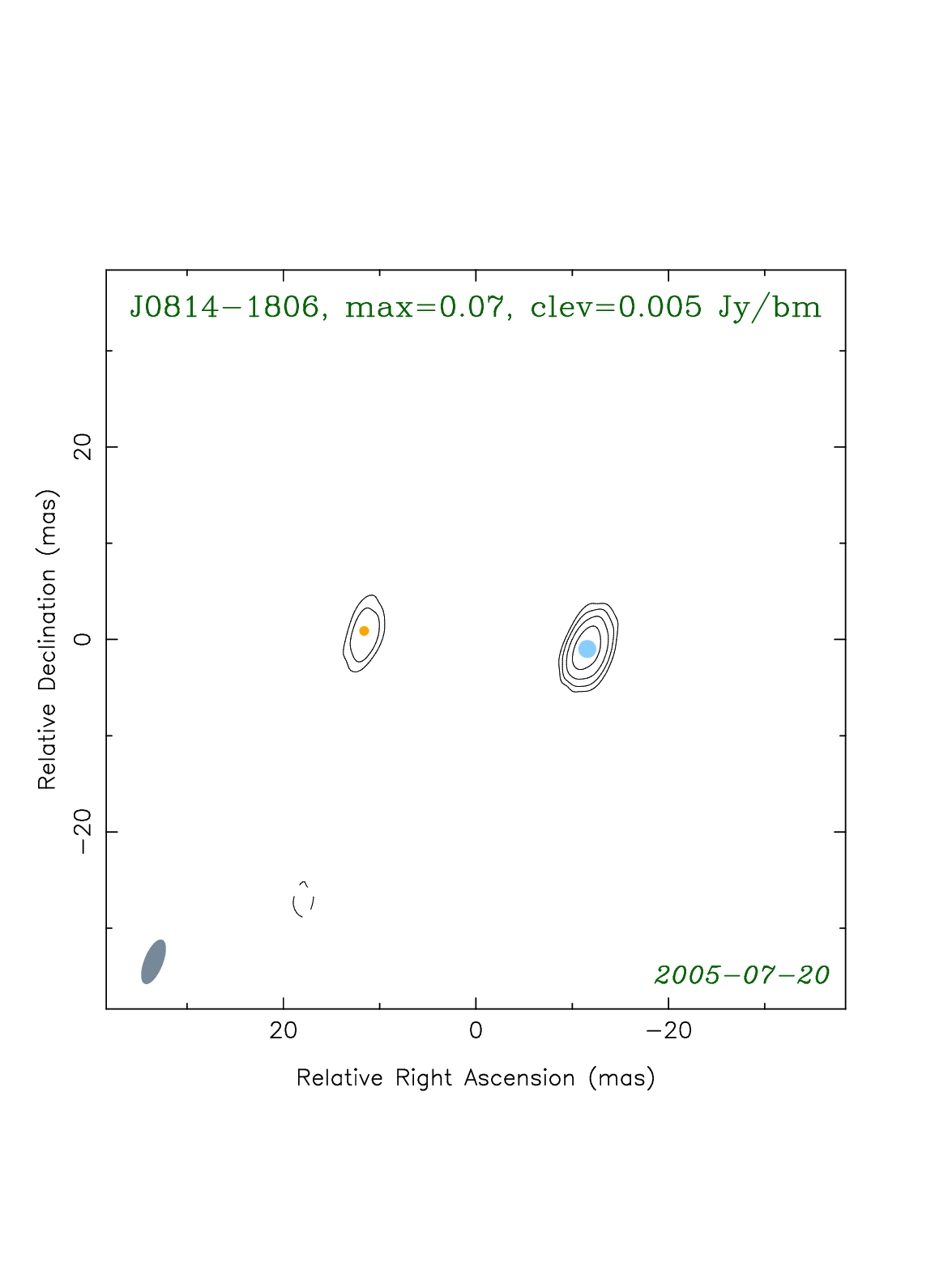}

 \includegraphics[width=0.44\textwidth,angle=0,trim=0 3cm 1cm 5cm,clip]{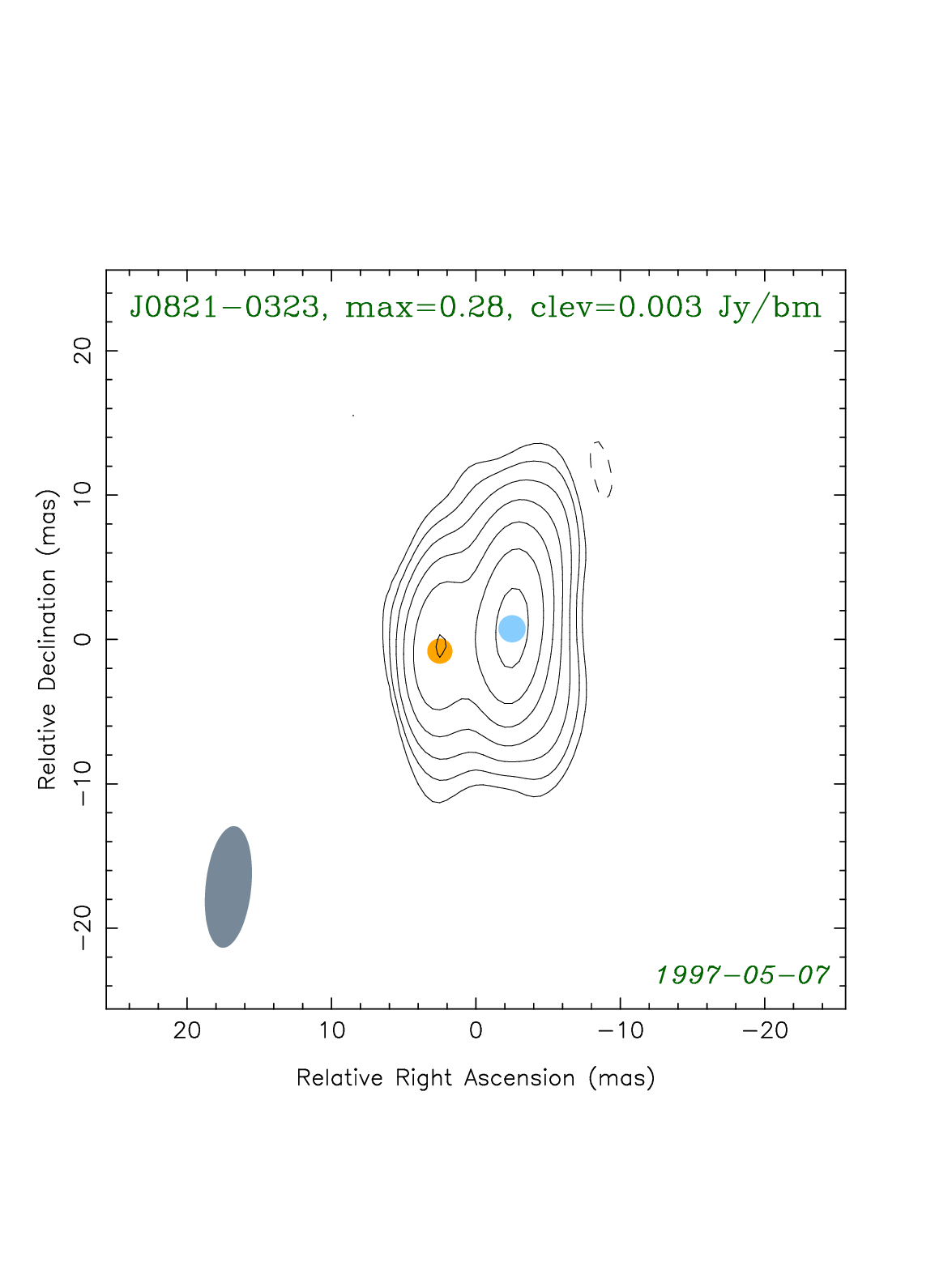}
 \includegraphics[width=0.44\textwidth,angle=0,trim=0 3cm 1cm 5cm,clip]{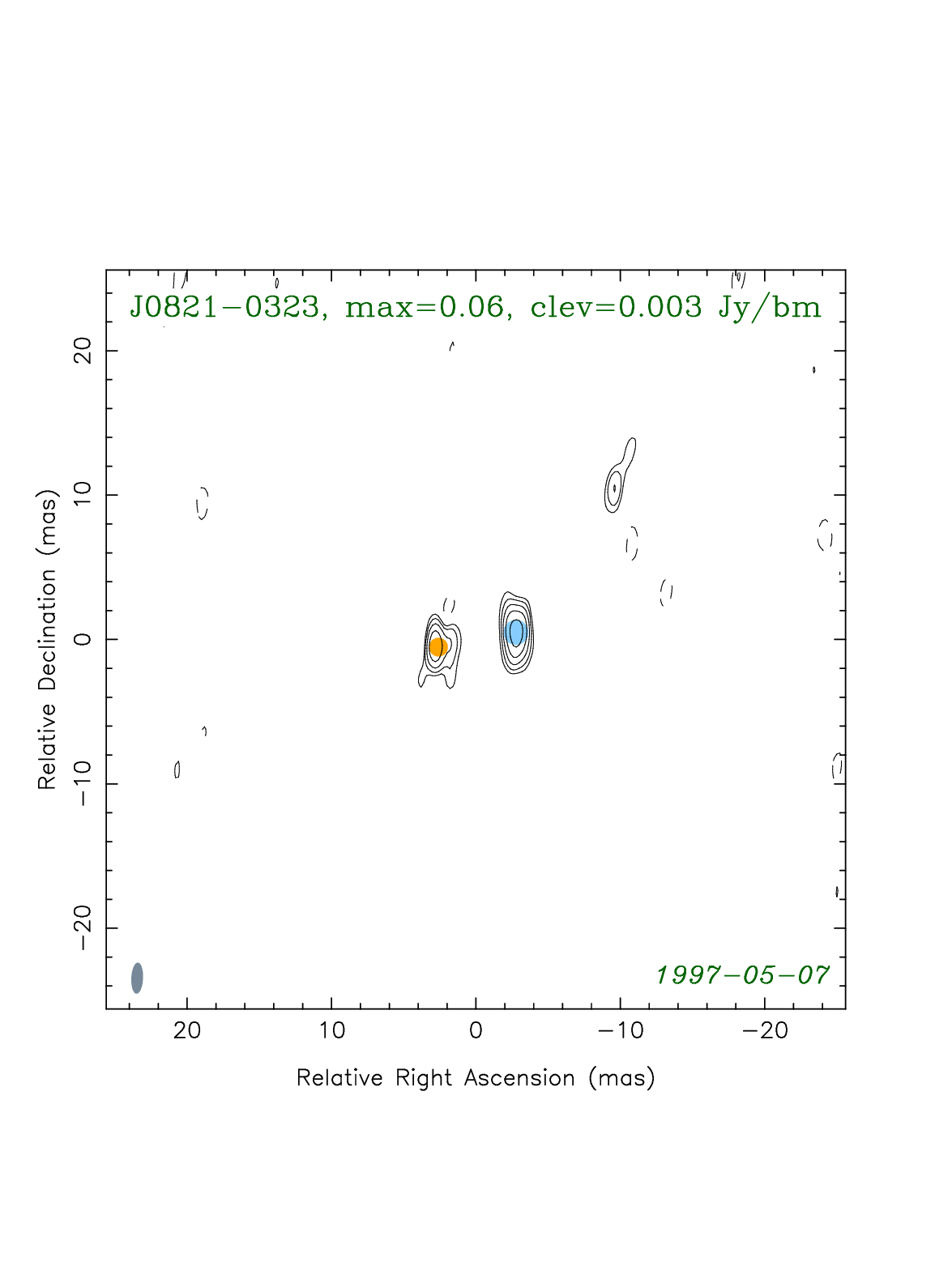}

 \caption{continued}
\end{figure*}

\clearpage
\addtocounter{figure}{-1}
\begin{figure*}
 \centering

 \includegraphics[width=0.44\textwidth,angle=0,trim=0 3cm 1cm 5cm,clip]{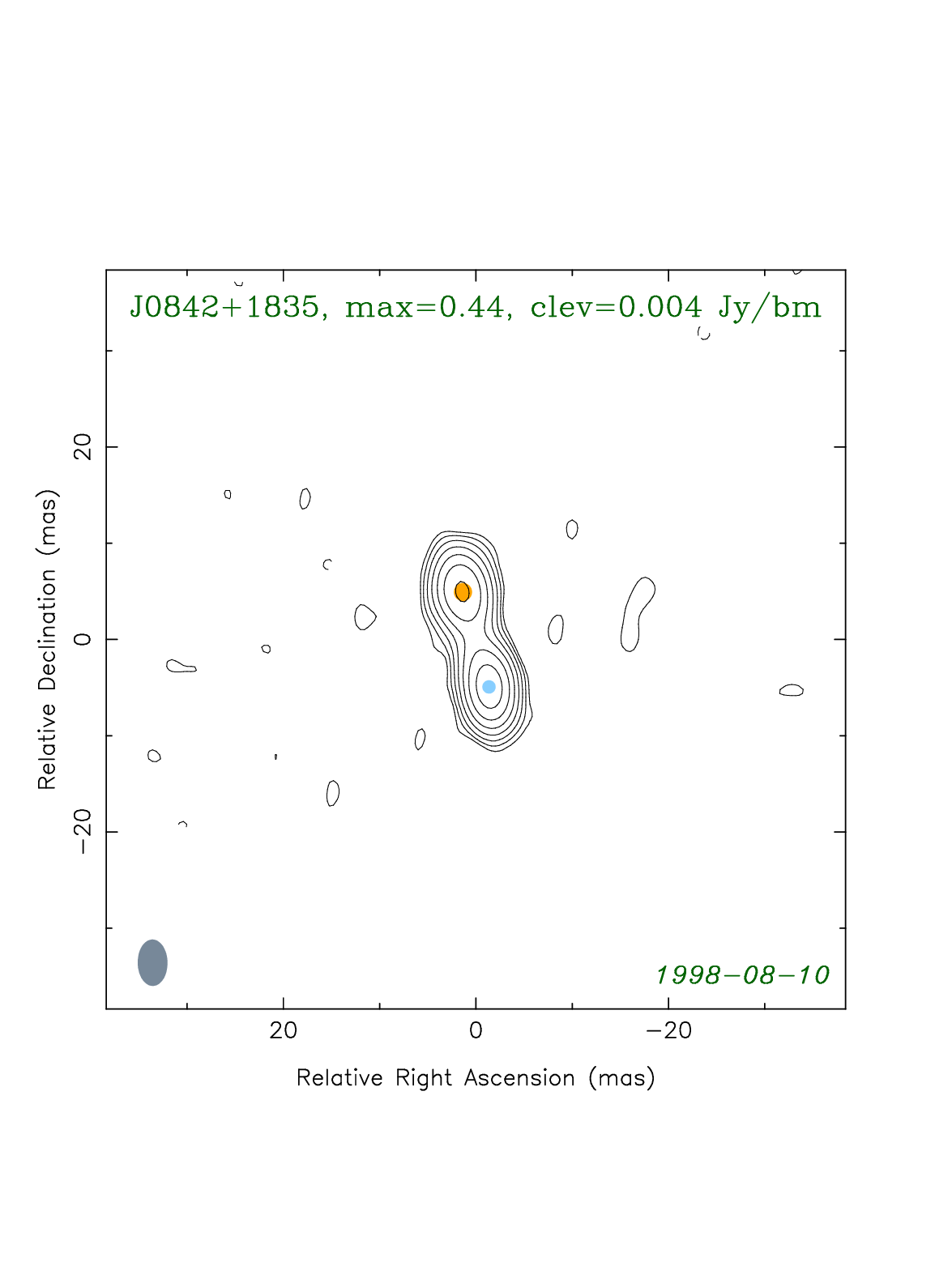}
 \includegraphics[width=0.44\textwidth,angle=0,trim=0 3cm 1cm 5cm,clip]{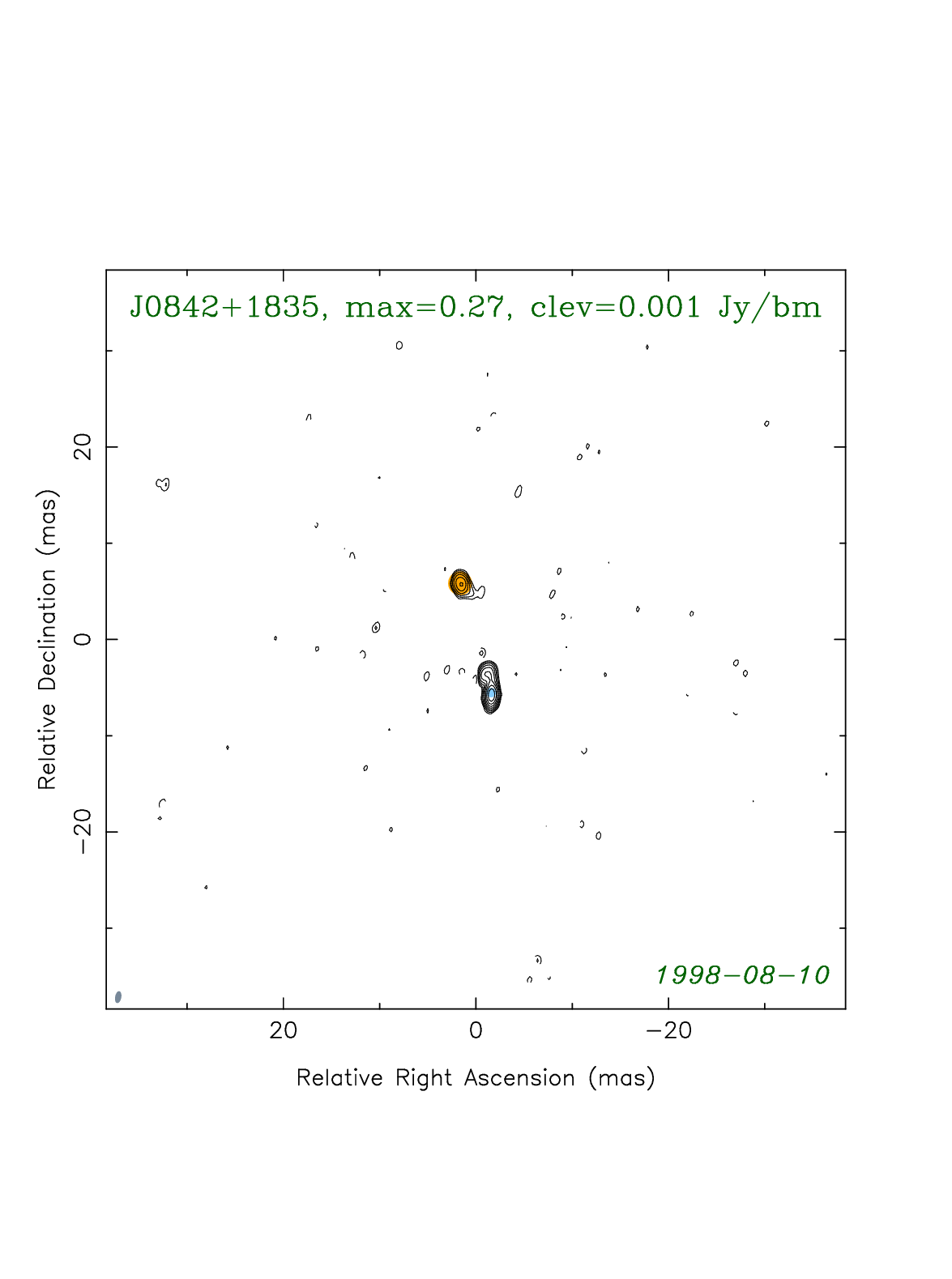}

 \includegraphics[width=0.44\textwidth,angle=0,trim=0 3cm 1cm 5cm,clip]{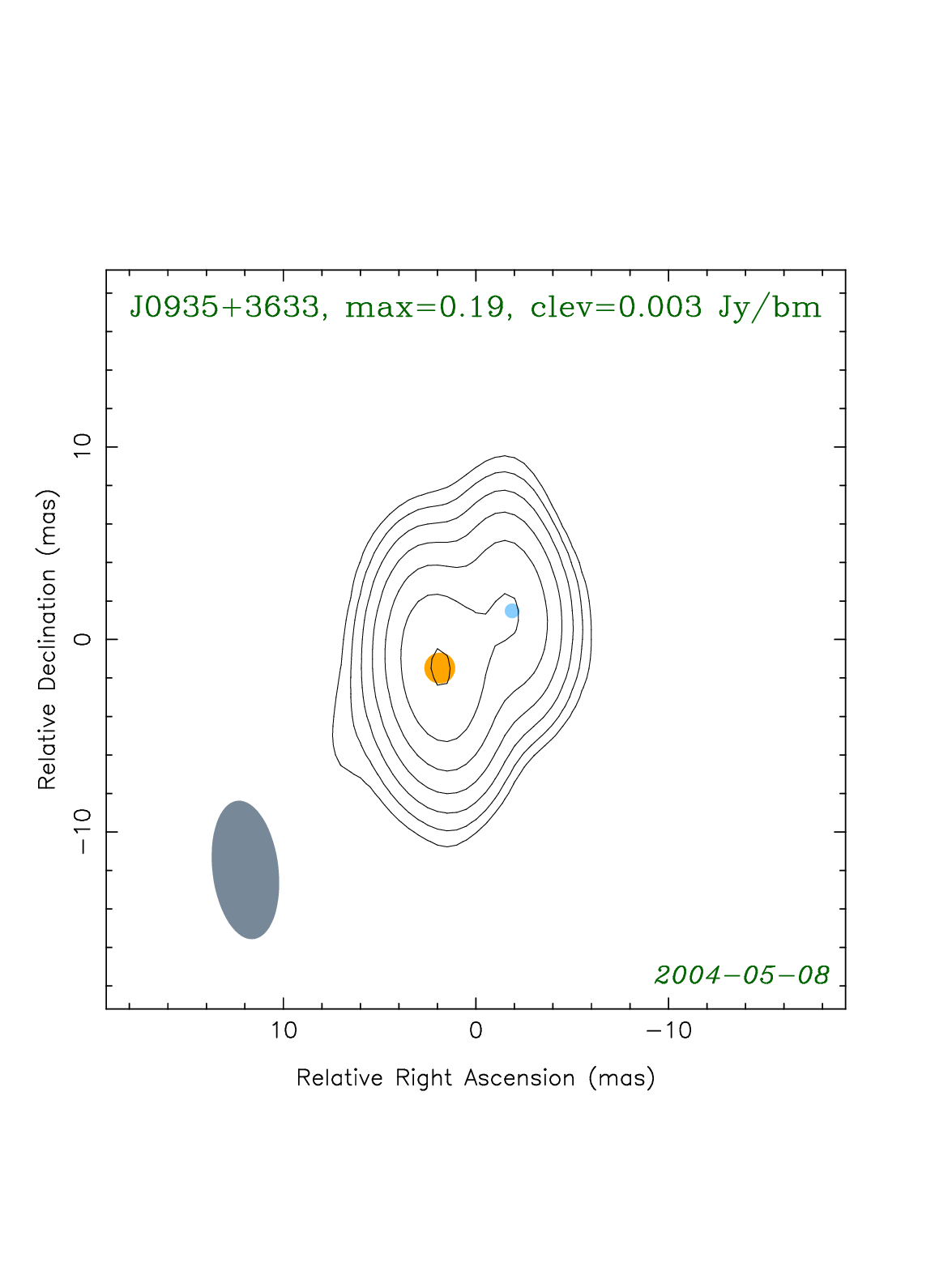}
 \includegraphics[width=0.44\textwidth,angle=0,trim=0 3cm 1cm 5cm,clip]{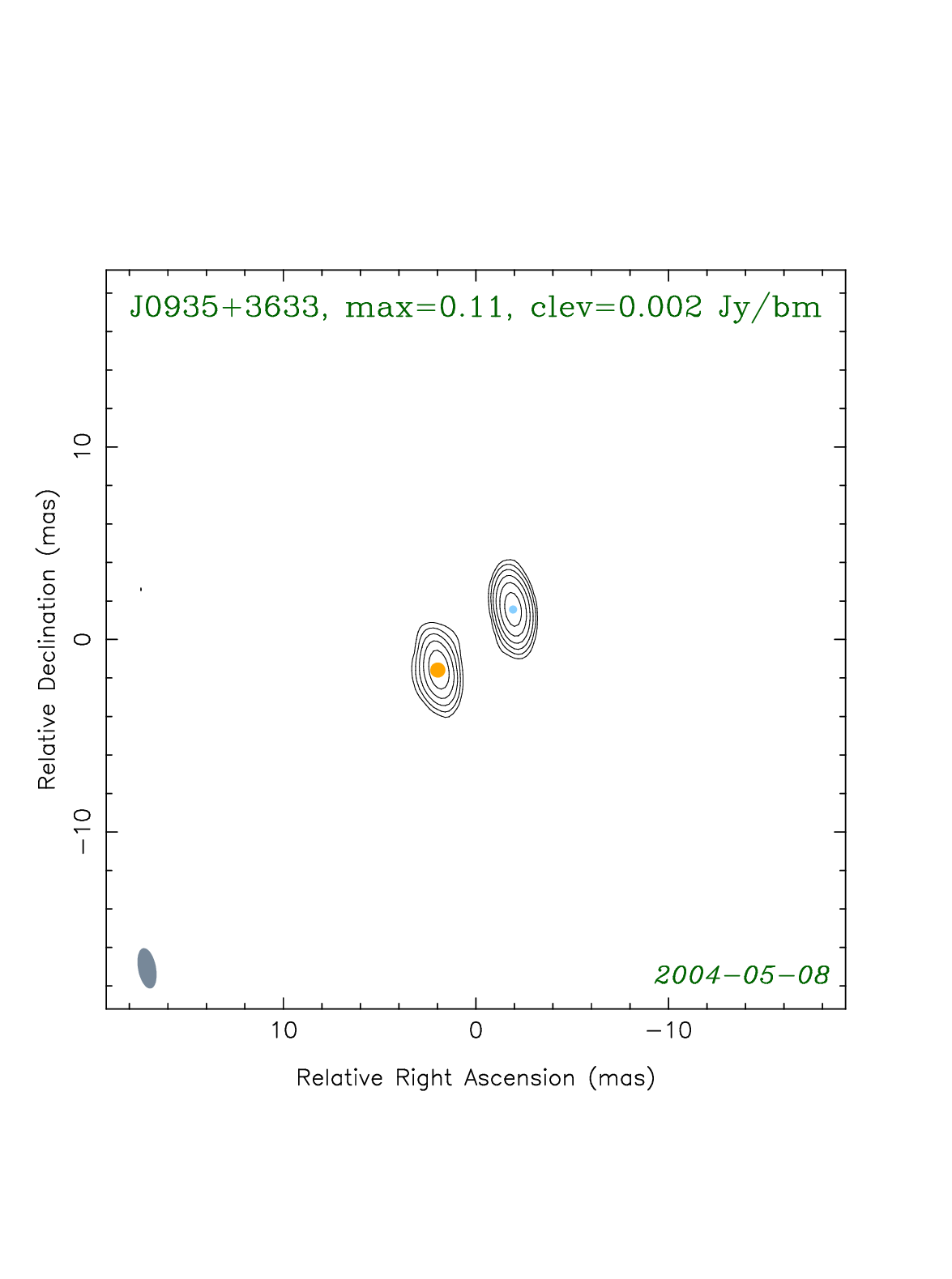}

 \includegraphics[width=0.44\textwidth,angle=0,trim=0 3cm 1cm 5cm,clip]{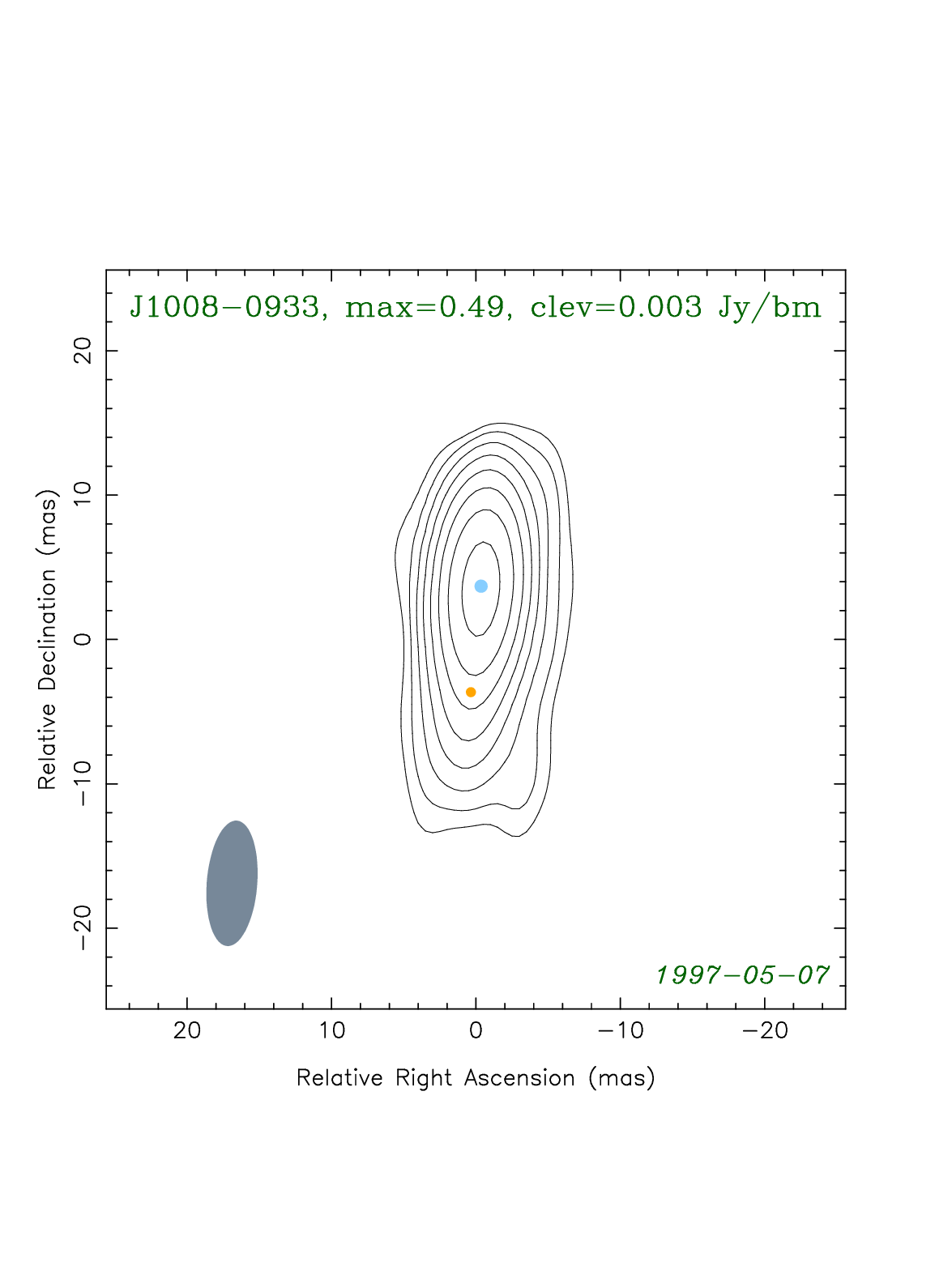}
 \includegraphics[width=0.44\textwidth,angle=0,trim=0 3cm 1cm 5cm,clip]{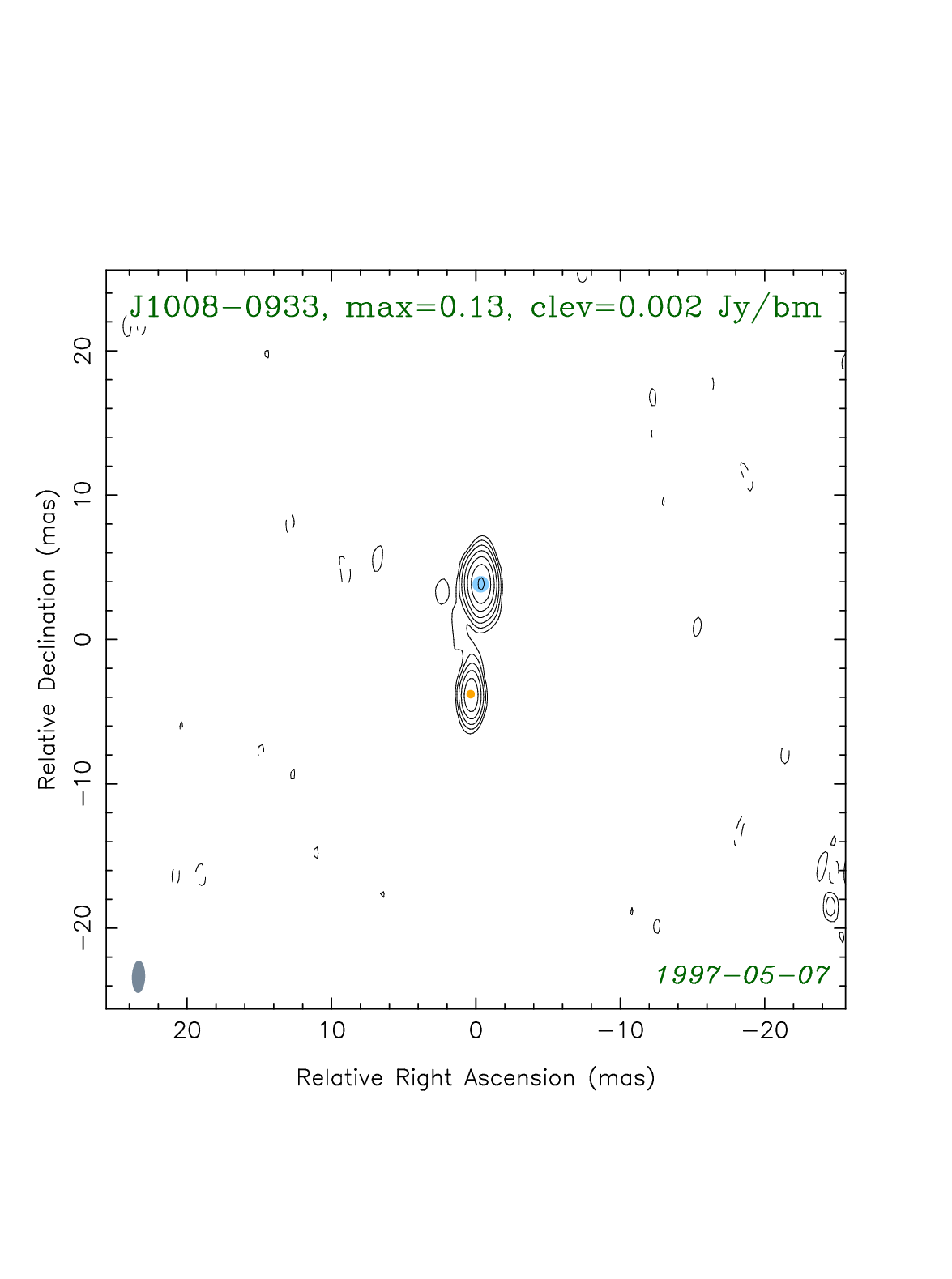}

 \caption{continued}
\end{figure*}

\clearpage
\addtocounter{figure}{-1}
\begin{figure*}
 \centering

 \includegraphics[width=0.44\textwidth,angle=0,trim=0 3cm 1cm 5cm,clip]{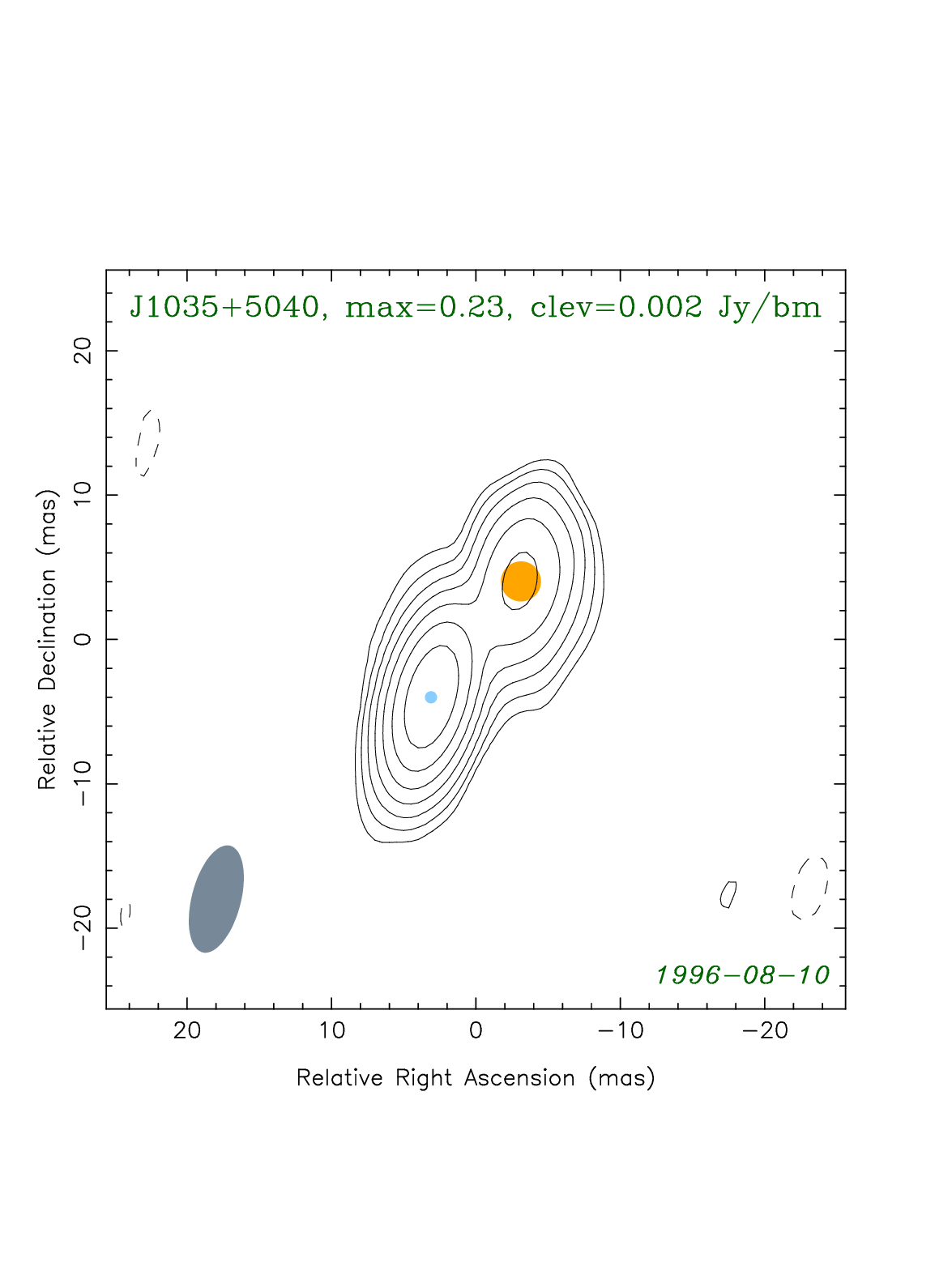}
 \includegraphics[width=0.44\textwidth,angle=0,trim=0 3cm 1cm 5cm,clip]{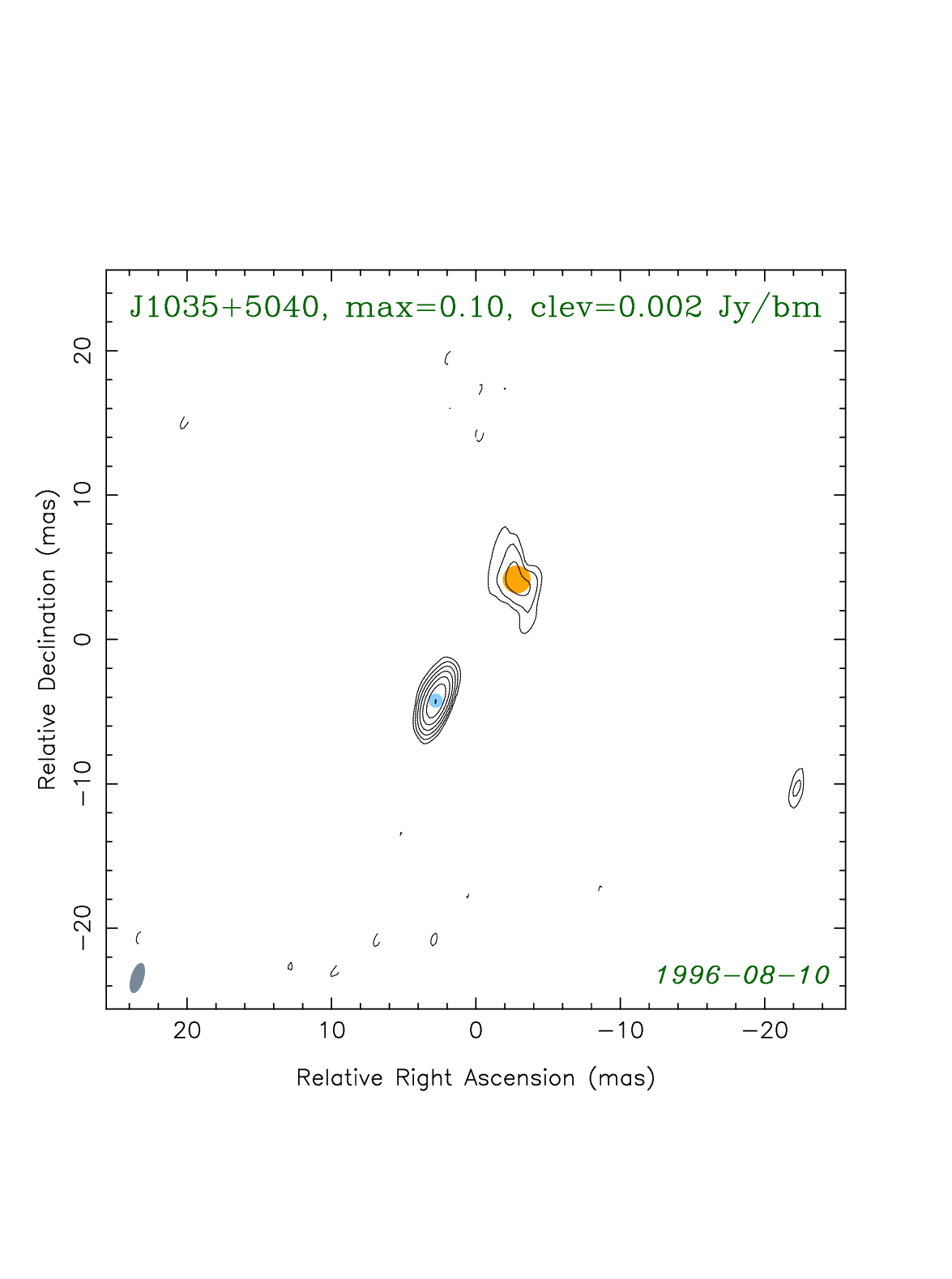}

 \includegraphics[width=0.44\textwidth,angle=0,trim=0 3cm 1cm 5cm,clip]{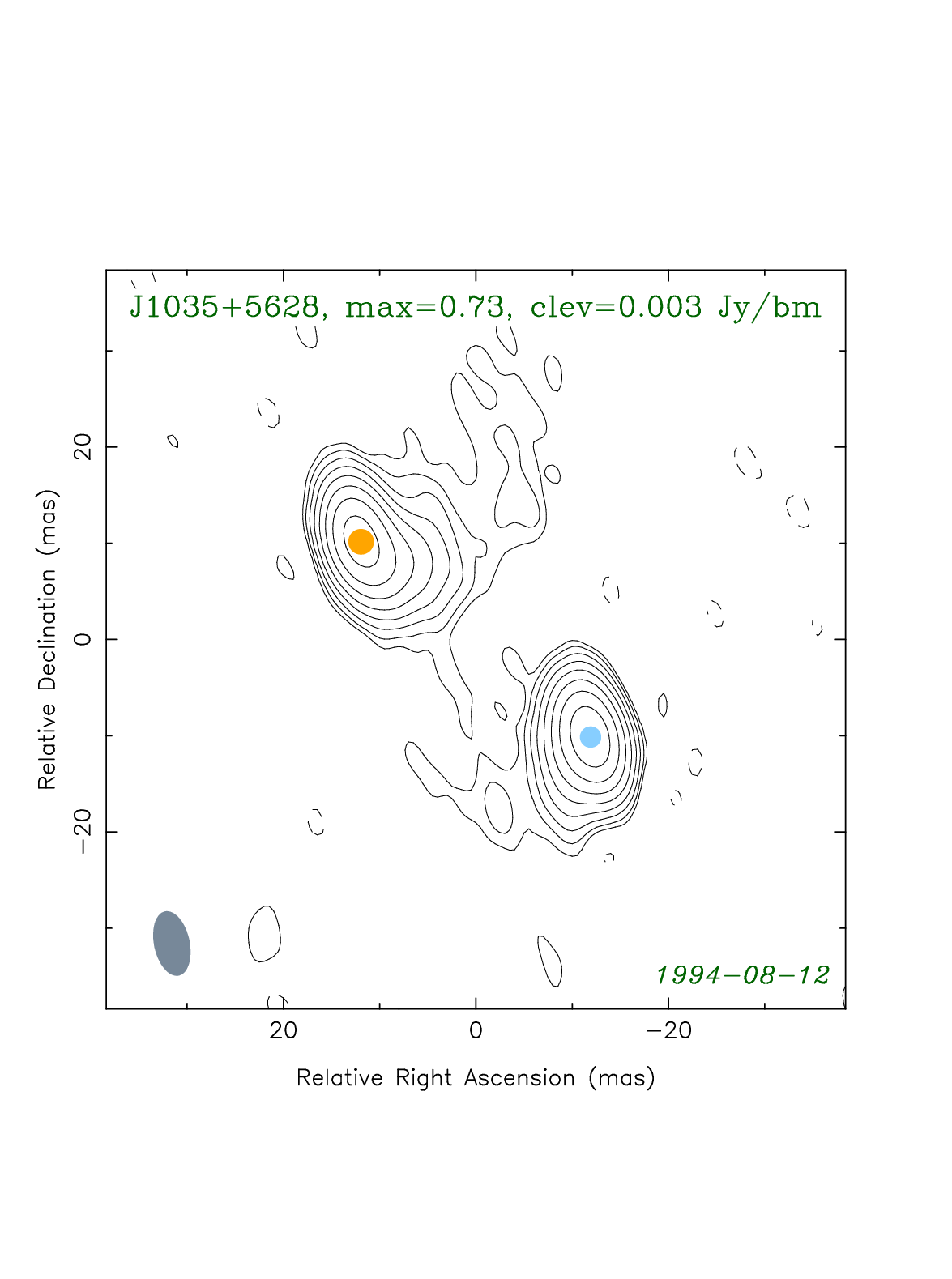}
 \includegraphics[width=0.44\textwidth,angle=0,trim=0 3cm 1cm 5cm,clip]{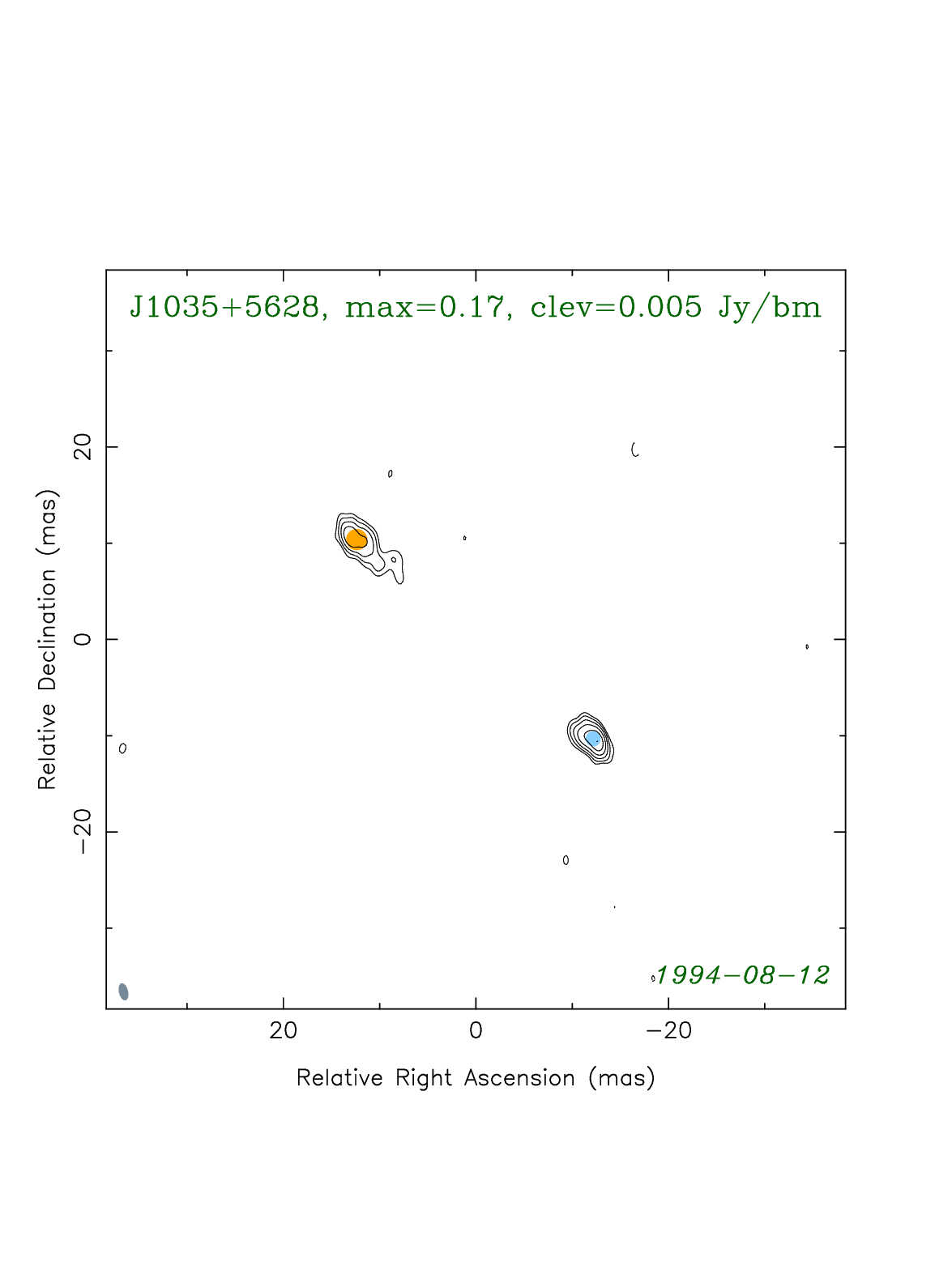}

 \includegraphics[width=0.44\textwidth,angle=0,trim=0 3cm 1cm 5cm,clip]{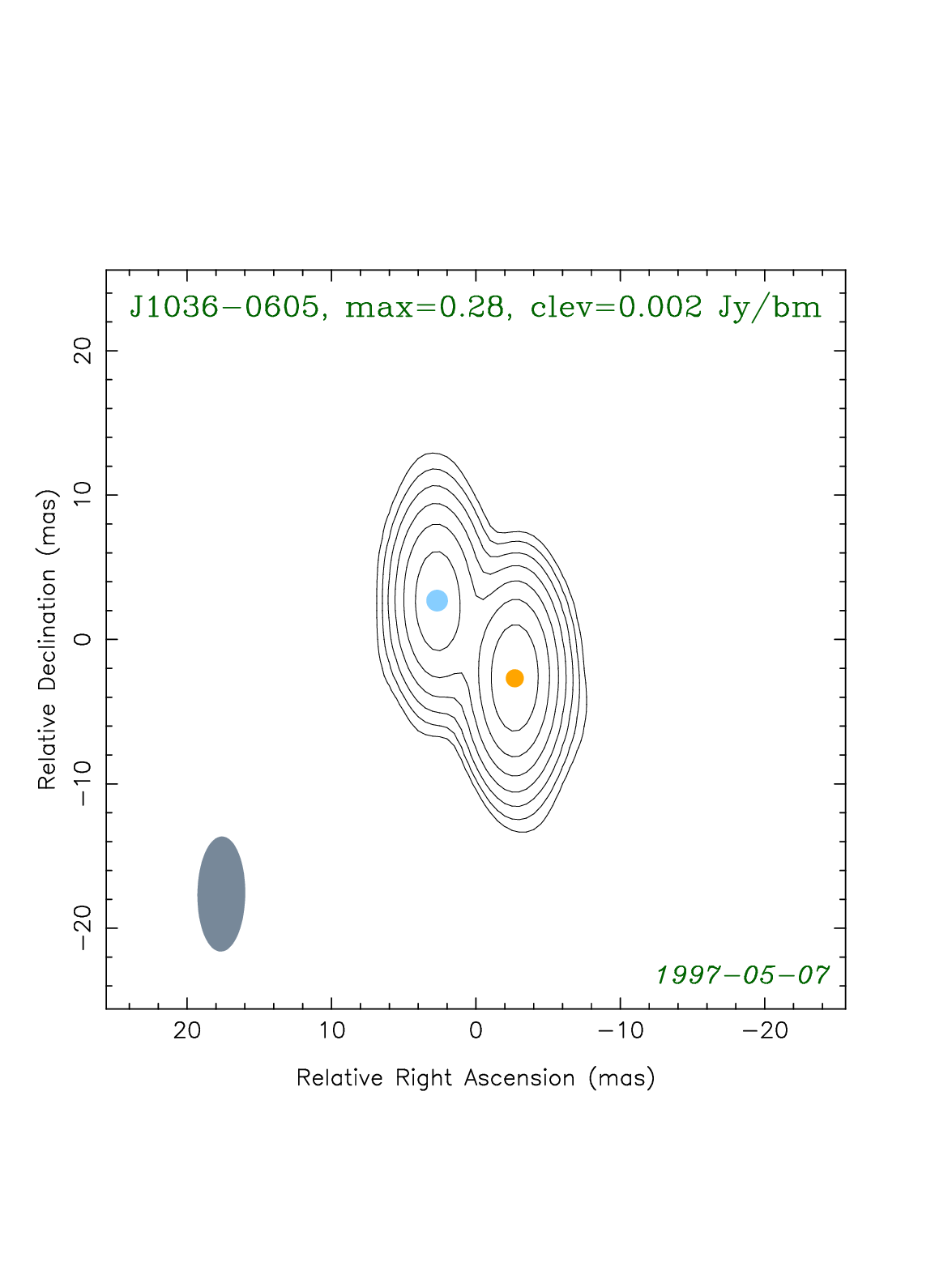}
 \includegraphics[width=0.44\textwidth,angle=0,trim=0 3cm 1cm 5cm,clip]{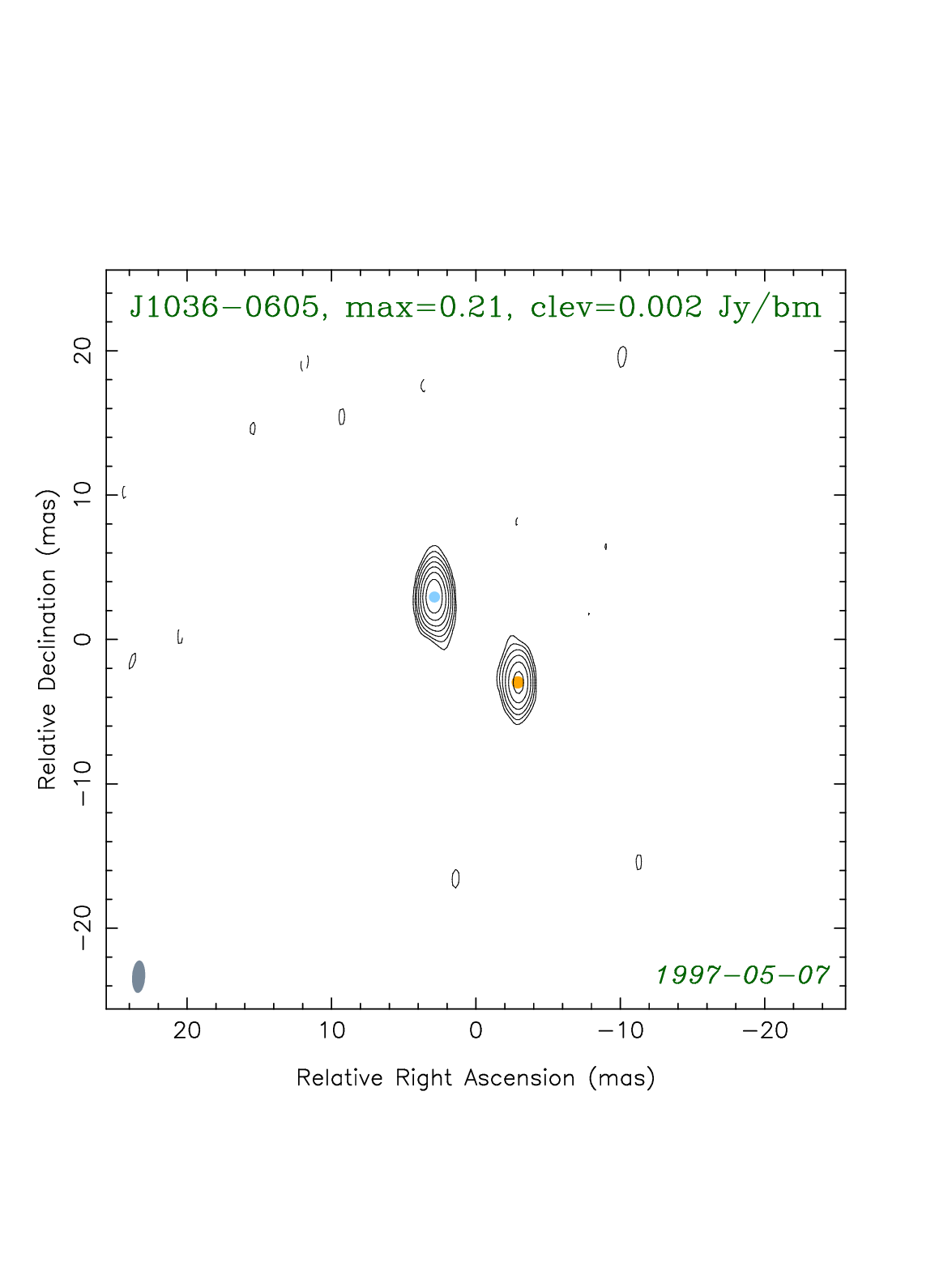}

 \caption{continued}
\end{figure*}

\clearpage
\addtocounter{figure}{-1}
\begin{figure*}
 \centering

 \includegraphics[width=0.44\textwidth,angle=0,trim=0 3cm 1cm 5cm,clip]{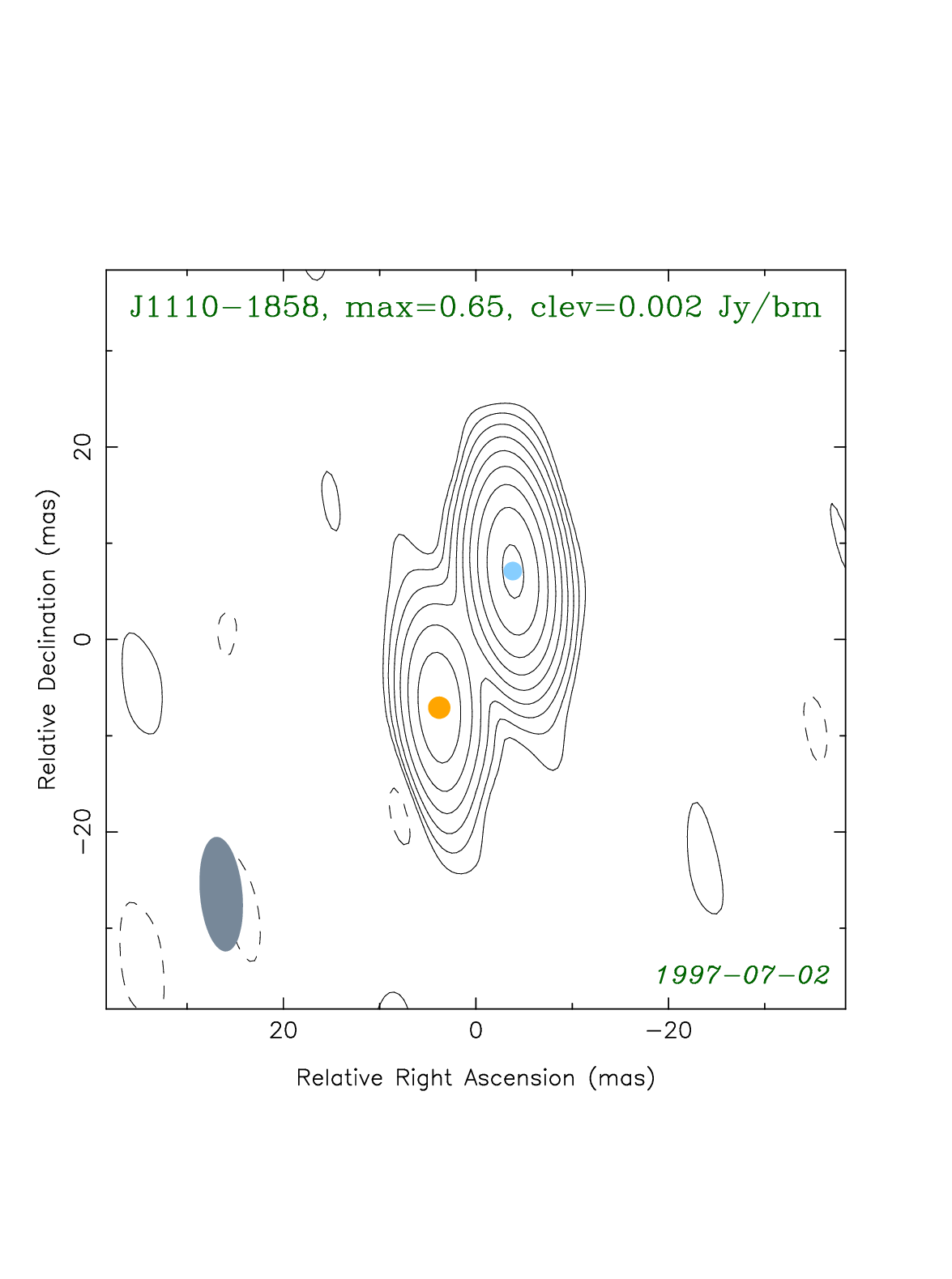}
 \includegraphics[width=0.44\textwidth,angle=0,trim=0 3cm 1cm 5cm,clip]{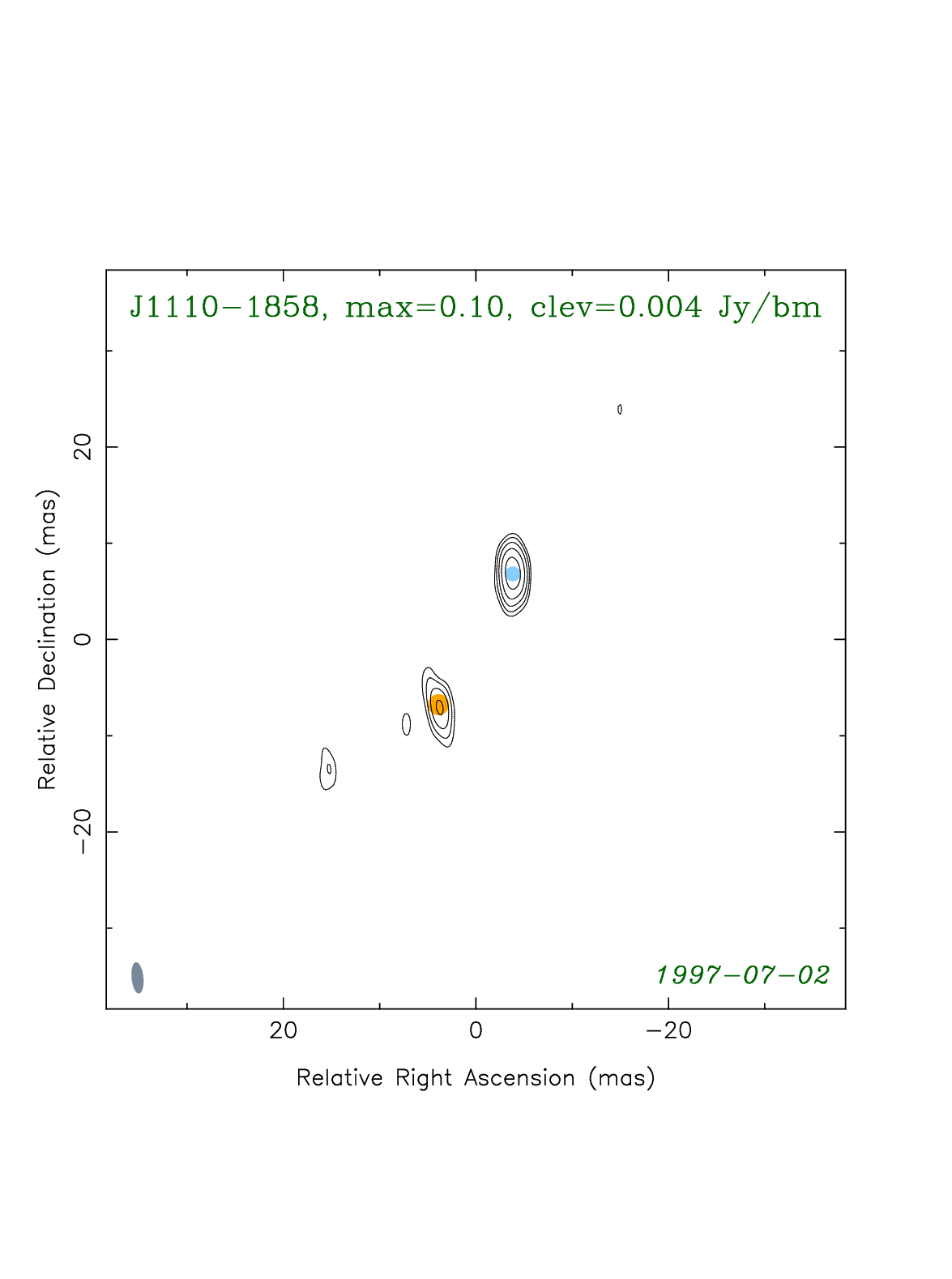}

 \includegraphics[width=0.44\textwidth,angle=0,trim=0 3cm 1cm 5cm,clip]{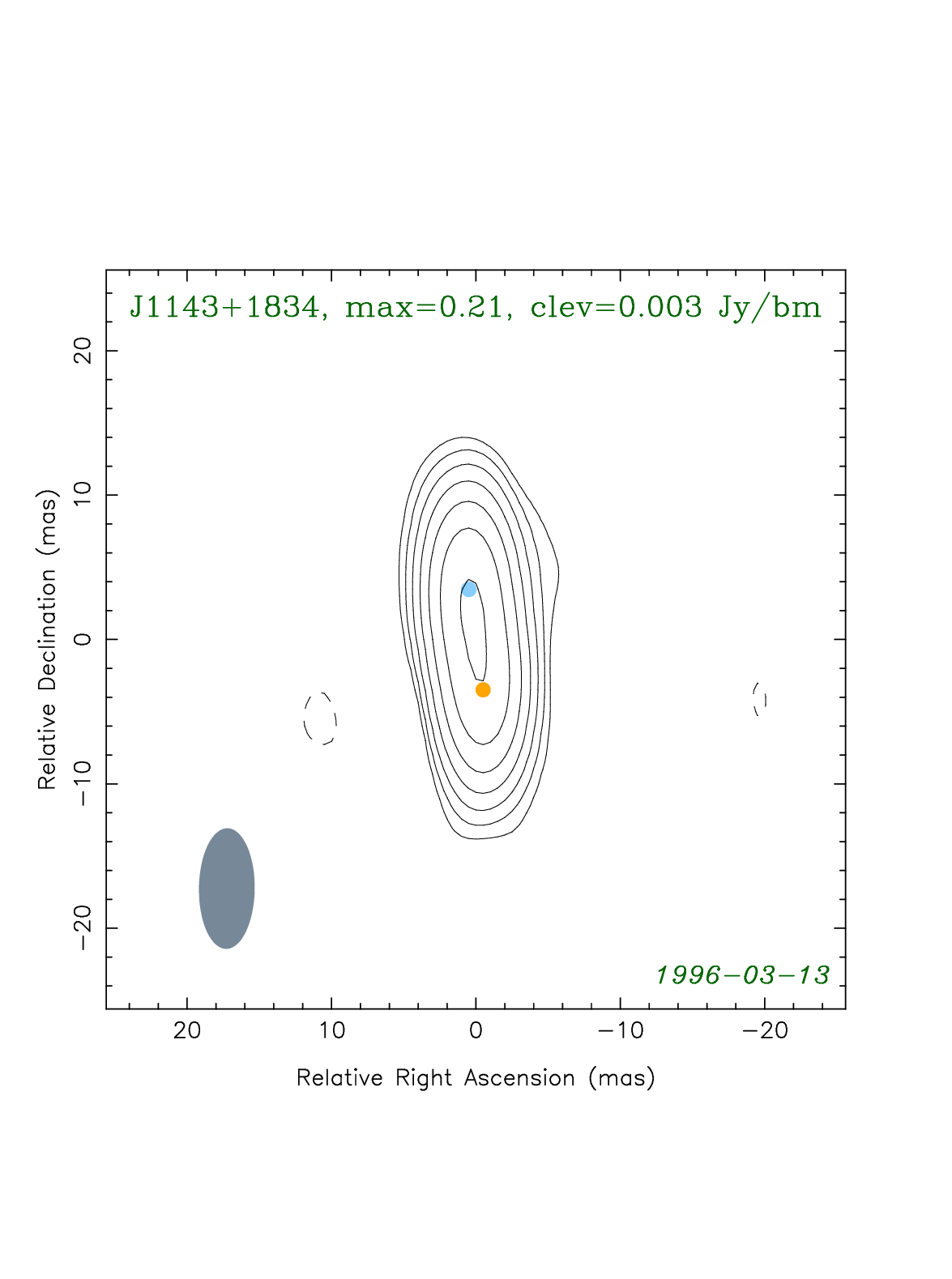}
 \includegraphics[width=0.44\textwidth,angle=0,trim=0 3cm 1cm 5cm,clip]{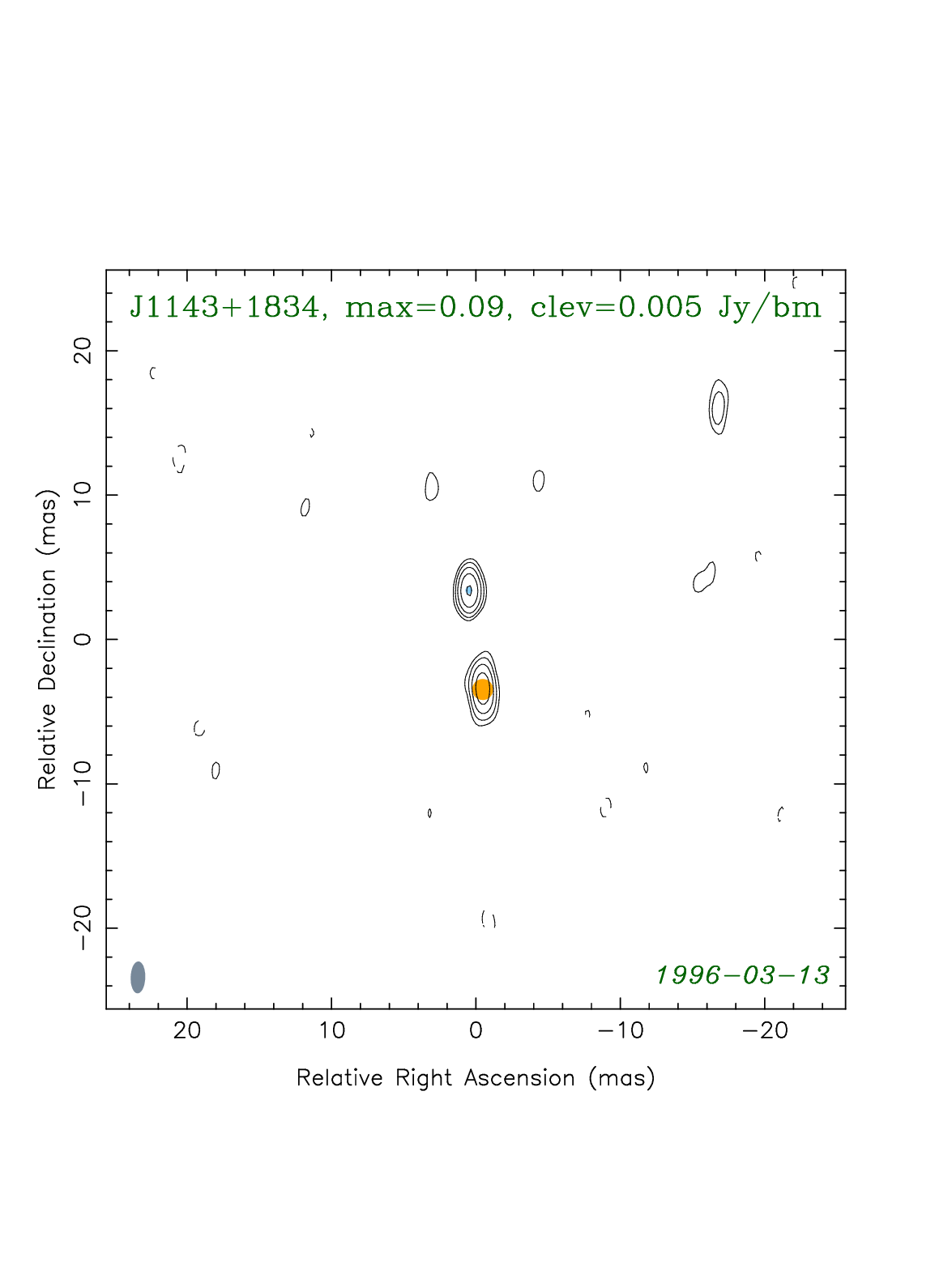}

 \includegraphics[width=0.44\textwidth,angle=0,trim=0 3cm 1cm 5cm,clip]{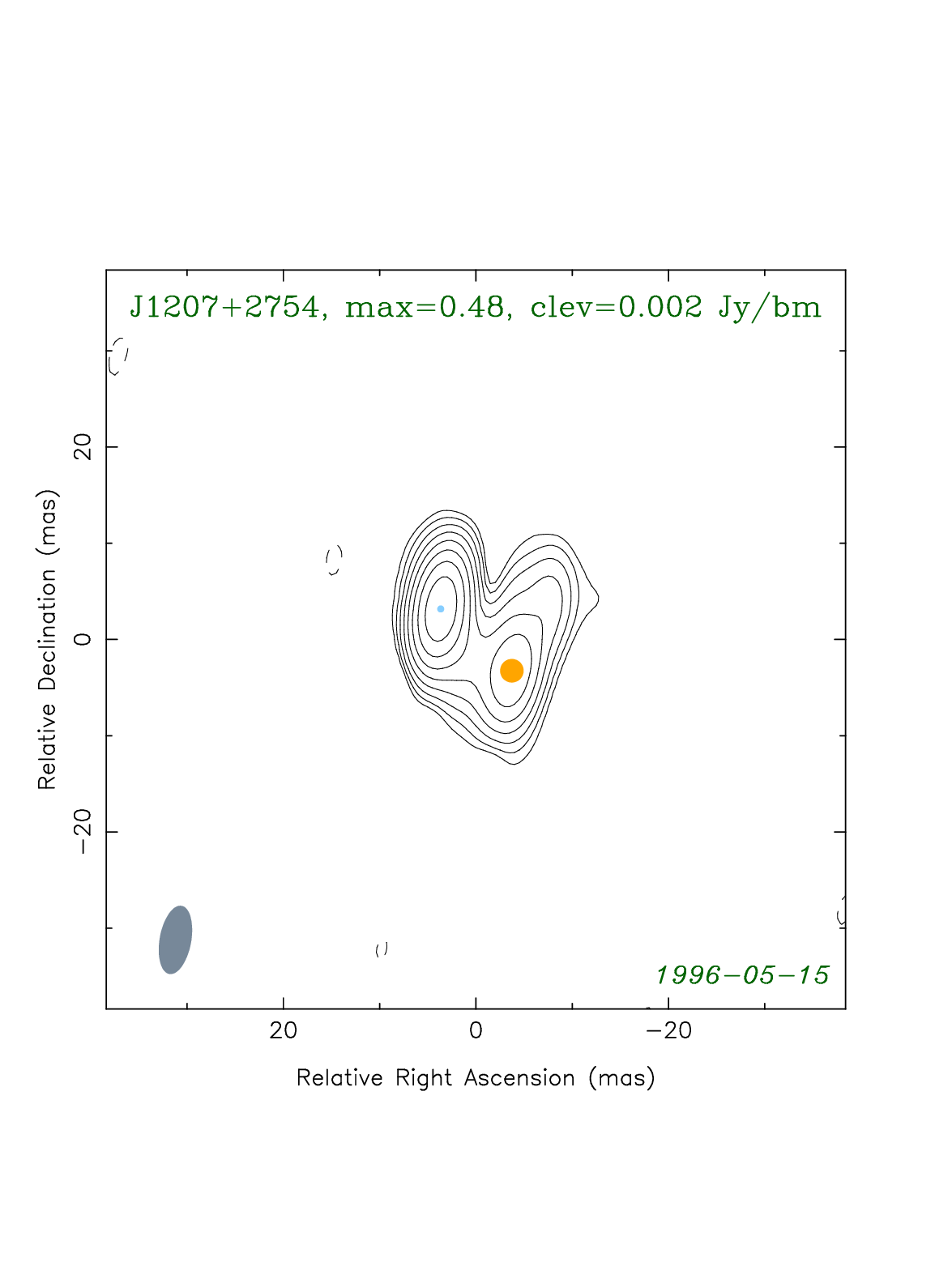}
 \includegraphics[width=0.44\textwidth,angle=0,trim=0 3cm 1cm 5cm,clip]{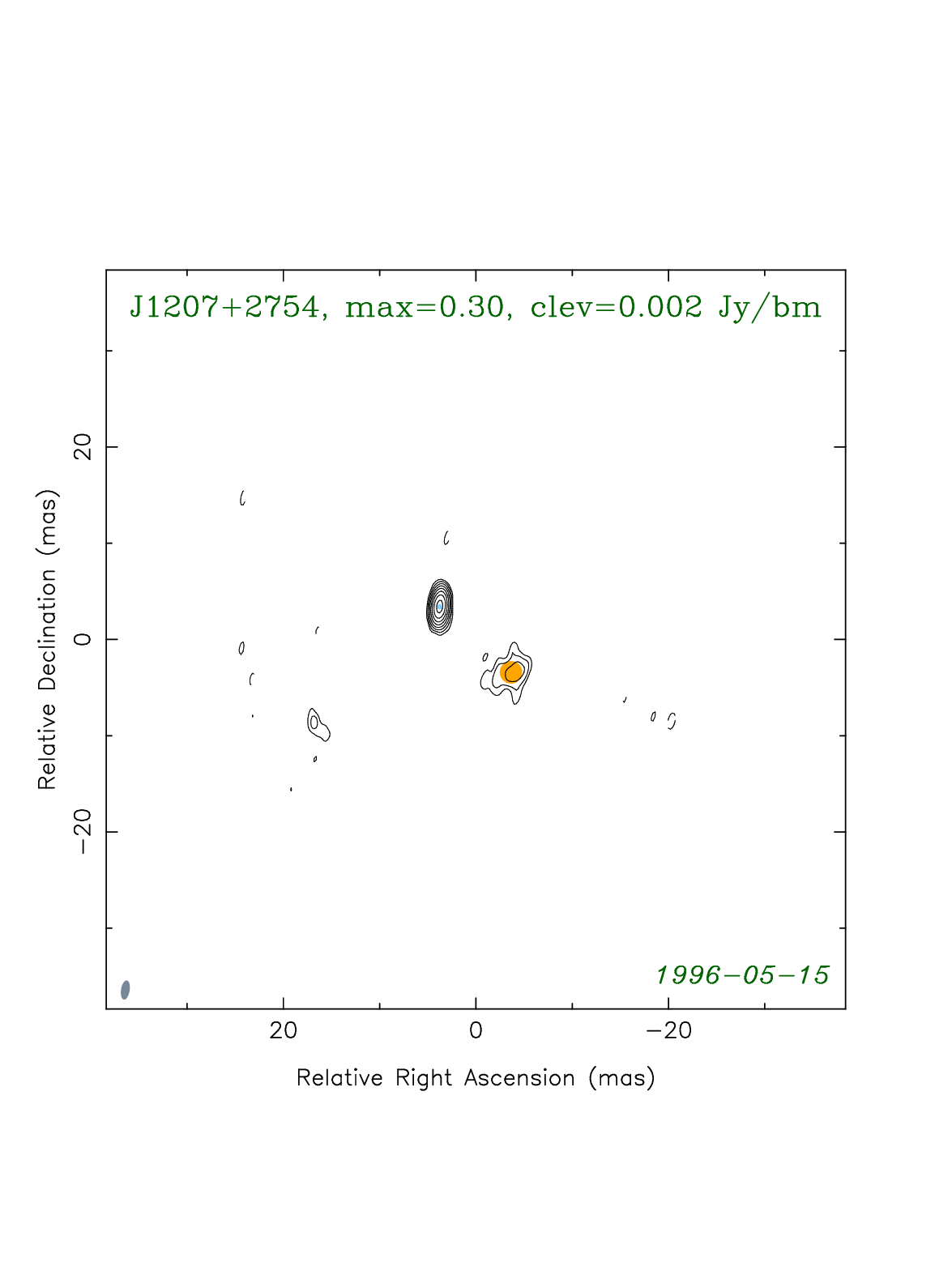}

 \caption{continued}
\end{figure*}

\clearpage
\addtocounter{figure}{-1}
\begin{figure*}
 \centering

 \includegraphics[width=0.44\textwidth,angle=0,trim=0 3cm 1cm 5cm,clip]{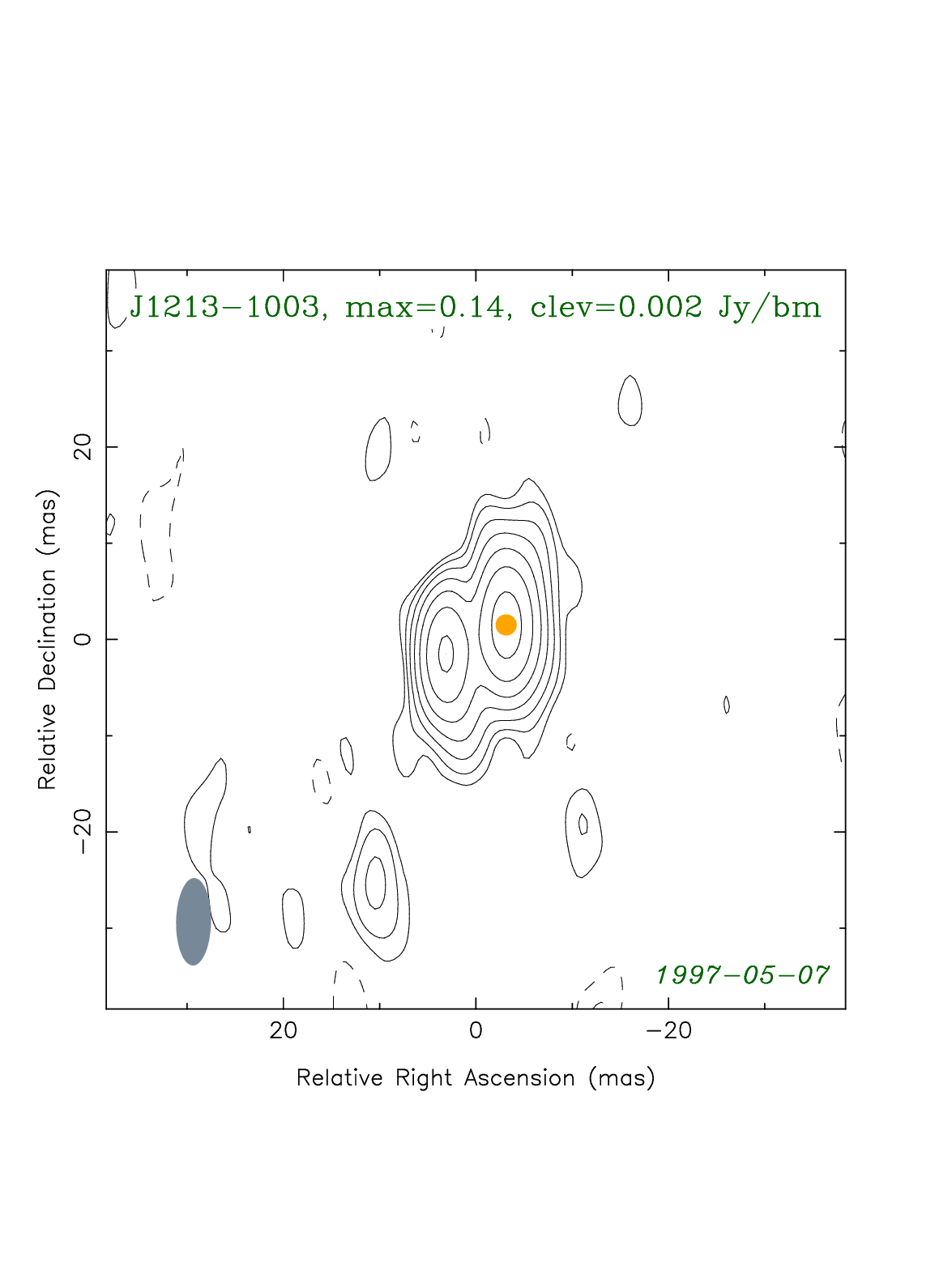}
 \includegraphics[width=0.44\textwidth,angle=0,trim=0 3cm 1cm 5cm,clip]{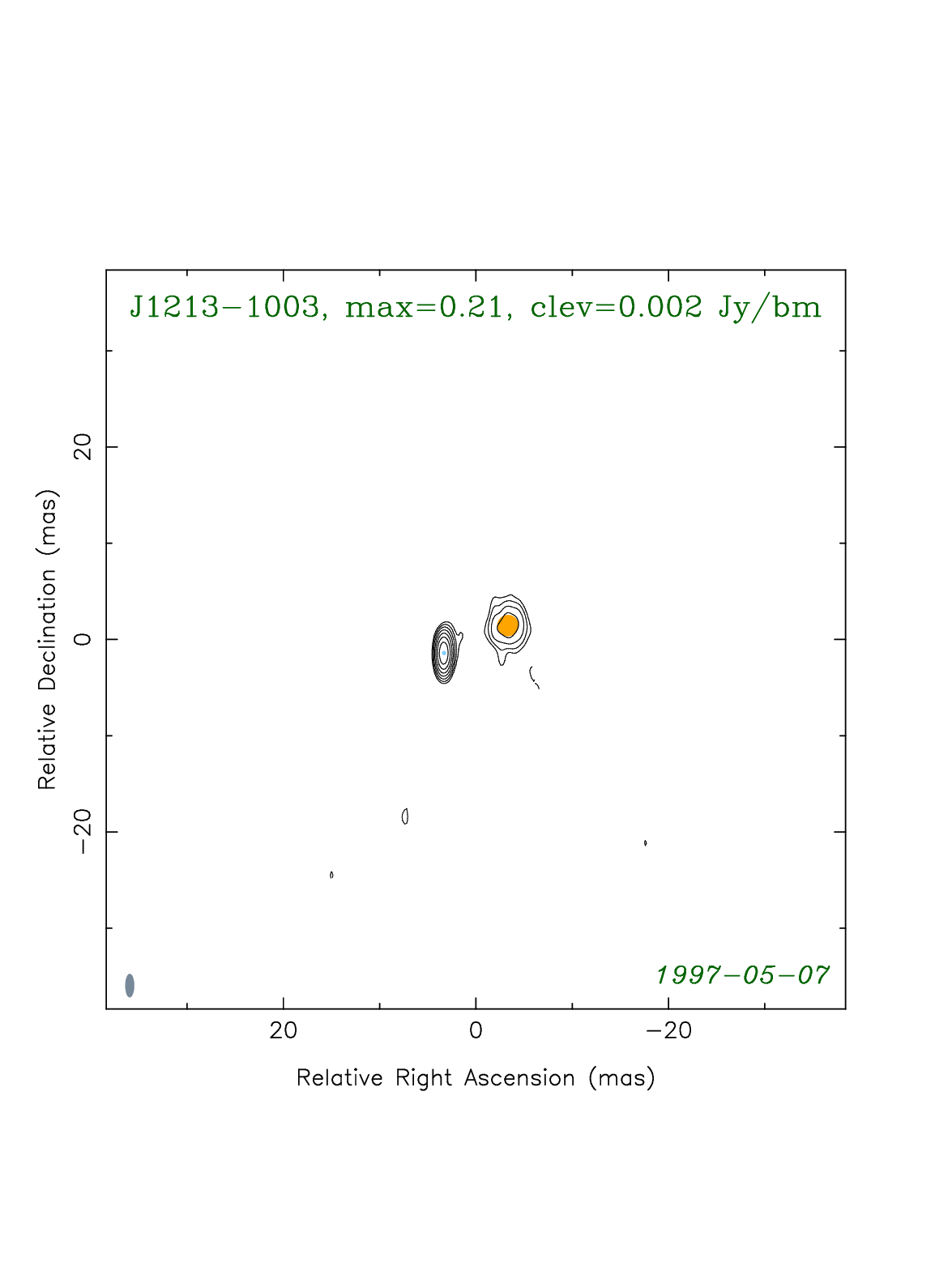}

 \includegraphics[width=0.44\textwidth,angle=0,trim=0 3cm 1cm 5cm,clip]{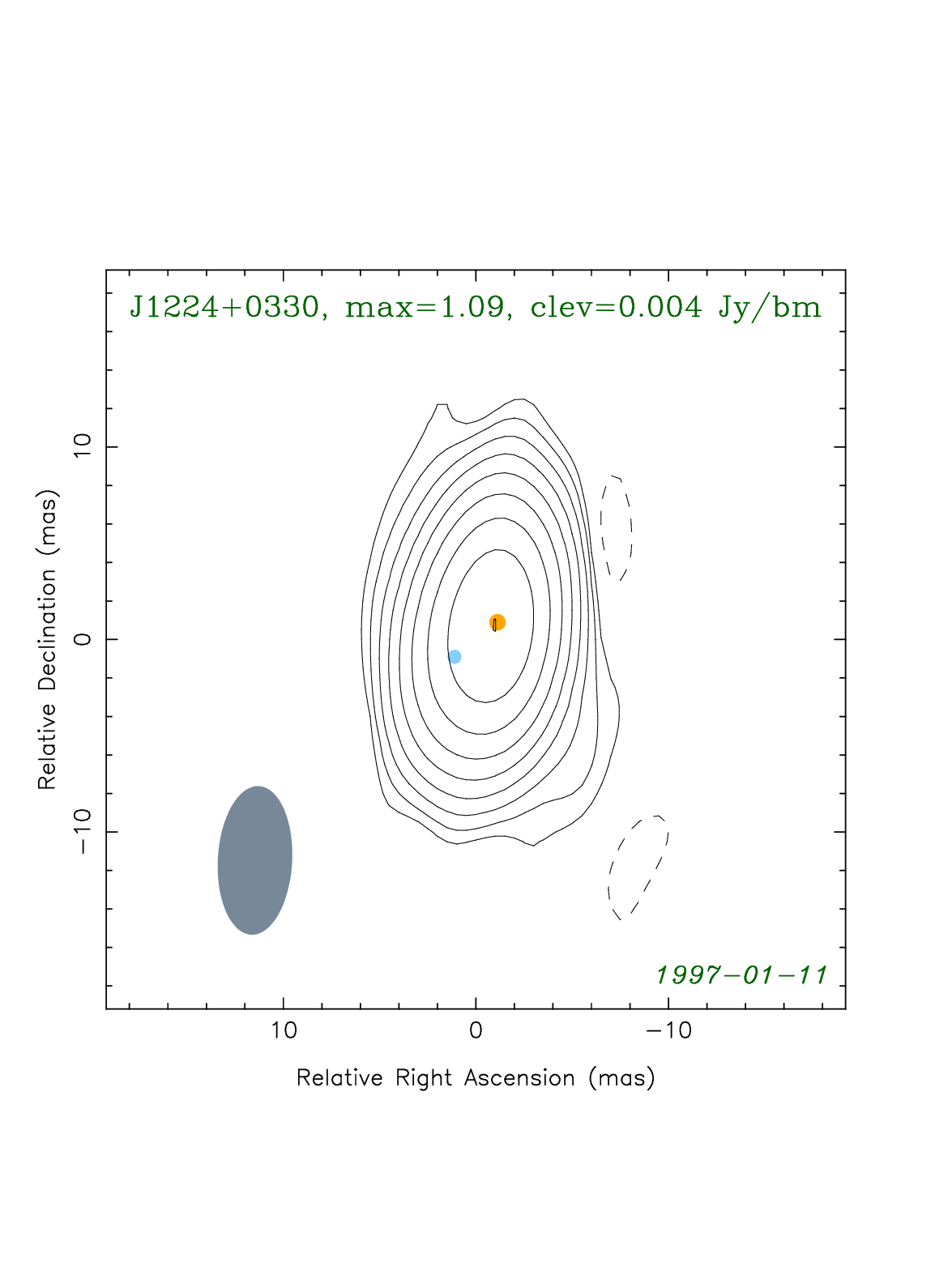}
 \includegraphics[width=0.44\textwidth,angle=0,trim=0 3cm 1cm 5cm,clip]{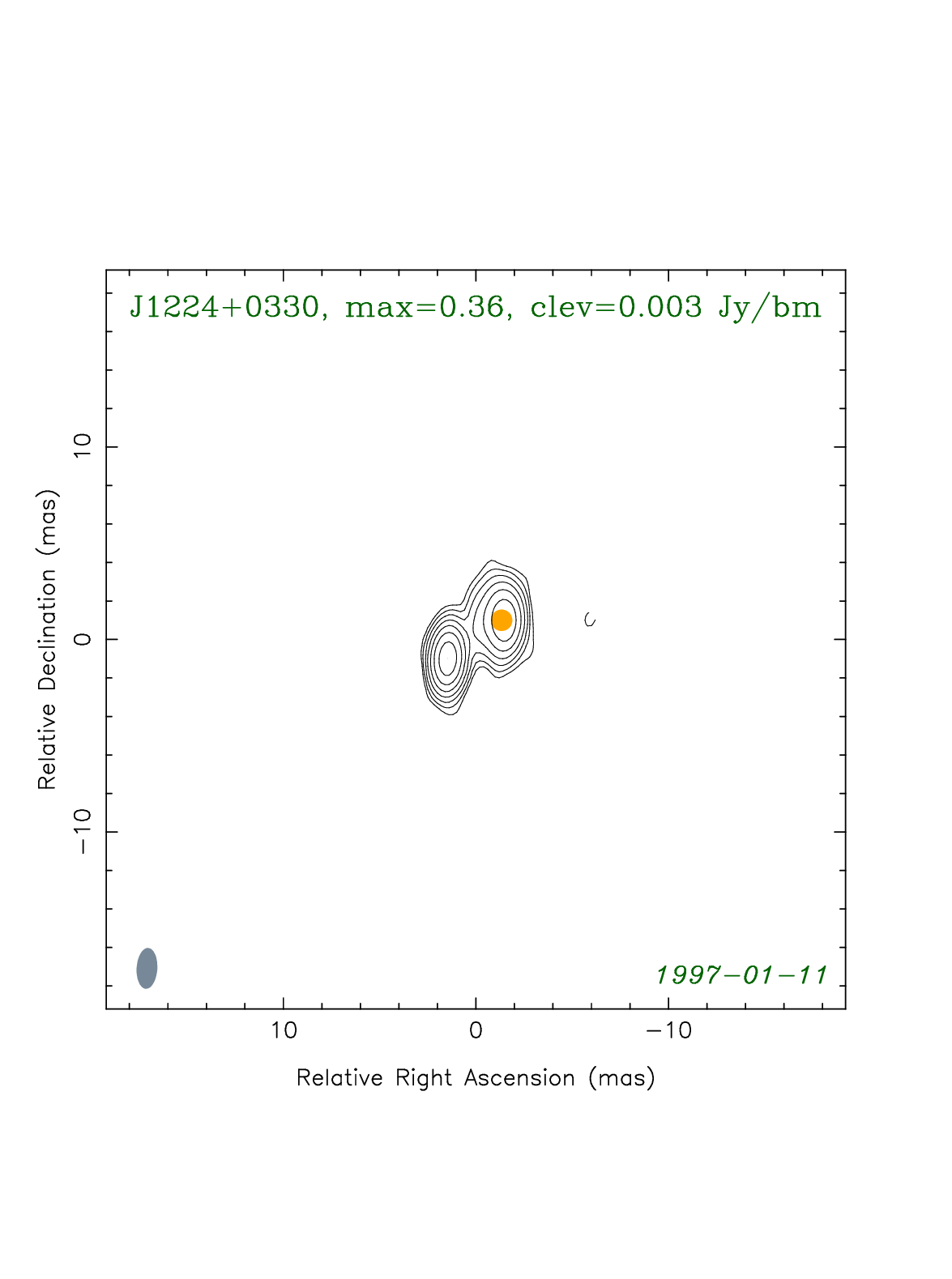}

 \includegraphics[width=0.44\textwidth,angle=0,trim=0 3cm 1cm 5cm,clip]{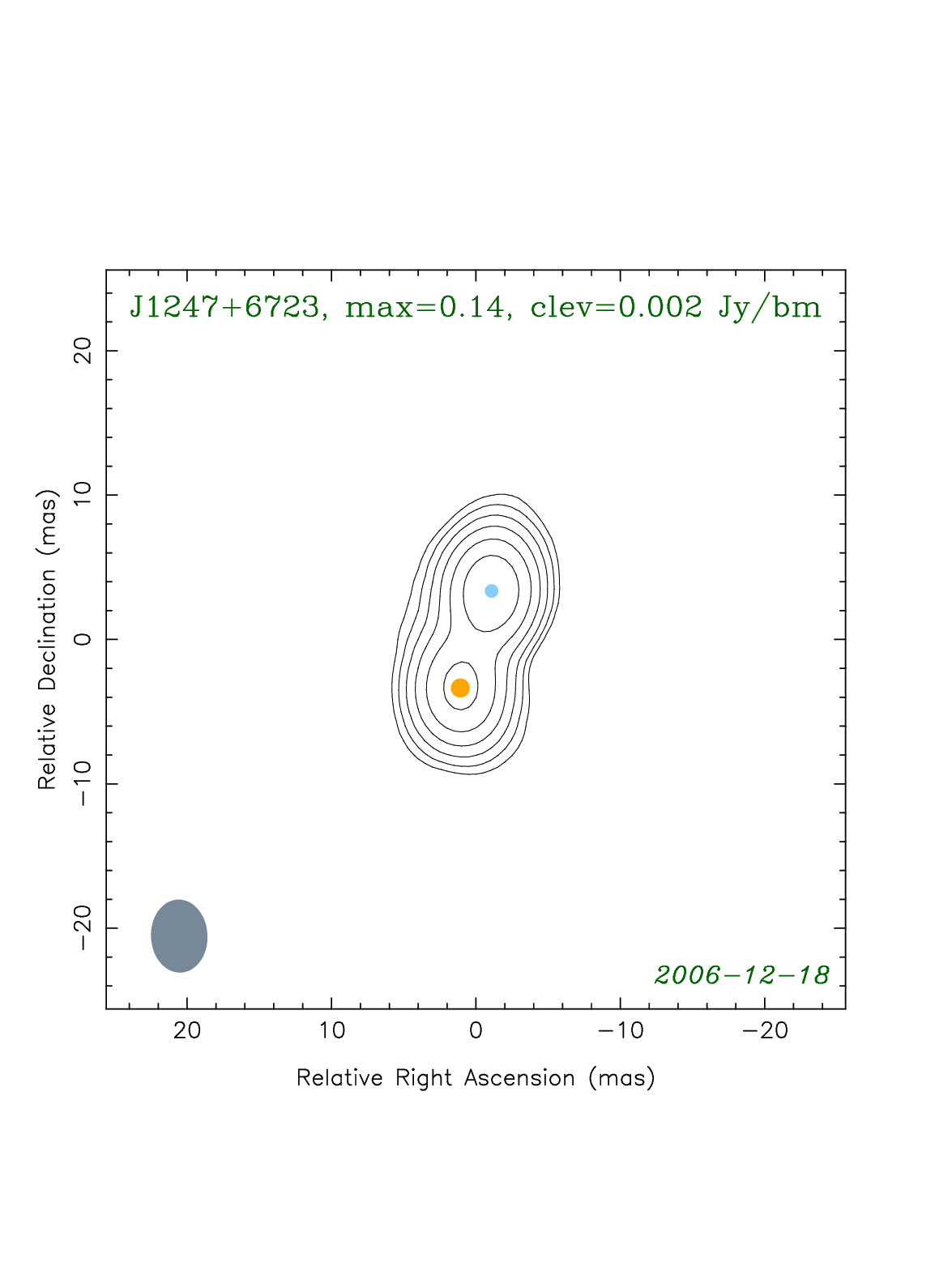}
 \includegraphics[width=0.44\textwidth,angle=0,trim=0 3cm 1cm 5cm,clip]{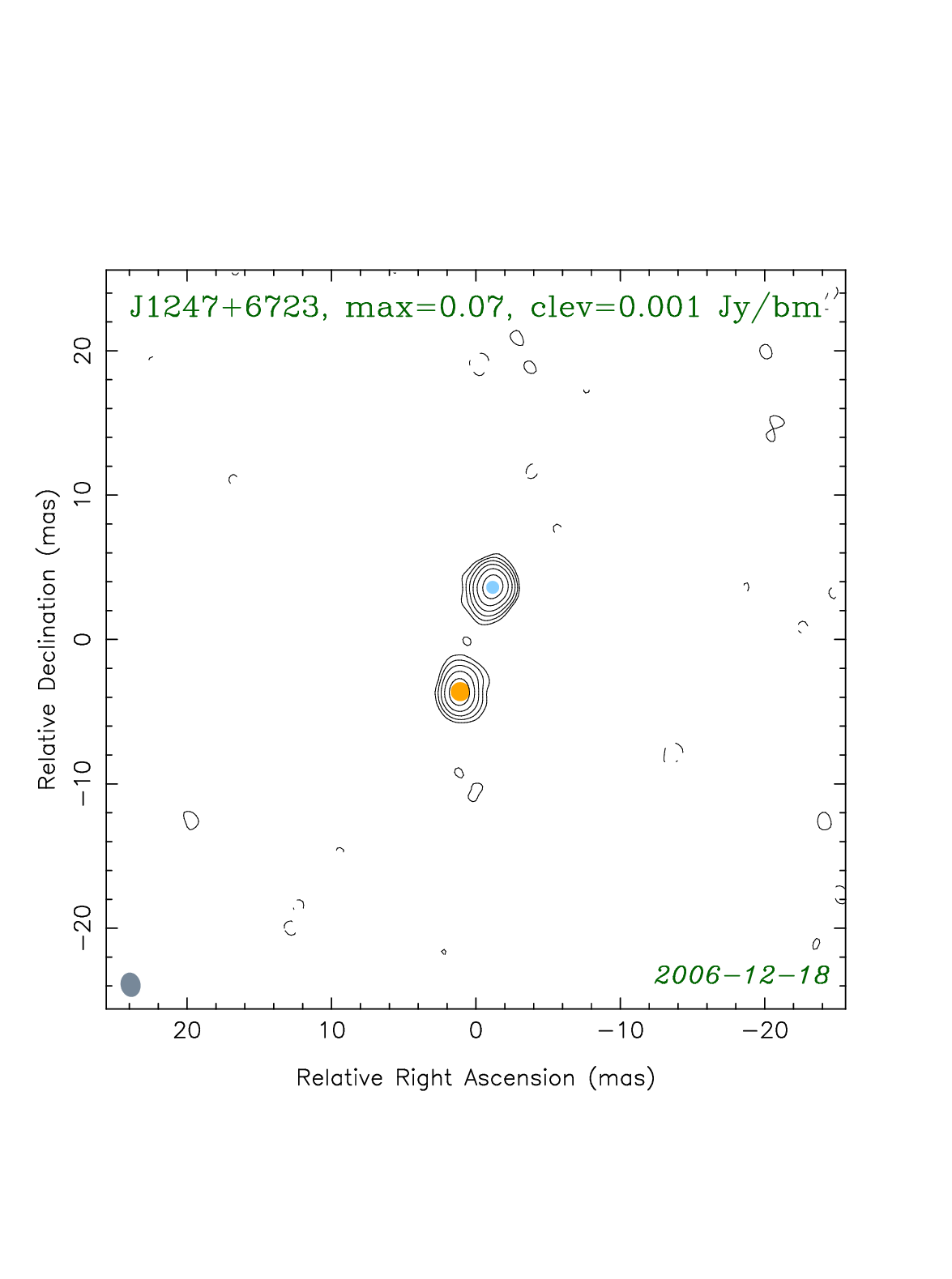}

 \caption{continued}
\end{figure*}

\clearpage
\addtocounter{figure}{-1}
\begin{figure*}
 \centering

 \includegraphics[width=0.44\textwidth,angle=0,trim=0 3cm 1cm 5cm,clip]{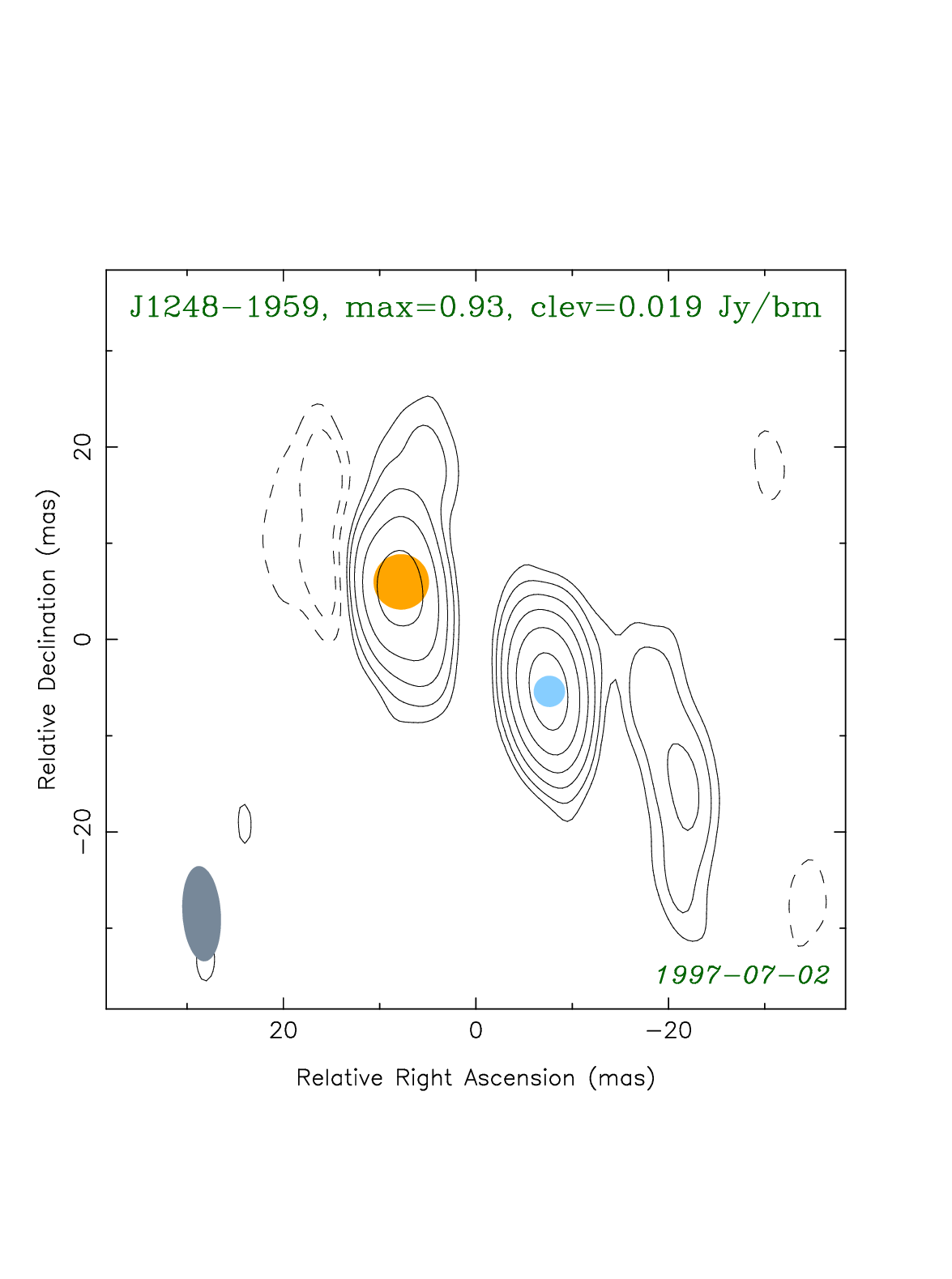}
 \includegraphics[width=0.44\textwidth,angle=0,trim=0 3cm 1cm 5cm,clip]{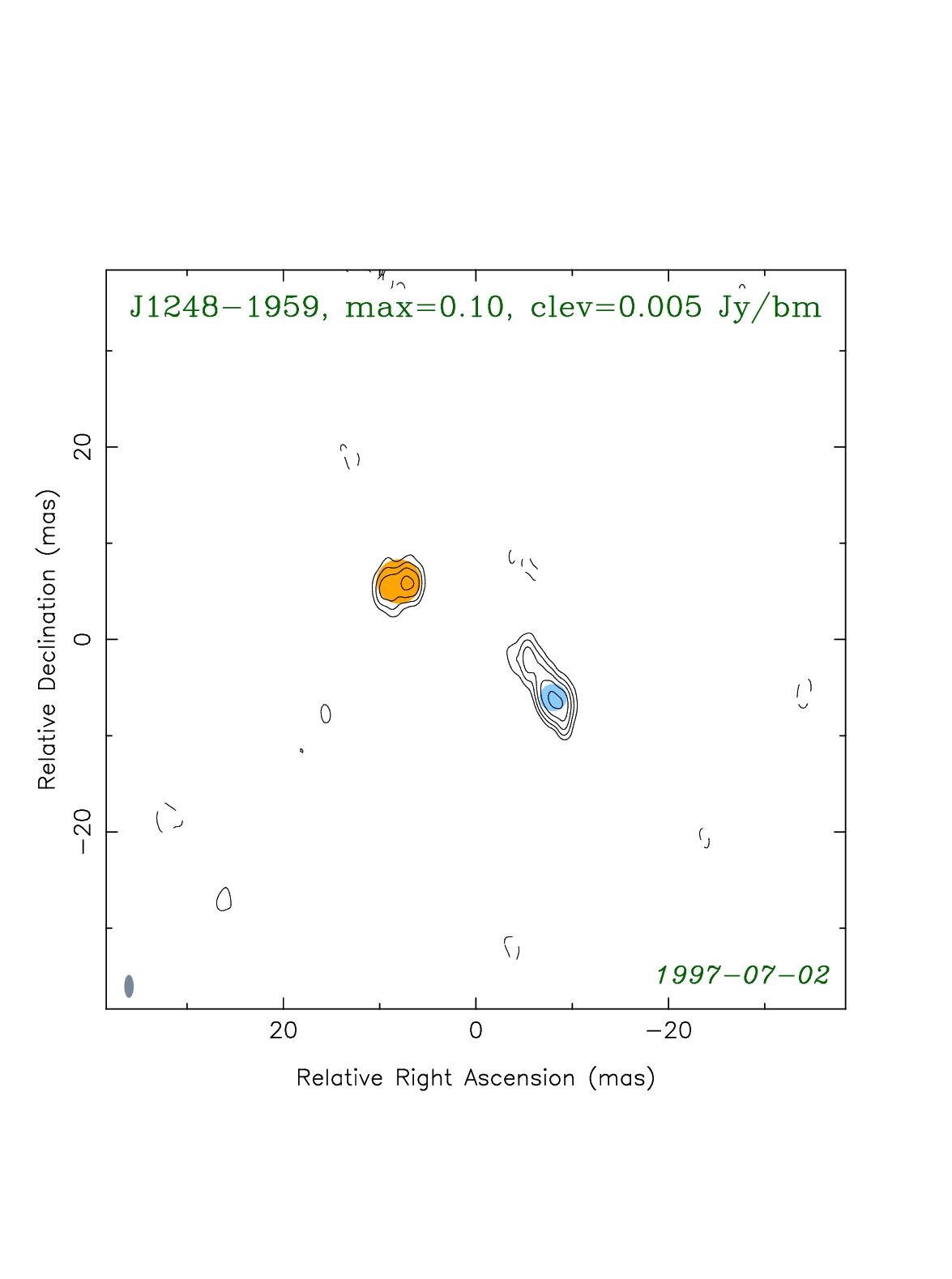}

 \includegraphics[width=0.44\textwidth,angle=0,trim=0 3cm 1cm 5cm,clip]{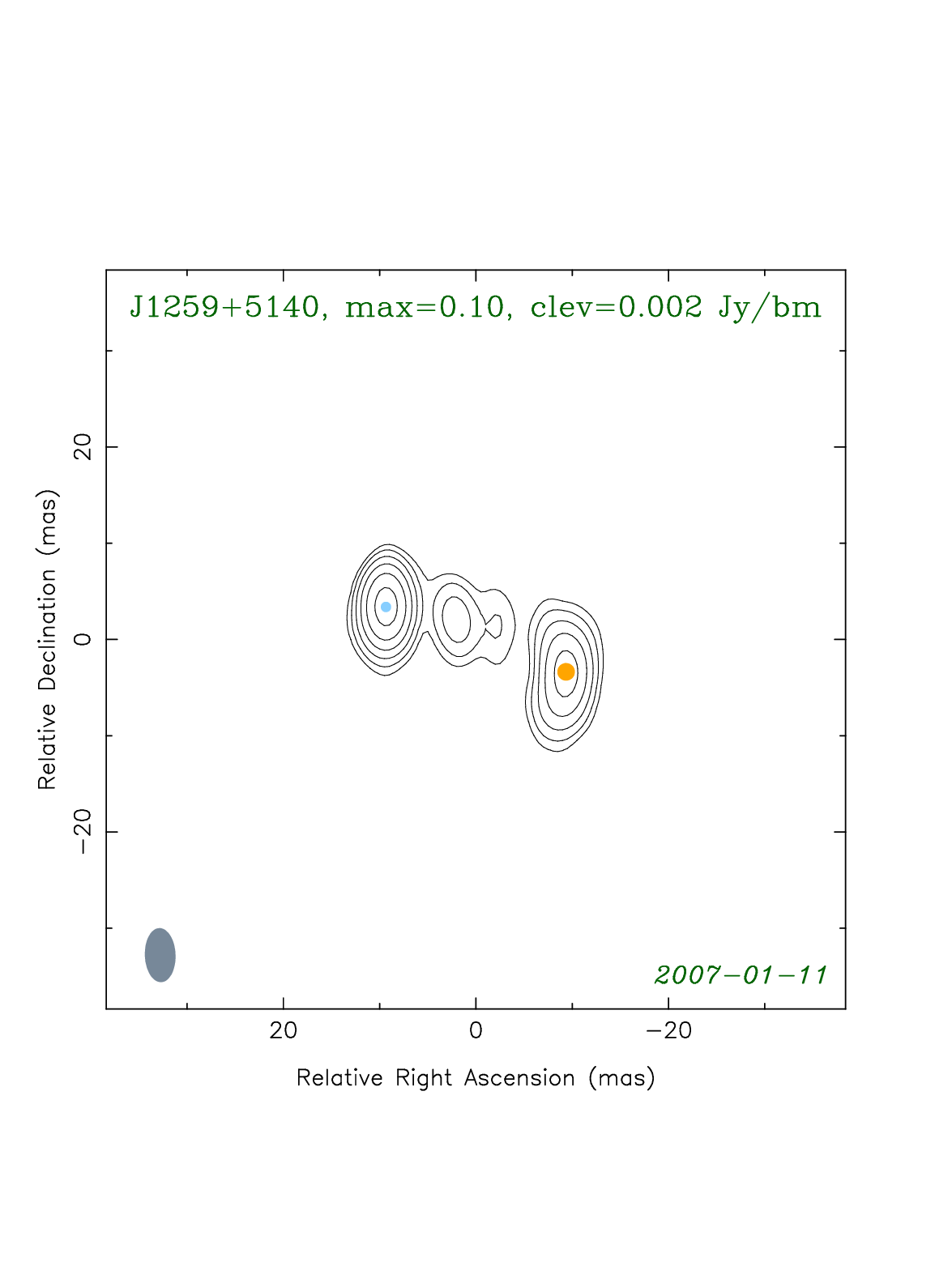}
 \includegraphics[width=0.44\textwidth,angle=0,trim=0 3cm 1cm 5cm,clip]{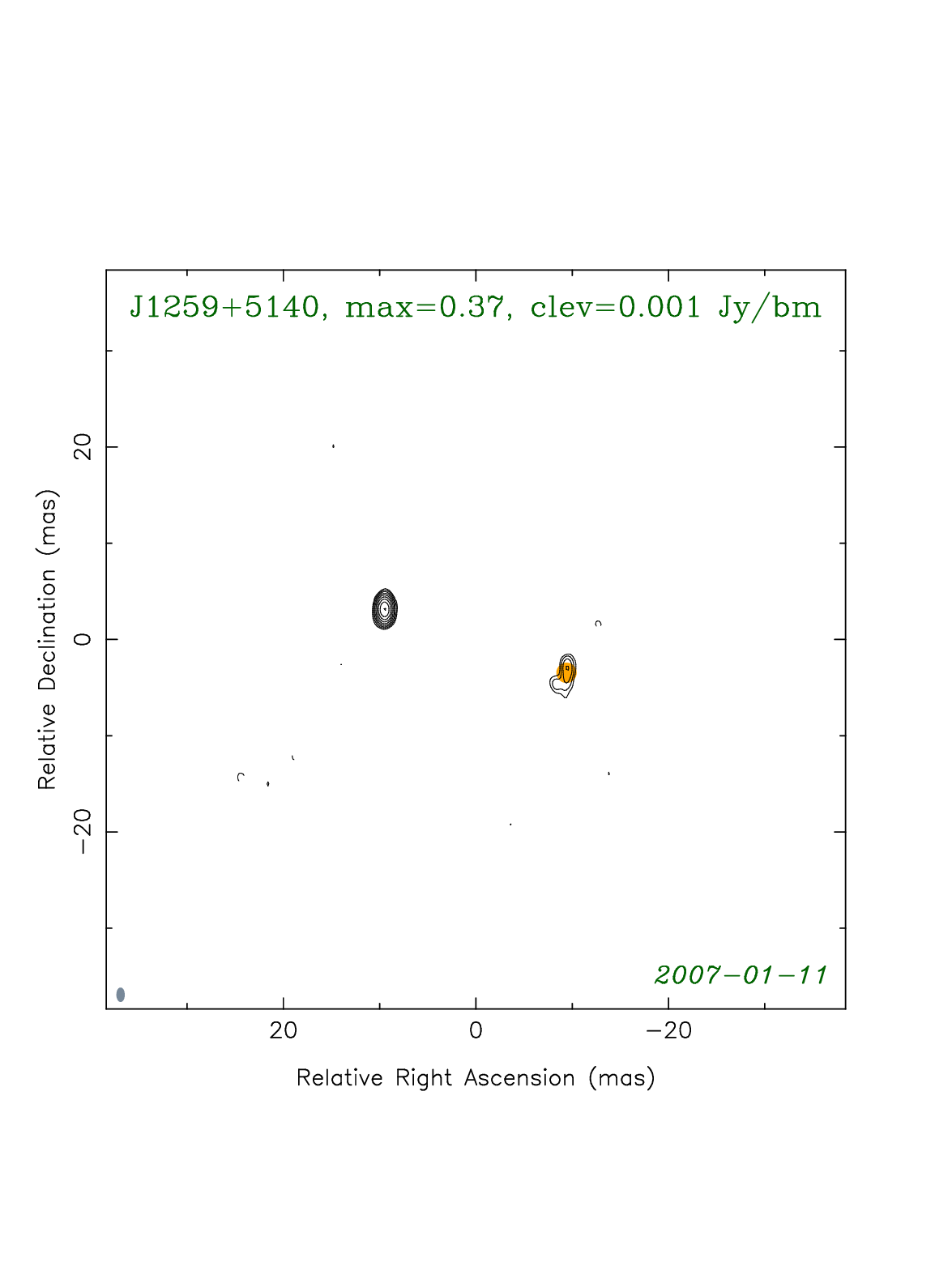}

 \includegraphics[width=0.44\textwidth,angle=0,trim=0 3cm 1cm 5cm,clip]{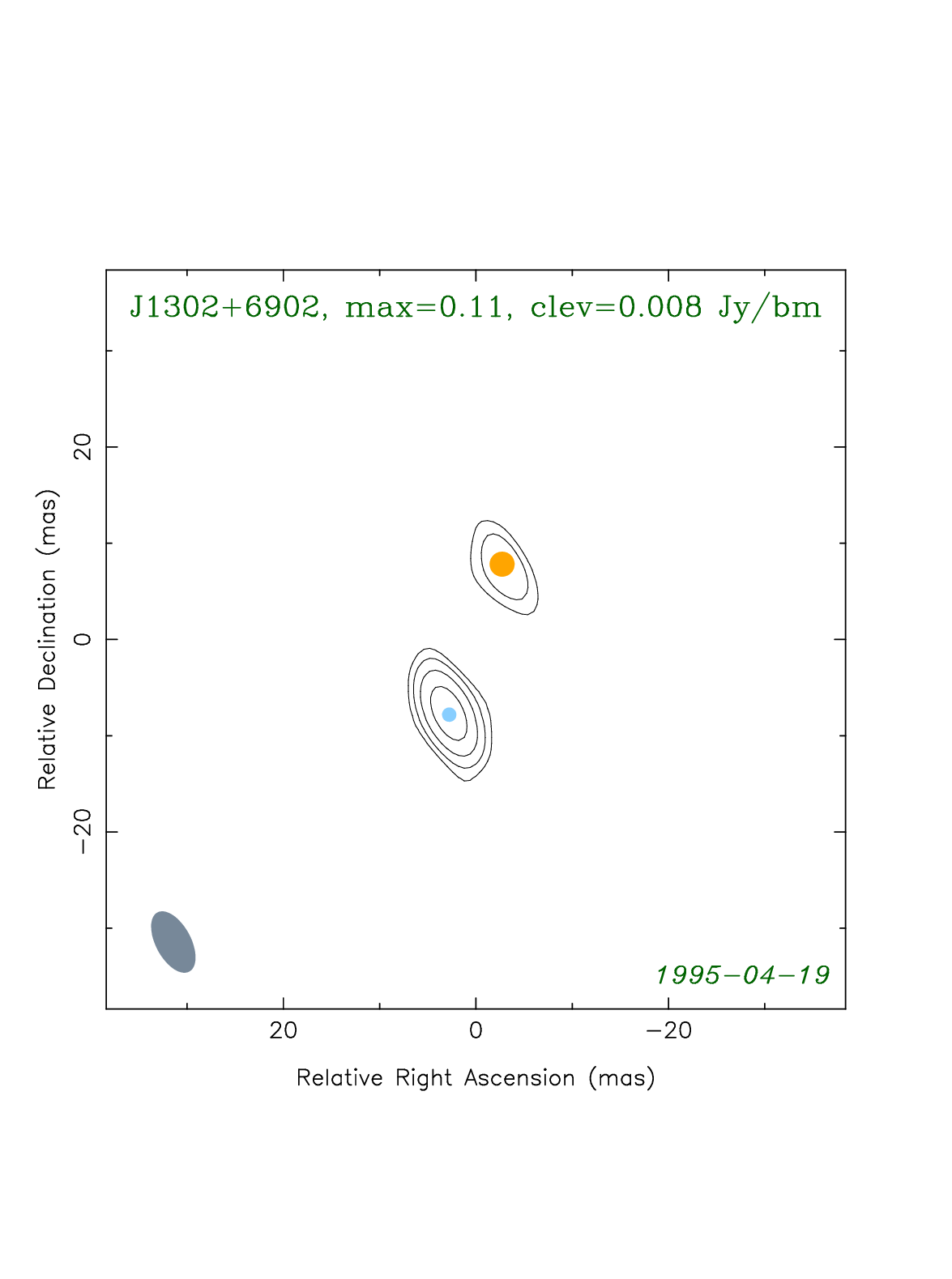}
 \includegraphics[width=0.44\textwidth,angle=0,trim=0 3cm 1cm 5cm,clip]{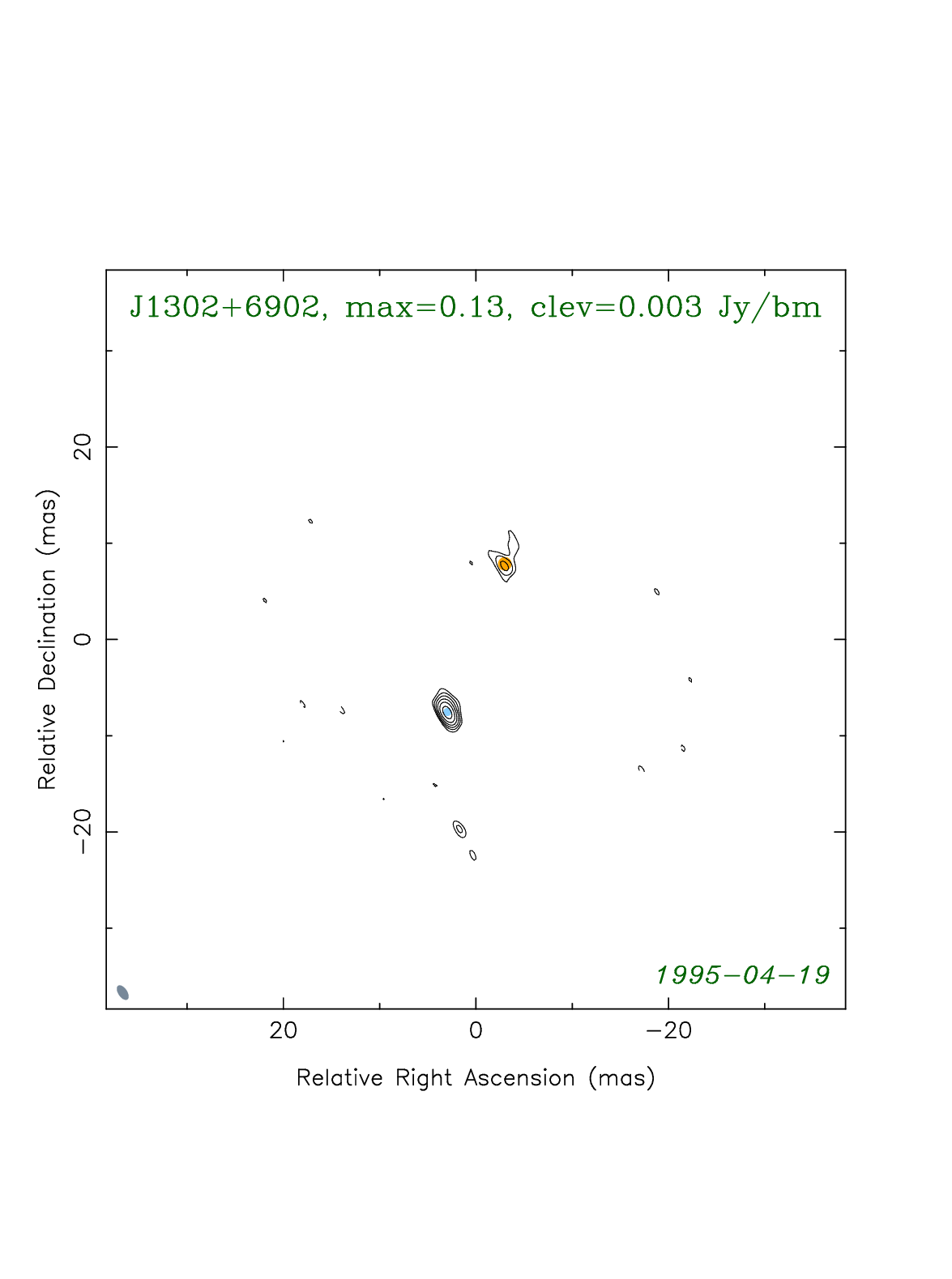}

 \caption{continued}
\end{figure*}

\clearpage
\addtocounter{figure}{-1}
\begin{figure*}
 \centering

 \includegraphics[width=0.44\textwidth,angle=0,trim=0 3cm 1cm 5cm,clip]{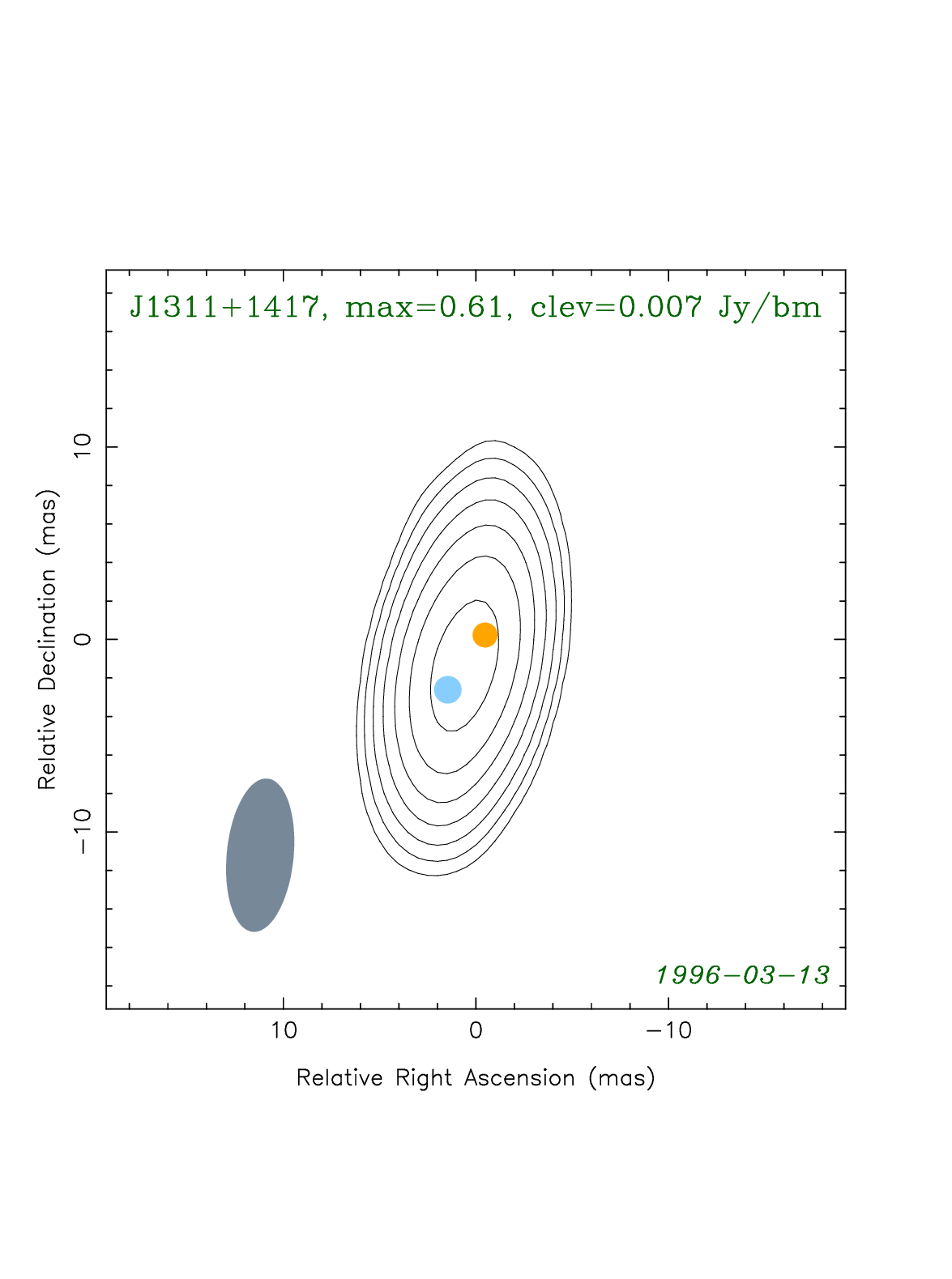}
 \includegraphics[width=0.44\textwidth,angle=0,trim=0 3cm 1cm 5cm,clip]{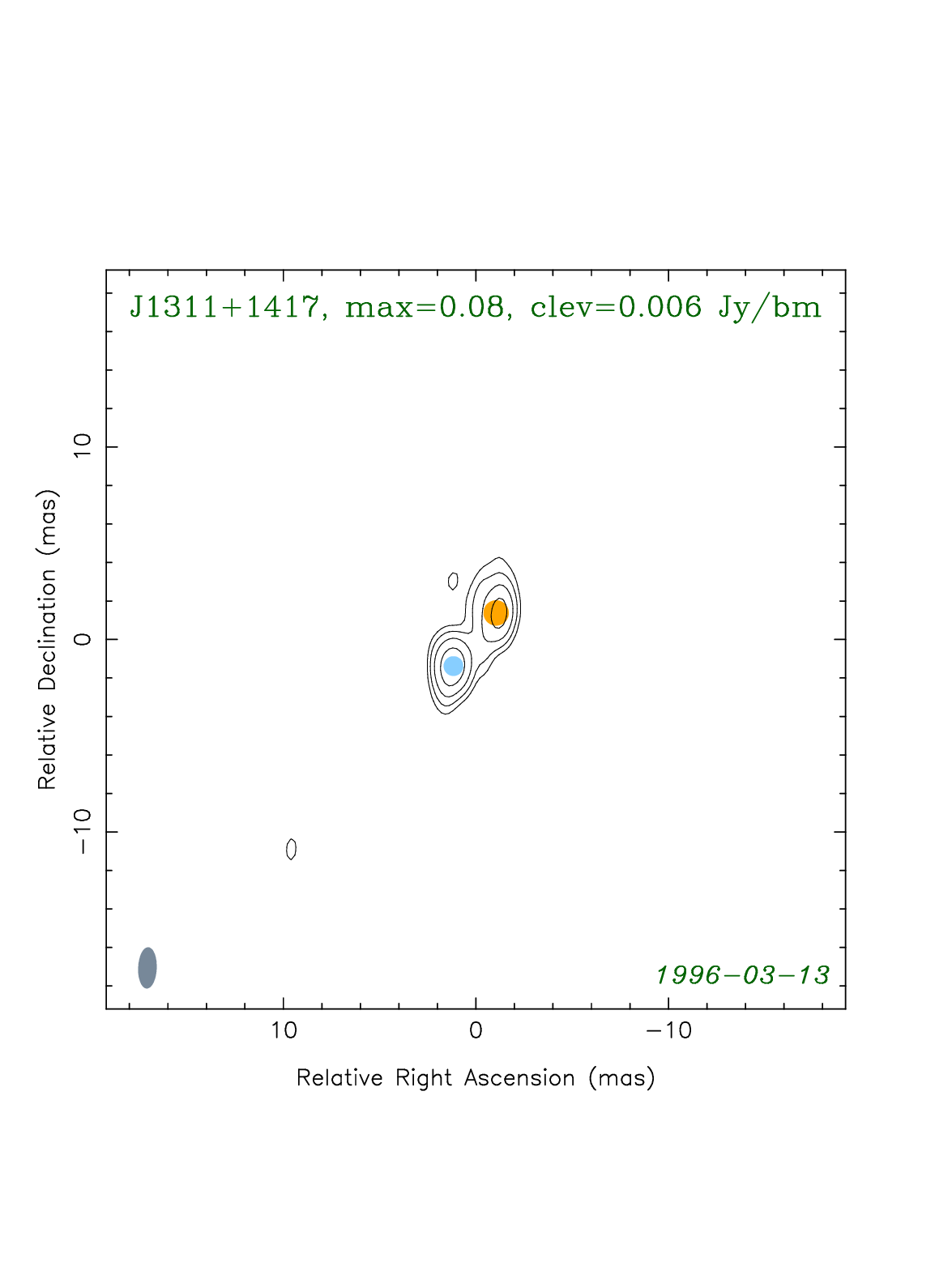}

 \includegraphics[width=0.44\textwidth,angle=0,trim=0 3cm 1cm 5cm,clip]{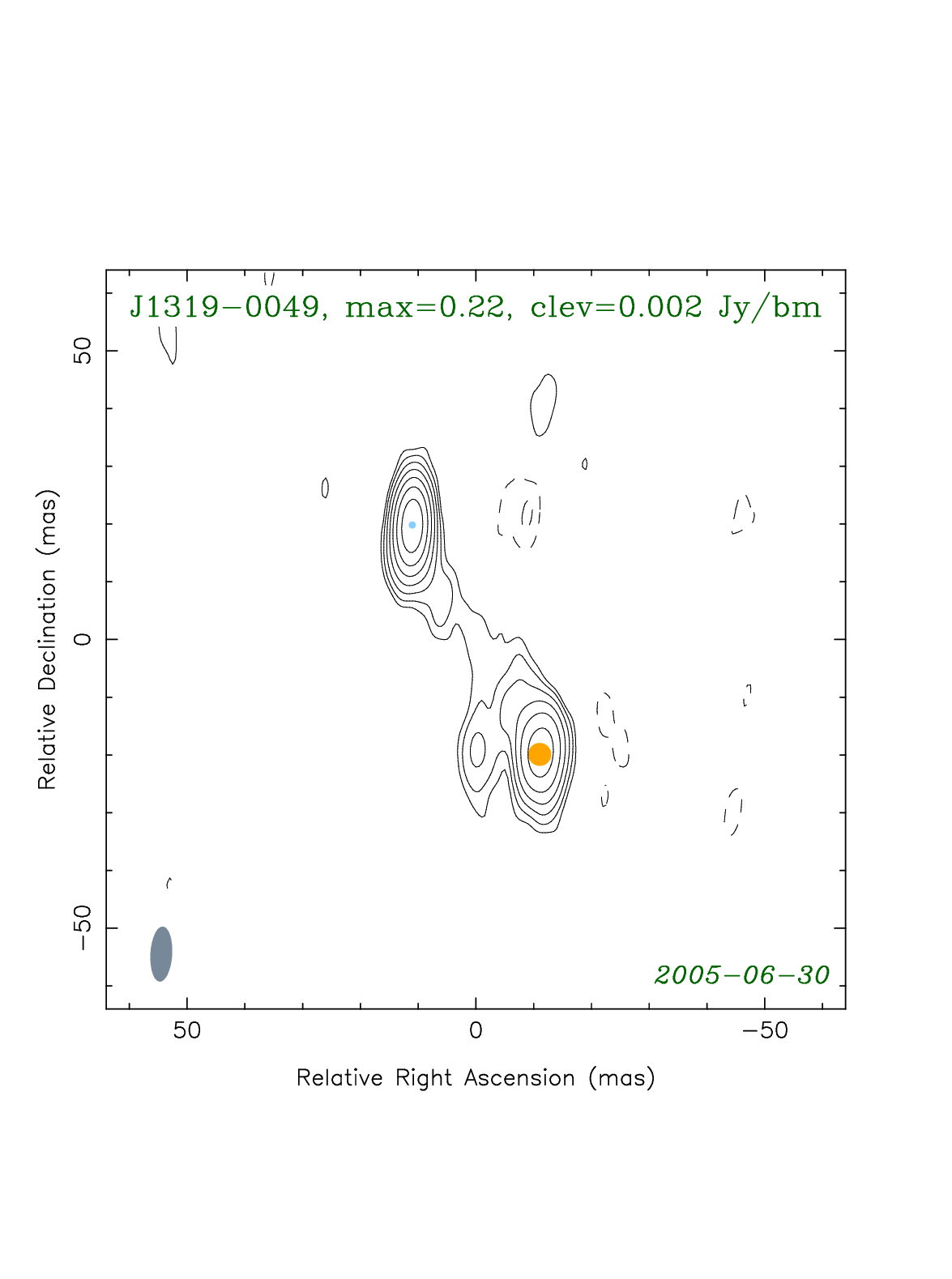}
 \includegraphics[width=0.44\textwidth,angle=0,trim=0 3cm 1cm 5cm,clip]{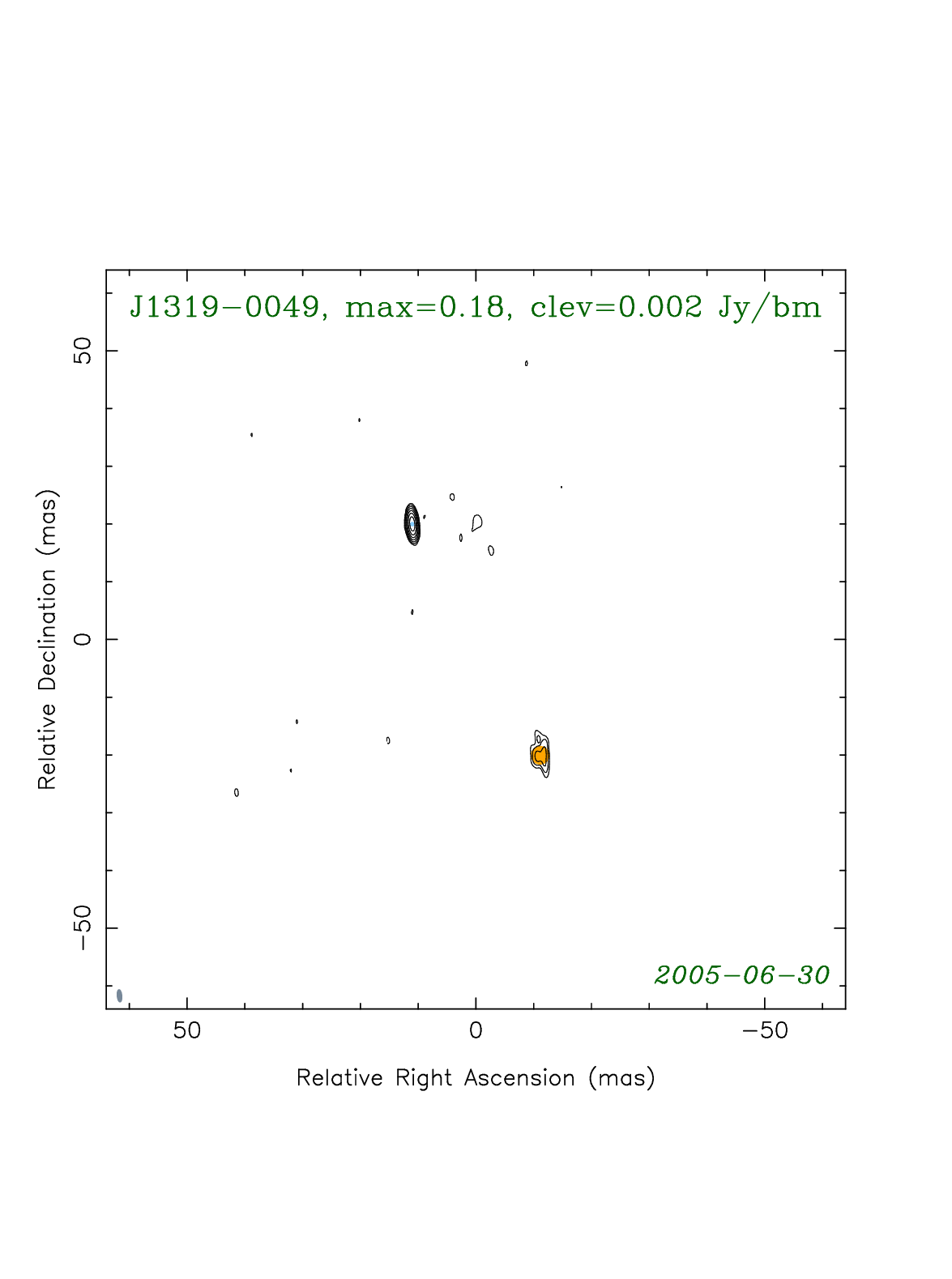}

 \includegraphics[width=0.44\textwidth,angle=0,trim=0 3cm 1cm 5cm,clip]{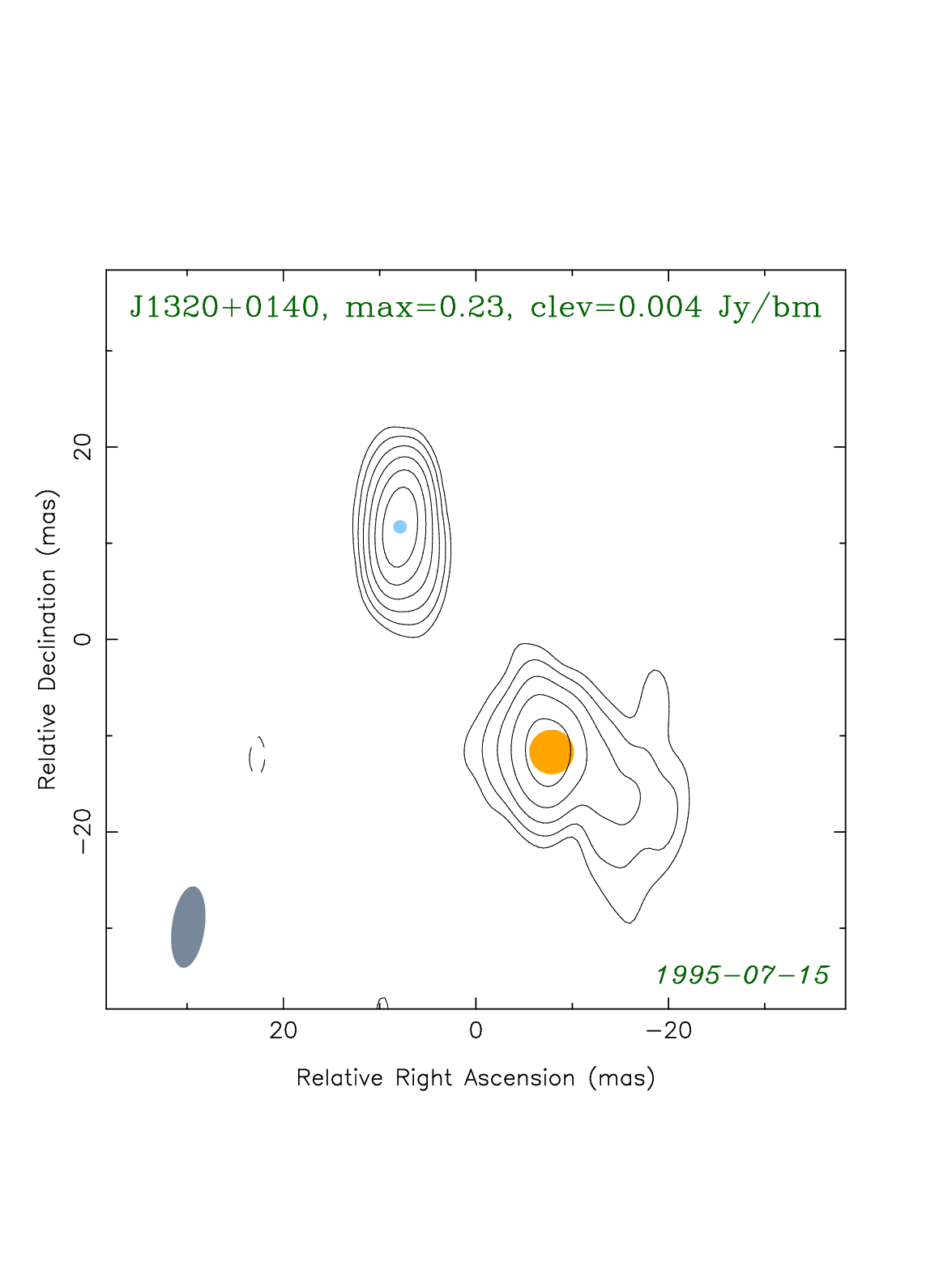}
 \includegraphics[width=0.44\textwidth,angle=0,trim=0 3cm 1cm 5cm,clip]{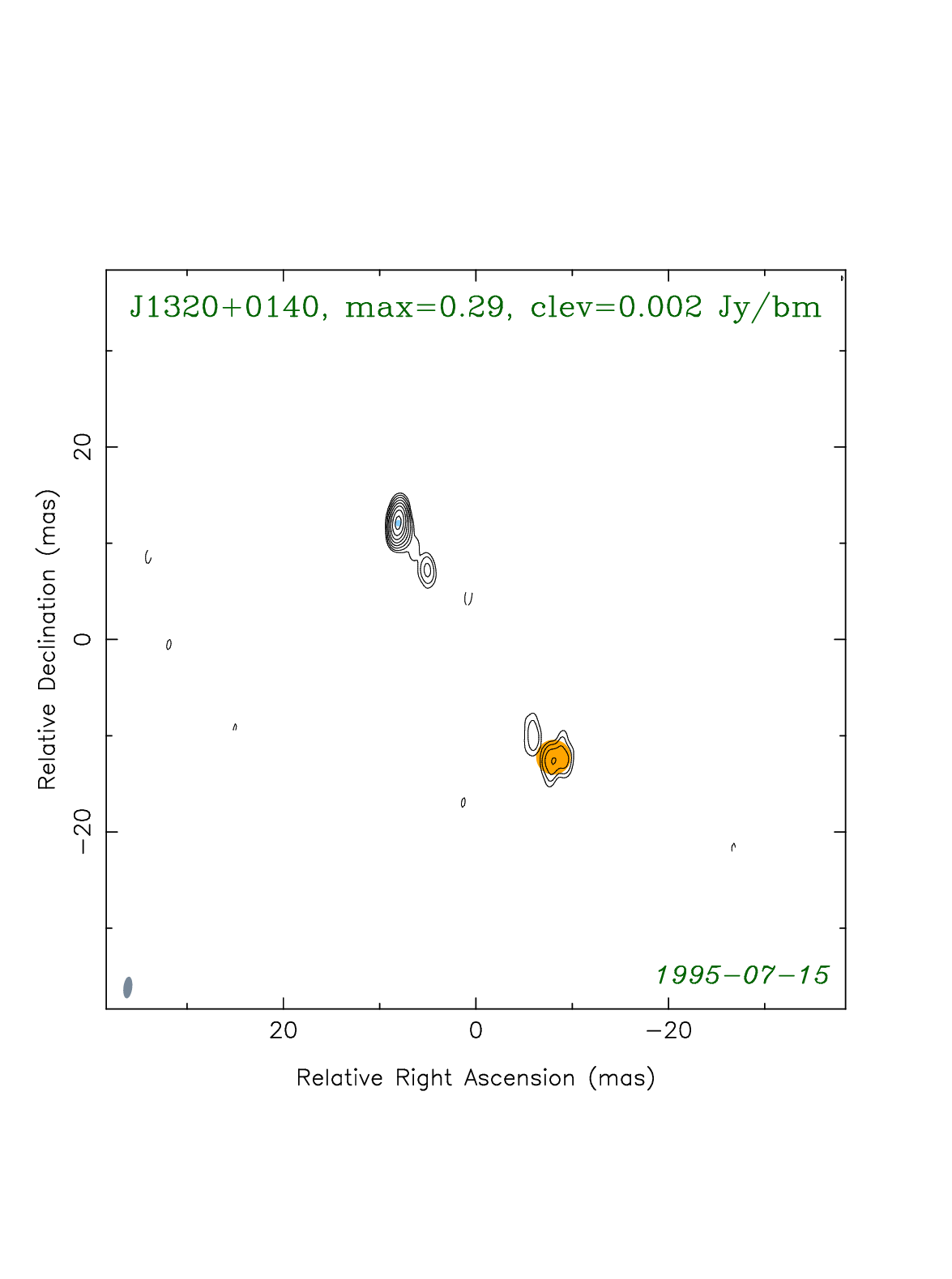}

 \caption{continued}
\end{figure*}

\clearpage
\addtocounter{figure}{-1}
\begin{figure*}
 \centering

 \includegraphics[width=0.44\textwidth,angle=0,trim=0 3cm 1cm 5cm,clip]{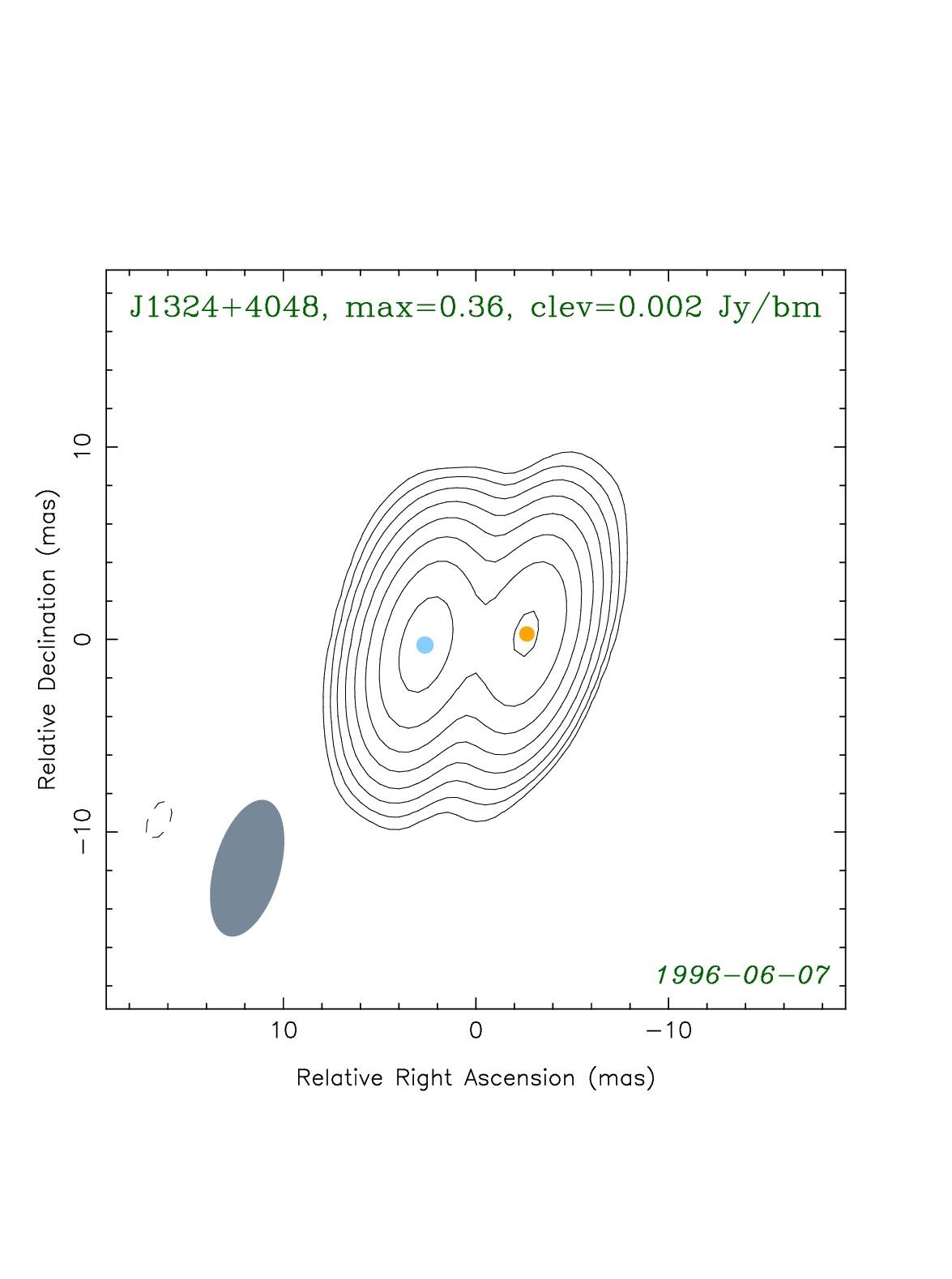}
 \includegraphics[width=0.44\textwidth,angle=0,trim=0 3cm 1cm 5cm,clip]{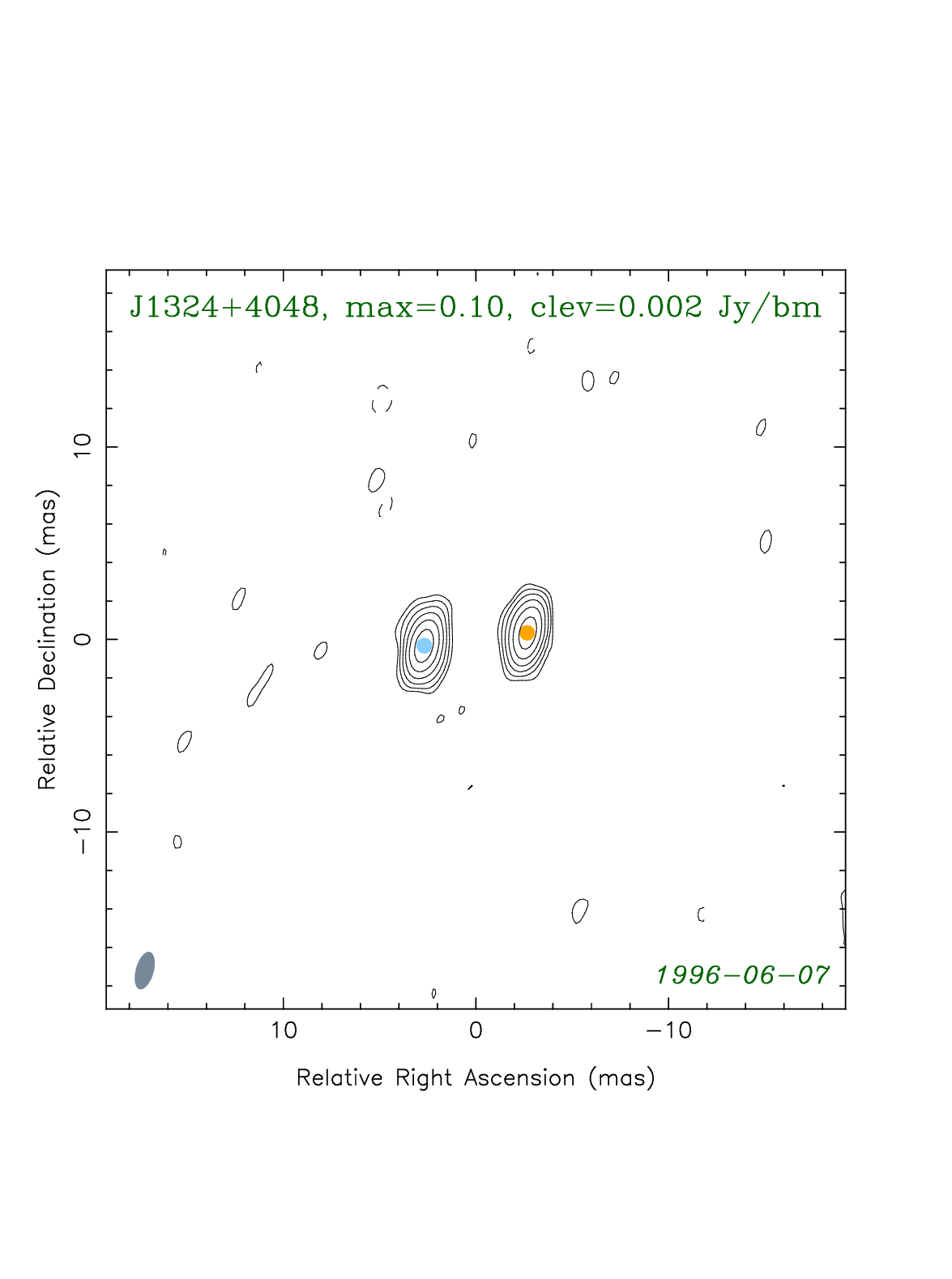}

 \includegraphics[width=0.44\textwidth,angle=0,trim=0 3cm 1cm 5cm,clip]{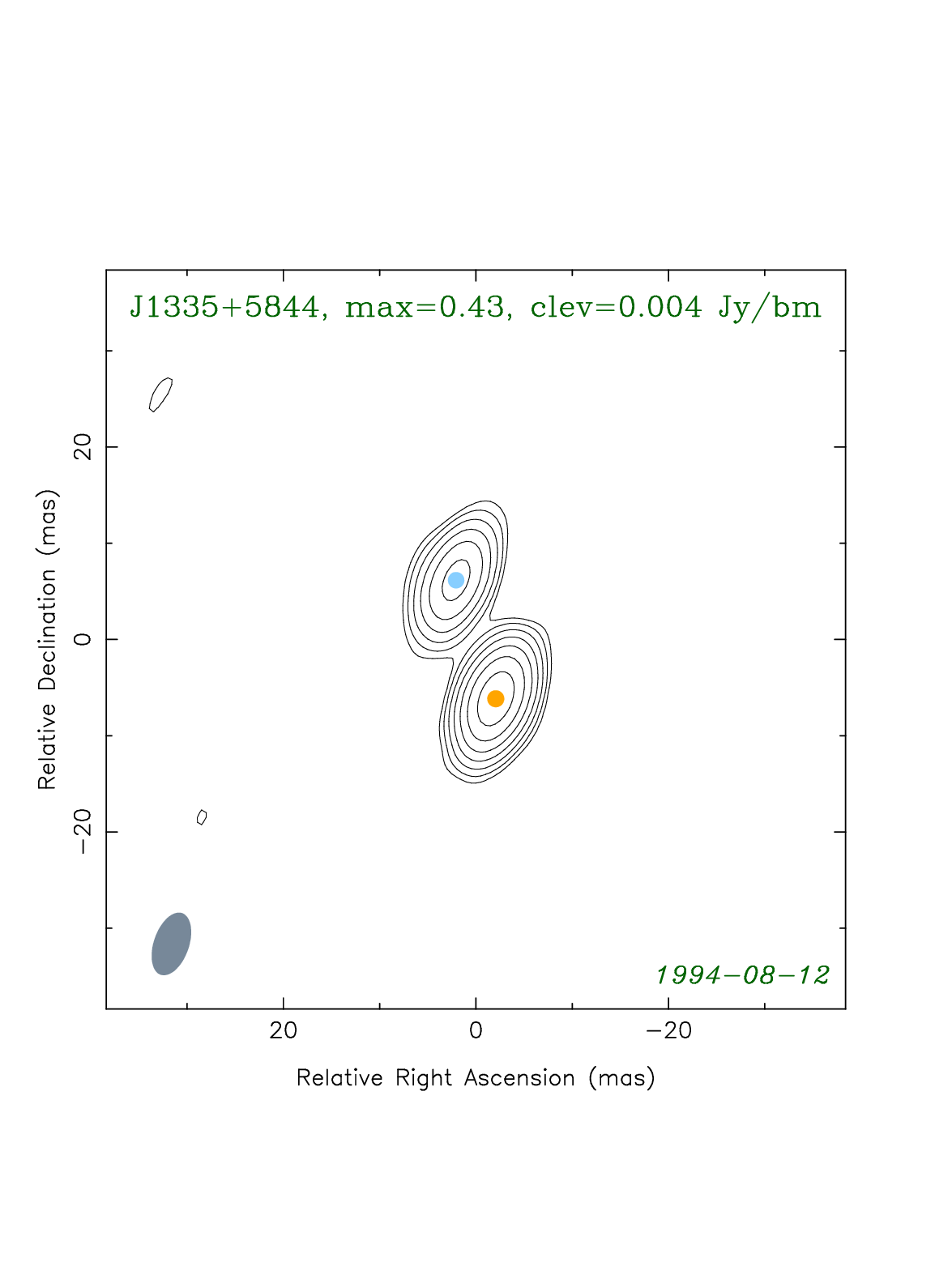}
 \includegraphics[width=0.44\textwidth,angle=0,trim=0 3cm 1cm 5cm,clip]{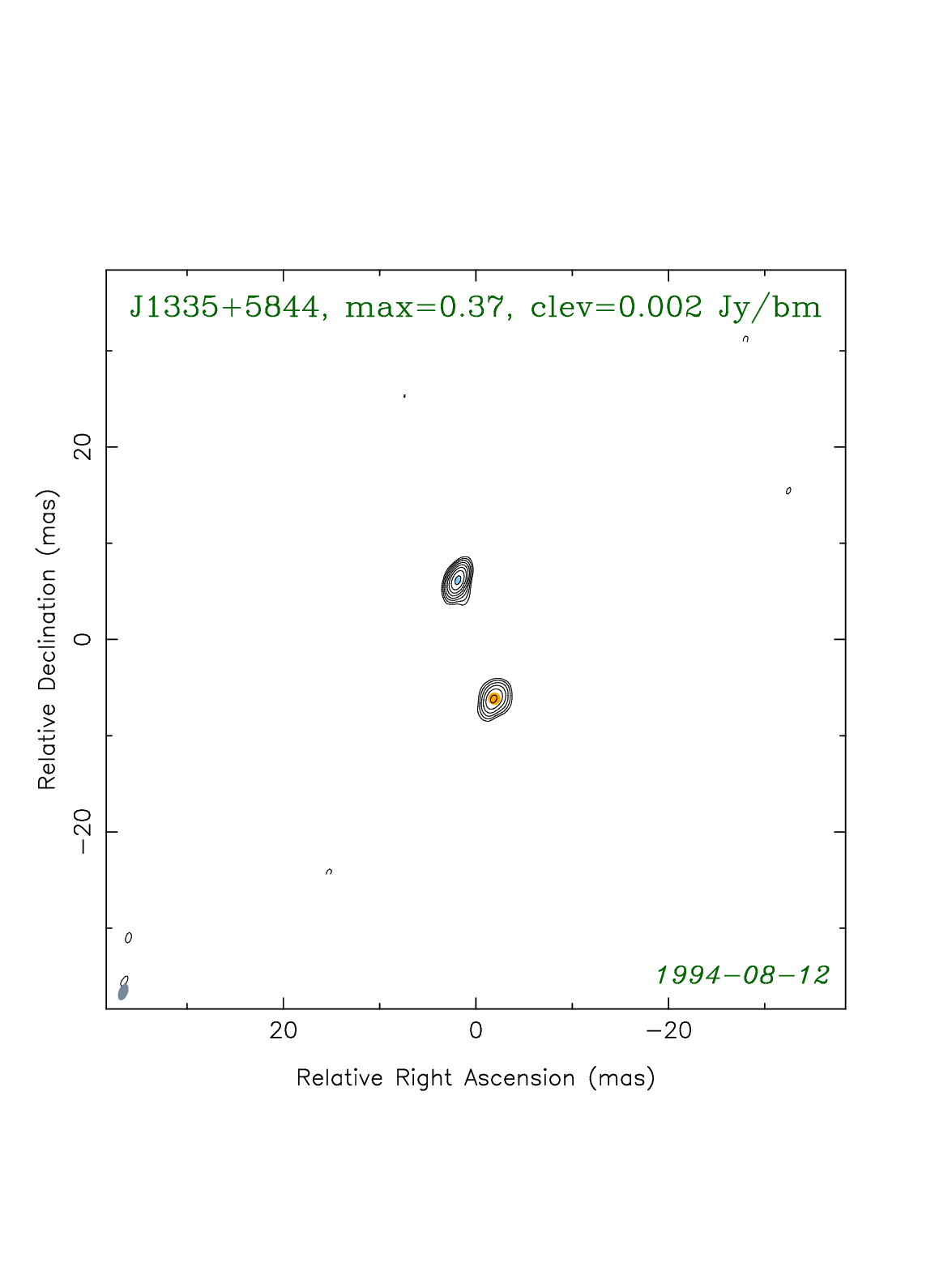}

 \includegraphics[width=0.44\textwidth,angle=0,trim=0 3cm 1cm 5cm,clip]{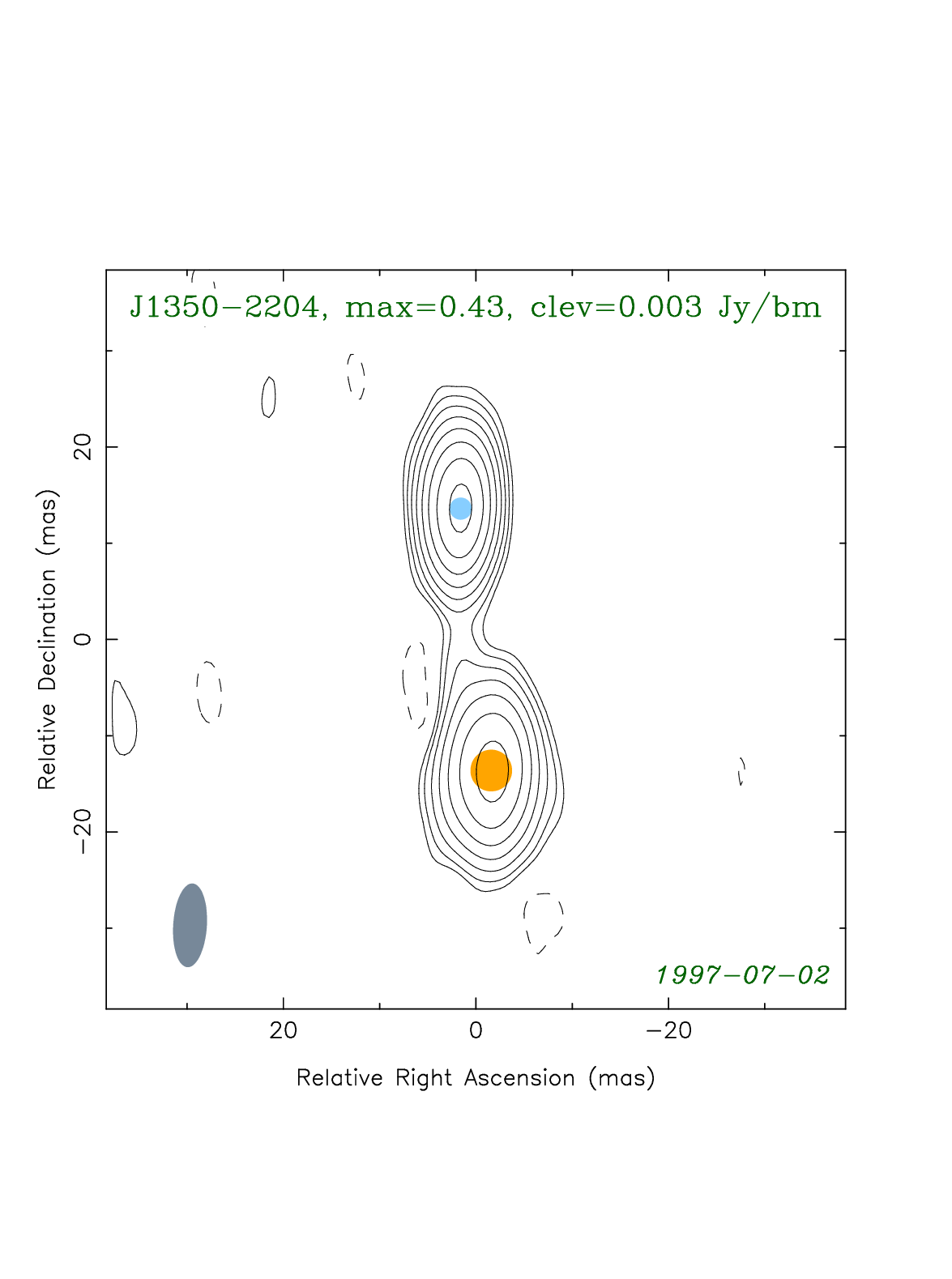}
 \includegraphics[width=0.44\textwidth,angle=0,trim=0 3cm 1cm 5cm,clip]{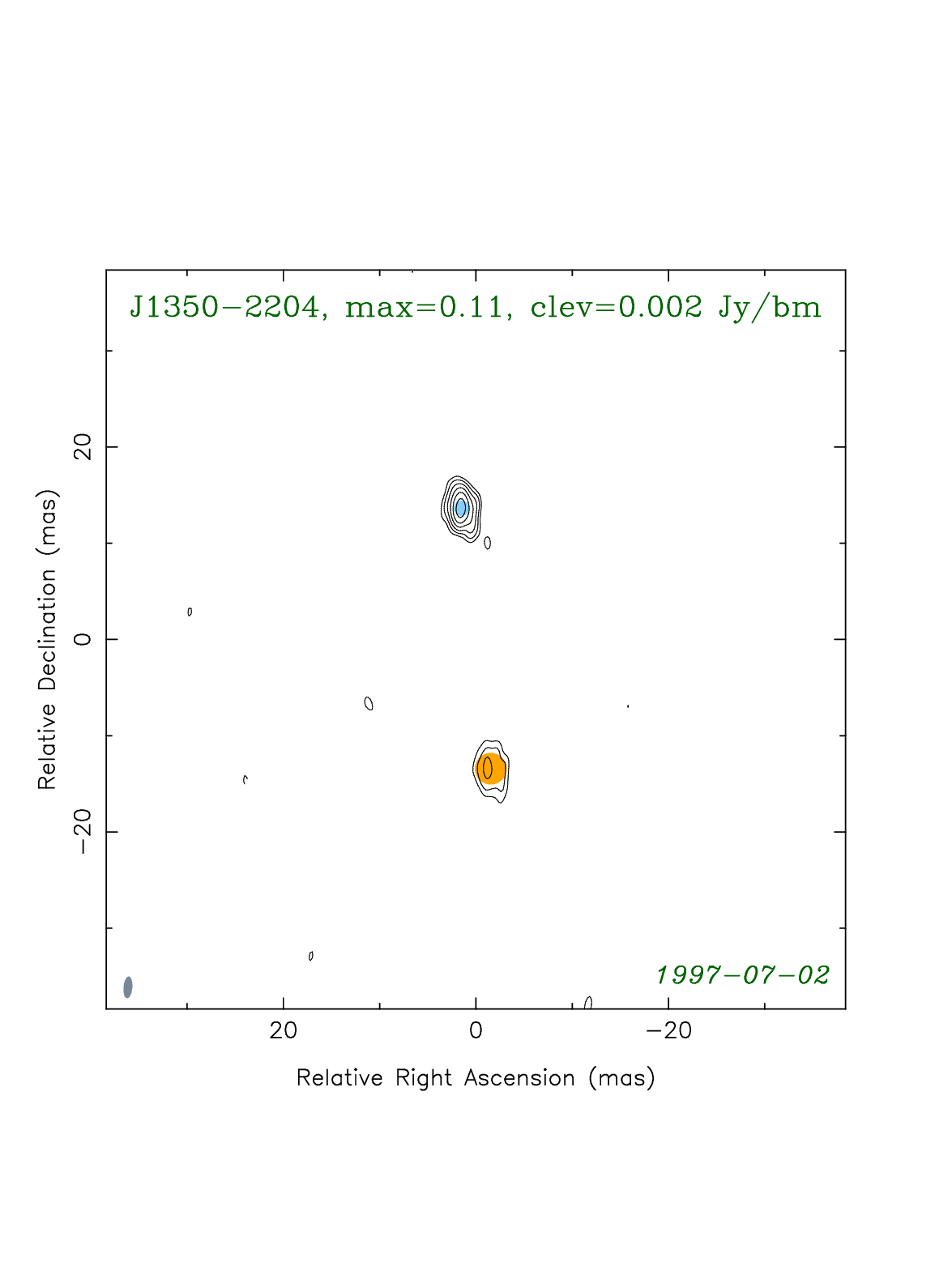}

 \caption{continued}
\end{figure*}

\clearpage
\addtocounter{figure}{-1}
\begin{figure*}
 \centering

 \includegraphics[width=0.44\textwidth,angle=0,trim=0 3cm 1cm 5cm,clip]{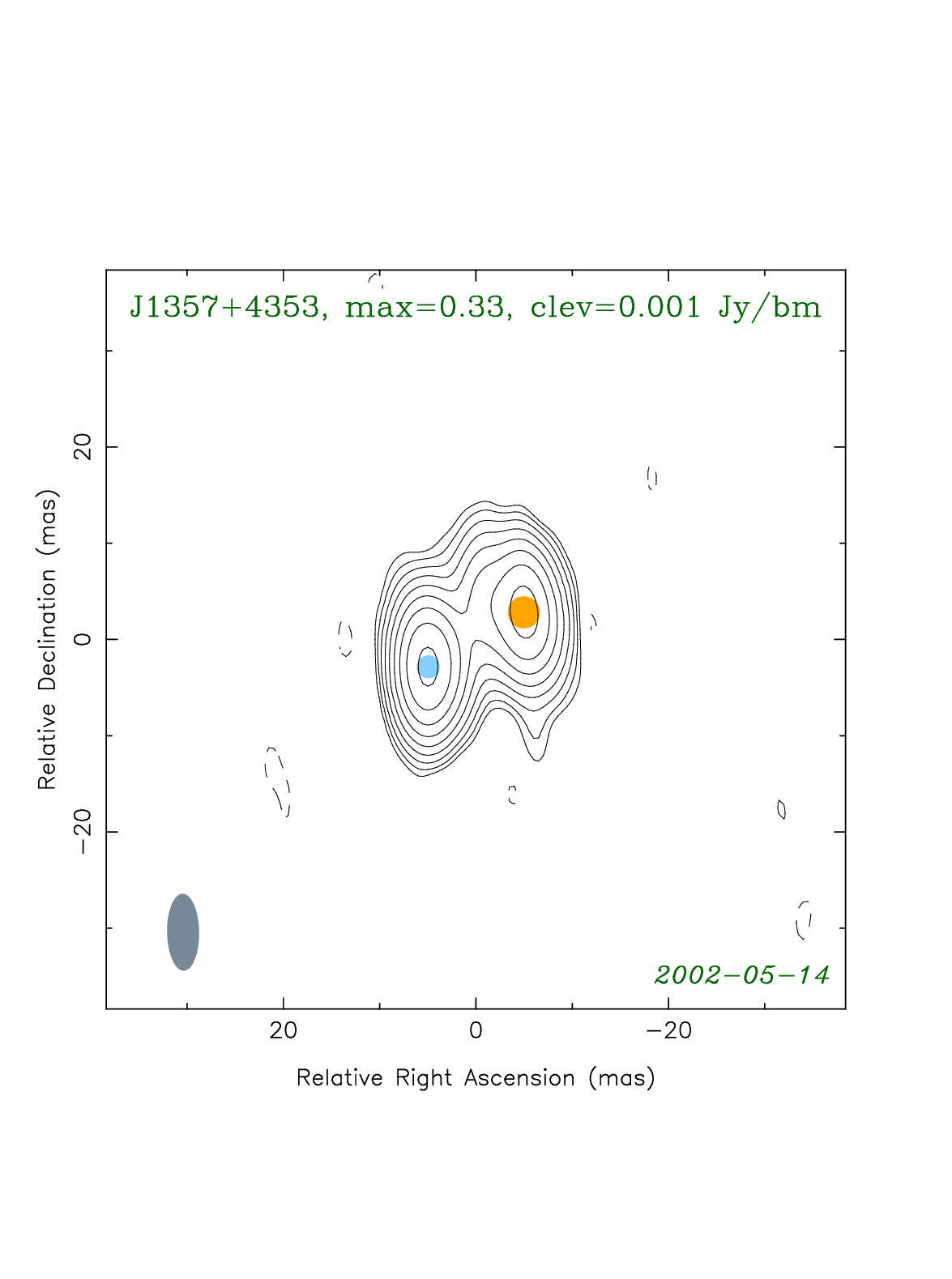}
 \includegraphics[width=0.44\textwidth,angle=0,trim=0 3cm 1cm 5cm,clip]{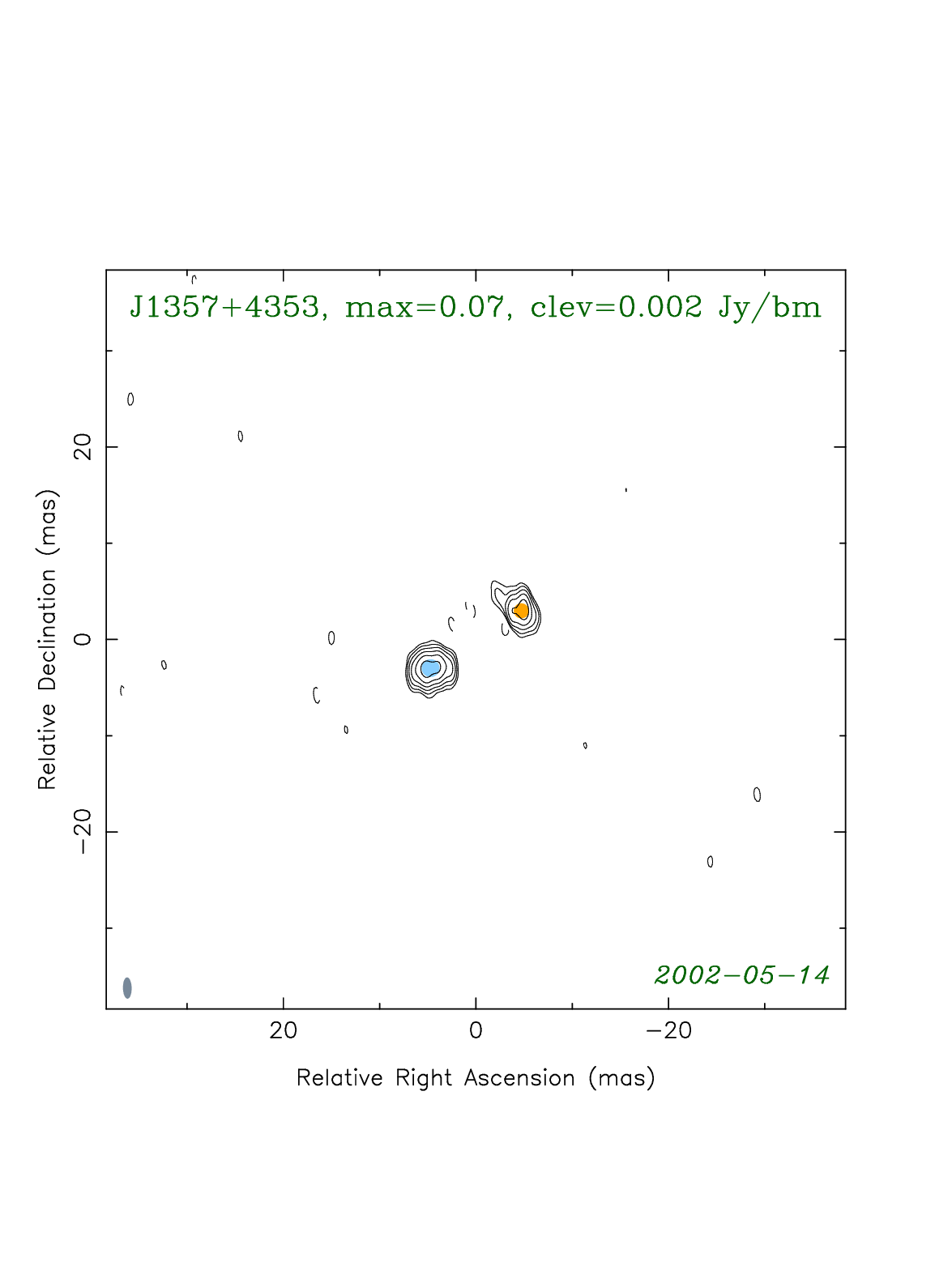}

 \includegraphics[width=0.44\textwidth,angle=0,trim=0 3cm 1cm 5cm,clip]{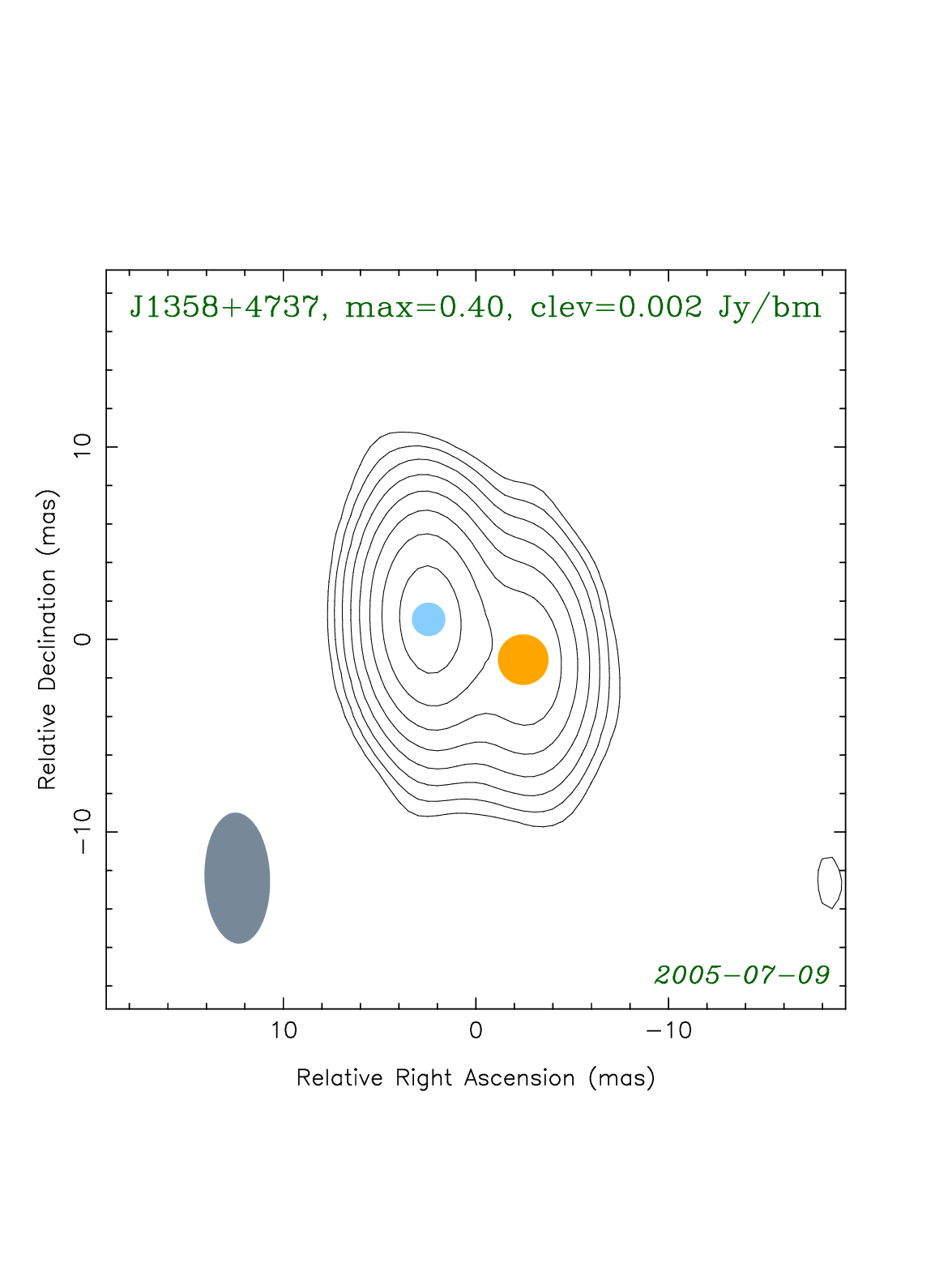}
 \includegraphics[width=0.44\textwidth,angle=0,trim=0 3cm 1cm 5cm,clip]{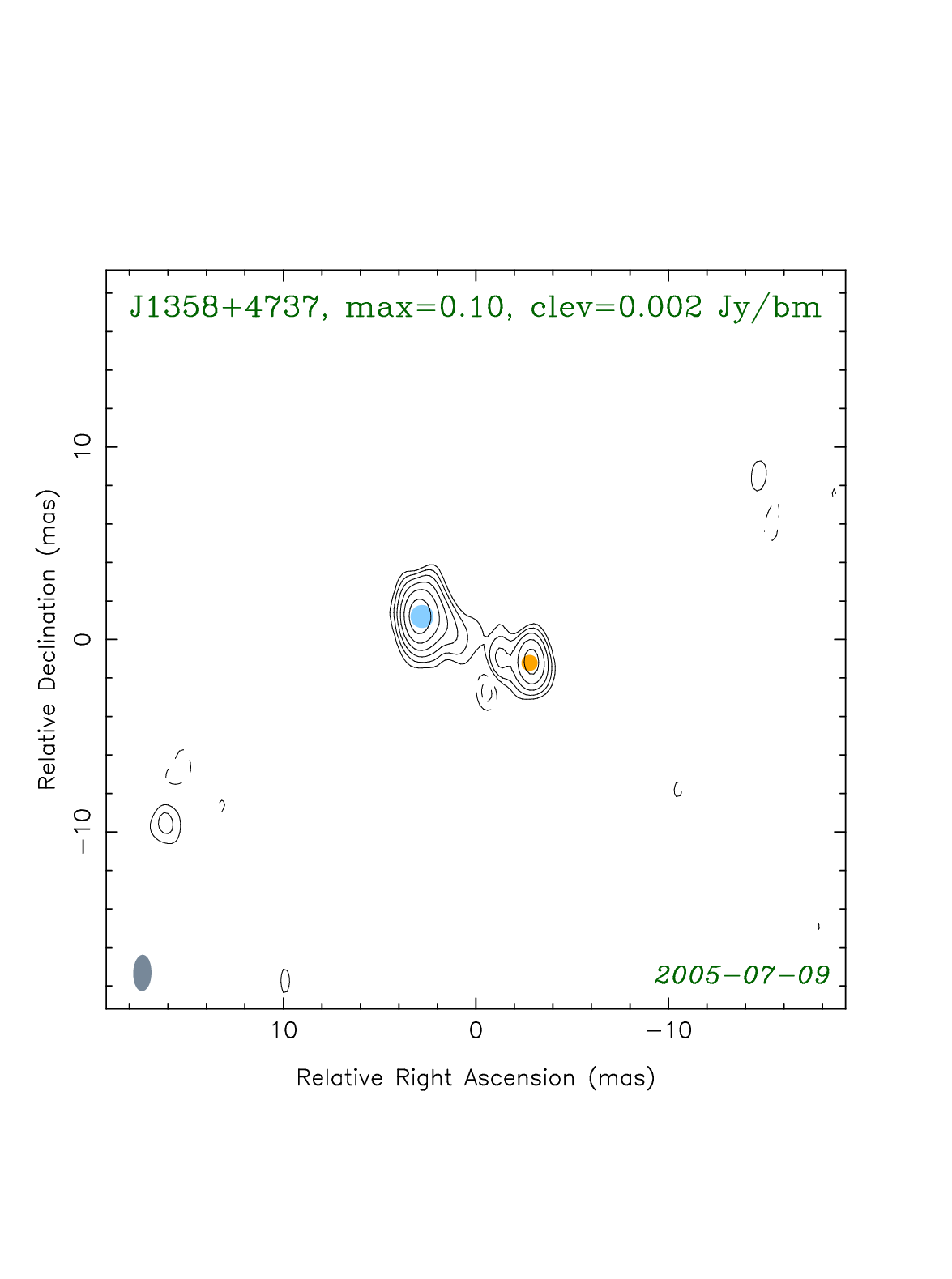}

 \includegraphics[width=0.44\textwidth,angle=0,trim=0 3cm 1cm 5cm,clip]{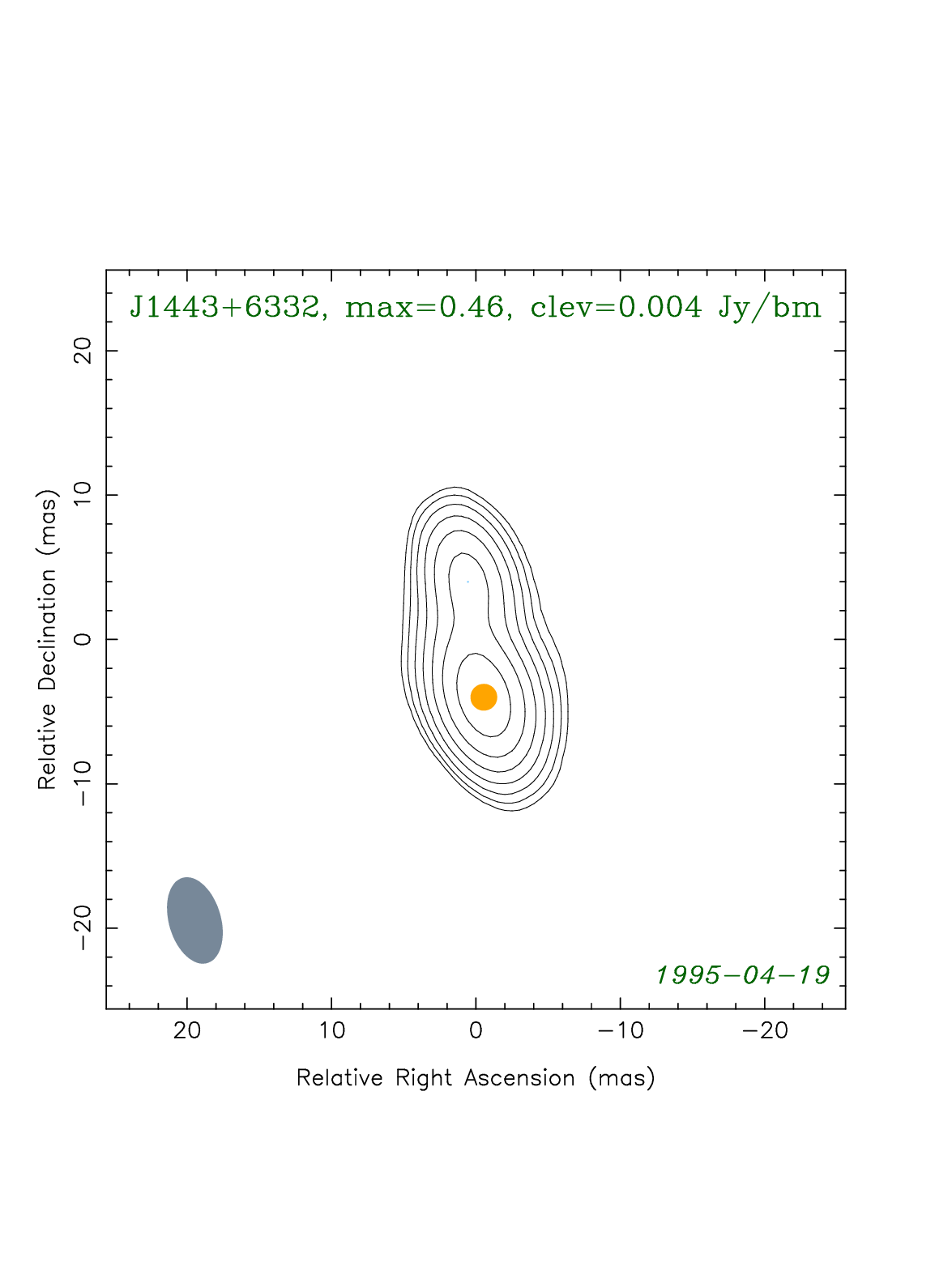}
 \includegraphics[width=0.44\textwidth,angle=0,trim=0 3cm 1cm 5cm,clip]{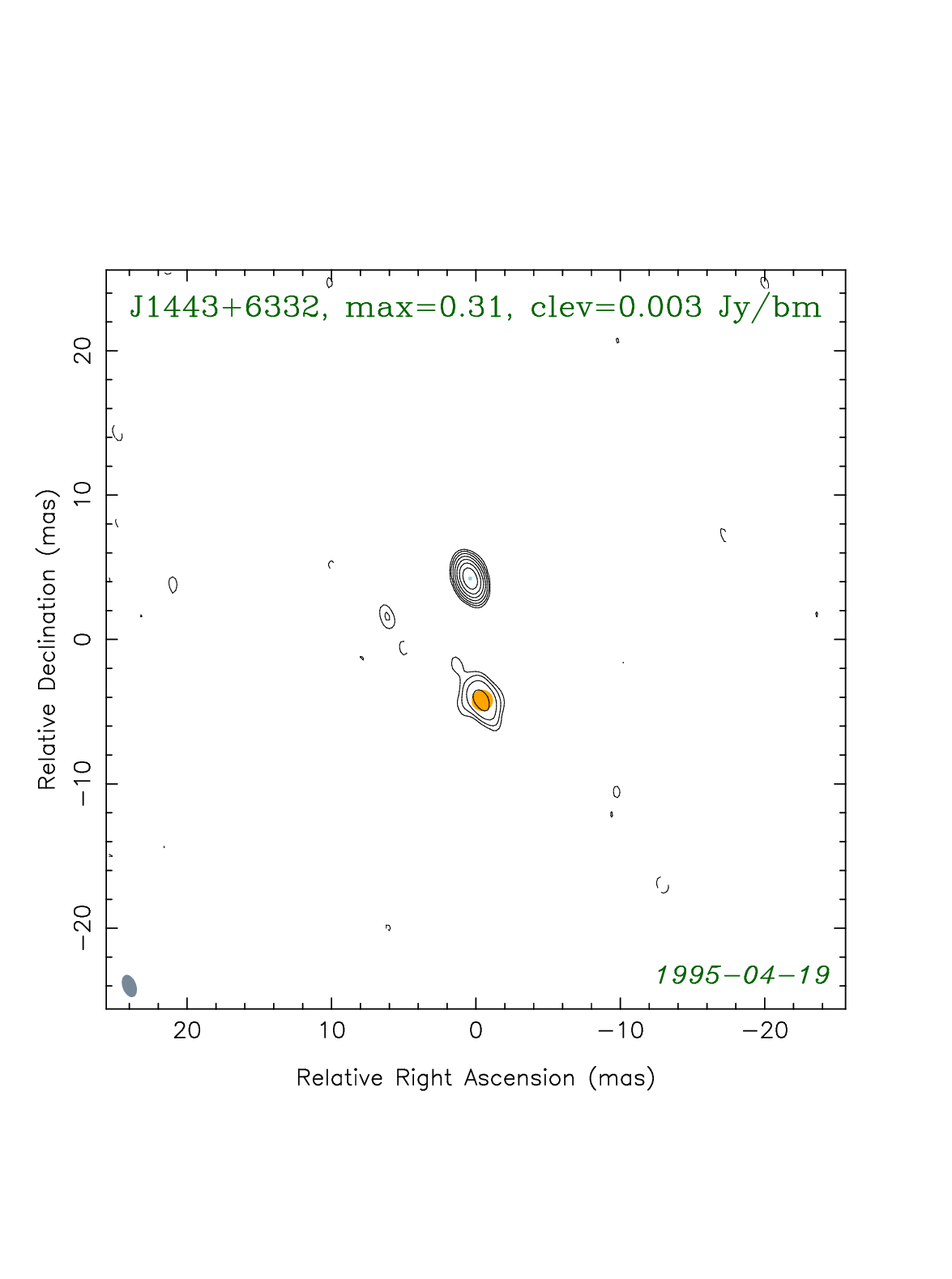}

 \caption{continued}
\end{figure*}

\clearpage
\addtocounter{figure}{-1}
\begin{figure*}
 \centering

 \includegraphics[width=0.44\textwidth,angle=0,trim=0 3cm 1cm 5cm,clip]{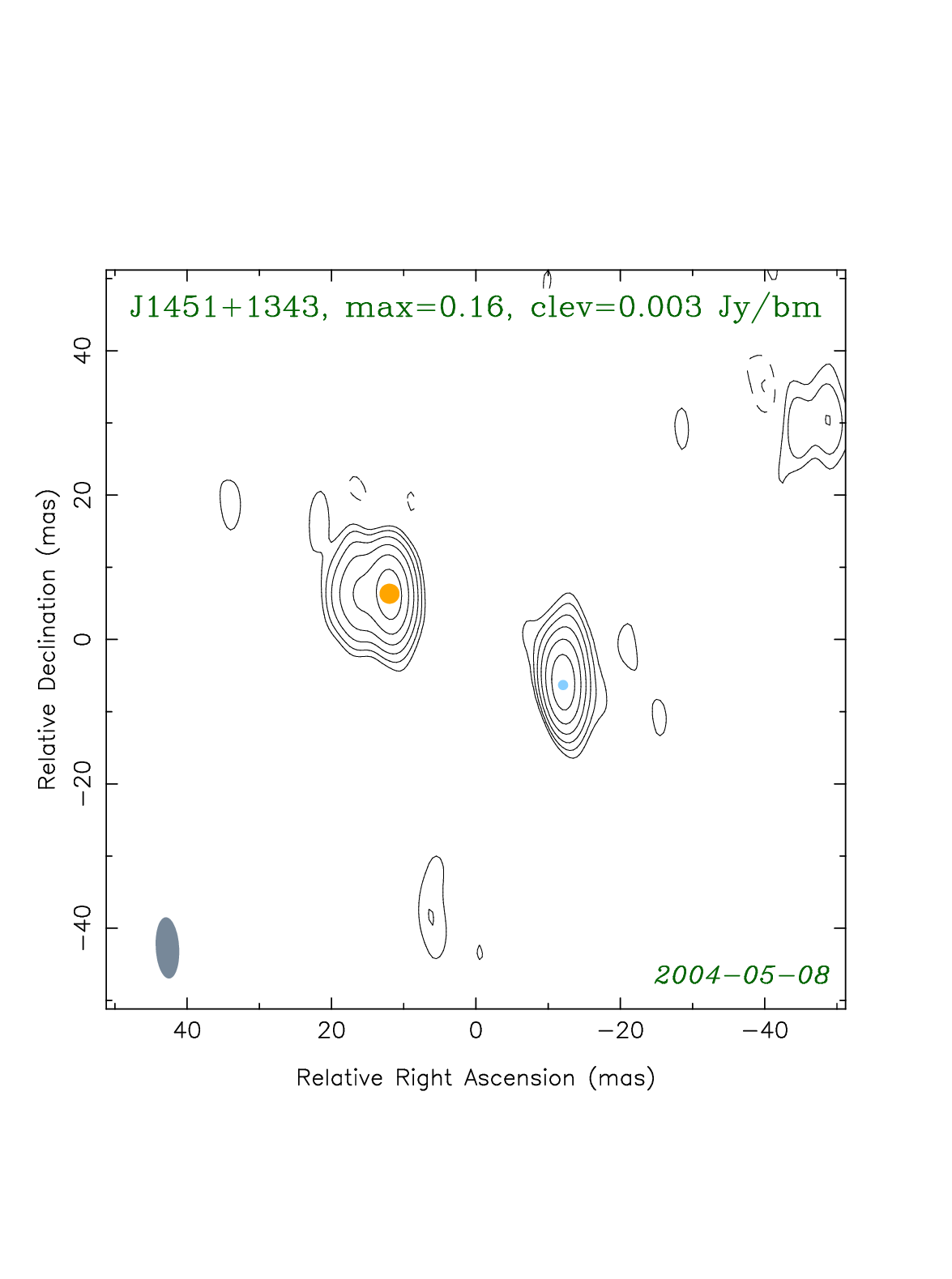}
 \includegraphics[width=0.44\textwidth,angle=0,trim=0 3cm 1cm 5cm,clip]{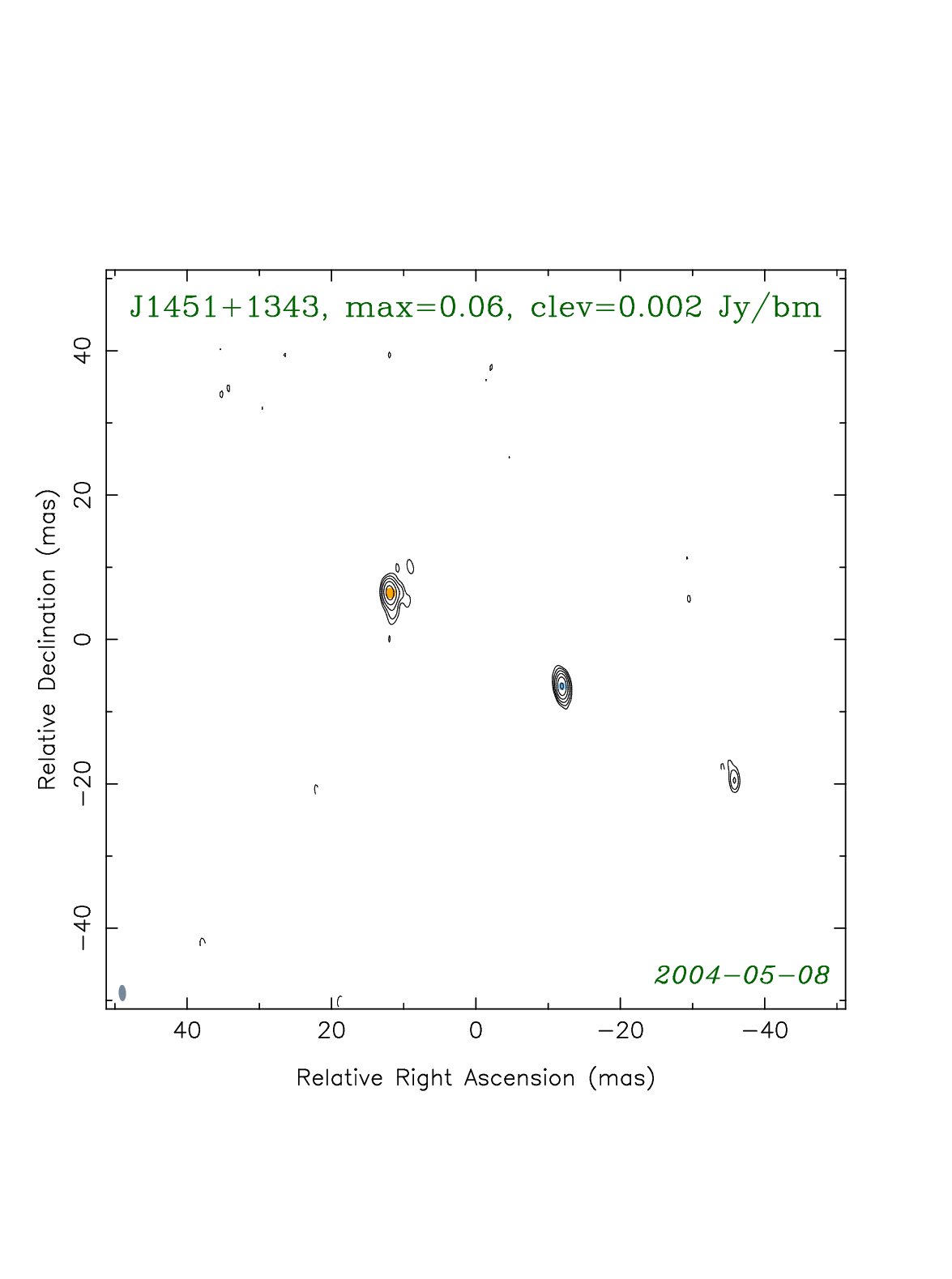}

 \includegraphics[width=0.44\textwidth,angle=0,trim=0 3cm 1cm 5cm,clip]{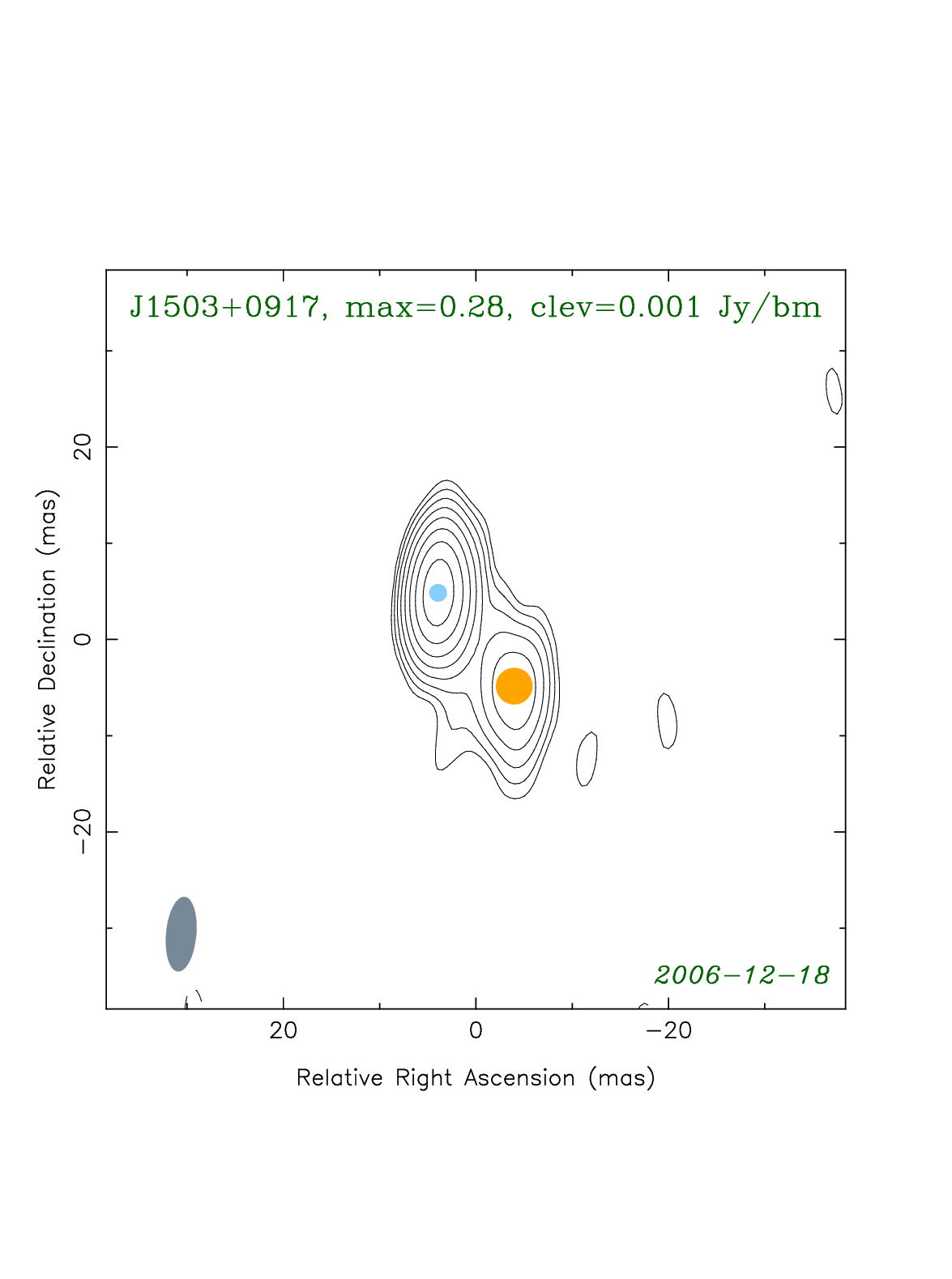}
 \includegraphics[width=0.44\textwidth,angle=0,trim=0 3cm 1cm 5cm,clip]{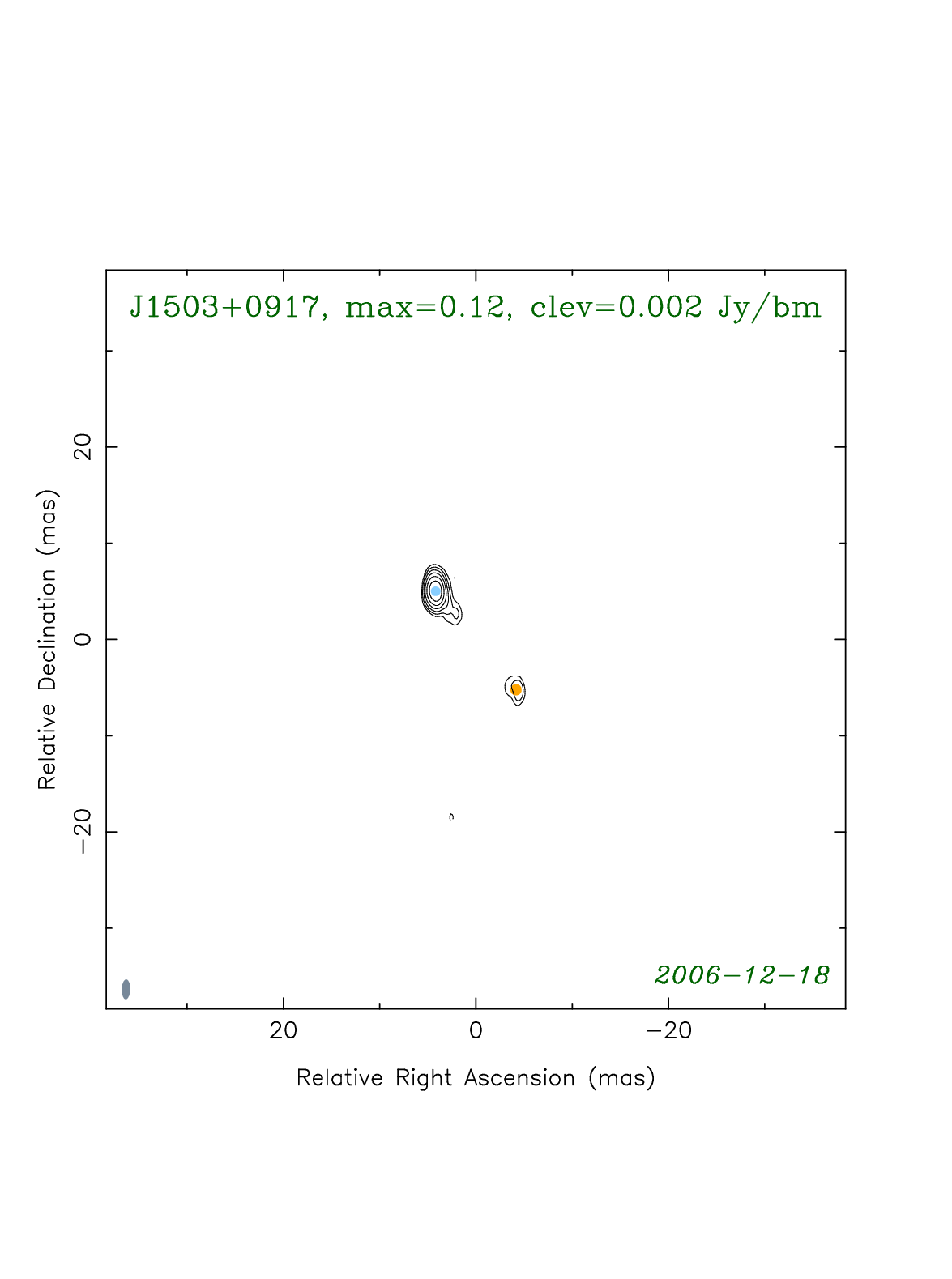}

 \includegraphics[width=0.44\textwidth,angle=0,trim=0 3cm 1cm 5cm,clip]{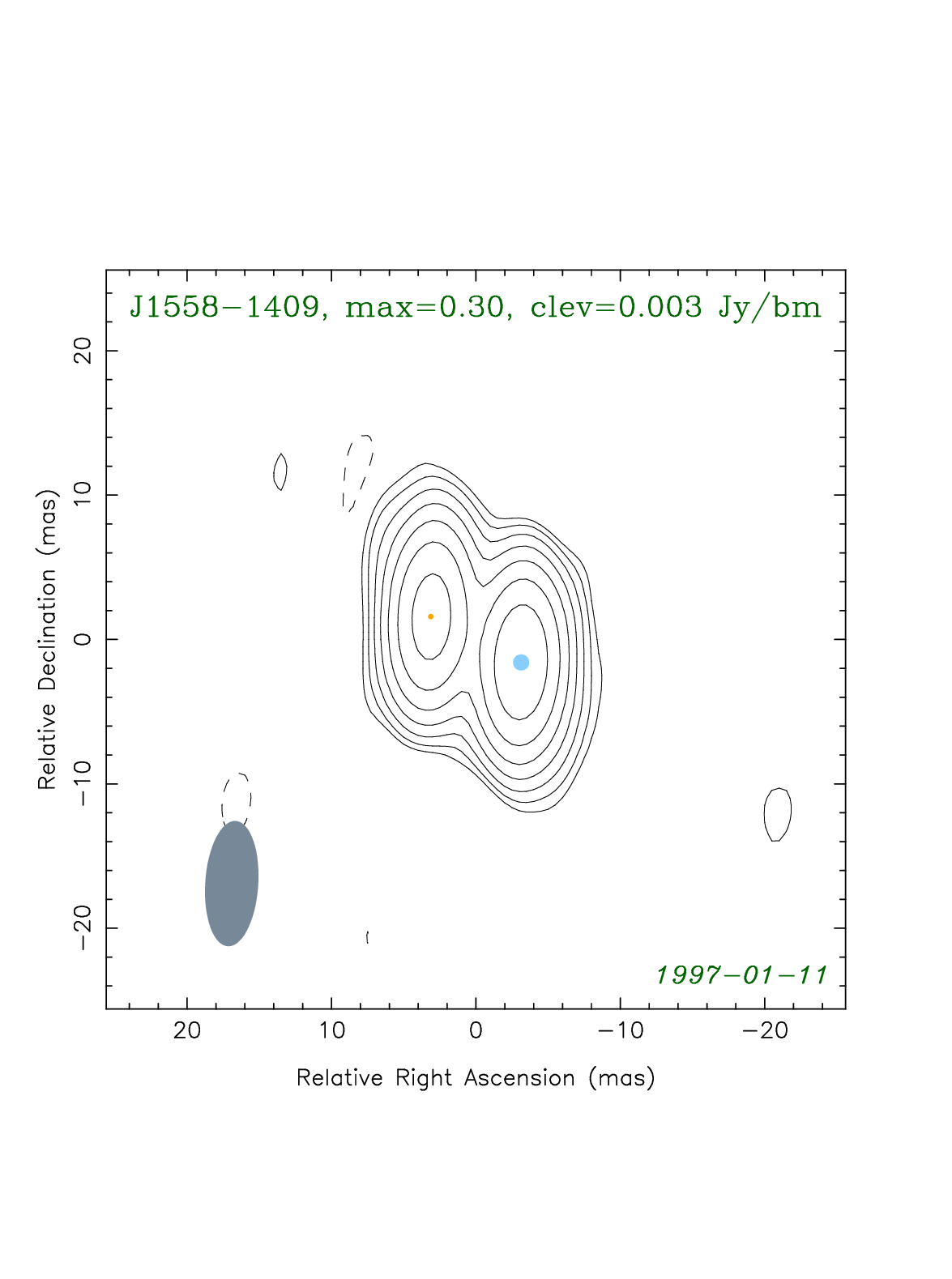}
 \includegraphics[width=0.44\textwidth,angle=0,trim=0 3cm 1cm 5cm,clip]{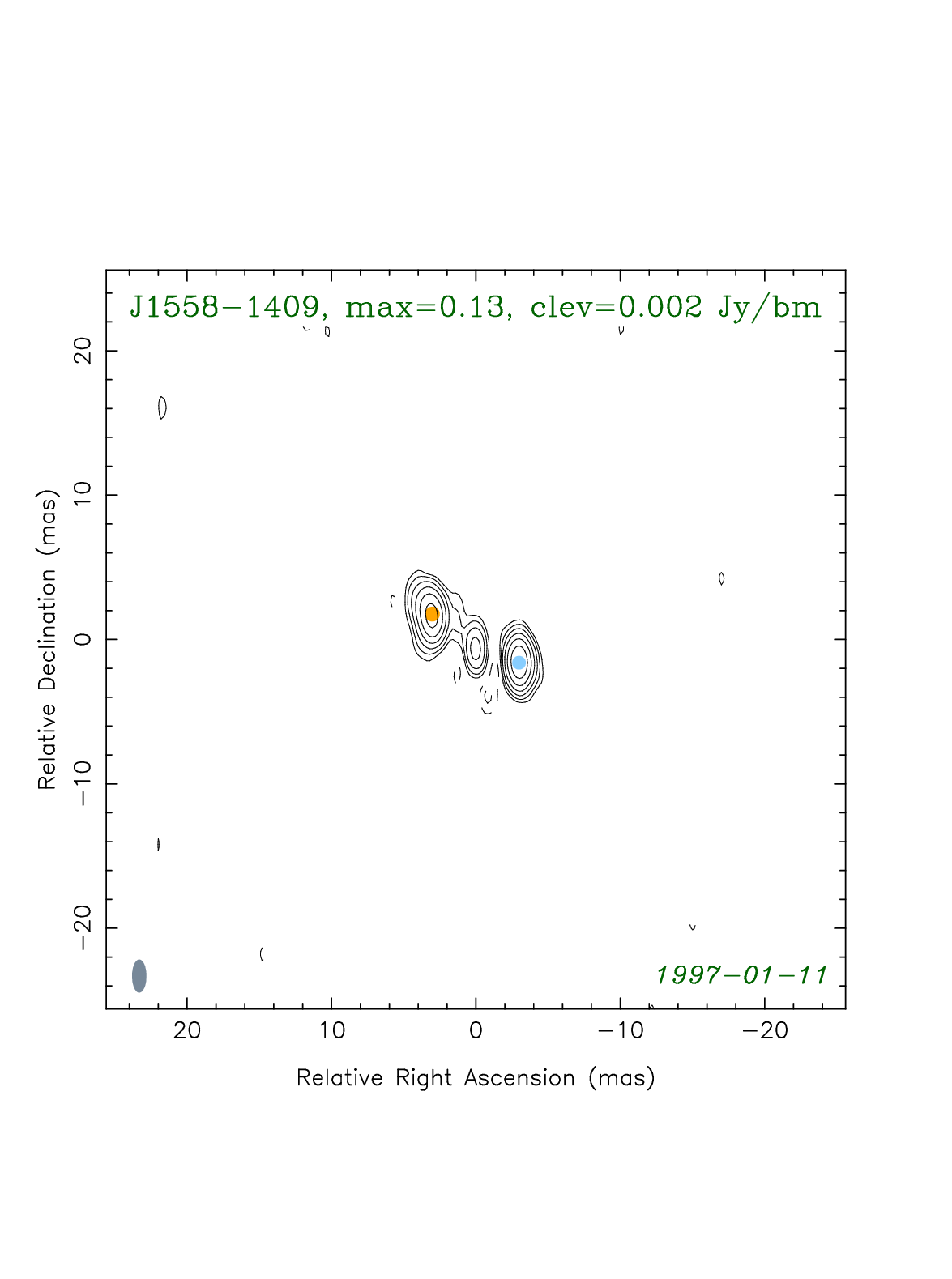}

 \caption{continued}
\end{figure*}

\clearpage
\addtocounter{figure}{-1}
\begin{figure*}
 \centering

 \includegraphics[width=0.44\textwidth,angle=0,trim=0 3cm 1cm 5cm,clip]{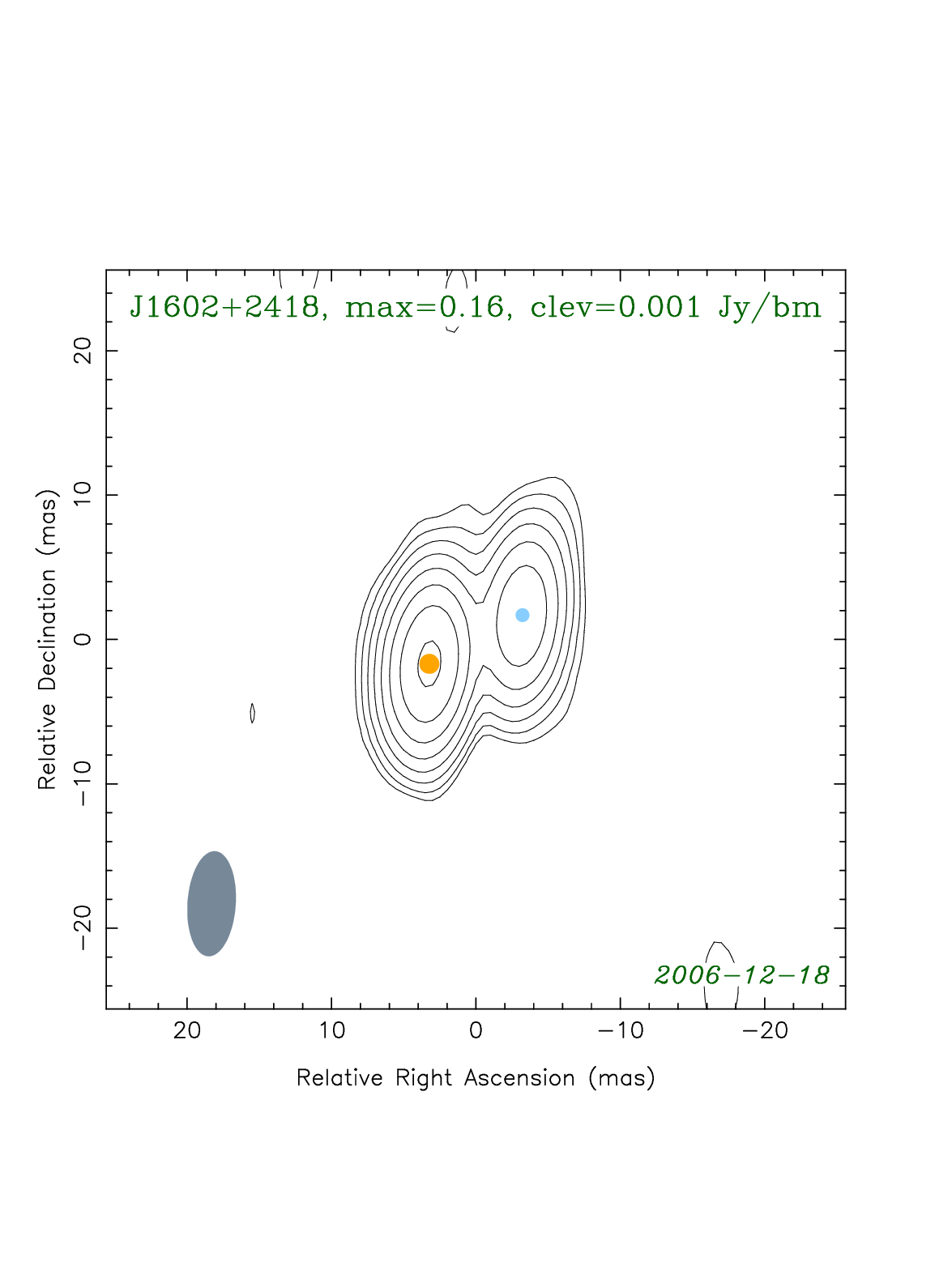}
 \includegraphics[width=0.44\textwidth,angle=0,trim=0 3cm 1cm 5cm,clip]{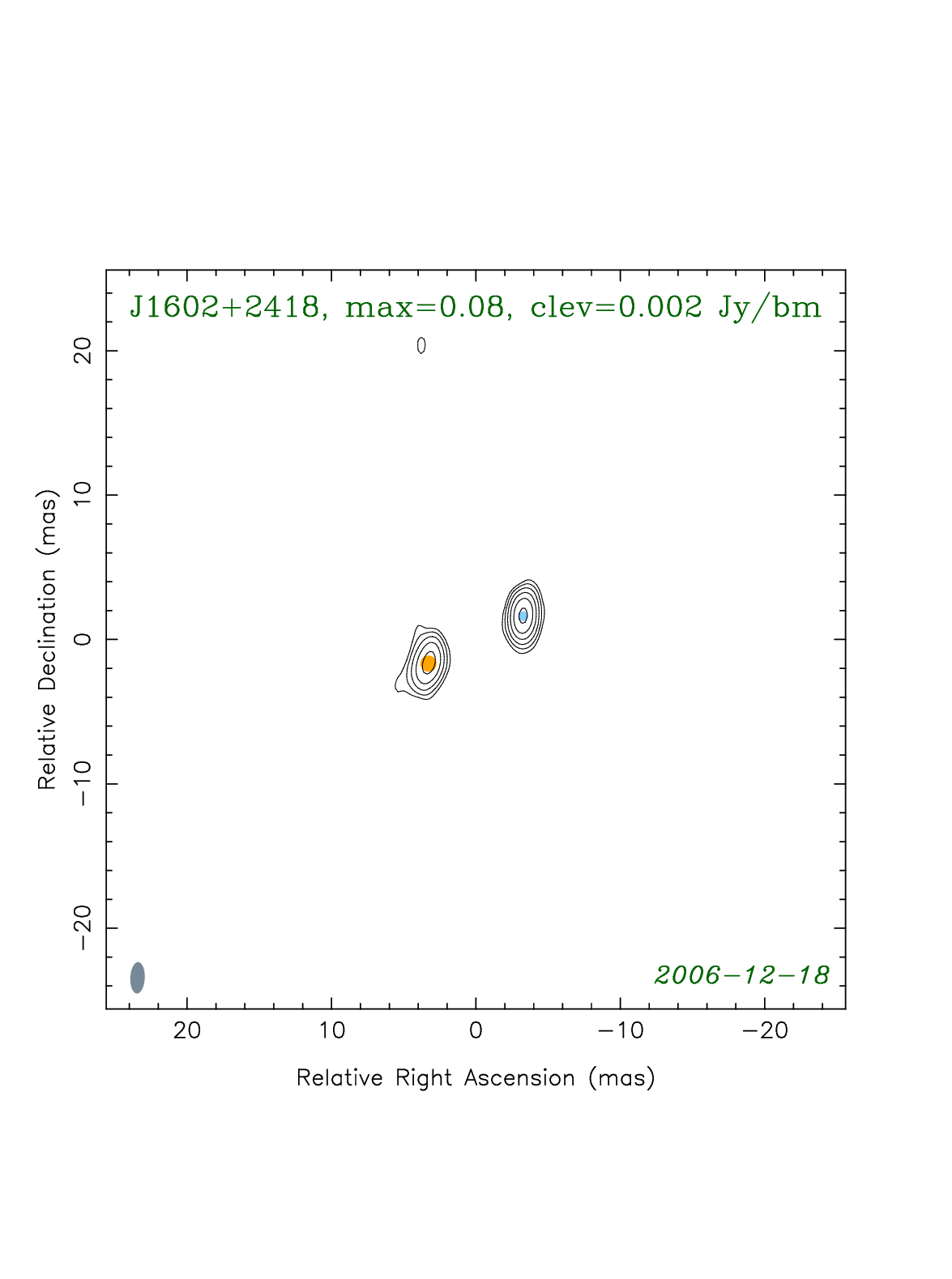}

 \includegraphics[width=0.44\textwidth,angle=0,trim=0 3cm 1cm 5cm,clip]{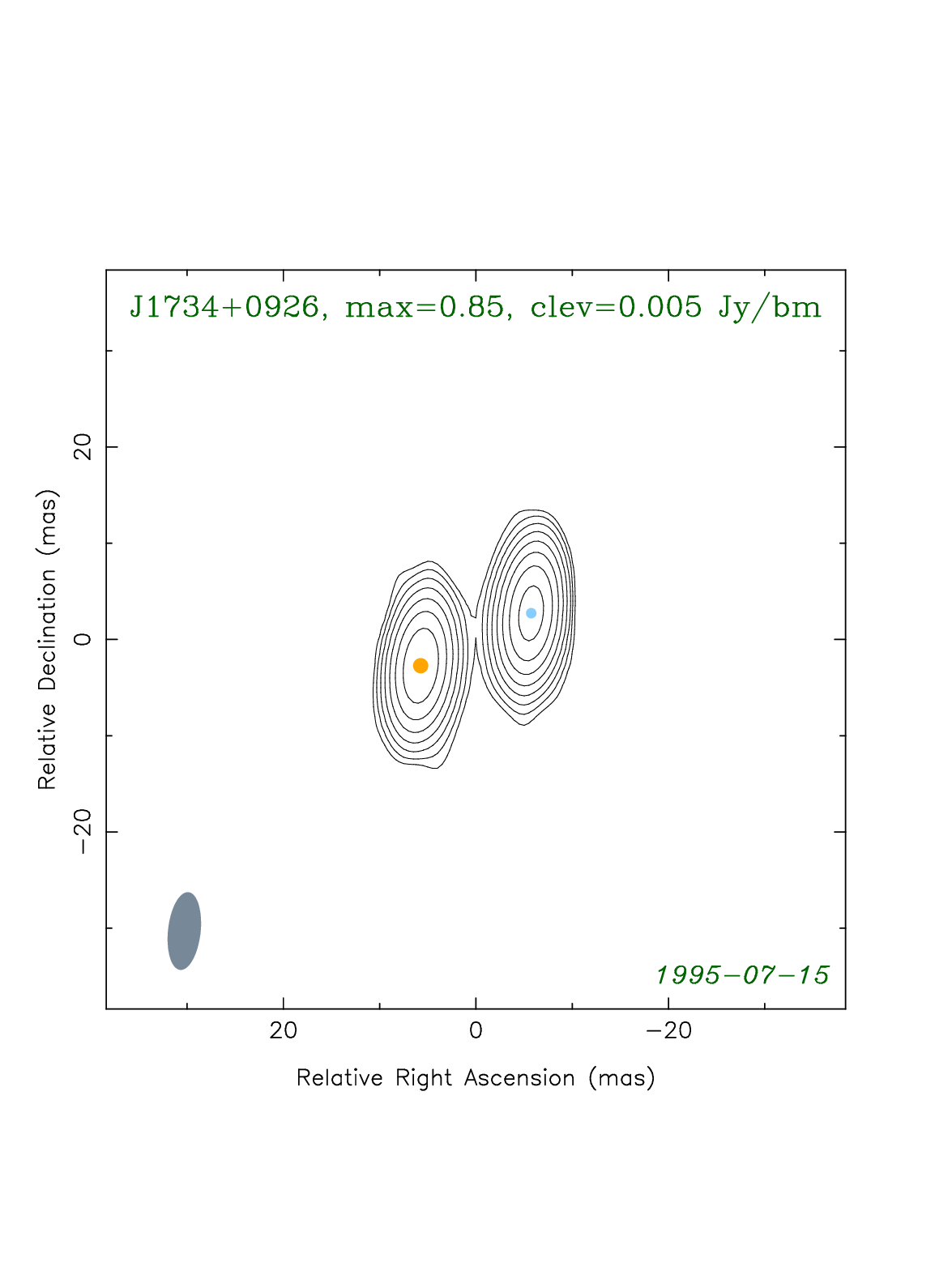}
 \includegraphics[width=0.44\textwidth,angle=0,trim=0 3cm 1cm 5cm,clip]{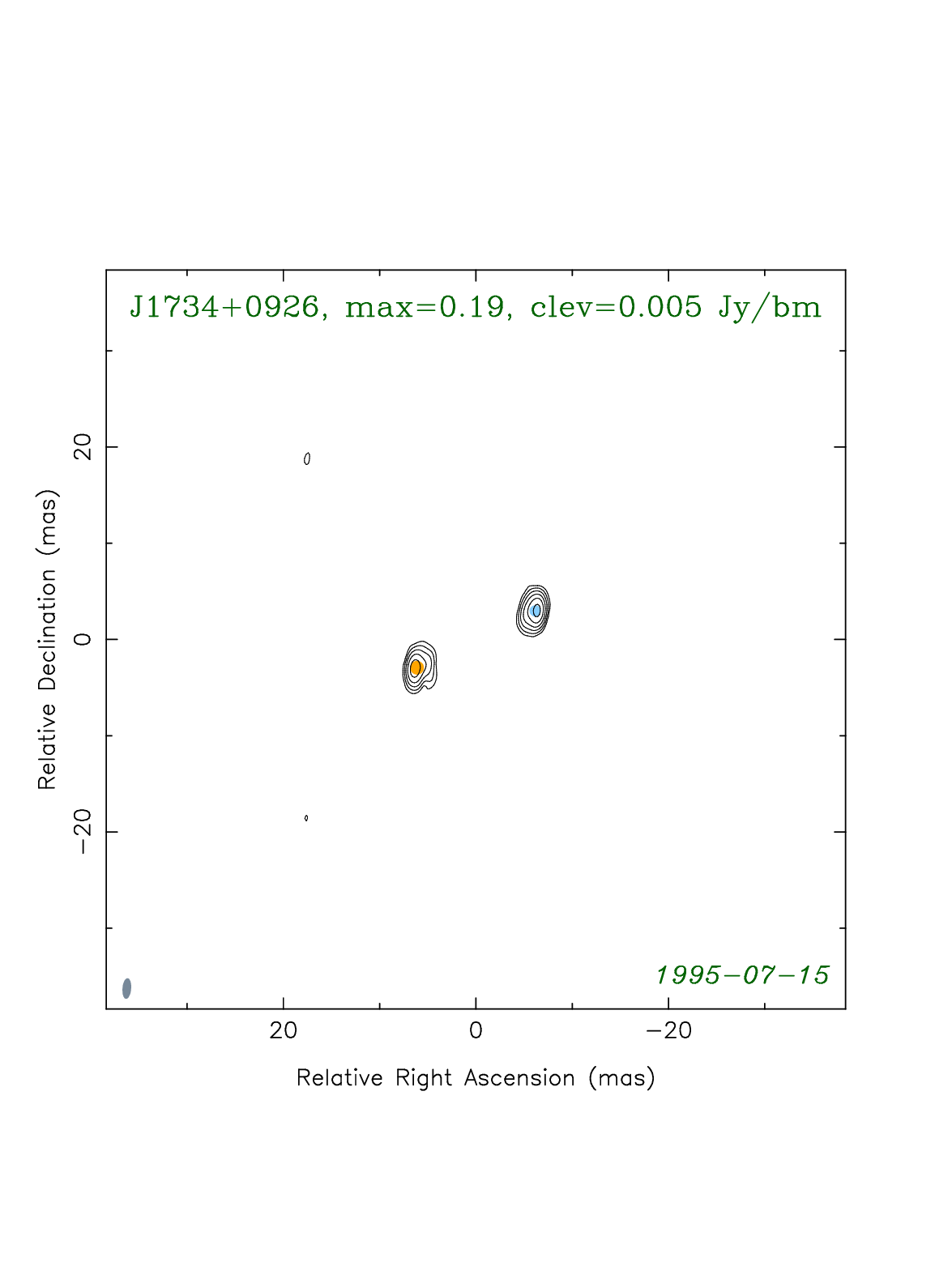}

 \includegraphics[width=0.44\textwidth,angle=0,trim=0 3cm 1cm 5cm,clip]{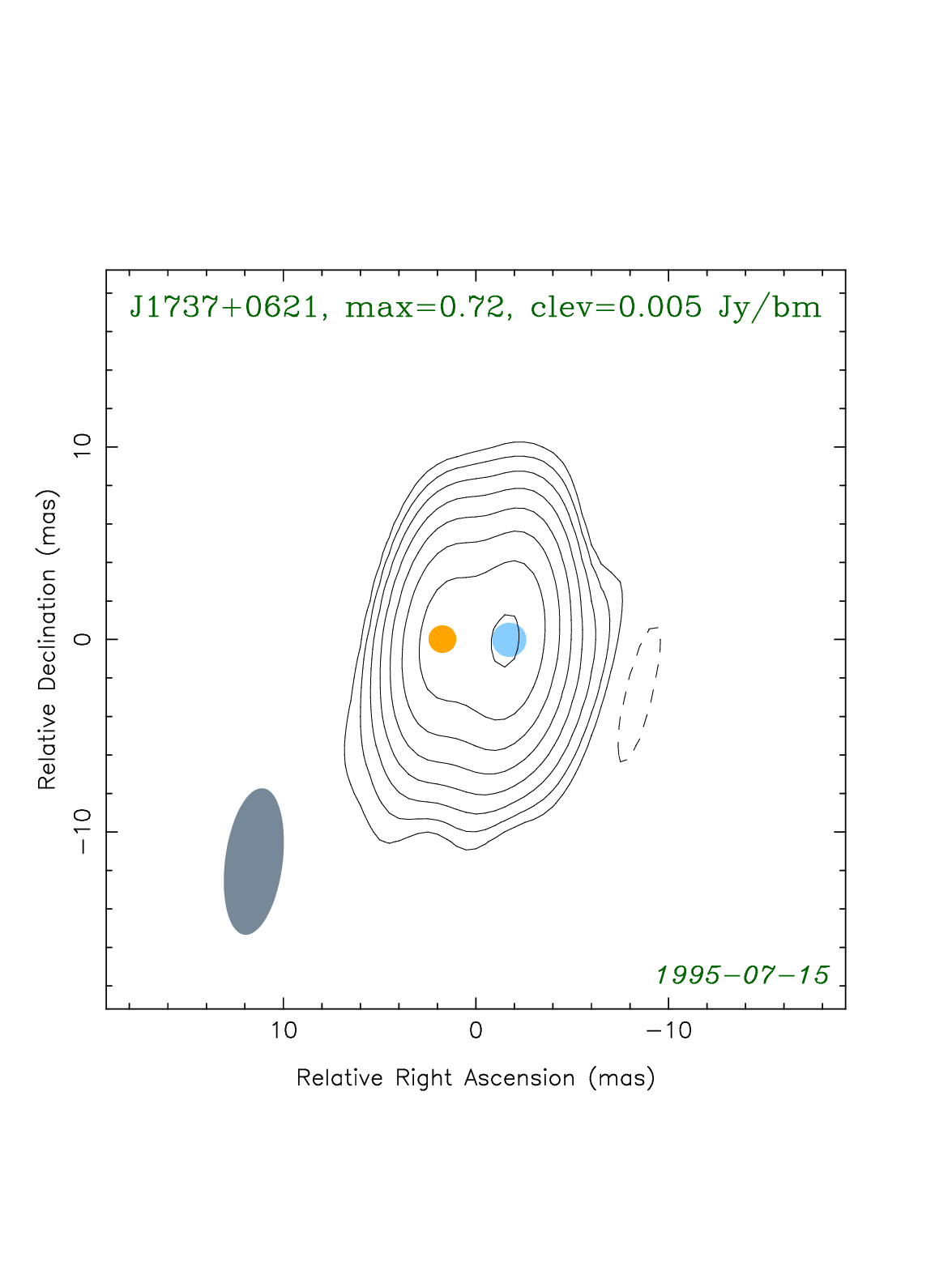}
 \includegraphics[width=0.44\textwidth,angle=0,trim=0 3cm 1cm 5cm,clip]{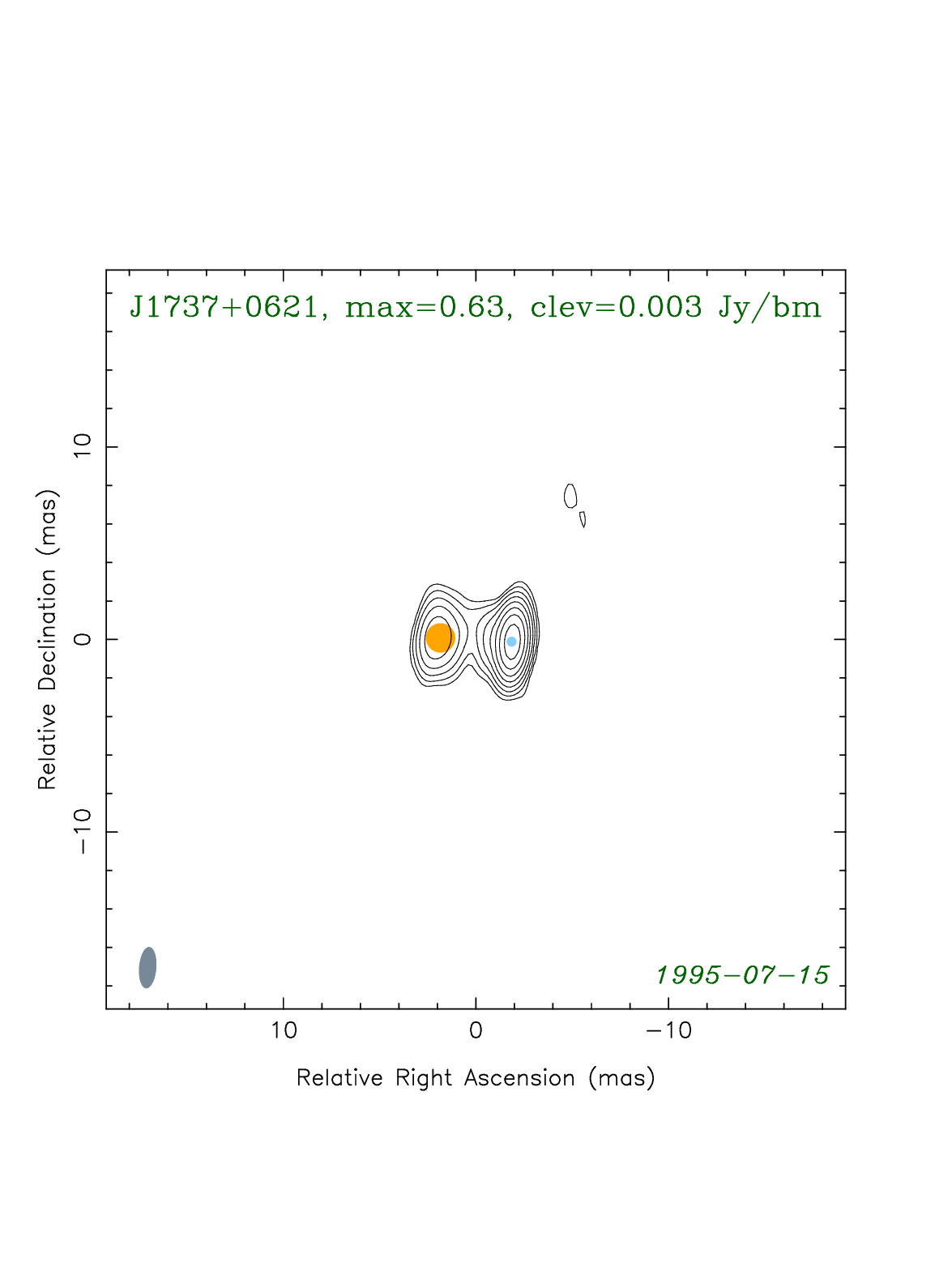}

 \caption{continued}
\end{figure*}

\clearpage
\addtocounter{figure}{-1}
\begin{figure*}
 \centering

 \includegraphics[width=0.44\textwidth,angle=0,trim=0 3cm 1cm 5cm,clip]{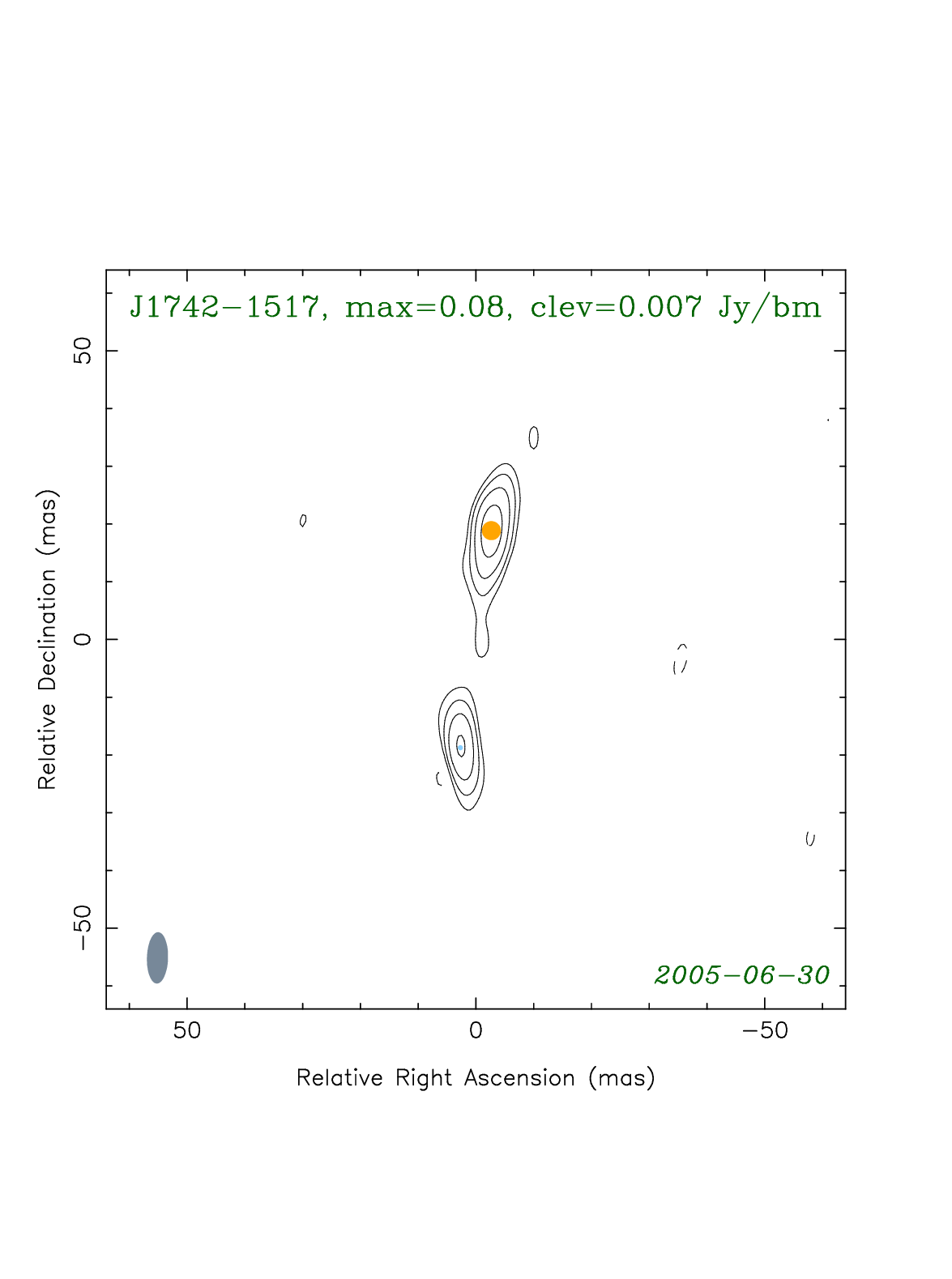}
 \includegraphics[width=0.44\textwidth,angle=0,trim=0 3cm 1cm 5cm,clip]{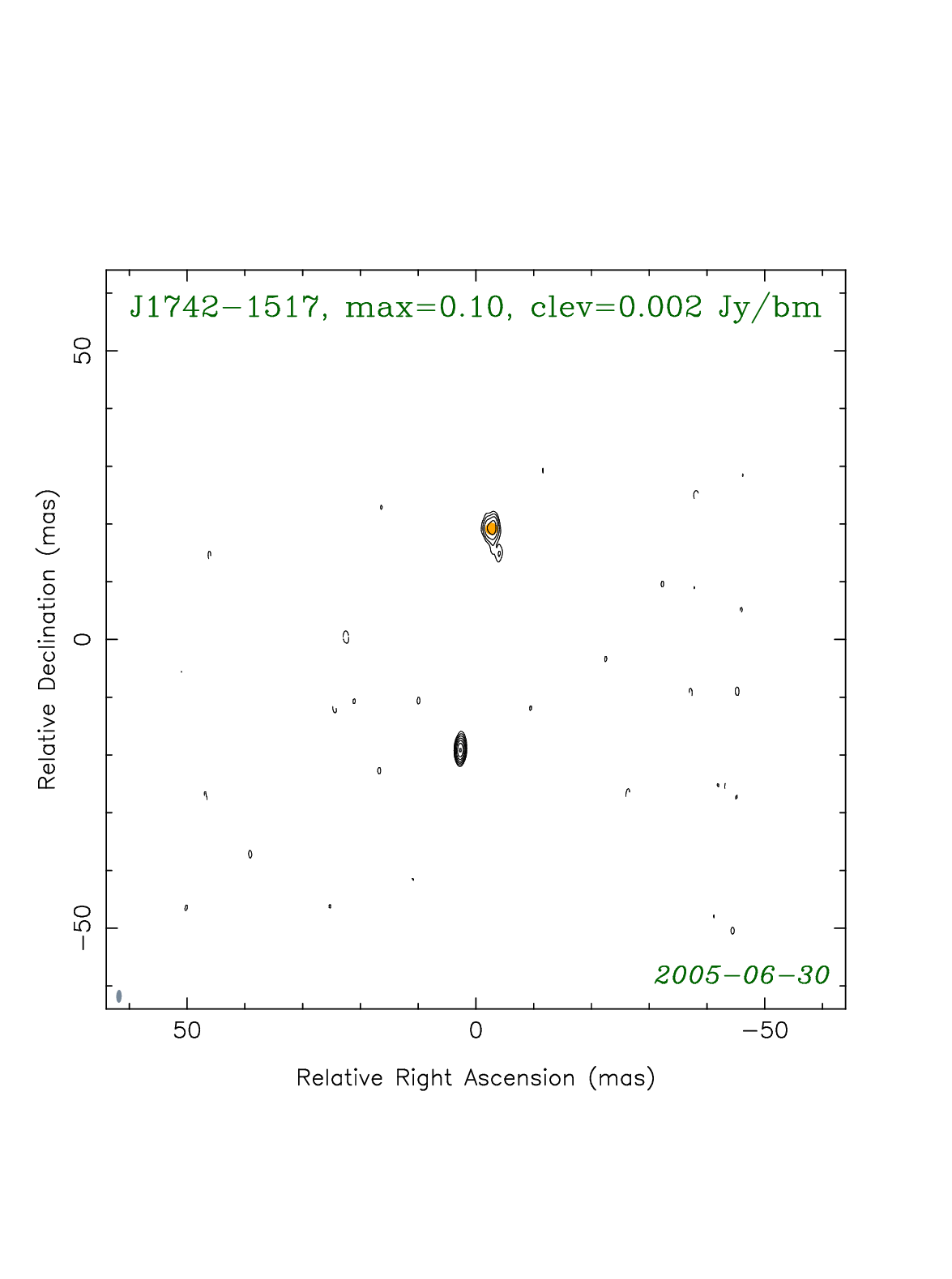}

 \includegraphics[width=0.44\textwidth,angle=0,trim=0 3cm 1cm 5cm,clip]{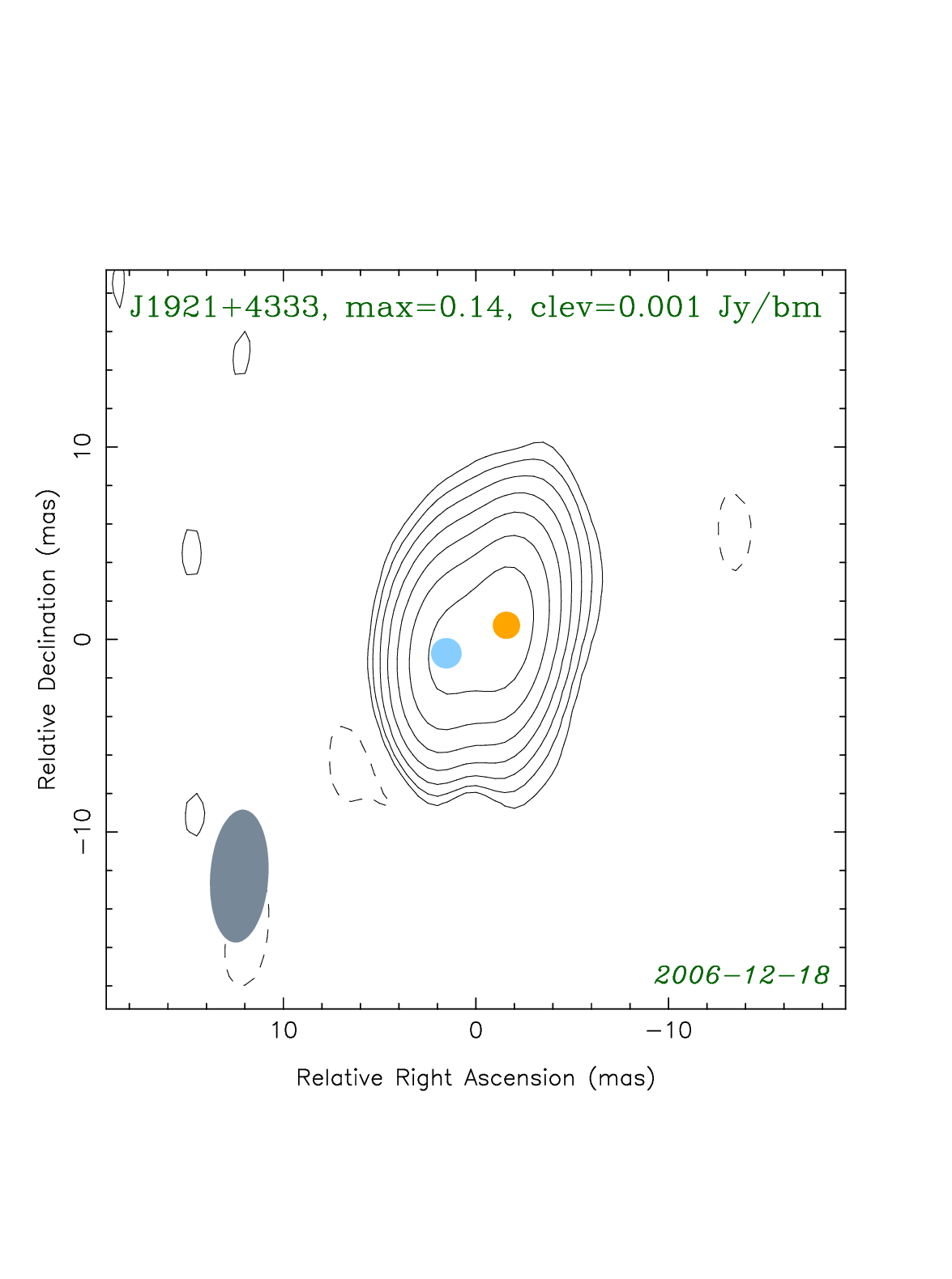}
 \includegraphics[width=0.44\textwidth,angle=0,trim=0 3cm 1cm 5cm,clip]{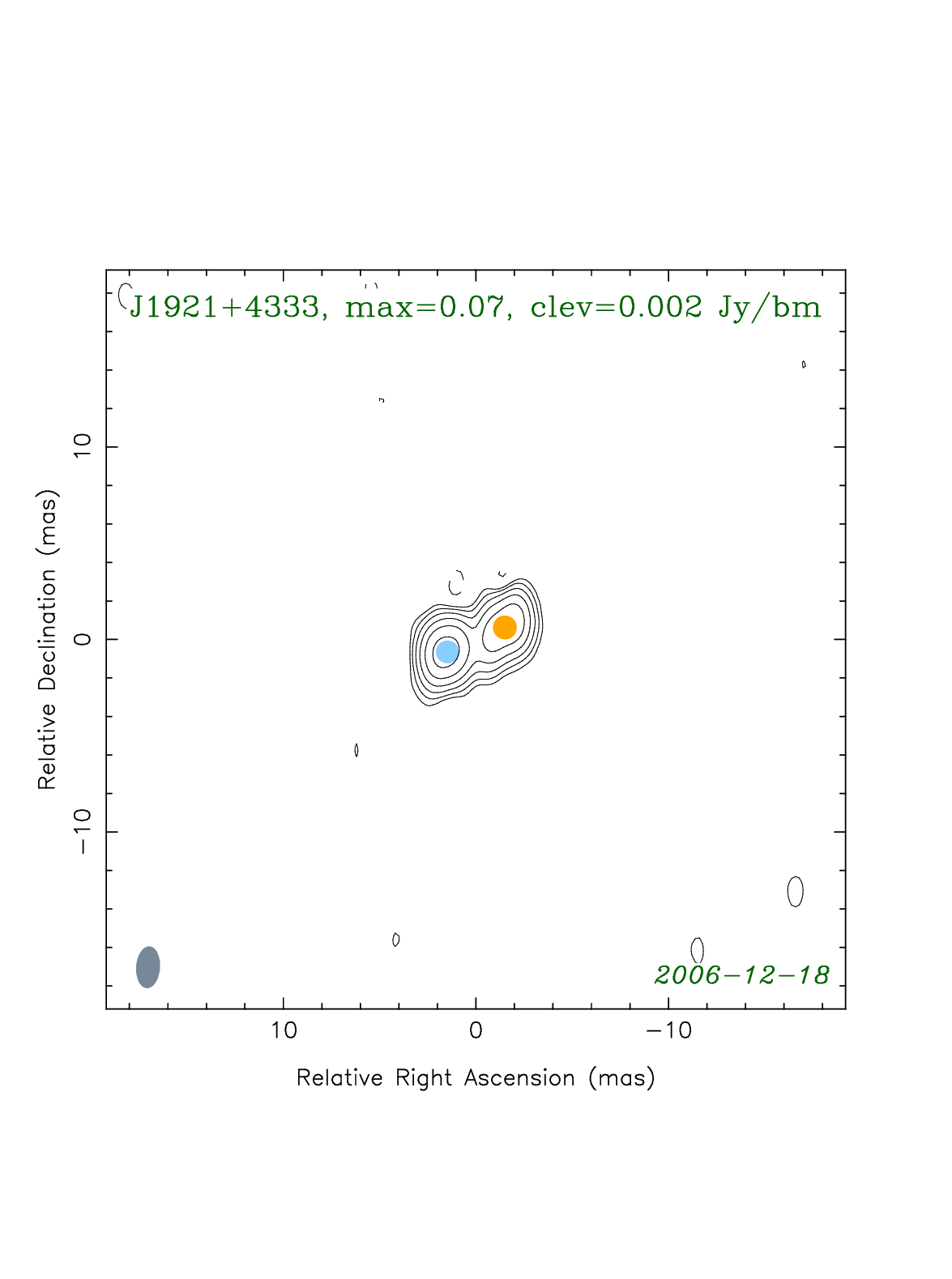}

 \includegraphics[width=0.44\textwidth,angle=0,trim=0 3cm 1cm 5cm,clip]{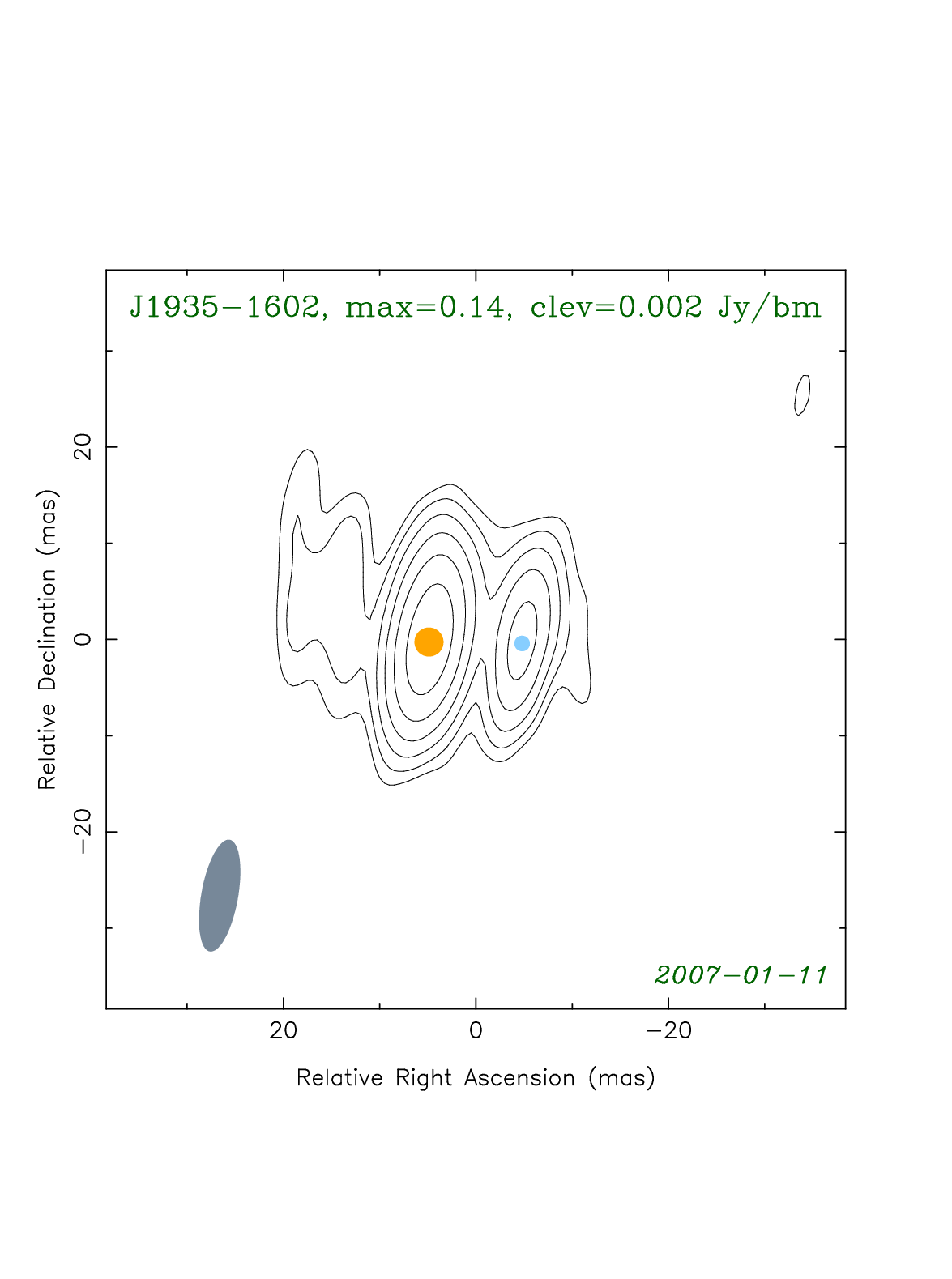}
 \includegraphics[width=0.44\textwidth,angle=0,trim=0 3cm 1cm 5cm,clip]{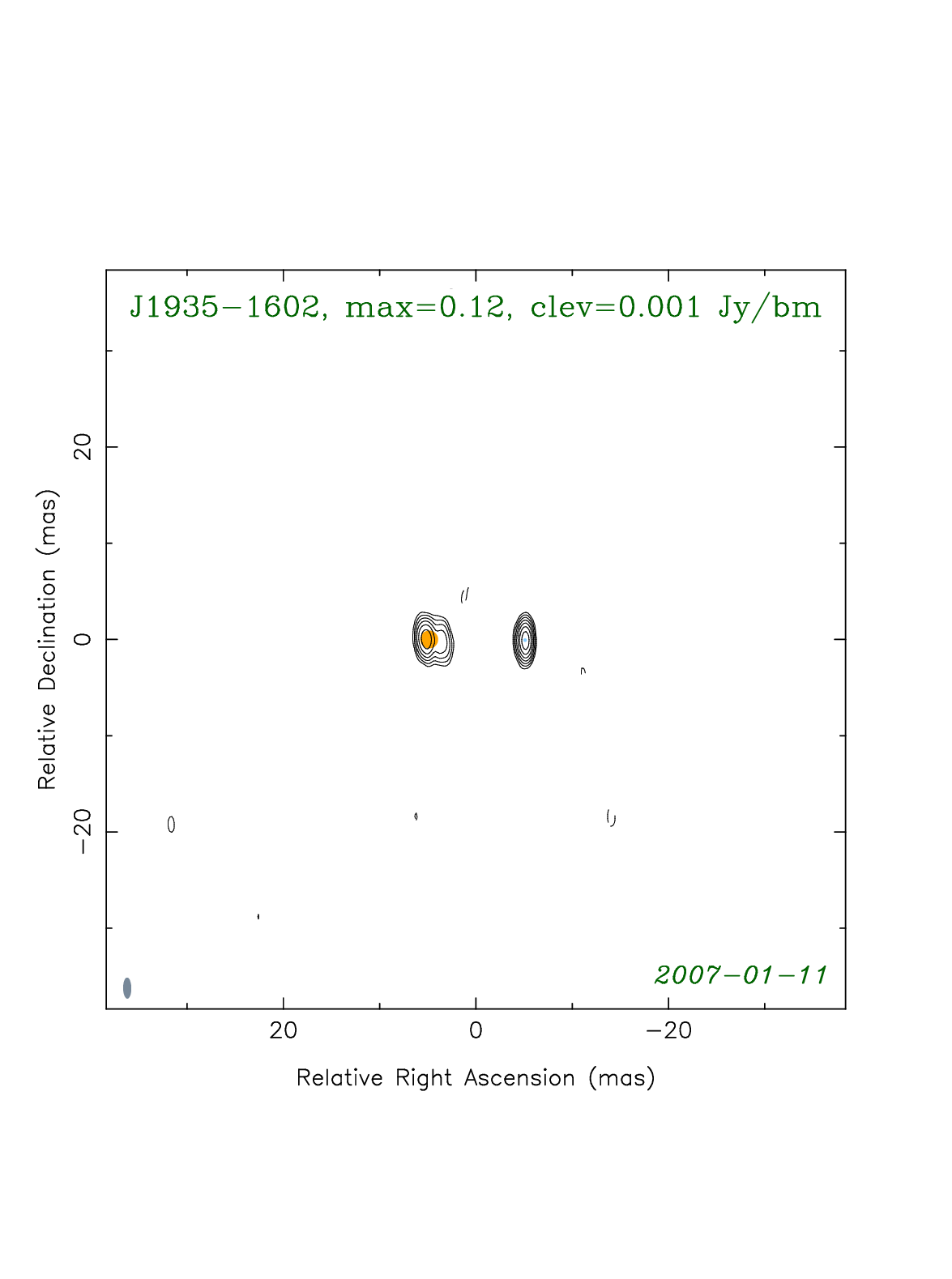}

 \caption{continued}
\end{figure*}

\clearpage
\addtocounter{figure}{-1}
\begin{figure*}
 \centering

 \includegraphics[width=0.44\textwidth,angle=0,trim=0 3cm 1cm 5cm,clip]{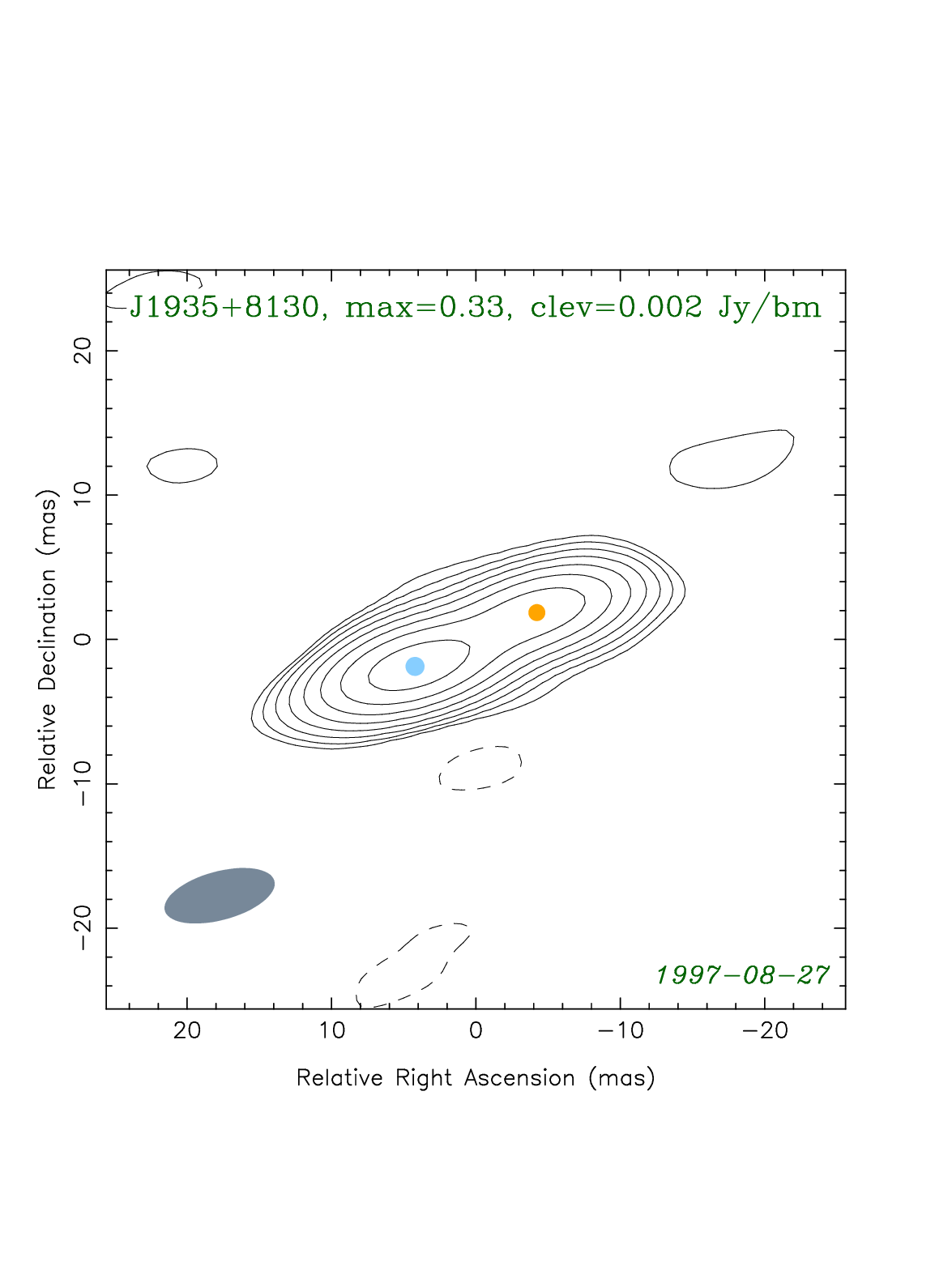}
 \includegraphics[width=0.44\textwidth,angle=0,trim=0 3cm 1cm 5cm,clip]{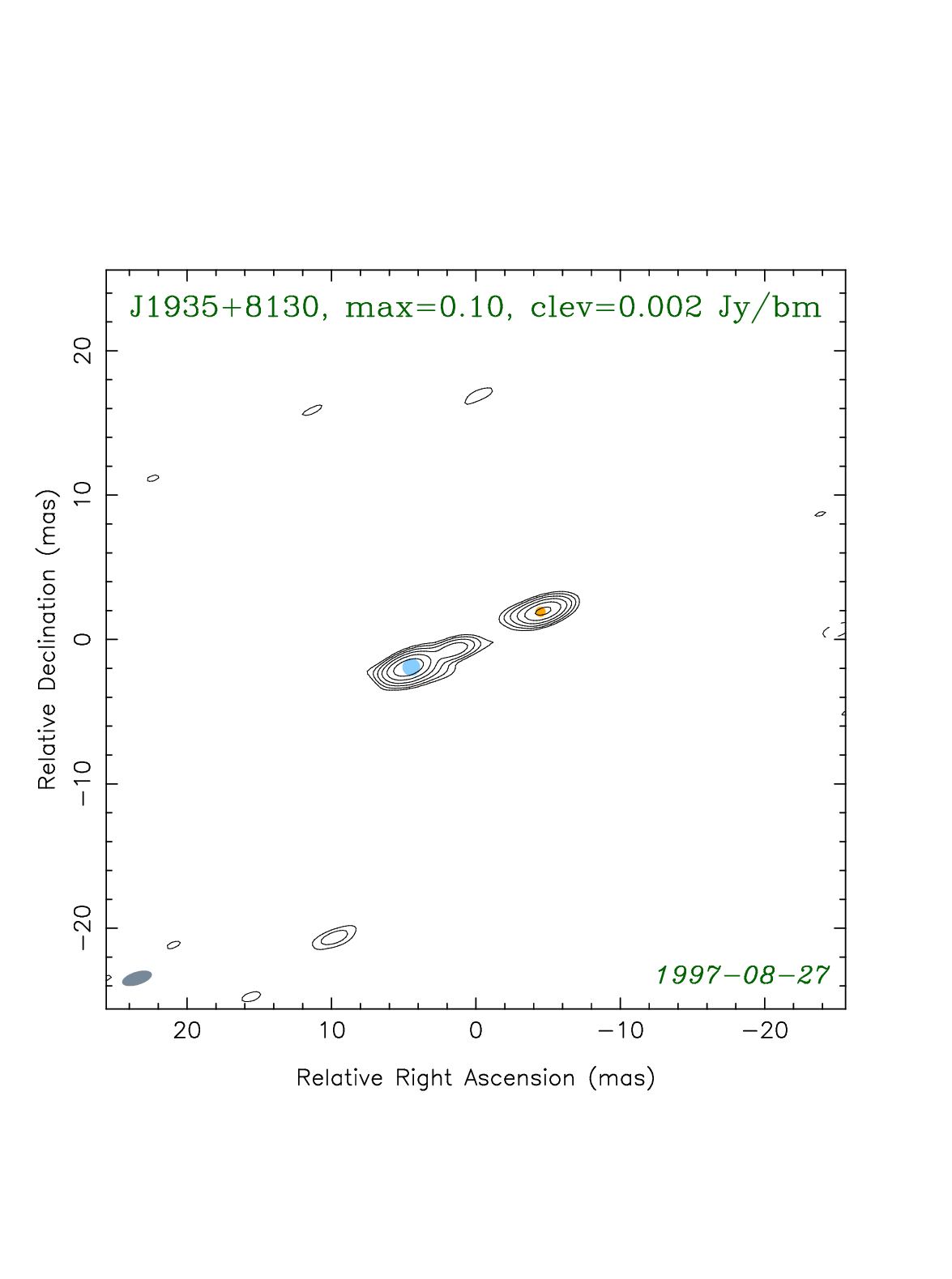}

 \includegraphics[width=0.44\textwidth,angle=0,trim=0 3cm 1cm 5cm,clip]{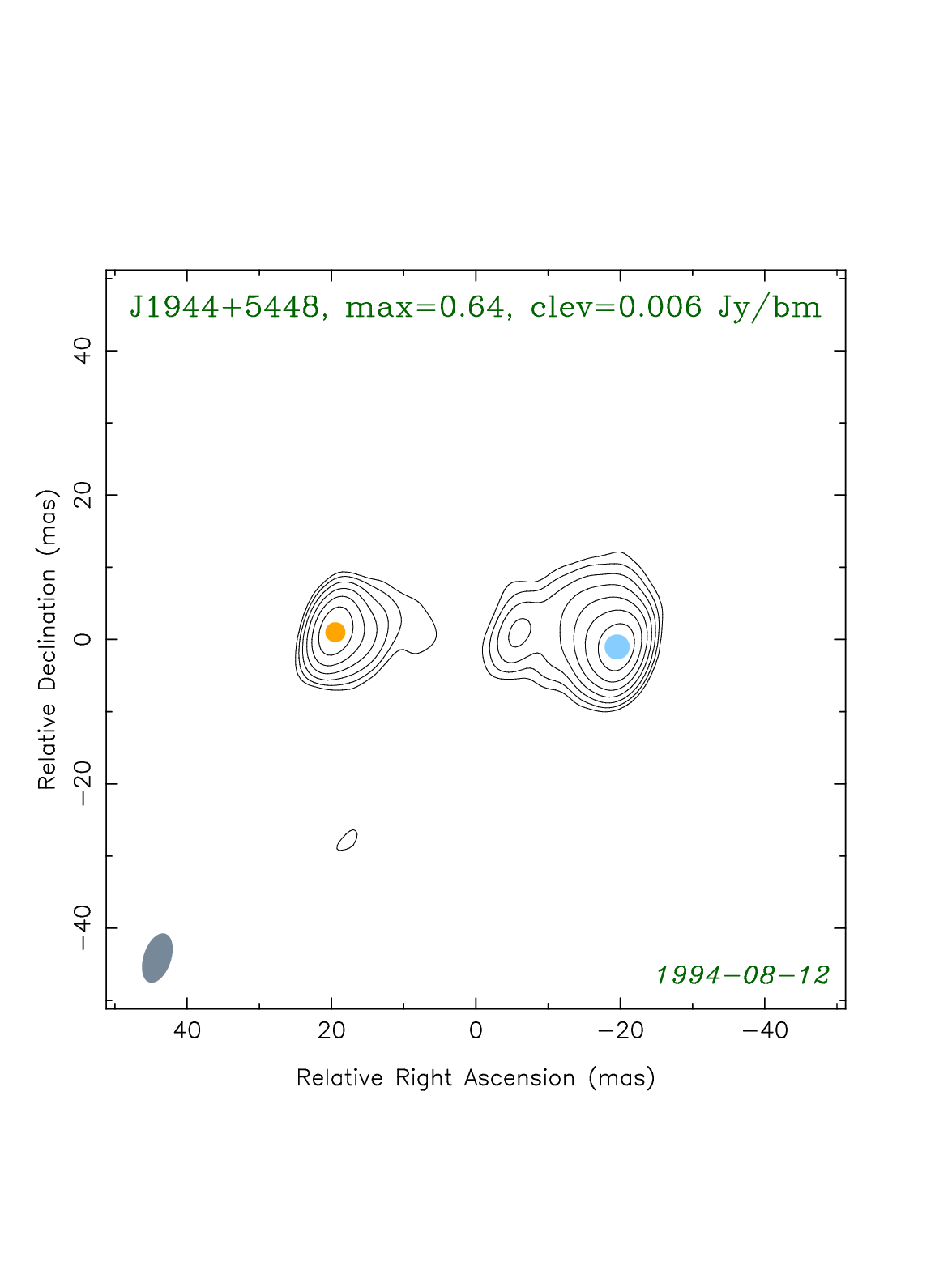}
 \includegraphics[width=0.44\textwidth,angle=0,trim=0 3cm 1cm 5cm,clip]{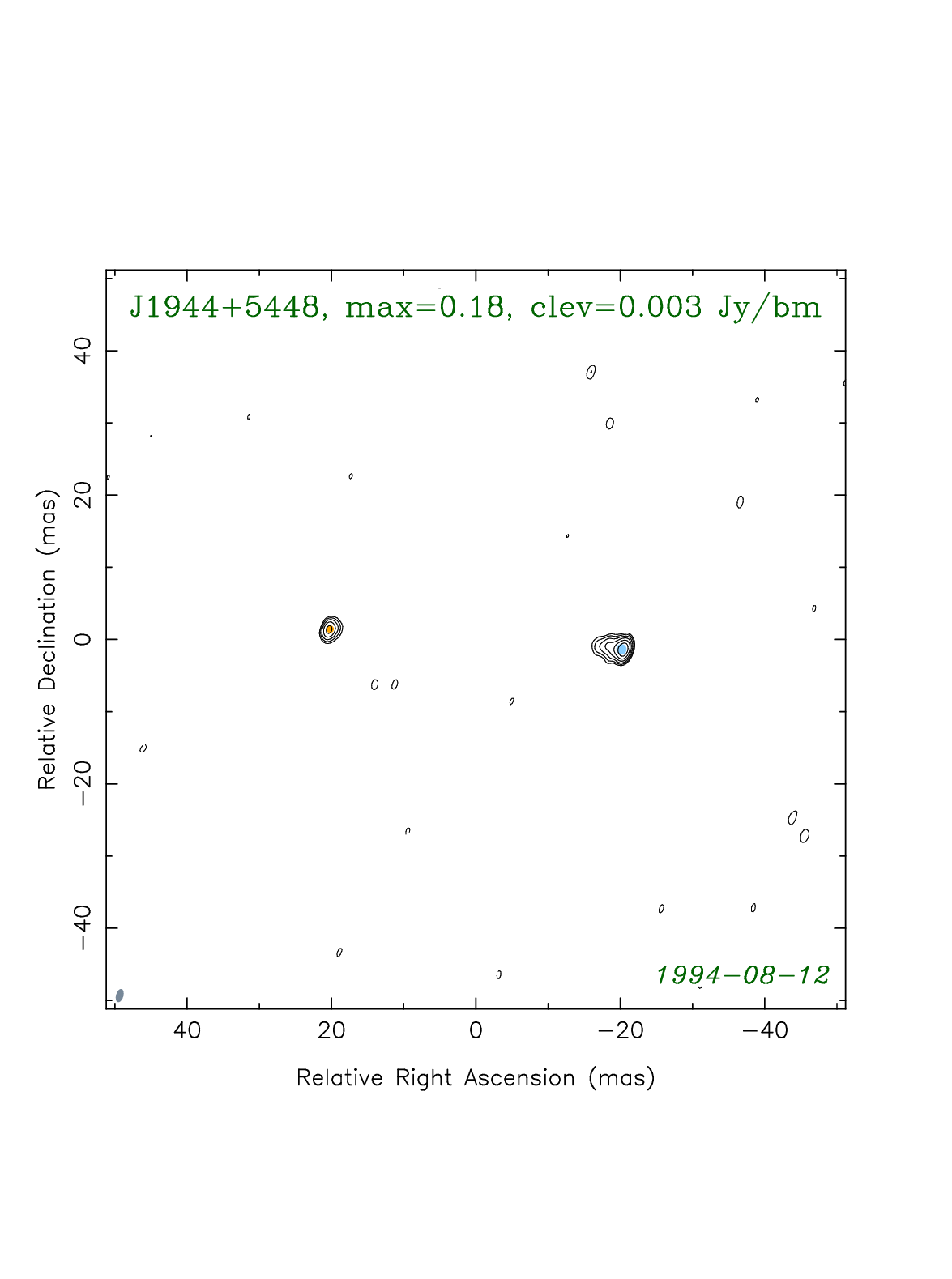}

 \includegraphics[width=0.44\textwidth,angle=0,trim=0 3cm 1cm 5cm,clip]{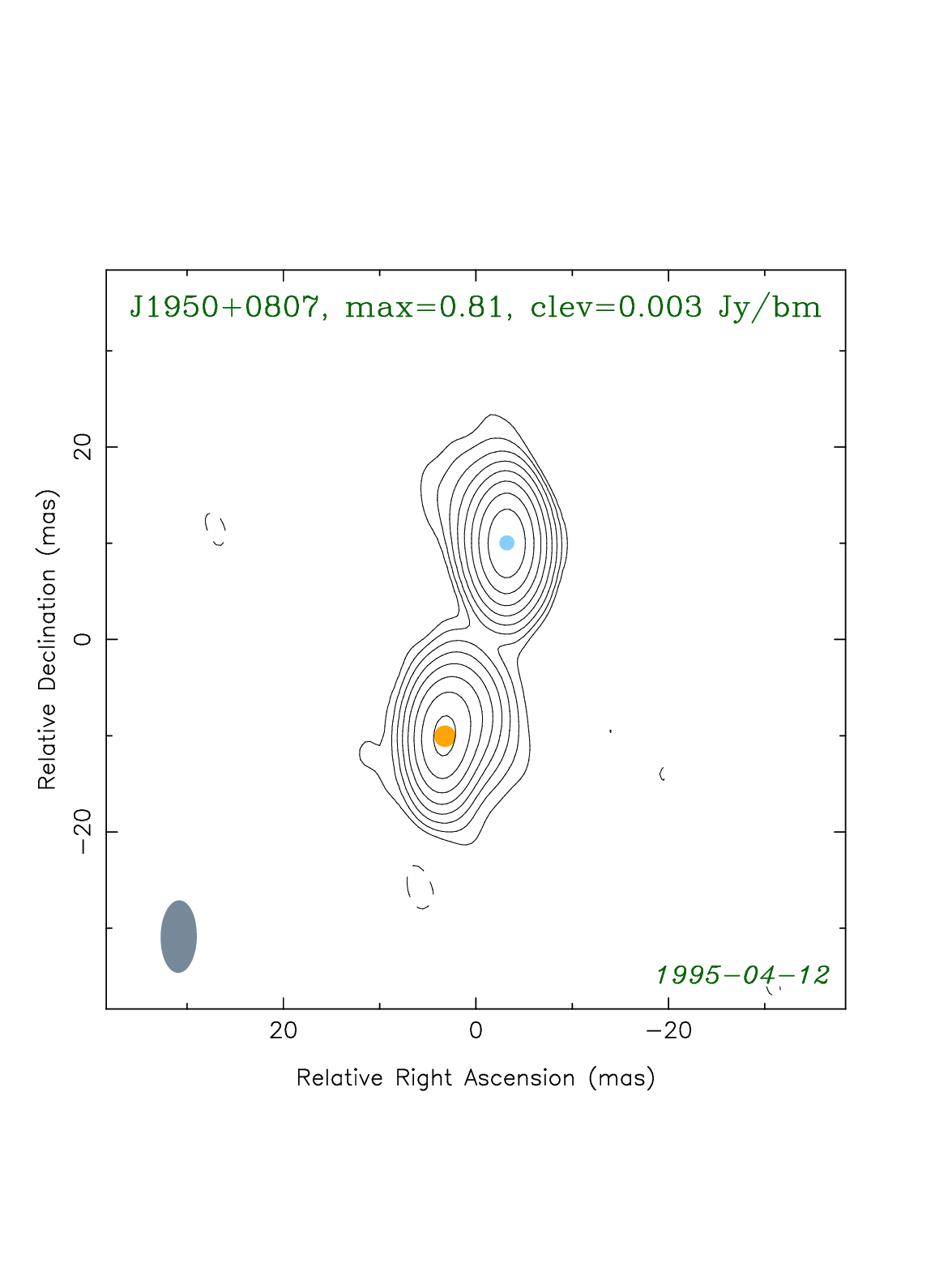}
 \includegraphics[width=0.44\textwidth,angle=0,trim=0 3cm 1cm 5cm,clip]{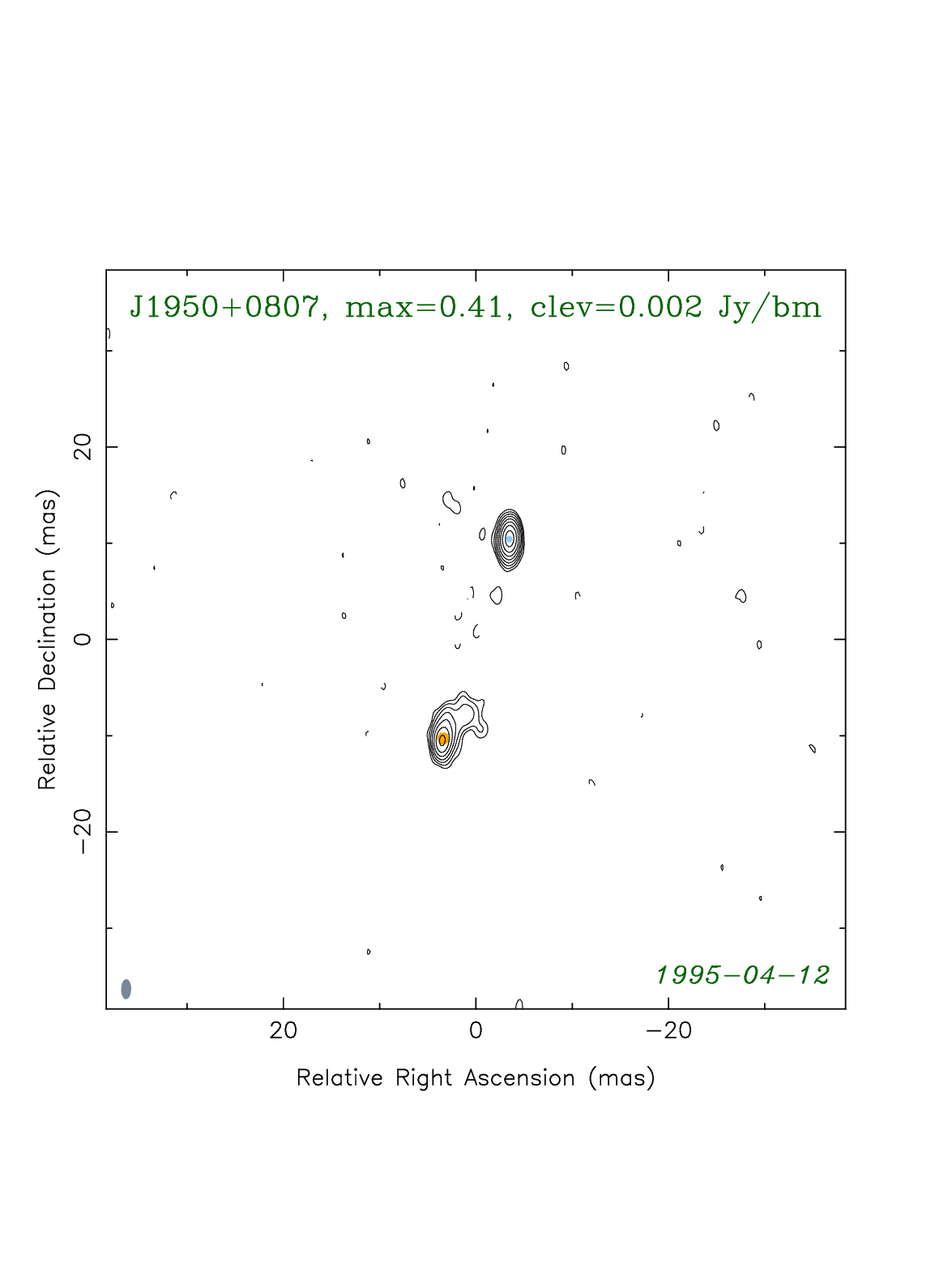}

 \caption{continued}
\end{figure*}

\clearpage
\addtocounter{figure}{-1}
\begin{figure*}
 \centering

 \includegraphics[width=0.44\textwidth,angle=0,trim=0 3cm 1cm 5cm,clip]{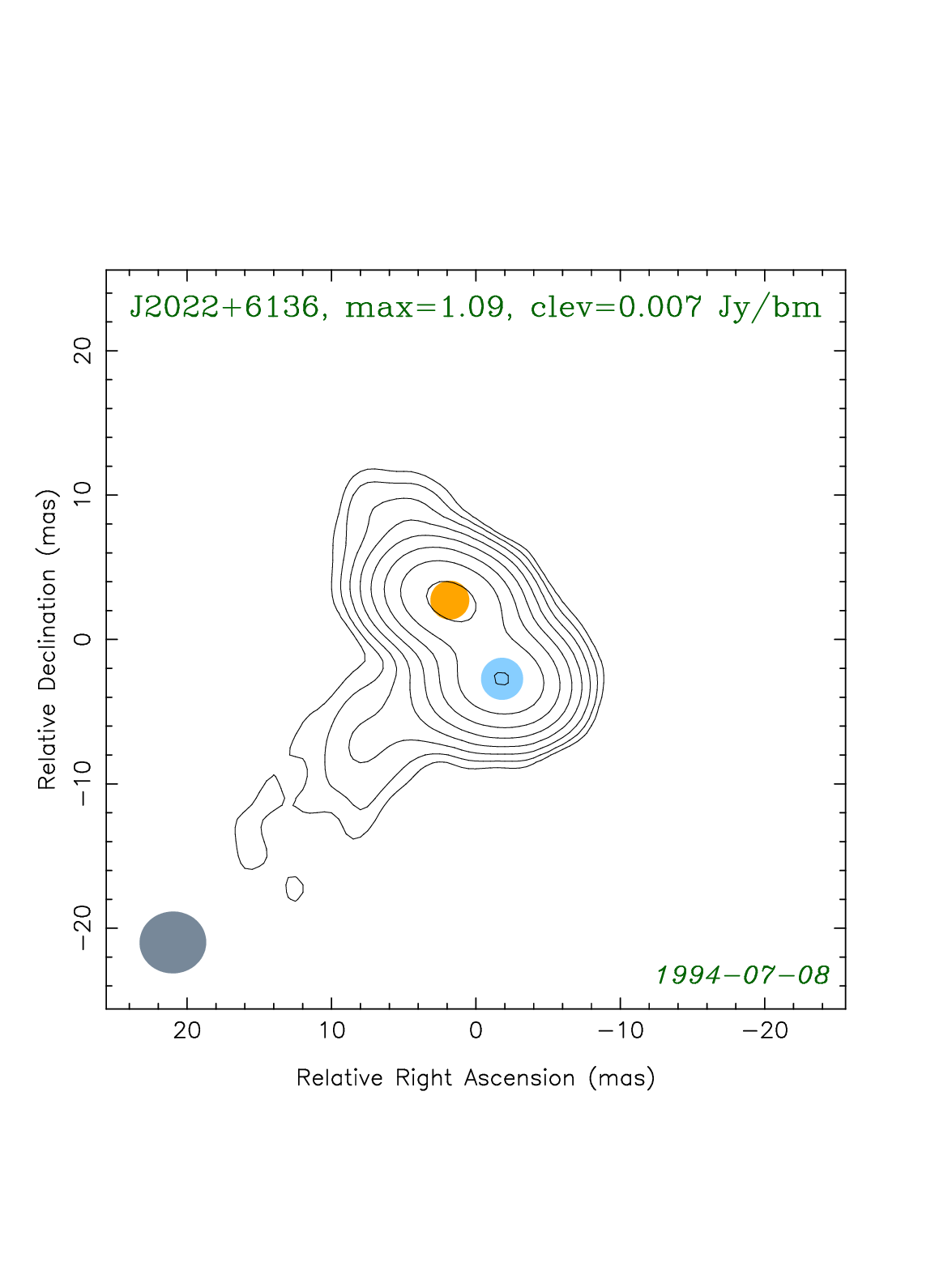}
 \includegraphics[width=0.44\textwidth,angle=0,trim=0 3cm 1cm 5cm,clip]{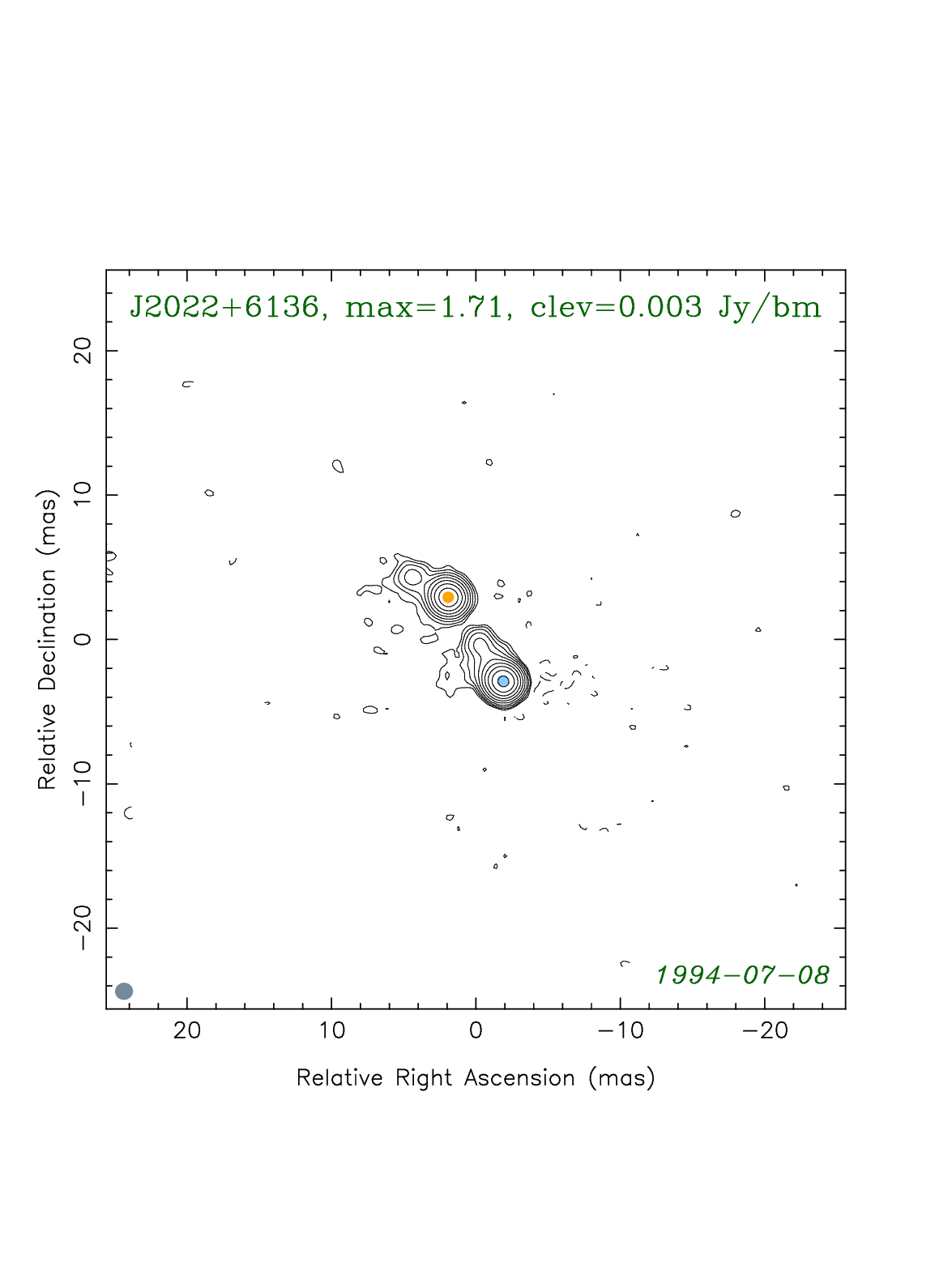}

 \includegraphics[width=0.44\textwidth,angle=0,trim=0 3cm 1cm 5cm,clip]{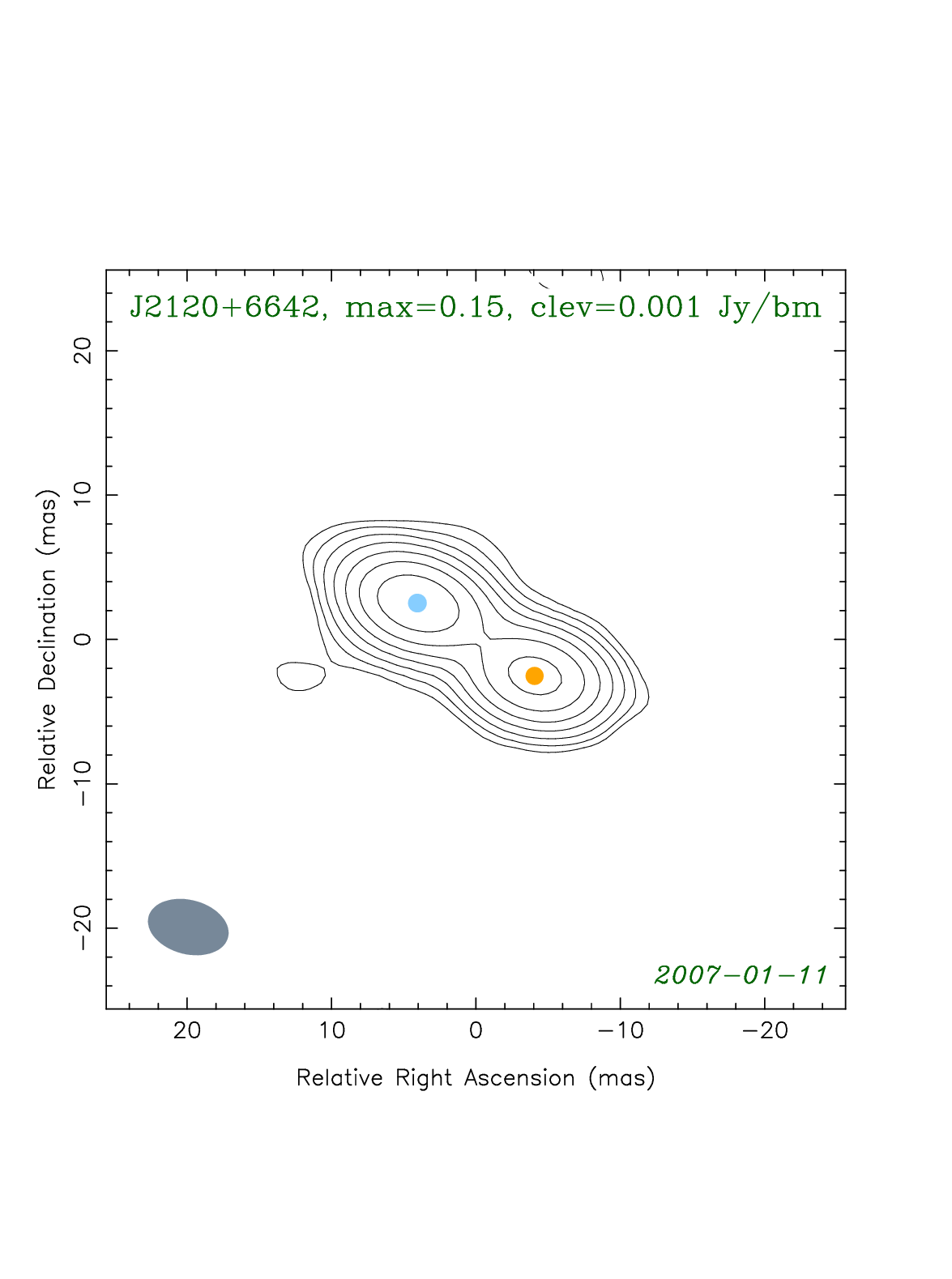}
 \includegraphics[width=0.44\textwidth,angle=0,trim=0 3cm 1cm 5cm,clip]{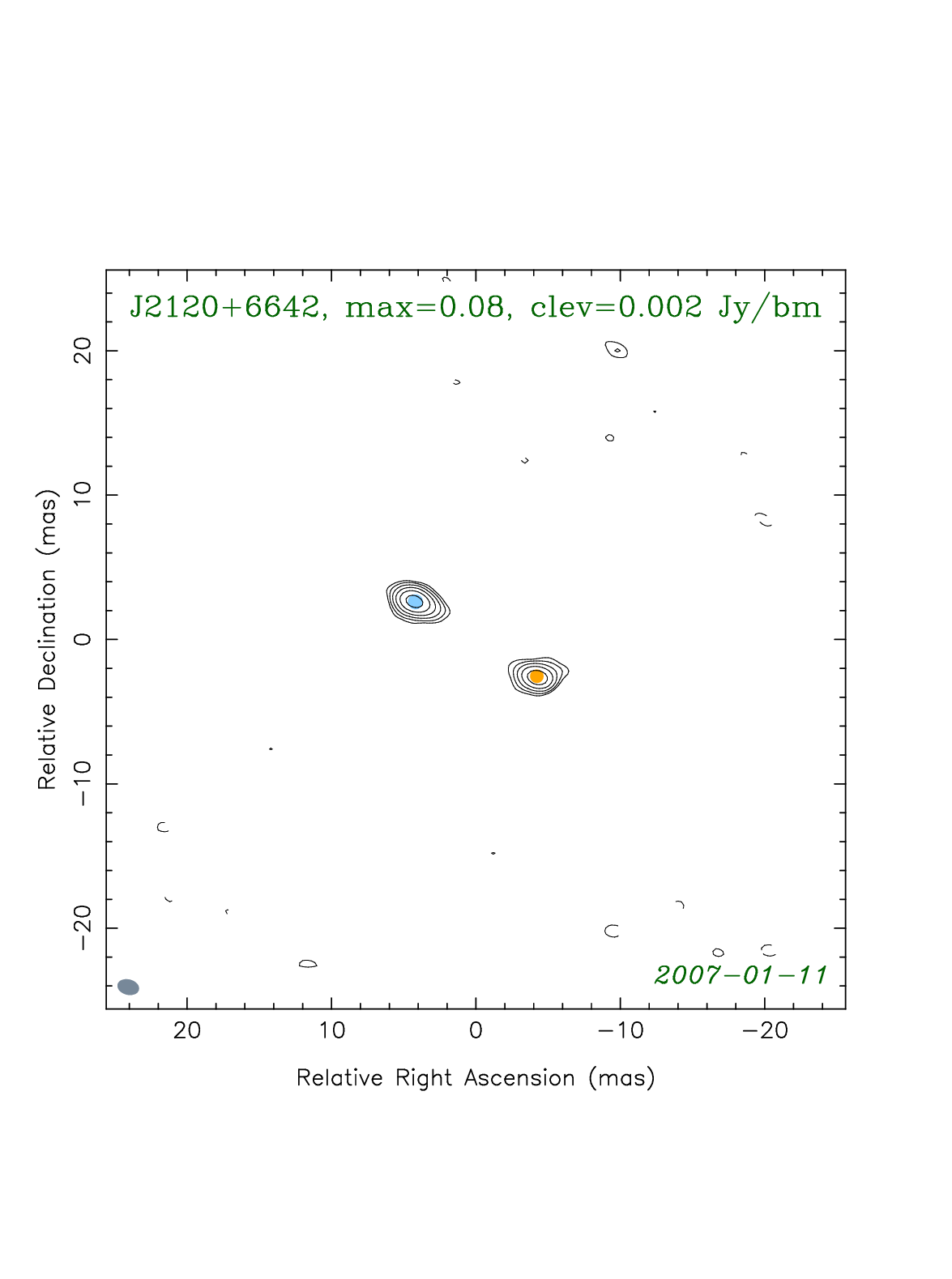}

 \includegraphics[width=0.44\textwidth,angle=0,trim=0 3cm 1cm 5cm,clip]{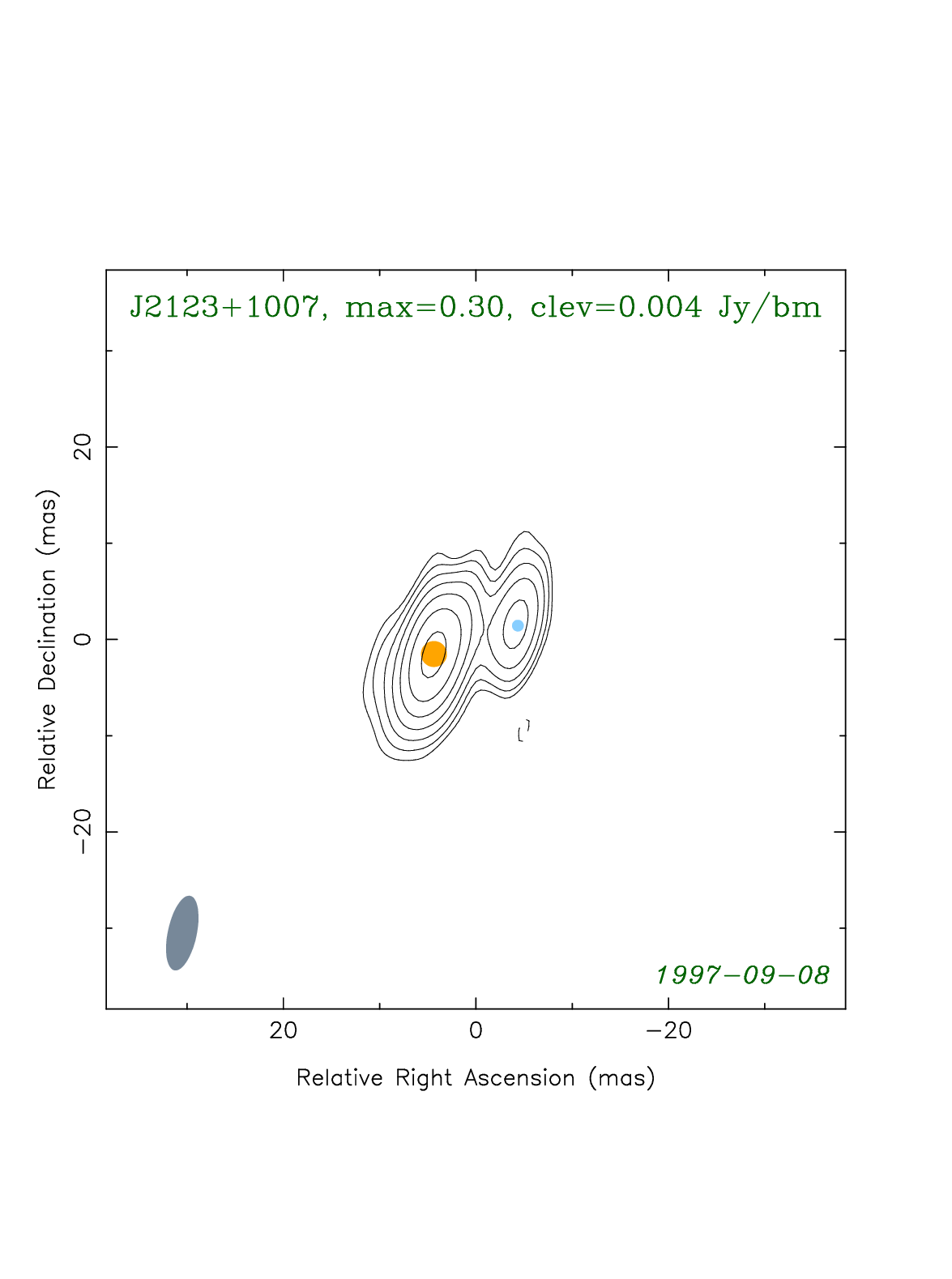}
 \includegraphics[width=0.44\textwidth,angle=0,trim=0 3cm 1cm 5cm,clip]{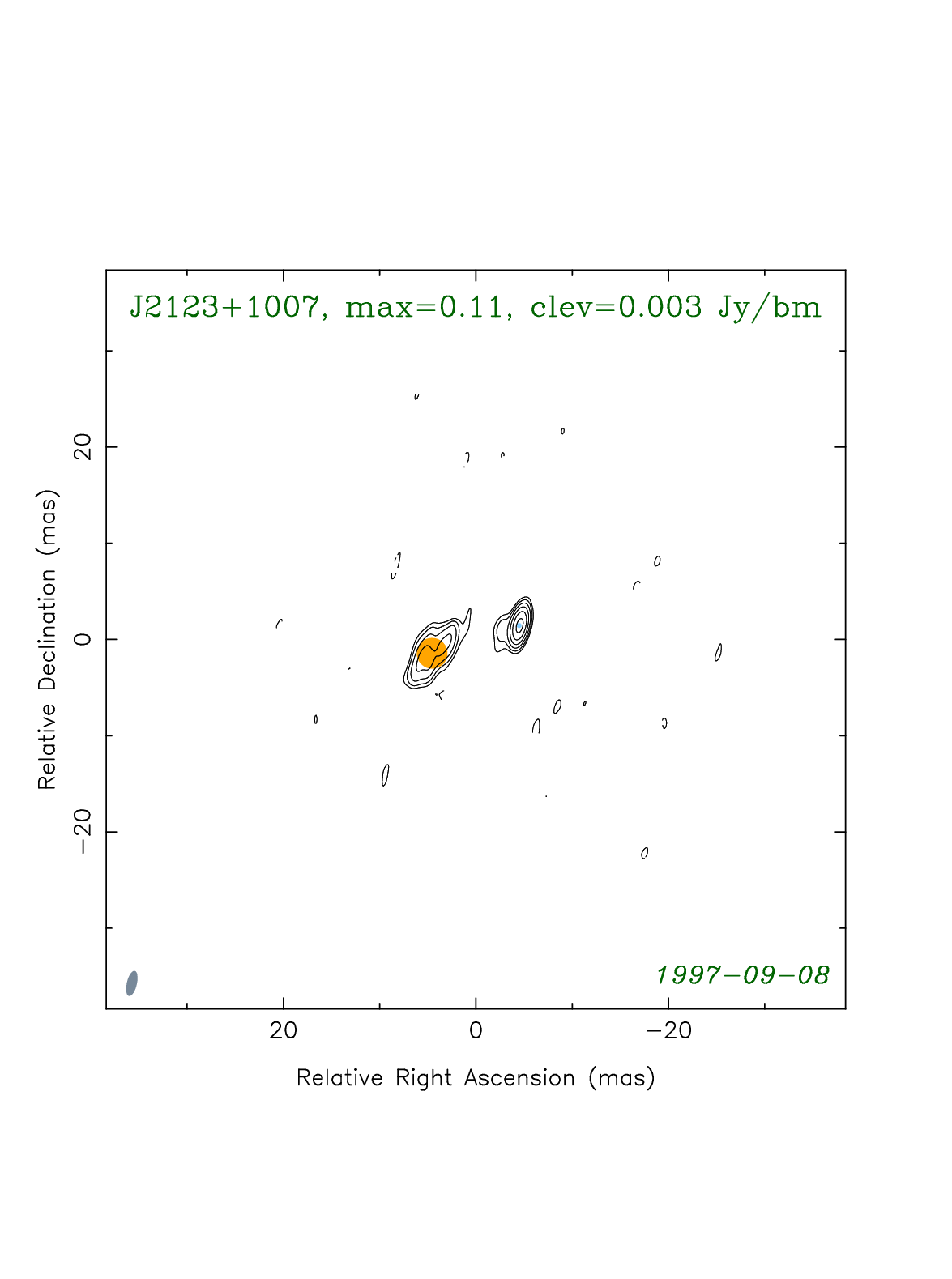}

 \caption{continued}
\end{figure*}

\clearpage
\addtocounter{figure}{-1}
\begin{figure*}
 \centering

 \includegraphics[width=0.44\textwidth,angle=0,trim=0 3cm 1cm 5cm,clip]{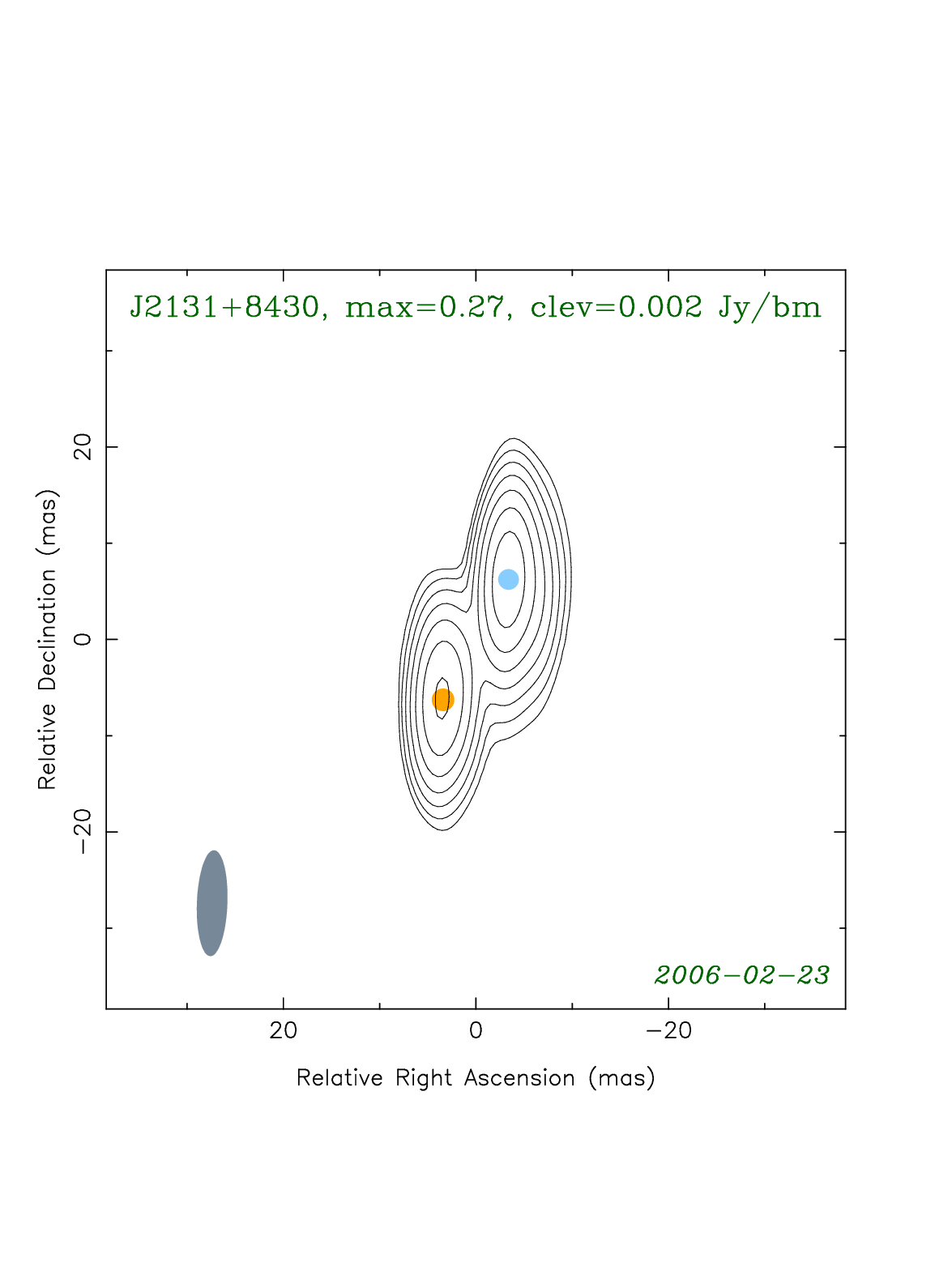}
 \includegraphics[width=0.44\textwidth,angle=0,trim=0 3cm 1cm 5cm,clip]{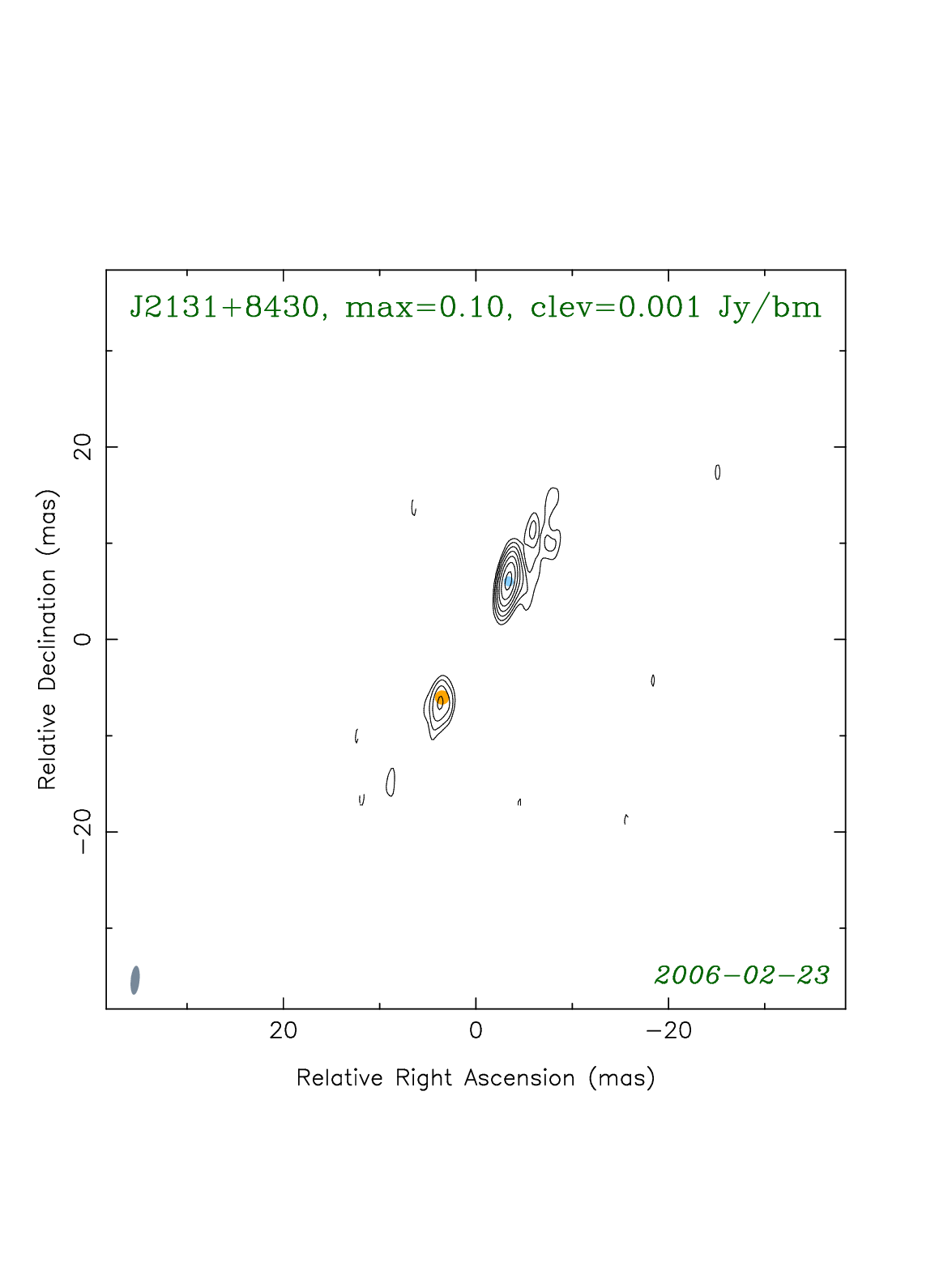}

 \includegraphics[width=0.44\textwidth,angle=0,trim=0 3cm 1cm 5cm,clip]{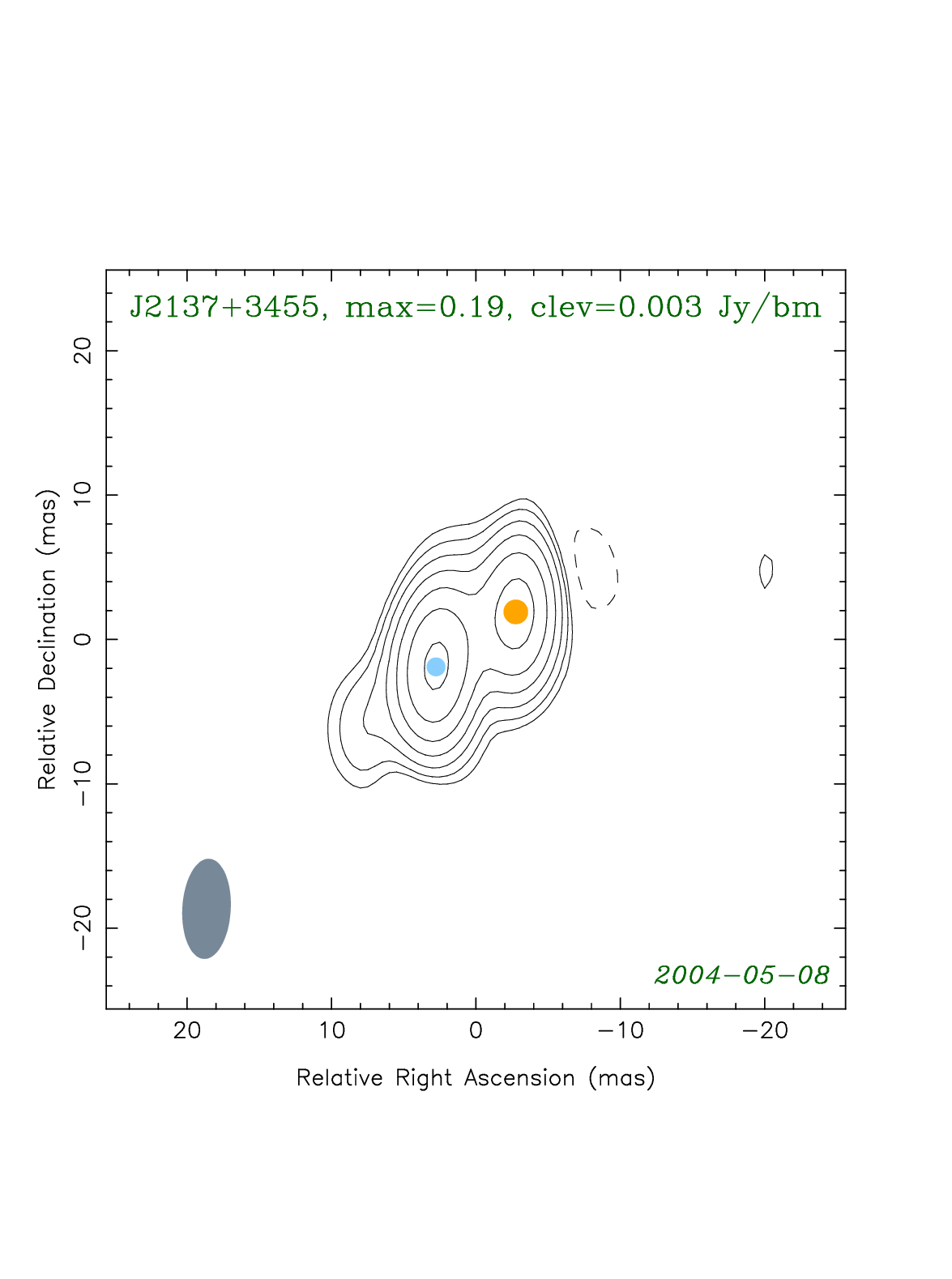}
 \includegraphics[width=0.44\textwidth,angle=0,trim=0 3cm 1cm 5cm,clip]{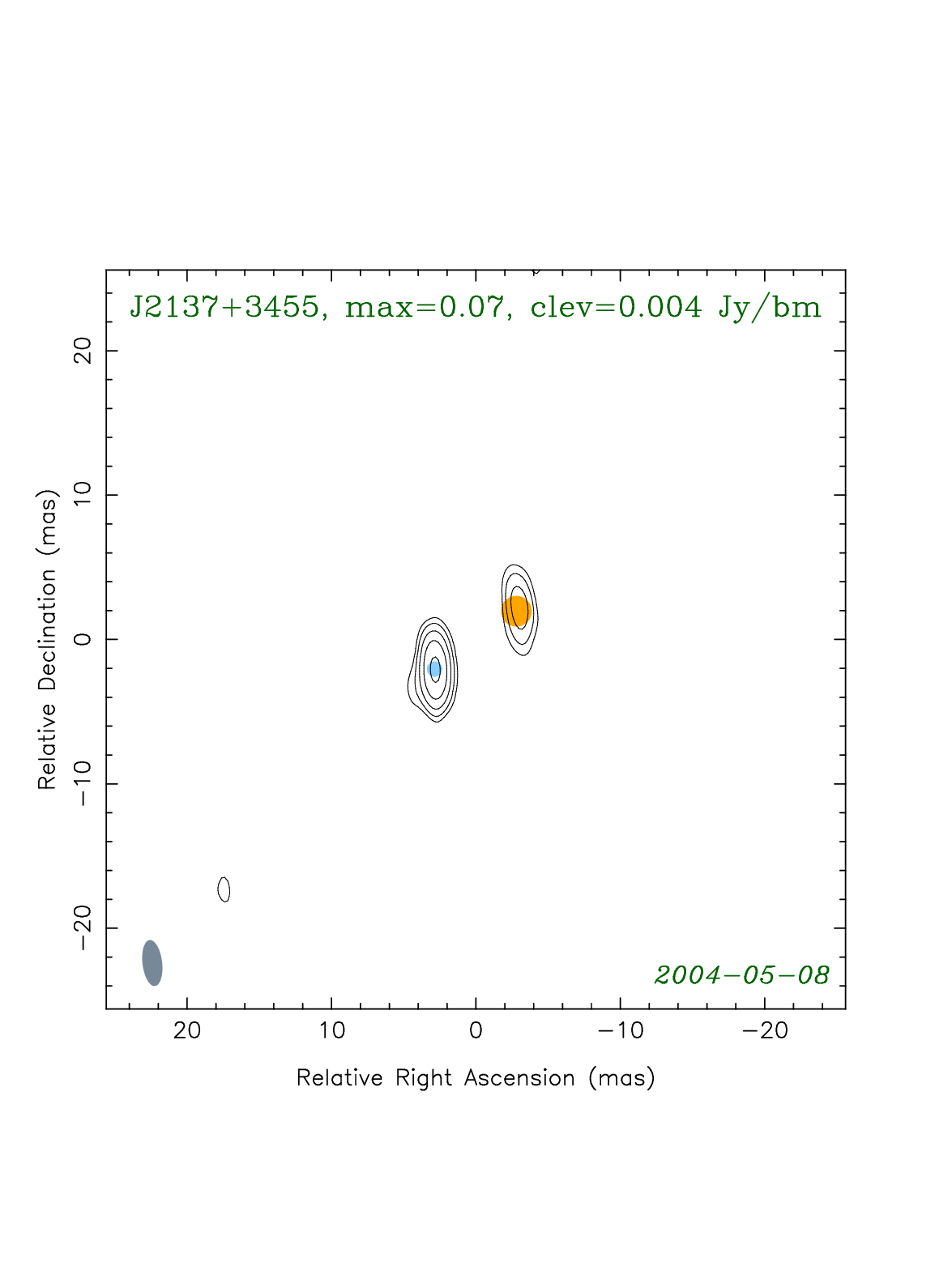}

 \includegraphics[width=0.44\textwidth,angle=0,trim=0 3cm 1cm 5cm,clip]{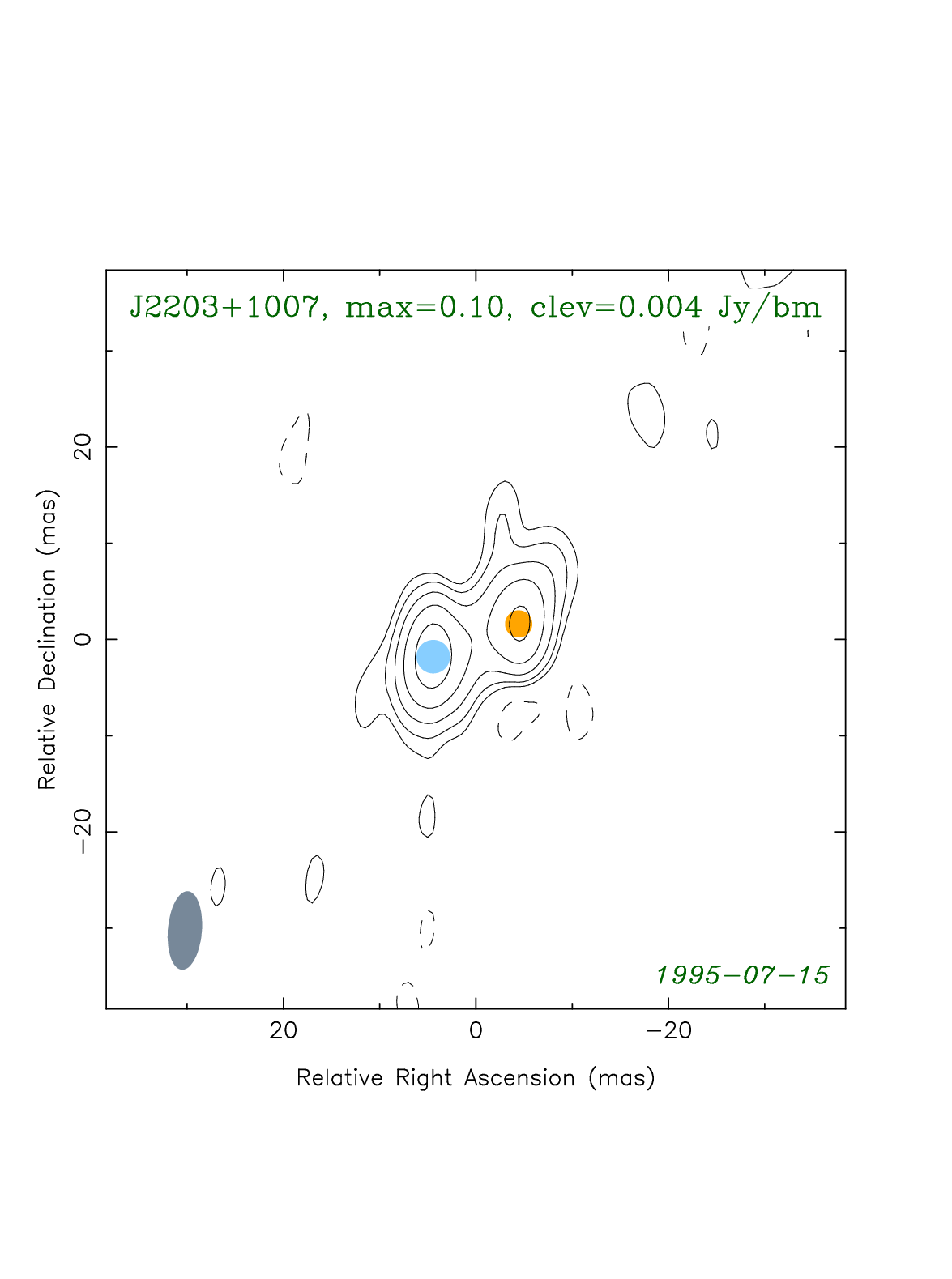}
 \includegraphics[width=0.44\textwidth,angle=0,trim=0 3cm 1cm 5cm,clip]{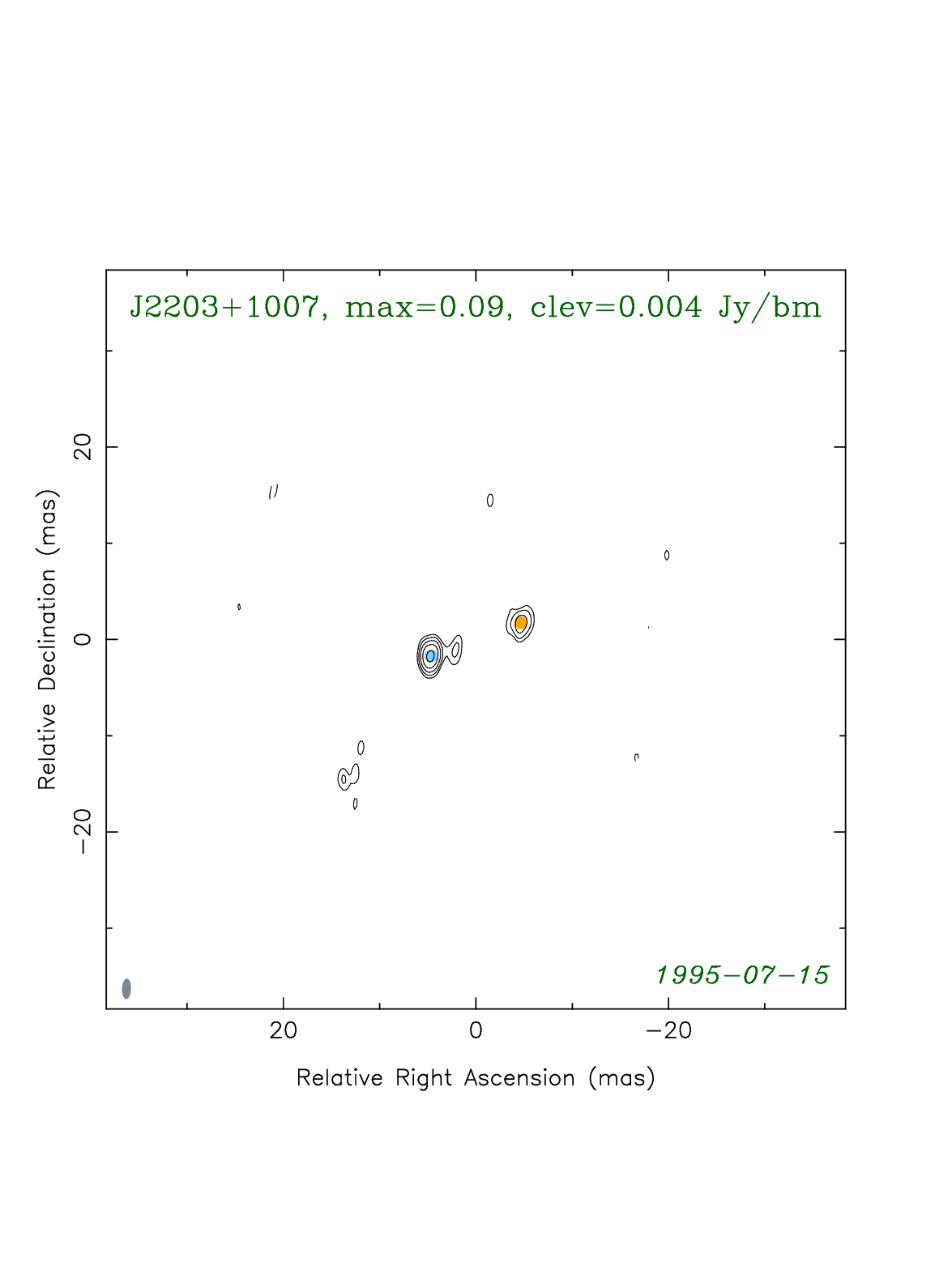}

 \caption{continued}
\end{figure*}

\clearpage
\addtocounter{figure}{-1}
\begin{figure*}
 \centering

 \includegraphics[width=0.44\textwidth,angle=0,trim=0 3cm 1cm 5cm,clip]{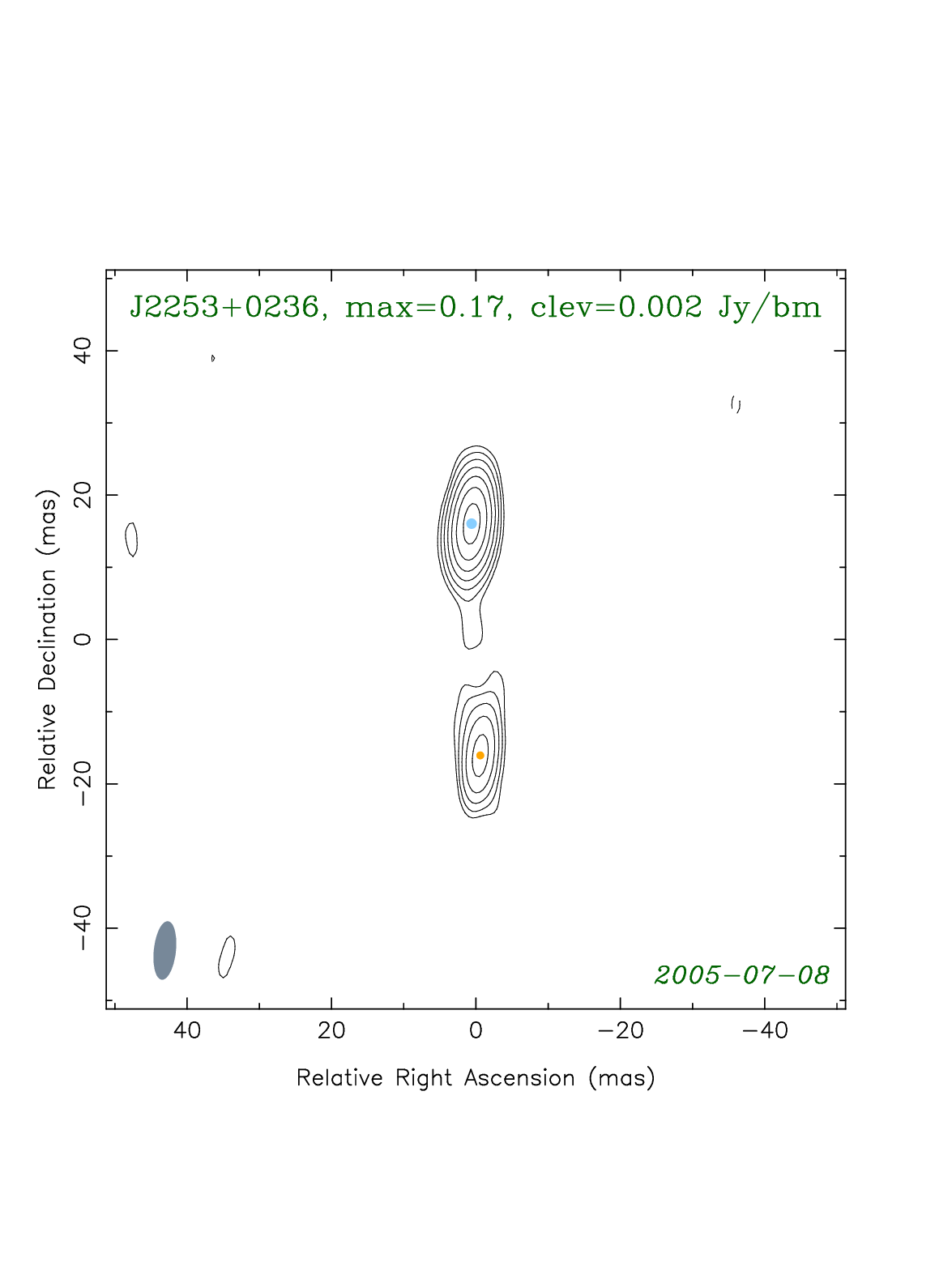}
 \includegraphics[width=0.44\textwidth,angle=0,trim=0 3cm 1cm 5cm,clip]{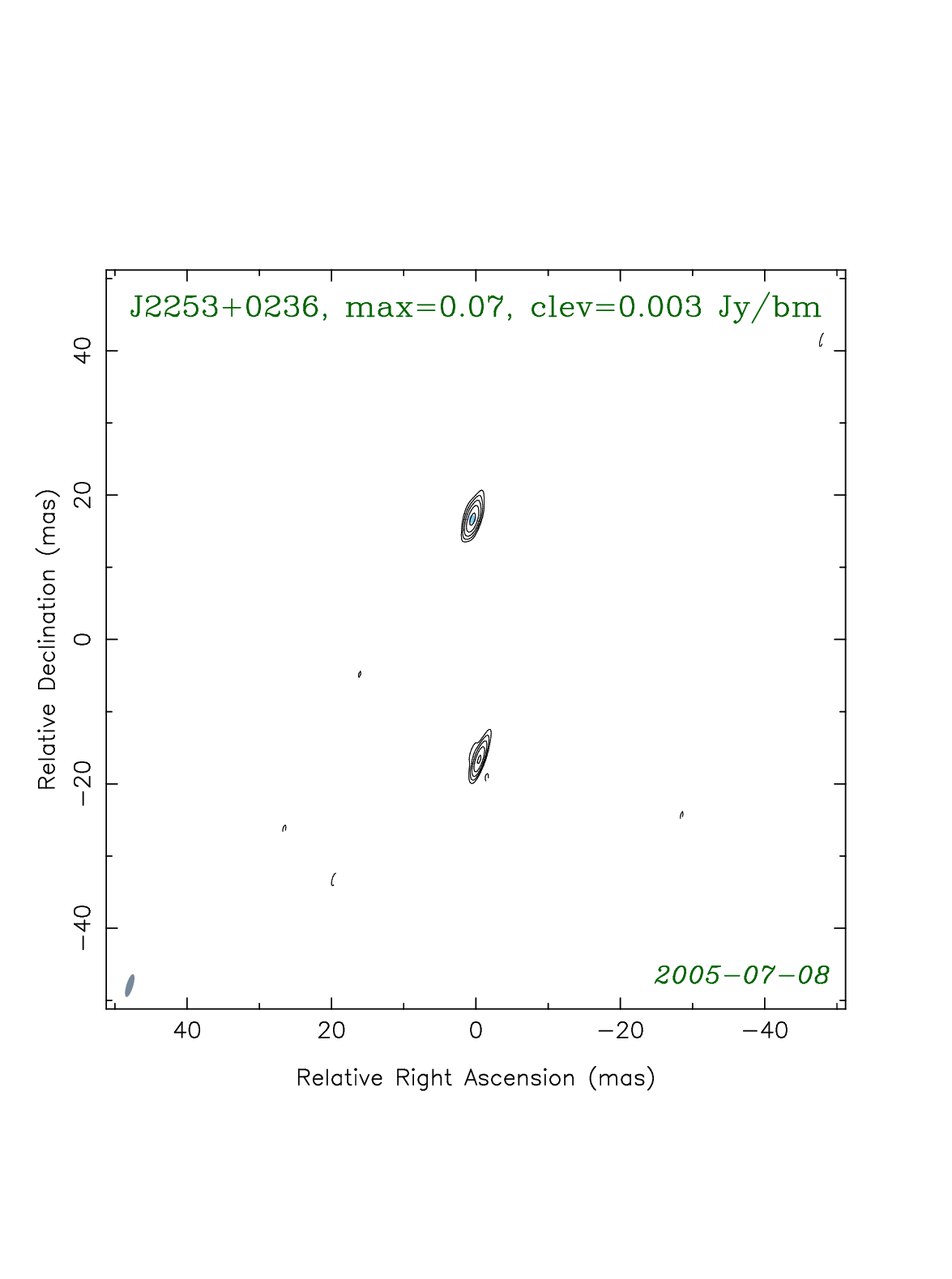}

 \includegraphics[width=0.44\textwidth,angle=0,trim=0 3cm 1cm 5cm,clip]{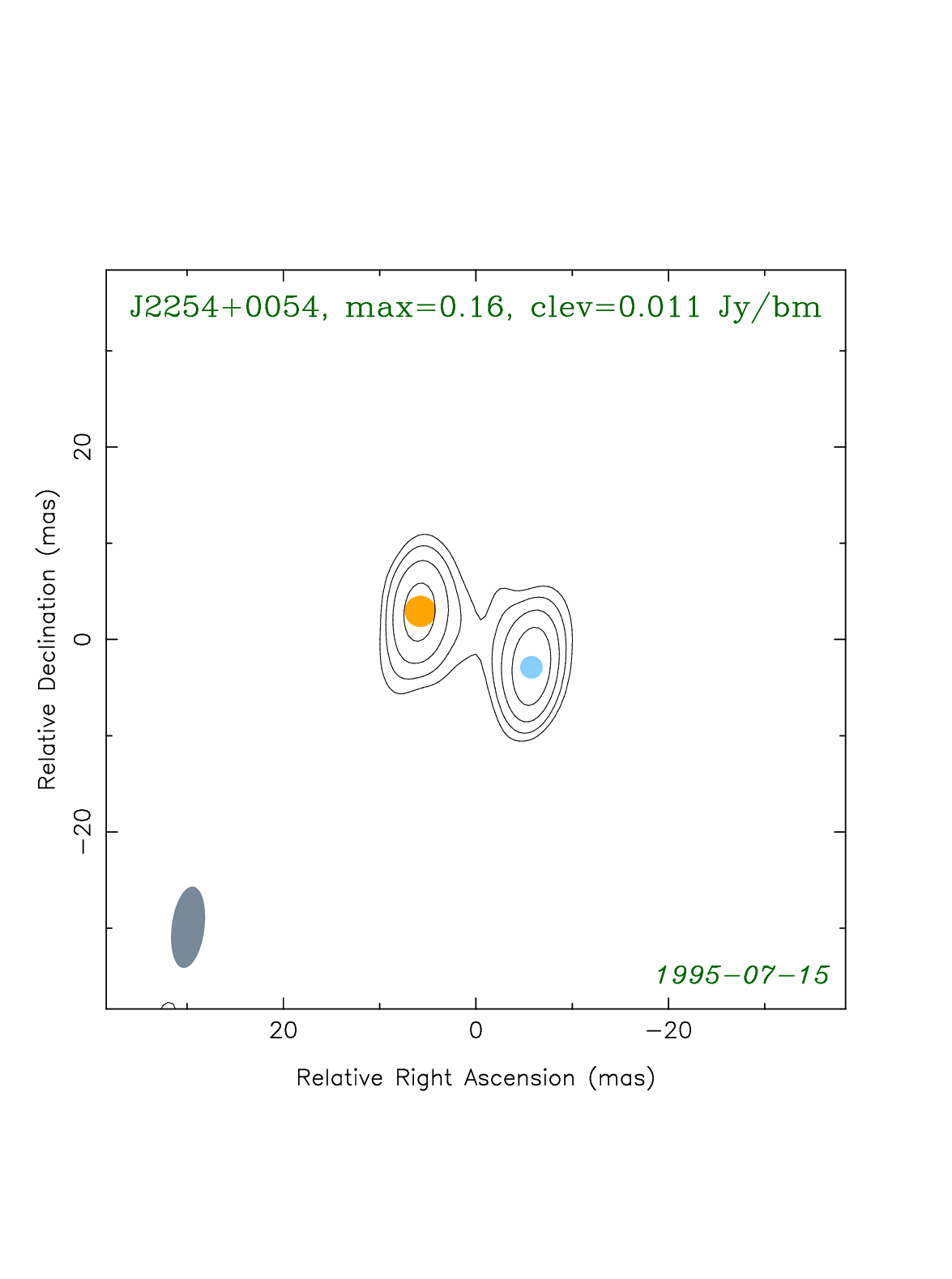}
 \includegraphics[width=0.44\textwidth,angle=0,trim=0 3cm 1cm 5cm,clip]{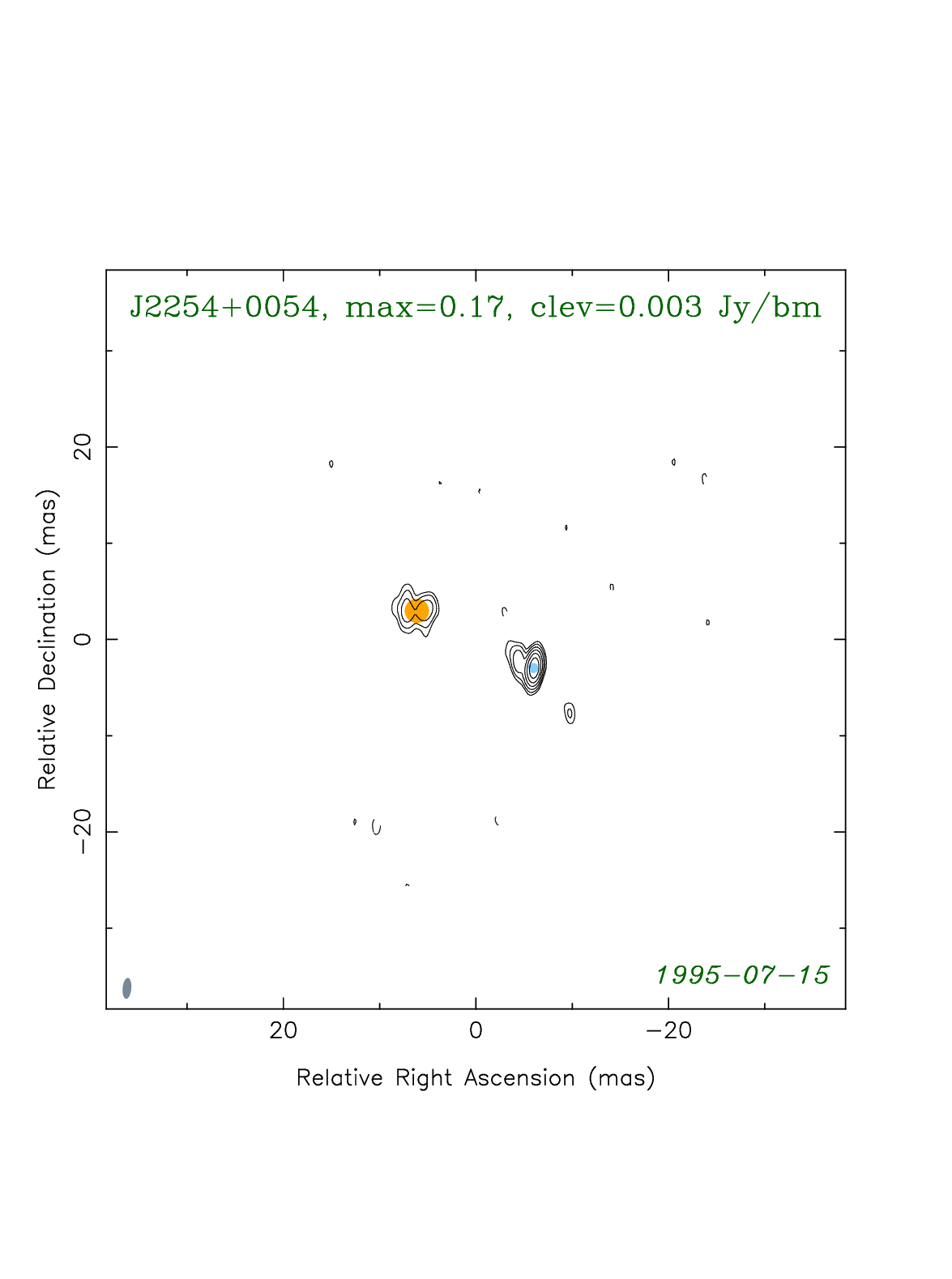}

 \includegraphics[width=0.44\textwidth,angle=0,trim=0 3cm 1cm 5cm,clip]{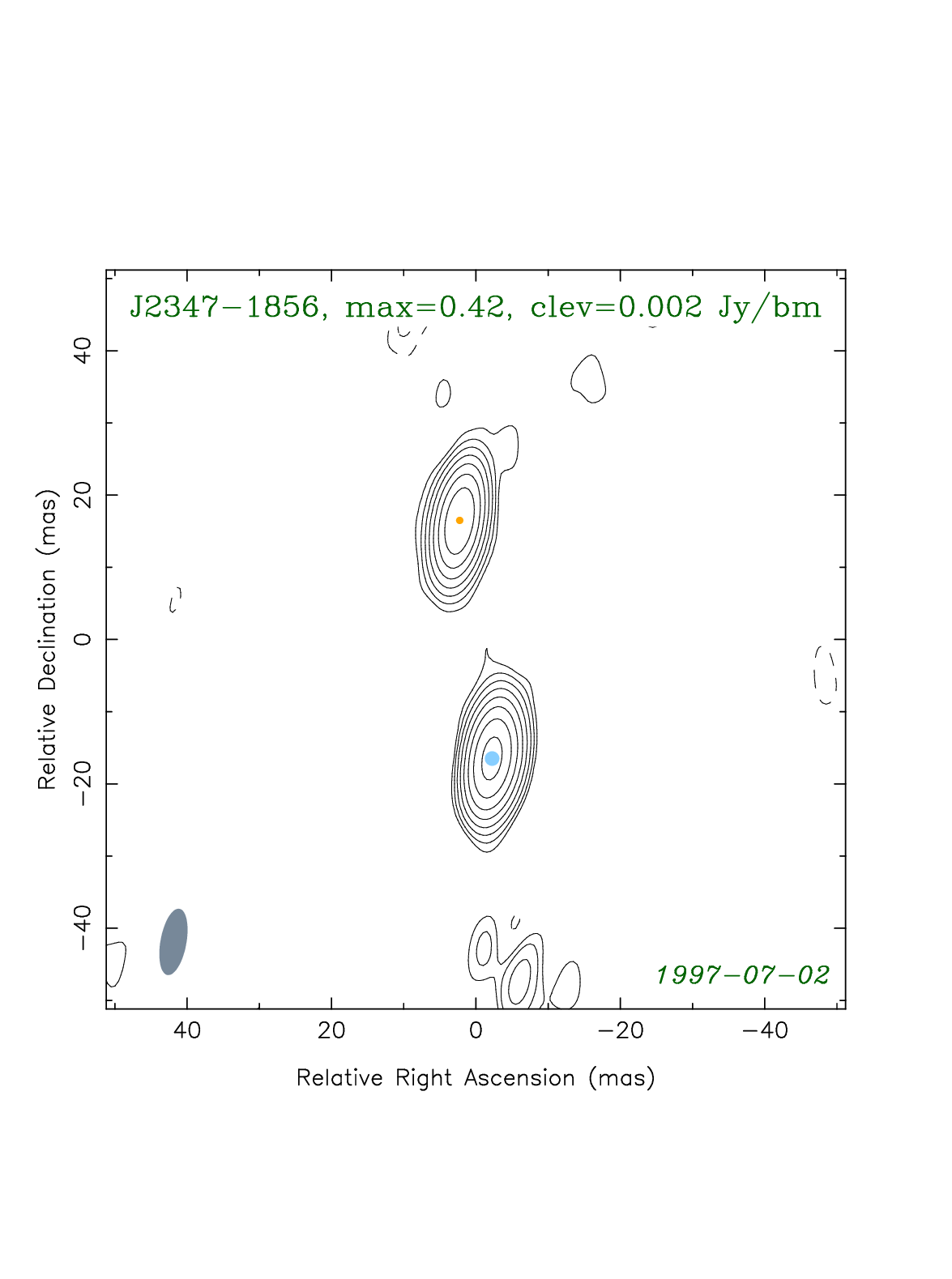}
 \includegraphics[width=0.44\textwidth,angle=0,trim=0 3cm 1cm 5cm,clip]{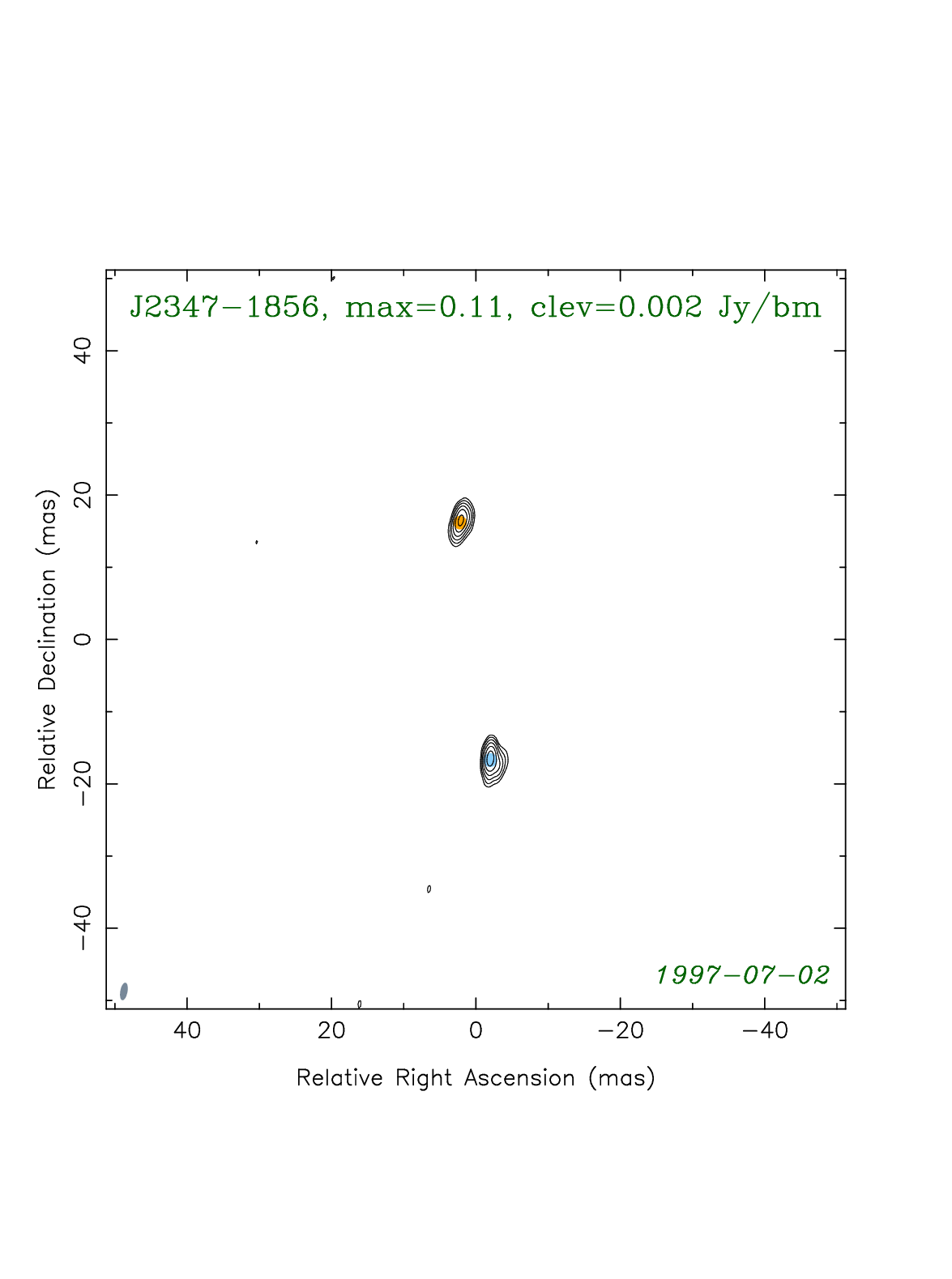}

 \caption{continued}
\end{figure*}

\clearpage
\addtocounter{figure}{-1}
\begin{figure*}
 \centering

 \includegraphics[width=0.44\textwidth,angle=0,trim=0 3cm 1cm 5cm,clip]{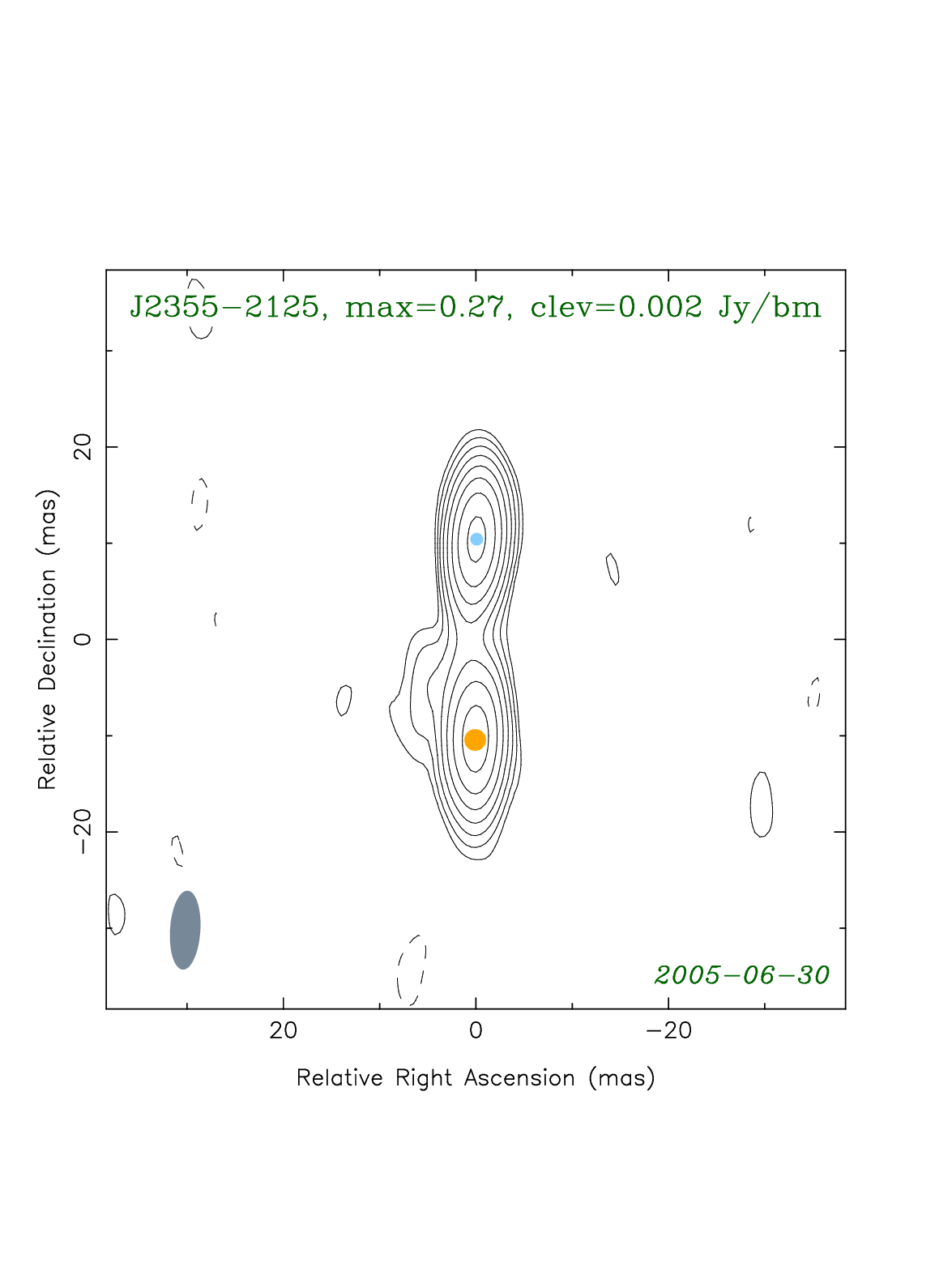}
 \includegraphics[width=0.44\textwidth,angle=0,trim=0 3cm 1cm 5cm,clip]{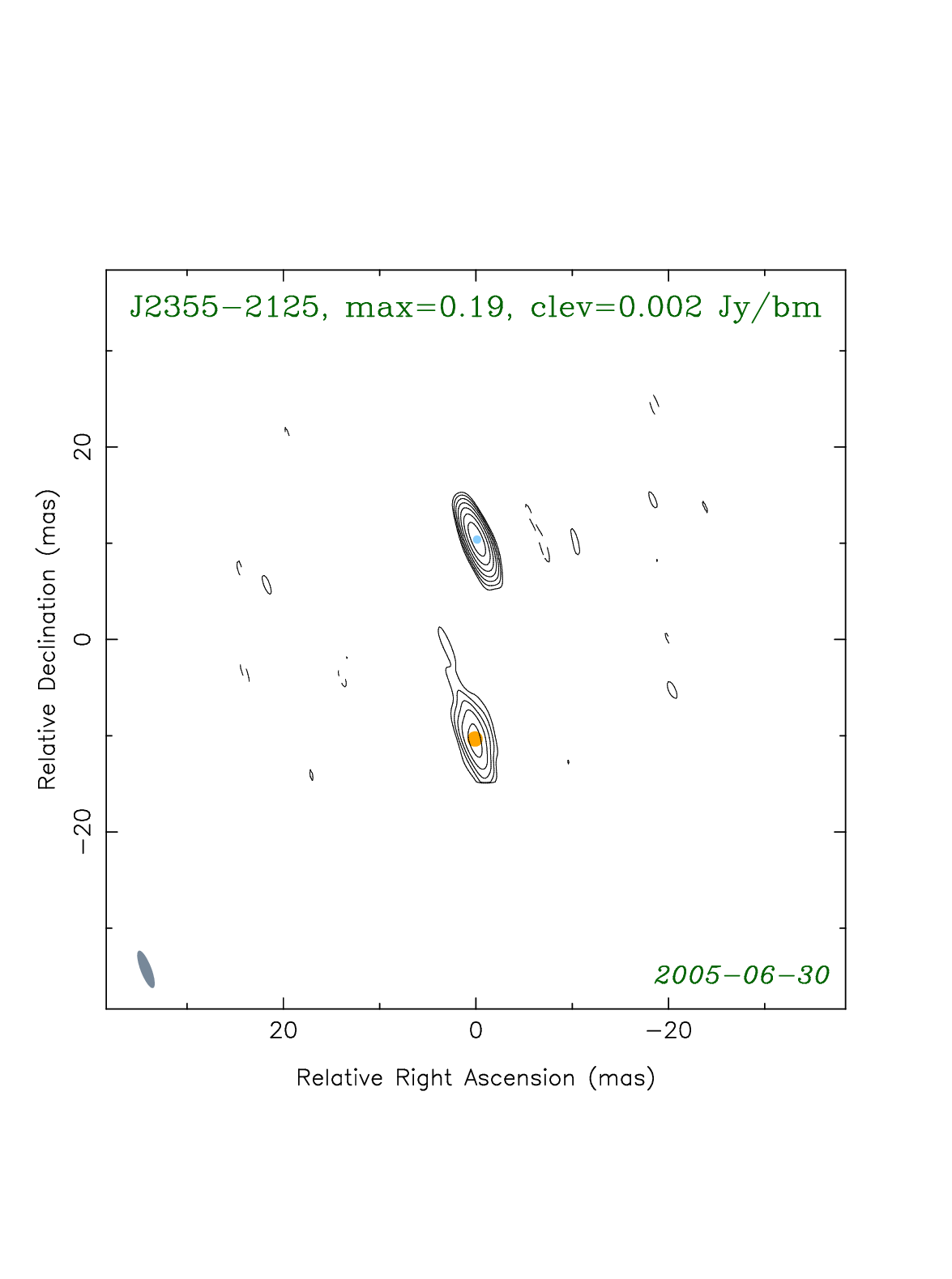}

 \caption{continued.}
\end{figure*}

\end{document}